\documentclass[a4paper,11pt]{article}
\pdfoutput=1
\usepackage{jheppub}
\usepackage{textcase}
\usepackage{graphicx,epsfig,psfrag,amssymb,hyperref}
\usepackage{multirow}
\usepackage{color,graphicx,epsfig,psfrag,amsmath,empheq}
\usepackage[dvipsnames]{xcolor}
\usepackage{bm}
\usepackage{mathrsfs,amsfonts,soul,color}
\usepackage{subfigure}
\usepackage{hepunits}
\usepackage{enumitem}
\usepackage{placeins}
\usepackage[stable]{footmisc}
\usepackage{mdframed}
\usepackage{bbm}
\usepackage{comment}
\usepackage{placeins}
\renewcommand\intercal{\mathsf{\scriptscriptstyle T}}
\newcommand{\bbm}[1]{\mathbf{#1}}
\newcommand{\bv}[1]{\bbm{#1}}
\newtheorem{theo}{Problem}[section]
\newenvironment{ftheo}
  {\begin{mdframed}\begin{theo}}
  {\end{theo}\end{mdframed}}
  
\usepackage{tikz}
\usetikzlibrary{arrows,shapes}
\usetikzlibrary{trees}
\usetikzlibrary{matrix,arrows} 				\usetikzlibrary{positioning}				\usetikzlibrary{calc,through}				\usetikzlibrary{decorations.pathreplacing}  \usepackage{pgffor}	
\usetikzlibrary{decorations.pathmorphing}	\usetikzlibrary{decorations.markings}
\usetikzlibrary{snakes}
\tikzset{
vector/.style={decorate, decoration={snake}, draw},
	provector/.style={decorate, decoration={snake,amplitude=2.5pt}, draw},
	antivector/.style={decorate, decoration={snake,amplitude=-2.5pt}, draw},
    fermion/.style={draw=black, postaction={decorate},
        decoration={markings,mark=at position .55 with {\arrow[draw=black]{>}}}},
    fermionbar/.style={draw=black, postaction={decorate},
        decoration={markings,mark=at position .55 with {\arrow[draw=black]{<}}}},
    fermionnoarrow/.style={draw=black},
    gluon/.style={decorate, draw=black,
        decoration={coil,amplitude=3pt, segment length=5pt}},
    scalar/.style={dashed,draw=black, postaction={decorate},
        decoration={markings,mark=at position .55 with {\arrow[draw=black]{>}}}},
    scalarbar/.style={dashed,draw=black, postaction={decorate},
        decoration={markings,mark=at position .55 with {\arrow[draw=black]{<}}}},
    scalarnoarrow/.style={dashed,draw=black},
    electron/.style={draw=black, postaction={decorate},
        decoration={markings,mark=at position .55 with {\arrow[draw=black]{>}}}},
	bigvector/.style={decorate, decoration={snake,amplitude=4pt}, draw},
}

\tikzstyle{block} = [draw, rectangle, 
    minimum height=3em, minimum width=6em]

\usepackage{float}
\usepackage[toc,page]{appendix}
\usepackage[ruled,vlined]{algorithm2e}
\usepackage{wrapfig}
\SetKwInOut{Input}{Input}
\SetKwInOut{Output}{Output}
\SetKwFunction{Fforward}{Forward}
\SetKwFunction{Fbackward}{Backward}
\SetKwProg{Fn}{function}{:}{}
\DontPrintSemicolon

\DeclareRobustCommand{\Eq}[1]{eq.~(\ref{#1})}

\DeclareRobustCommand{\Fig}[1]{figure~\ref{#1}}

\newcommand{\pt}{\ensuremath{p_{\mathrm{T}}}\xspace}
\newcommand{\PZpr}{\ensuremath{\mathrm{Z}^{\prime}}\xspace}  \newcommand{\PWpr}{\ensuremath{\mathrm{W}^{\prime}}\xspace} \newcommand{\PX}{\ensuremath{\mathrm{X}}\xspace} \newcommand{\PY}{\ensuremath{\mathrm{Y}}\xspace} \newcommand{\Pq}{\ensuremath{\mathrm{q}}\xspace}

\begin{document}

\title{\sffamily\bfseries {\Huge The LHC Olympics 2020} \\ \vspace{3mm}{\Large A Community Challenge for Anomaly} \\\vspace{1mm} {\Large Detection in High Energy Physics} }

\author{Gregor Kasieczka (ed),$^1$}
\author{Benjamin Nachman (ed),$^{2,3}$}
\author{David Shih (ed),$^4$}

\author{Oz Amram,$^{5}$}
\author{Anders Andreassen,$^{6}$}
\author{Kees Benkendorfer,$^{2,7}$}
\author{Blaz Bortolato,$^{8}$}
\author{Gustaaf Brooijmans,$^{9}$}
\author{Florencia Canelli,$^{10}$}
\author{Jack H. Collins,$^{11}$}
\author{Biwei Dai,$^{12}$}
\author{Felipe F. De Freitas,$^{13}$}
\author{Barry M. Dillon,$^{8,14}$}
\author{Ioan-Mihail Dinu,$^5$}
\author{Zhongtian Dong,$^{15}$}
\author{Julien Donini,$^{16}$}
\author{Javier Duarte,$^{17}$}
\author{D. A. Faroughy$^{10}$}
\author{Julia Gonski,$^{9}$}
\author{Philip Harris,$^{18}$} 
\author{Alan Kahn,$^{9}$}
\author{Jernej F. Kamenik,$^{8,19}$}
\author{Charanjit K. Khosa,$^{20,30}$}
\author{Patrick Komiske,$^{21}$}
\author{Luc Le Pottier,$^{2,22}$}
\author{Pablo Mart\'in-Ramiro,$^{2,23}$}
\author{Andrej Matevc,$^{8,19}$}
\author{Eric Metodiev,$^{21}$}
\author{Vinicius Mikuni,$^{10}$}
\author{In\^{e}s Ochoa,$^{24}$}
\author{Sang Eon Park,$^{18}$}
\author{Maurizio Pierini,$^{25}$}
\author{Dylan Rankin,$^{18}$}
\author{Veronica Sanz,$^{20,26}$}
\author{Nilai Sarda,$^{27}$}
\author{Uro\u{s} Seljak,$^{2,3,12}$}
\author{Aleks Smolkovic,$^{8}$}
\author{George Stein,$^{2,12}$}
\author{Cristina Mantilla Suarez,$^{5}$}
\author{Manuel Szewc,$^{28}$}
\author{Jesse Thaler,$^{21}$}
\author{Steven Tsan,$^{17}$}
\author{Silviu-Marian Udrescu,$^{18}$}
\author{Louis Vaslin,$^{16}$}
\author{Jean-Roch Vlimant,$^{29}$}
\author{Daniel Williams,$^{9}$}
\author{Mikaeel Yunus$^{18}$}

\affiliation{
{$^1$Institut f\"{u}r Experimentalphysik, Universit\"{a}t Hamburg, Germany} \\
{$^2$Physics Division, Lawrence Berkeley National Laboratory, Berkeley, CA 94720, USA} \\
{$^3$Berkeley Institute for Data Science, University of California, Berkeley, CA 94720, USA} \\
{$^4$NHETC, Department of Physics \& Astronomy, Rutgers University, Piscataway, NJ 08854, USA} \\
{$^{5}$Department of Physics \& Astronomy, The Johns Hopkins University, Baltimore, MD 21211, USA}\\
{$^{6}$Google, Mountain View, CA 94043, USA}\\
{$^{7}$Physics Department, Reed College, Portland, OR 97202, USA}\\
{$^{8}$Jo\v zef Stefan Institute, Jamova 39, 1000 Ljubljana, Slovenia}\\
{$^{9}$Nevis Laboratories, Columbia University, 136 S Broadway, Irvington NY, USA}\\
{$^{10}$Physik Institut, University of Zurich, Winterthurerstrasse 190, 8057 Zurich, Switzerland}\\
{$^{11}$SLAC National Accelerator Laboratory, Stanford University, Stanford, CA 94309, USA} \\
{$^{12}$Berkeley Center for Cosmological Physics, University of California, Berkeley}\\
{$^{13}$Departamento de F\'\i sica da Universidade de Aveiro and CIDMA Campus de Santiago, 3810-183 Aveiro, Portugal}\\
{$^{14}$Institute for Theoretical Physics, University of Heidelberg, Heidelberg, Germany}\\
{$^{15}$Department of Physics \& Astronomy, University of Kansas, 1251 Wescoe Hall Dr., Lawrence, KS 66045, USA}\\
{$^{16}$Laboratoire de Physique de Clermont, Universit\'{e} Clermont Auvergne, France}\\
{$^{17}$University of California San Diego, La Jolla, CA 92093, USA}\\
{$^{18}$Laboratory for Nuclear Science, MIT, 77 Massachusetts Ave, Cambridge, MA 02139}\\
{$^{19}$Faculty of Mathematics and Physics, University of Ljubljana, Jadranska 19, 1000 Ljubljana, Slovenia}\\
{$^{20}$Department of Physics and Astronomy, University of Sussex, Brighton BN1 9QH, UK}\\
{$^{21}$Center for Theoretical Physics, MIT, 77 Massachusetts Ave, Cambridge, MA 02139}\\
{$^{22}$Physics Department, University of Michigan, Ann Arbor, MI 48109, USA}\\
{$^{23}$Instituto de F\'isica Te\'orica, IFT-UAM/CSIC, Universidad Aut\'onoma de Madrid, 28049 Madrid, Spain}\\
{$^{24}$Laboratory of Instrumentation and Experimental Particle Physics, Lisbon, Portugal}\\
{$^{25}$European Organization for Nuclear Research (CERN), CH-1211, Geneva 23, Switzerland}\\
{$^{26}$Instituto de F\'isica Corpuscular (IFIC), Universidad de Valencia-CSIC, E-46980, Valencia, Spain}\\
{$^{27}$CSAIL, Massachusetts Institute of Technology, 32 Vassar Street, Cambridge, MA 02139, USA}\\
{$^{28}$International Center for Advanced Studies and CONICET, UNSAM, CP1650, Buenos Aires, Argentina}\\
{$^{29}$Division Office Physics, Math and Astronomy, California Institute of Technology, Pasadena, CA 91125, USA}\\
{$^{30}$Department of Physics, University of Genova, Via Dodecaneso 33, 16146 Genova, Italy}\\
}

\emailAdd{shih@physics.rutgers.edu}
\emailAdd{bpnachman@lbl.gov}
\emailAdd{gregor.kasieczka@uni-hamburg.de}

\abstract{
A new paradigm for data-driven, model-agnostic new physics searches at colliders is emerging, and aims to leverage recent breakthroughs in anomaly detection and machine learning. 
In order to develop and benchmark new anomaly detection methods within this framework, it is essential to have standard datasets.  To this end, we have created the LHC Olympics 2020, a community challenge accompanied by a set of simulated collider events.  Participants in these Olympics have developed their methods using an R\&D dataset and then tested them on black boxes: datasets with an unknown anomaly (or not).  This paper will review the LHC Olympics 2020 challenge, including an overview of the competition, a description of methods deployed in the competition, lessons learned from the experience, and implications for data analyses with future datasets as well as future colliders.
}

\maketitle

\clearpage

\section{Introduction}
\label{sec:intro}

The field of high energy physics (HEP) has reached an exciting stage in its development.  After many decades of searching, the Standard Model (SM) of particle physics was completed in 2012 with the discovery of the Higgs boson~\cite{Aad:2012tfa,Chatrchyan:2012ufa}.
Meanwhile, there are strong motivations for physics beyond the Standard Model (BSM).  For example, the nature of dark matter and dark energy, the mass of neutrinos, the minuteness of the neutron dipole moment, and the baryon-anti-baryon asymmetry in the universe are all well-established problems that do not have solutions in the Standard Model.  Furthermore, the Higgs boson mass is unstable with respect to quantum corrections, and a consistent theory of quantum gravity remains mysterious. The Large Hadron Collider (LHC) at CERN has the potential to shed light on all of these fundamental challenges.

Searching for BSM physics is a major part of the research program at the LHC across experiments~\cite{atlasexoticstwiki,atlassusytwiki,atlashdbspublictwiki,cmsexoticstwiki,cmssusytwiki,cmsb2gtwiki,lhcbtwiki}. The current dominant search paradigm is top-down, meaning searches target specific models. Nearly all of the existing BSM searches at the LHC pick a signal model that addresses one or more of the above experimental or theoretical motivations for BSM physics.  Then, high-fidelity synthetic or simulated data are generated using this signal model.  These signal events are then often combined with synthetic background events to develop an analysis strategy which is ultimately applied to data.  An analysis strategy requires a proposal for selecting signal-like events as well as a method for calibrating the background rate to ensure that the subsequent statistical analysis is unbiased. Many searches provide ``model-independent" results, in the form of a limit on cross-section or cross-section times acceptance ungoverned by any theoretical calculation. However, the event selection and background estimation are still strongly model-dependent.

These search efforts are constantly improving and are important to continue and expand with new data.  However, it is also becoming clear that a complementary search paradigm is critical for fully exploring the complex LHC data.  One possible explanation for why we have not discovered new physics yet is that the model dependence of the current search paradigm has created blind spots to unconventional new physics signatures. In fact, despite thousands of BSM searches to date, much of phase space and countless possible signals remain unexplored at present (for many examples just in the realm of 2-body resonances, see~\cite{Craig:2016rqv,1907.06659}). 

Model independent searches for new particles have a long history in high energy physics.  
With a venerable history dating back at least to the discovery of the $\rho$ meson~\cite{Button:1962bnf},
generic searches like the \textit{bump hunt}\footnote{This is a search where signal events present as a localized enhancement on top of a smoothly falling background distribution.} assume little about the signal and have been used to discover many new particles, including the Higgs boson~\cite{Aad:2012tfa,Chatrchyan:2012ufa}.  While generic, the bump hunt is not particularly sensitive because it usually does not involve other event properties aside from the resonant feature.  More differential signal model independent searches have been performed by D0~\cite{sleuth,Abbott:2000fb,Abbott:2000gx,Abbott:2001ke}, H1~\cite{Aaron:2008aa,Aktas:2004pz}, ALEPH~\cite{Cranmer:2005zn}, CDF~\cite{Aaltonen:2007dg,Aaltonen:2007ab,Aaltonen:2008vt}, CMS~\cite{CMS-PAS-EXO-14-016,CMS-PAS-EXO-10-021,CMS:2020ohc,Sirunyan:2020jwk}, and ATLAS~\cite{Aaboud:2018ufy,ATLAS-CONF-2014-006,ATLAS-CONF-2012-107}.  The general strategy in these analyses is to directly compare data with simulation in a large number of exclusive final states (bins).  Aside from the feature selection, these approaches are truly signal model independent.  The cost for signal model independence is sensitivity if there are a large number of bins because of the look elsewhere effect~\cite{Gross:2010qma}. Also, given the extreme reliance on simulation (a form of background model dependence) in these approaches, and they are typically only as sensitive as the simulation is accurate, and characterizing systematic uncertainties across thousands of final states can be challenging. 

Machine learning offers great potential to enhance and extend model independent searches.  In particular, semi-, weak-, or un-supervised training can be used to achieve sensitivity to weak or complex signals with fewer model assumptions than traditional searches.   \textit{Anomaly detection} is an important topic in applied machine learning research, but HEP challenges require dedicated approaches.   In particular, single events often contain no useful information --- it is only when considering a statistical ensemble that an anomaly becomes apparent.  This is a contrast between anomaly detection that is common in industry (``off manifold" or ``out-of-sample" anomalies) and that which is the target of searches in high energy physics (``over-densities").  Furthermore, HEP data are systematically different than natural images and other common data types used for anomaly detection in applied machine learning.  In order to test the resulting tailored methods, it is essential to have public datasets for developing and benchmarking new approaches. 

For this purpose, we have developed the LHC Olympics 2020 challenge and corresponding datasets~\cite{lhco}.  The name of this community effort is inspired by the first LHC Olympics that took place over a decade ago before the start of the LHC~\cite{lhco_old1,lhco_old2,lhco_old3,lhco_old4}.   In those Olympics, researchers prepared `black boxes' (BBs) of simulated signal events and contestants had to examine these simulations to infer the underlying signal process.  These boxes were nearly all signal events and many of the signatures were dramatic (e.g. dilepton mass peaks) and all were findable with simple analysis procedures.  While this was an immensely useful exercise, we are now faced with the reality that the new physics is rare or at least hard to find, and characterizing the BSM properties will not be our biggest challenge.

The LHC Olympics 2020 challenge is also composed of black boxes.  In contrast to the previous Olympics, these contain mostly simulated SM events.  The goal of the challenge is to determine if there is new physics in the box and then to identify its properties. As stressed above, calibrating the background prediction is an essential aspect of BSM searches and so we have restricted this search to high energy hadronic final states where sideband methods can be used to estimate the background.  We provide lists of the detector-reconstructed final state particles
in order to allow contestants to test methods that can process low-level event information.  To aid in method development and testing, we also provide a simulated dataset (with no anomalies) and a benchmark signal dataset. These two are combined to form the R\&D dataset, and provided in addition to the black box datasets.  The goal of this paper is to describe the Winter~\cite{winterolympics} and Summer~\cite{summerolympics} Olympics 2020 competitions.  Well over one hundred researchers participated in these events, with over a dozen teams submitting their results for the black boxes.

This paper is organized as follows.  Section~\ref{sec:challenge} introduces the LHC Olympics competition, including the R\&D and black box datasets.  A brief description of methods deployed in the competition are provided in Secs.~\ref{sec:unsupervised},~\ref{sec:weaklysupervised}, and~\ref{sec:supervised}.  Each contribution includes an introduction to the method, a concise statement of the results, as well as lessons learned before, during, and/or after the challenge.   The results and lessons learned are synthesized in Sec.~\ref{sec:discussion}.  Implications for data analyses with future datasets and as well as future colliders are discussed in Sec.~\ref{sec:outlook} and the paper concludes in Sec.~\ref{sec:conclusion}. 

\clearpage

\section{Dataset and Challenge}
\label{sec:challenge}

The portal for the LHC Olympics dataset can be found at the challenge website~\cite{lhco}.  The datasets described below are all publicly available and downloadable from Zenodo~\cite{lhc_bb1}.  Contestants entered their results in a Google form. On the form, participants were asked to state:

\begin{itemize}
    \item The black box number (1-3) corresponding to their submission.
    \item A short abstract describing their method.
    \item A $p$-value associated with the dataset having no new particles (null hypothesis).
    \item As complete a description of the new physics as possible. For example: the masses and decay modes of all new particles (and uncertainties on those parameters).
    \item How many signal events (with the associated uncertainty) are in the dataset (before any selection criteria).
\end{itemize}

Additionally, contestants were encouraged to submit plots or a \textsc{Jupyter} notebook~\cite{PER-GRA:2007}. The LHC Olympics website includes a basic \textsc{Jupyter} notebook for reading in the data and running basic preprocessing using the \textsc{pyjet} software~\cite{noel_dawe_2020_4289190,Cacciari:2011ma,Cacciari:2005hq}.  Further details of the R\&D and three black box datasets can be found below.

\subsection{R\&D Dataset}
\label{sec:challenge_rnd}

\begin{figure}[h!]
\centering

\begin{tikzpicture}[line width=1.5 pt,scale=2]
	\draw[fermion] (-2,1) -- (-1,0);
	\draw[fermion] (-1,0) -- (-2,-1);
	\draw[vector] (-1,0) -- (1,0);
	\draw[vector] (2,1) -- (1,0);
	\draw[vector] (1,0) -- (2,-1);
	\draw[fermion] (2,1) -- (3,1.5);
	\draw[fermion] (3,0.5) -- (2,1);
	\draw[fermion] (2,-1) -- (3,-1.5);
	\draw[fermion] (3,-0.5) -- (2,-1);
	\node at (-1.5, -1.1) {\Huge $q$};
	\node at (-1.5, 1.1) {\Huge $q$};
	\node at (1.5, -1.1) {\Huge $X$};
	\node at (1.5, 1.1) {\Huge $Y$};
	\node at (2.5, 0.3) {\Huge $q$};
	\node at (2.5, 1.8) {\Huge $q$};
	\node at (2.5, -0.3) {\Huge $q$};
	\node at (2.5, -1.8) {\Huge $q$};
	\node at (0,0.4) {\Huge $Z'$};
 \end{tikzpicture}

\caption{Feynman diagram for signals of R\&D dataset and Black Box 1.}
\label{fig:bb1sig}
\end{figure}
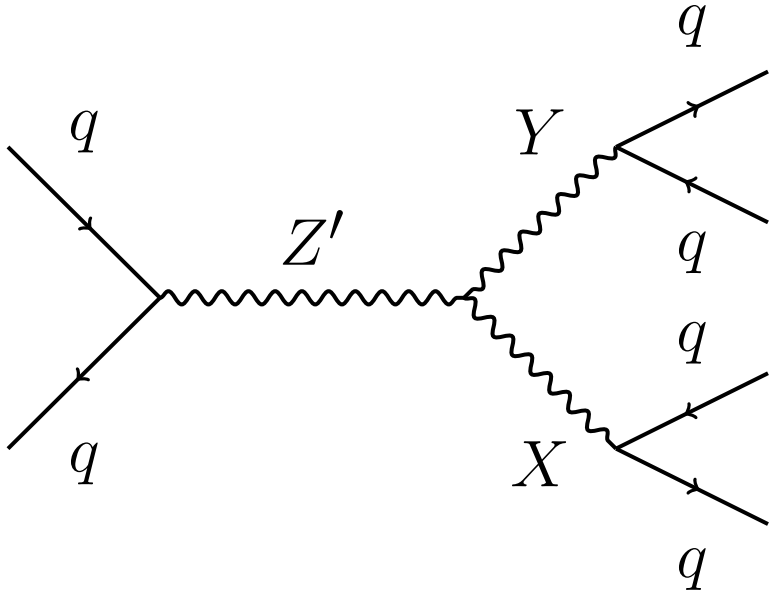

The R\&D dataset consisted of one million SM events each comprised of two jets produced through the strong interaction, referred to as quantum chromodynamics (QCD) dijet events, and 100,000 $Z'\to XY$ events, with $X\to q\bar q$ and $Y\to q\bar q$, as shown in Fig.~\ref{fig:bb1sig} for the topology.  The masses of the new BSM particles $Z'$, $X$, and $Y$ are 3.5~TeV, 500~GeV and 100~GeV, respectively. The events were produced using \textsc{Pythia}~8.219~\cite{Sjostrand:2006za,Sjostrand:2014zea} and \textsc{Delphes}~3.4.1~\cite{deFavereau:2013fsa,Mertens:2015kba,Selvaggi:2014mya}, with default settings, and with no pileup or multiparton interactions included. They are selected using a single large-radius ($R=1$) anti-$k_\mathrm{T}$~\cite{Cacciari:2008gp} jet trigger with a $p_\text{T}$ threshold of 1.2 TeV. 

The signal model was discussed in Ref.~\cite{1907.06659} and has the feature that existing generic searches for dijet resonances or targeted searches for diboson resonances may not be particularly sensitive.  For example, existing searches may register a low significance ($<2\,\sigma$) while automated methods may be able to identify the signal with a high significance.

These events are stored as pandas dataframes~\cite{mckinney-proc-scipy-2010} saved to compressed HDF5~\cite{koranne2011hierarchical} format. For each event, all \textsc{Delphes} reconstructed particles in the event are assumed to be massless and are recorded in detector coordinates ($p_\text{T}$, $\eta$, $\phi$). More detailed information such as particle charge is not included. Events are zero padded to constant size arrays of 700 particles, with the truth bit appended at the end to dictate whether the event is signal or background. The array format is therefore ($N_\text{events}$=1.1~M, 2101).

\subsection{Black Box 1}
\label{sec:challenge_bb1}

\begin{table}
\begin{centering}
\begin{tabular}{|c|c|c|c|}
\hline 
Setting & R\&D & BB1  & BB3\\
\hline
Tune:pp & 14 & 3 &     10                   \\
PDF:pSet &  13 & 12 & 5\\
TimeShower:alphaSvalue &  0.1365 & 0.118 & 0.16\\
SpaceShower:alphaSvalue & 0.1365 & 0.118 & 0.16\\
TimeShower:renormMultFac & 1 & 0.5 & 2\\
SpaceShower:renormMultFac & 1 & 0.5 &2\\
TimeShower:factorMultFac & 1 & 1.5 &0.5\\
SpaceShower:factorMultFac & 1 & 1.5 &0.5\\
TimeShower:pTmaxMatch & 1 & 2 & 1 \\
SpaceShower:pTmaxMatch & 0 & 2     &1\\    
\hline 
\end{tabular} 
\caption{\textsc{Pythia} settings for the different datasets. For R\&D the settings were the \textsc{Pythia} defaults while for BB1 and BB3 they were modified. BB2 is not shown here because it was produced using \textsc{Herwig}++ with default settings.}
\label{tab:pythiasettings}
\end{centering}
\end{table}

This box contained the same signal topology as the R\&D dataset (see Fig.~\ref{fig:bb1sig}) but with masses $m_{Z'}=3.823$~TeV, $m_{X}=732$~GeV and $m_Y=378$~GeV. A total of 834 signal events were included (out of a total of 1M events in all).  This number was chosen so that the approximate local significance inclusively is not significant.  In order to emulate reality, the background events in Black Box 1 are different to the ones from the R\&D dataset.  The background still uses the same generators as for the R\&D dataset, but a number of \textsc{Pythia} and \textsc{Delphes} settings were changed from their defaults. For the \textsc{Pythia} settings, see Table\footnote{Setting \texttt{pTmaxMatch = 2} in \textsc{Pythia} invokes a ``power shower'', where emissions are allowed to occur all the way to the kinematical limit.  With a phase space cut on the hard scattering process, this sculpts a bump-like feature in the multijet background, which was flagged as anomalous by the authors of Section~\ref{sec:factorizedtopics}.  Identification of this bump is labeled as ``Human NN'' in Figure~\ref{fig:bb1res}.}~\ref{tab:pythiasettings}. For the \textsc{Delphes} settings, we changed \textsc{EfficiencyFormula} in the \textsc{ChargedHadronTrackingEfficiency} module, \textsc{ResolutionFormula} in the \textsc{ChargedHadronMomentumSmearing} module, and \textsc{HCalResolutionFormula} in the hadronic calorimeter (HCal) module.  The tracking variations are approximated using the inner-detector measurements from Ref~\cite{Aad:2016jkr} and the calorimeter energy resolutions are varied by 10\% inspired by measurements from Ref.~\cite{Aaboud:2016hwh}.

\subsection{Black Box 2}
\label{sec:challenge_bb2}

This sample of 1M events was background only. The background was produced using Herwig++~\cite{Bahr:2008pv} instead of \textsc{Pythia}, and used a modified \textsc{Delphes} detector card that is different from Black Box 1 but with similar modifications on top of the R\&D dataset card. 

\subsection{Black Box 3}
\label{sec:challenge_bb3}

The signal was inspired by Ref.~\cite{Agashe:2016rle,Agashe:2016kfr} and consisted of a heavy resonance (the KK graviton) with mass $m=4.2$~TeV which had two different decay modes. The first is just to dijets (gluons), while the second is to a lighter neutral resonance $R$ (the IR radion) of mass $m_{R}=2.217$~TeV plus a gluon, with $R\to gg$. 1200 dijet events and 2000 trijet events were included along with QCD backgrounds in Black Box 3.  These numbers were chosen so that an analysis that found only one of the two modes would not observe a significant excess.  The background events were produced with modified \textsc{Pythia} and \textsc{Delphes} settings (different than the R\&D and the other black boxes). For the \textsc{Pythia} settings, see Table~\ref{tab:pythiasettings}.

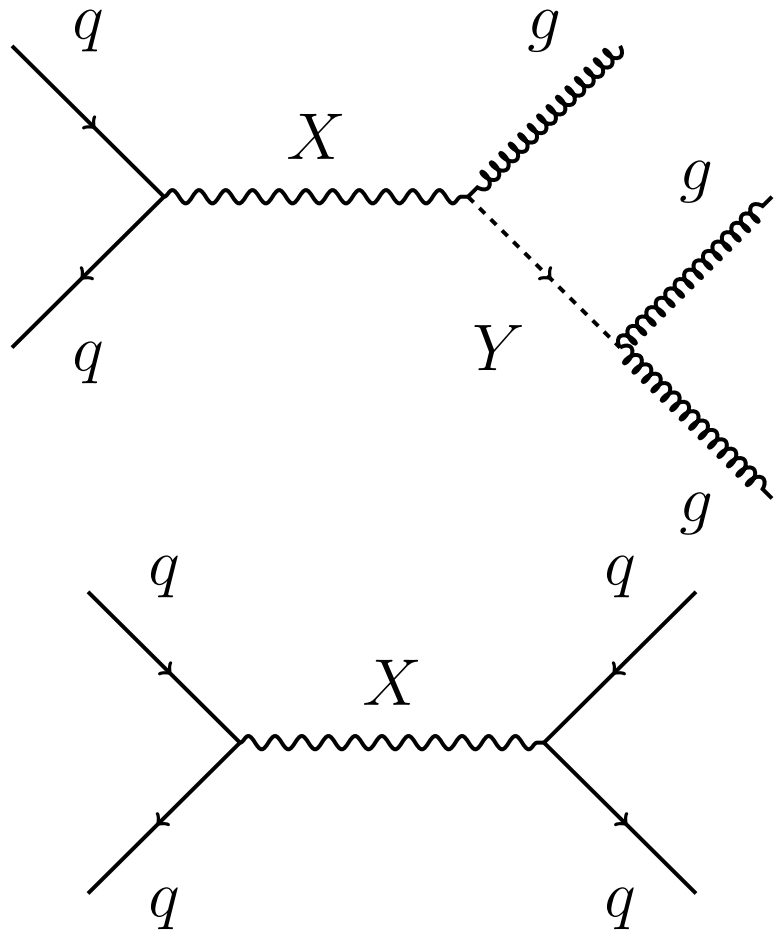
\begin{figure}[h!]
\centering

\begin{tikzpicture}[line width=1.5 pt,scale=2]
	\draw[fermion] (-2,1) -- (-1,0);
	\draw[fermion] (-1,0) -- (-2,-1);
	\draw[vector] (-1,0) -- (1,0);
	\draw[gluon] (2,1) -- (1,0);
	\draw[scalar] (1,0) -- (2,-1);
	\draw[gluon] (2,-1) -- (3,-2);
	\draw[gluon] (2,-1) -- (3,0);
	\node at (-1.5, -1.1) {\Huge $q$};
	\node at (-1.5, 1.1) {\Huge $q$};
	\node at (1.2, -1.) {\Huge $Y$};
	\node at (1.5, 1.1) {\Huge $g$};
	\node at (2.5, .1) {\Huge $g$};
	\node at (2.5, -2.1) {\Huge $g$};
	\node at (0,0.4) {\Huge $X$};
 \end{tikzpicture}

\begin{tikzpicture}[line width=1.5 pt,scale=2]
	\draw[fermion] (-2,1) -- (-1,0);
	\draw[fermion] (-1,0) -- (-2,-1);
	\draw[vector] (-1,0) -- (1,0);
	\draw[fermion] (2,1) -- (1,0);
	\draw[fermion] (1,0) -- (2,-1);
	\node at (-1.5, -1.1) {\Huge $q$};
	\node at (-1.5, 1.1) {\Huge $q$};
	\node at (1.5, -1.1) {\Huge $q$};
	\node at (1.5, 1.1) {\Huge $q$};
	\node at (0,0.4) {\Huge $X$};
 \end{tikzpicture}

\caption{Feynman diagrams for signal of Black Box 3.}
\label{fig:bb3sig}
\end{figure}

\clearpage

\addcontentsline{toc}{section}{Individual Approaches}

\noindent {\huge\textbf{ Individual Approaches}}\\

\noindent The following sections describes a variety of approaches to anomaly detection.  In addition to an explanation of the method, each section includes a set of results on the LHC Olympics datasets as well as a brief description of lessons learned.  

We have grouped the various methods into three loose categories: Unsupervised (Sec.~\ref{sec:unsupervised}), Weakly Supervised (Sec.~\ref{sec:weaklysupervised}), and (Semi)-Supervised (Sec.~\ref{sec:supervised}).  \textit{Supervision} refers to the type of label information provided to the machine learning algorithms during training.  Unsupervised methods do not provide any label information and learn directly from background-dominated data.  Typically, these methods try to look for events with low $p(\text{background})$. (Exceptions exist, see e.g.\  ANODE in Sec.~\ref{sec:ANODE} and GIS in Sec.~\ref{sec:gis} which use likelihood ratios.) Weakly supervised methods have noisy labels.\footnote{Such a categorisation is not unique, see e.g.~\cite{zhou2018brief} for an alternative way of defining weak supervision.  We follow the established usage in applications of machine learning for particle physics.}
Many of these approaches operate by comparing two datasets with different amounts of a potential signal.  The labels are noisy because instead of being pure `signal' and `background', the labels are `possibly signal-depleted' and `possibly signal-enriched'.  The goal of these methods is to look for events with high $p(\text{possibly signal-depleted})/p(\text{possibly signal-enriched})$.  Supervised methods have labels for each event.  Semi-supervised methods have labels for some events.  Methods that are labeled as (Semi-)Supervised use signal simulations in some way to build signal sensitivity.  These three categories are not exact and the boundaries are not rigid.  However, this categorization may help to identify similarities and differences between approaches.  Within each category, the methods are ordered alphabetically by title.

Furthermore, the results on the datasets can be grouped into three types: (i) blinded contributions using the black boxes, (ii) unblinded results or updates on blinded results (and thus, also unblinded) on the black boxes, and (iii) results only using the R\&D dataset.  All three of these contribution types provide valuable insight, but each serves a different purpose.  The first category (i) corresponds to the perspective of a pure challenge that is analogous to a real data analysis.  The organizers of the LHCO challenge could not participate in this type of analysis.  Section~\ref{sec:discussion_overall} provides an overview of the challenge results.  The LHC Olympics datasets have utility beyond the initial blinded challenge as well and so contributions of type (ii) and (iii) are also important.  Some of the results of these types came from collaborations with the challenge organizers and some came from other groups as well who did not manage (for whatever reason) to deploy their results on the blinded black boxes. 

A summary of all of the methods and results can be found in Table~\ref{tab:overview}. Note that in some cases, blinded results (of type (i)) were presented at the LHC Olympics workshops, but only a subset (sometimes of type (iii)) appear in the subsequent sections.  The table gives precedence to the workshops results, which are also discussed in Sec.~\ref{sec:discussion_overall}. 

\begin{table}[h!]
\begin{center}
\begin{tabular}{|c|c|c|c|} 
 \hline
 Section&Short Name&Method Type&Results Type \\
  \hline
\ref{sec:vrnn}& VRNN& Unsupervised & (i) (BB2,3) and (ii) (BB1)\\
\ref{sec:ANODE}& ANODE& Unsupervised& (iii)\\
\ref{sec:buhula}& BuHuLaSpa& Unsupervised& (i) (BB2,3) and (ii) (BB1)\\
\ref{sec:GAN-AE}& GAN-AE& Unsupervised& (i) (BB2-3) and (ii) (BB1)\\
\ref{sec:gis}& GIS &Unsupervised &(i) (BB1)\\
\ref{sec:lda}& LDA& Unsupervised& (i) (BB1-3)\\
\ref{sec:pga}& PGA& Unsupervised &(ii) (BB1-2)\\
\ref{sec:regualrizedlikes}& Reg. Likelihoods & Unsupervised& (iii)\\
\ref{sec:ucluster}& UCluster& Unsupervised& (i) (BB2-3)\\
\ref{sec:cwola}& CWoLa& Weakly Supervised &(ii) (BB1-2)\\
\ref{sec:cwolaaecompare}& CWoLa AE Compare& Weakly/Unsupervised &(iii)\\
\ref{sec:tnt}& Tag N' Train & Weakly Supervised &(i) (BB1-3)\\
\ref{sec:salad}& SALAD &Weakly Supervised &(iii)\\
\ref{sec:sacwola}& SA-CWoLa &Weakly Supervised &(iii)\\
\ref{sec:CNNBDT}& Deep Ensemble& Semisupervised& (i) (BB1)\\
\ref{sec:factorizedtopics}& Factorized Topics &Semisupervised &(iii)\\
\ref{sec:quak}& QUAK& Semisupervised& (i) (BB2,3) and (ii) (BB1)\\
\ref{sec:LSTM}& LSTM& Semisupervised & (i) (BB1-3)  \\
 \hline
\end{tabular}
\end{center}
\caption{A categorization in terms of method and result type for all of the results presented in the Sec.~\ref{sec:unsupervised}, Sec.~\ref{sec:weaklysupervised}, and Sec.~\ref{sec:supervised}.  }
\label{tab:overview}
\end{table}

\clearpage

\section{Unsupervised}
\label{sec:unsupervised}

\subsection[Anomalous Jet Identification via Variational Recurrent Neural Network]{Anomalous Jet Identification via Variational Recurrent Neural Network\footnote{Authors: Alan Kahn, Julia Gonski, In\^{e}s Ochoa, Daniel Williams, and Gustaaf Brooijmans.}}

\label{sec:vrnn}

\subsubsection{Method}
\label{sec:methodvnn}

The method described here employs a Variational Recurrent Neural Network (VRNN) to perform jet-level anomaly detection by modeling jets as a sequence of constituents. A VRNN is a sequence-modeling architecture which replaces the standard encoder-decoder architecture of a Recurrent Neural Network with a Variational Autoencoder (VAE)~\cite{chung2016recurrent}. 
This allows the VRNN to perform both sequence modeling in addition to variational inference, which has been shown to be a very powerful tool for anomaly detection~\cite{An2015VariationalAB}.
A sequence-modeling architecture is well-motivated as it is capable of accommodating variable-length inputs, such as lists of constituent four-vectors in a jet, while suppressing the ability of the model to learn correlations with the jet's constituent multiplicity.
By contrast, fixed-length architectures such as VAEs rely on a loss function that is computed between the input layer and the reconstructed output layer. As a result, zero-padded inputs directly affect the value of the loss function, leading to correlations that are difficult to remove when using inputs that are naturally variable in length, but forced to work in a fixed-length framework. 

\begin{figure}
  \begin{center}
  
\includegraphics[width=0.8\textwidth]{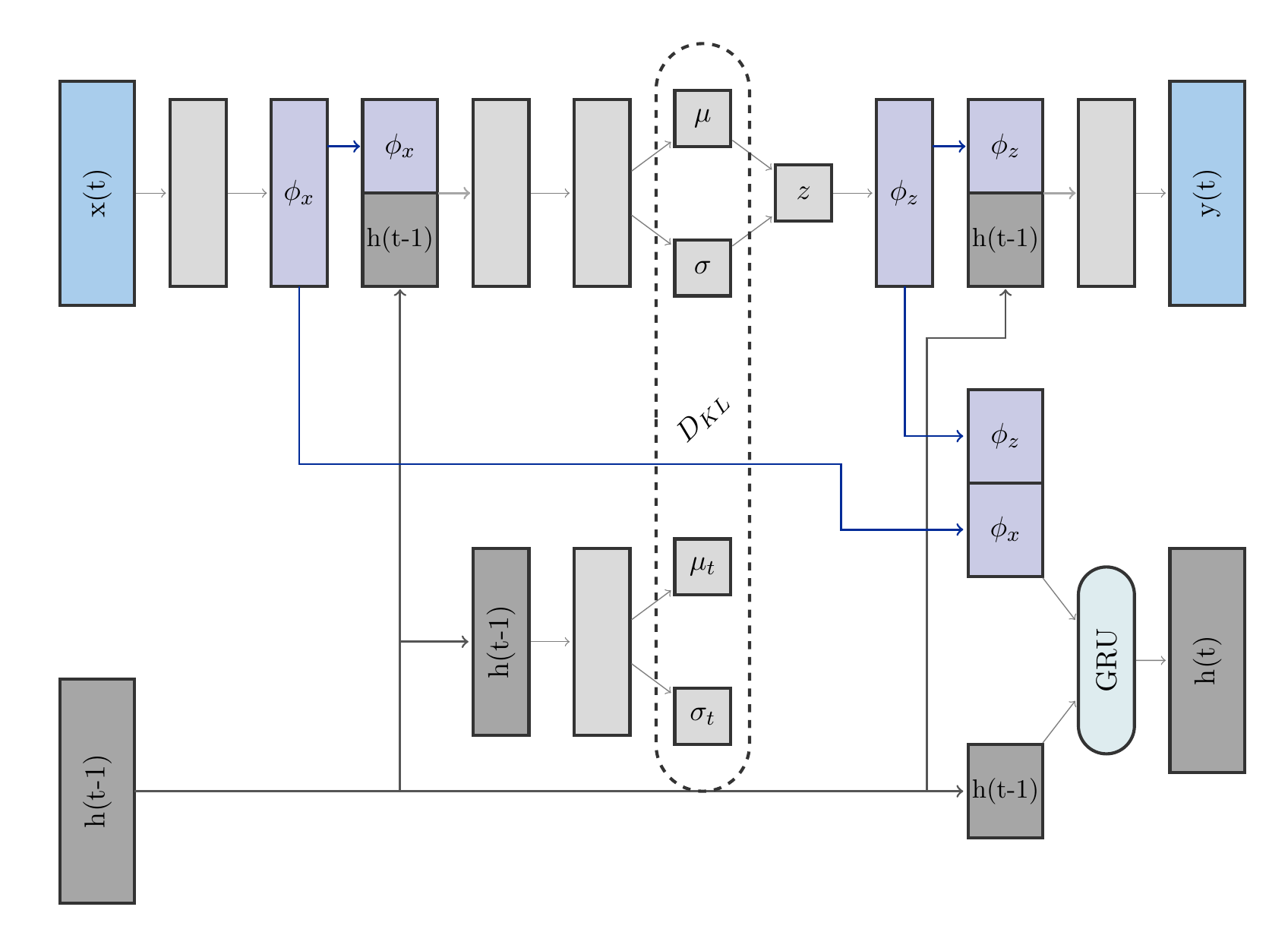}
  
  \end{center}
  \caption{A Variational Recurrent Neural Network cell. The $x(t)$ and $y(t)$ layers represent respectively the input constituent and reconstructed constituents' four-momentum components $p_\text{T}$, $\eta$, and $\phi$. The $\phi_{x}$ and $\phi_{z}$ layers are \textit{feature-extracting layers} which encode a representation of the features in the input layer $x(t)$ and latent space $z$ respectively. $h(t-1)$ represents the current time-step's hidden state, which is updated each iteration via a transition function between $h(t-1)$, $\phi_{x}$, and $\phi_{z}$ carried out by a Gated Recurrent Unit (GRU). At each time-step, the prior distribution defined by $\mu_{t}$ and $\sigma_{t}$ is determined from the current hidden state.}
  \label{fig:VRNN}
\end{figure}

Figure \ref{fig:VRNN} shows a diagram of one VRNN cell. The VAE portion of the architecture is displayed on the top row of layers in the diagram, where a constituent's four-momentum components are input as a vector $x(t)$, which is encoded into a multivariate Gaussian distribution in the latent space $z$, and then decoded to produce a reconstruction of the same input constituent's components $y(t)$. The variable $t$ refers to the \textit{time-step}, which advances as the sequence is processed, and can be interpreted as the constituent number currently being processed by the model.

Inputs to the VRNN consist of sequences of jet four-vector constituent components $p_\text{T}$, $\eta$, and $\phi$, where constituents are assumed to be massless.
Jets are reconstructed with FastJet~\cite{Cacciari:2011ma,Cacciari:2005hq} using the anti-$k_t$ algorithm with a radius parameter of 1.0~\cite{Cacciari:2008gp}.
Before training, a pre-processing method is applied which boosts each jet to the same reference mass, energy, and orientation in $\eta-\phi$ space, such that all input jets differ only by their substructure.
In addition, our pre-processing method includes a choice of \textit{sequence ordering}, in which the constituent sequence input into the model is sorted by $k_{t}$-distance instead of by the typical constituent $p_\text{T}$. 
In more detail, the $n^\text{th}$ constituent in the list, $c_{n}$, is determined by Eq.~\ref{eq:seq} to be the constituent with the highest $k_{t}$-distance relative to the previous constituent, with the first constituent in the list being the highest $p_\text{T}$ constituent. 

\begin{equation}
\label{eq:seq}
c_{n} = \max(p_{Tn}\Delta R_{n, n-1})
\end{equation}

This ordering is chosen such that non-QCD-like substructure, characterized by two or more separate prongs of constituents within the jet, is more easily characterized by the sequence. 
When compared to $p_{T}$-sorted constituent ordering, the $k_{t}$-sorted sequence consistently travels back and forth between each prong, making their existence readily apparent and easy to model. As a result, a significant boost in performance is observed.

The loss function $\mathcal{L}(t)$ for each constituent, defined in Eq.~\ref{eq:lossVAE}, is very similar to that of an ordinary VAE. 
It consists of a mean-squared-error (MSE) loss between input constituents and generated output constituents as a reconstruction loss, as well as a weighted KL-Divergence from the learned latent space prior to the encoded approximate posterior distribution. 
Since softer constituents contribute less to the overall classification of jet substructure, each KL-Divergence term, computed constituent-wise, is weighted by the constituent's $p_{T}$-fraction with respect to the jet's total $p_\text{T}$, averaged over all jets in the dataset to avoid correlations with constituent multiplicity. 
The weight coefficient of the KL-Divergence term is enforced as a hyperparameter, and has been optimized to a value of 0.1 in dedicated studies. 

\begin{equation}
\label{eq:lossVAE}
\mathcal{L}(t)=\text{MSE}+0.1 \times \overline{p_T}(t)D_\text{KL}
\end{equation}

After a jet is fully processed by the VRNN, a total loss function $\mathcal{L}$ is computed as the average of the individual constituent losses over the jet: $\mathcal{L} = \frac{\Sigma \mathcal{L}(t)}{N}$.

The architecture is built with 16 dimensional hidden layers, including the hidden state, with a two-dimensional latent space. All hyperparameters used are determined by a hyperparameter optimization scan. 

The model is trained on the leading and sub-leading jets of each event, where events are taken from the LHC Olympics datasets. 
After training, each jet in the dataset is assigned an \textit{Anomaly Score}, defined in Eq.~\ref{eq:as}, where $D_\text{KL}$ is the KL-Divergence from the learned prior distribution to the encoded posterior distribution.

\begin{equation}
\label{eq:as}
\text{Anomaly Score} = 1 - e^{-\overline{D_\text{KL}}}
\end{equation}

Since the LHC Olympics challenge entails searching for a signal on the event level instead of the jet level, an overall \textit{Event Score} is determined by choosing the most anomalous score between the leading and sub-leading jets in an event. 
To ensure consistency between training scenarios, Event Scores are subject to a transformation in which the mean of the resulting distribution is set to a value of 0.5, and Event Scores closer to 1 correspond to more anomalous events.

\subsubsection{Results on LHC Olympics}
\label{sec:results}

The performance of the VRNN was first assessed with the LHC Olympics R\&D dataset, which includes a known signal of a beyond-the-Standard-Model $Z'$ boson with a mass of 3500 GeV which decays to two hadronically decaying $X$ and $Y$ particles, each reconstructed by a $R=1.0$ jet.
This study was used as a validation of the method, with a goal of directly investigating the ability of the Event Score selection to reconstruct the $Z'$ mass. 
Therefore, no selections beyond those described in Section~\ref{sec:methodvnn} are applied.

The VRNN was trained over a contaminated dataset consisting of 895113 background events and 4498 signal events, corresponding to a signal contamination level of 0.5\%.
A selection on the Event Score is applied as the sole discriminator, and the invariant mass $m_{JJ}$ of the two jets is then scanned to assess the prominence of the $Z'$ mass peak.
In this validation analysis, the Event Score is required to exceed a value of 0.65. 
This value is chosen to significantly enrich the fraction of anomalous jet events over the background, while retaining enough statistics in the background to display its smoothly falling behavior.

\begin{figure}[h!]
	\begin{center}
		\includegraphics[width=0.47\textwidth]{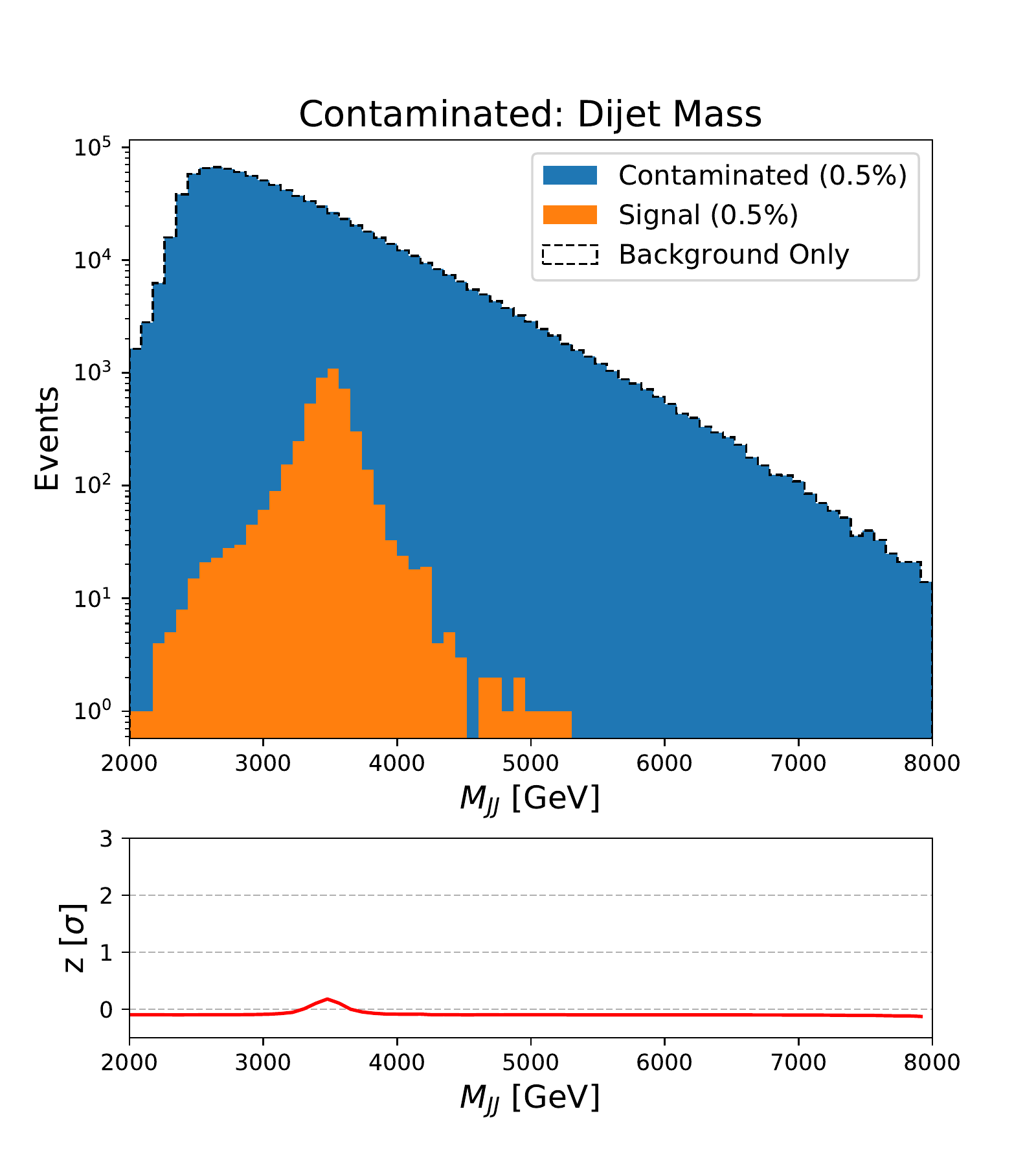}
		\includegraphics[width=0.47\textwidth]{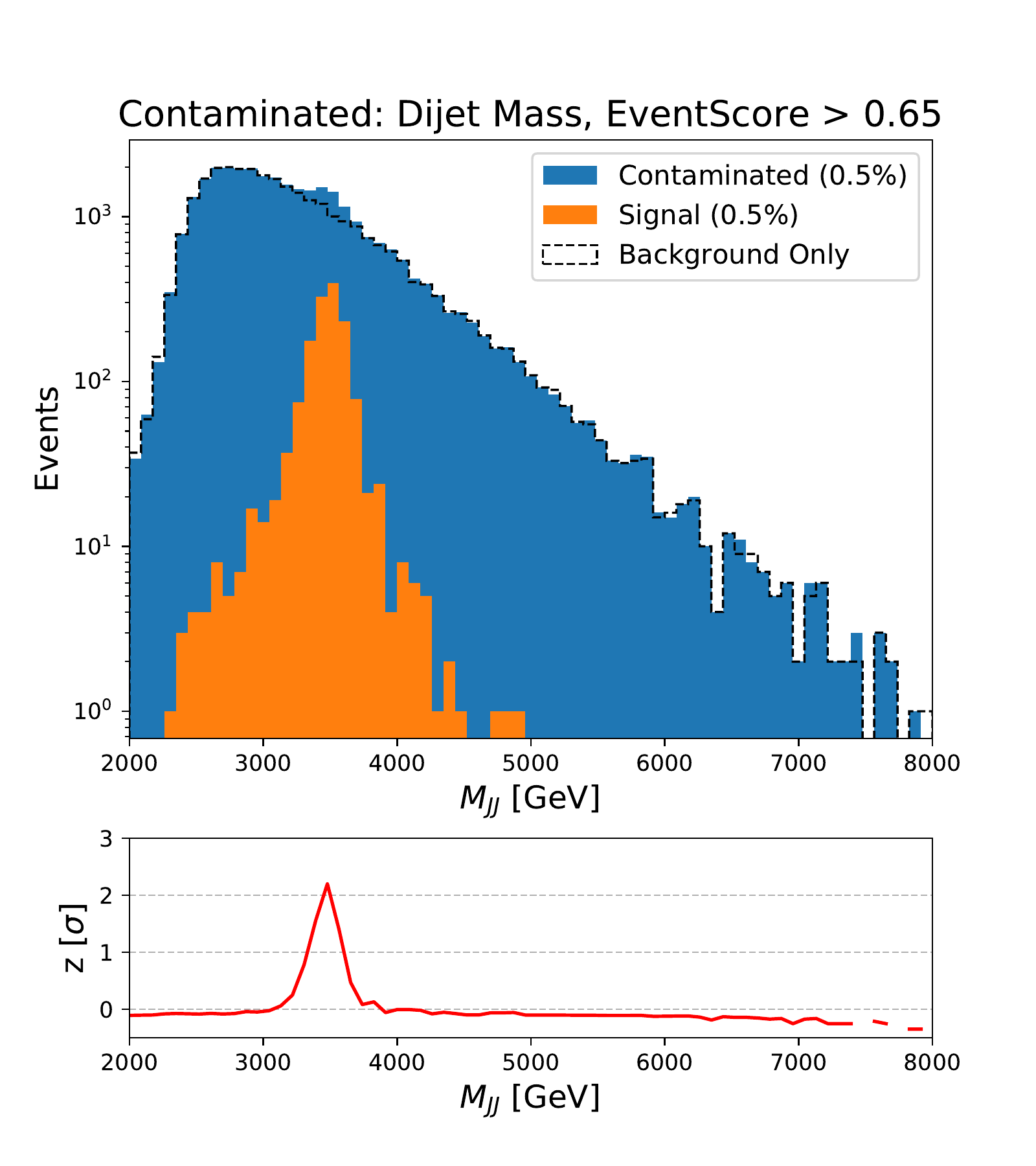}
	\end{center}
	\caption{Dijet invariant mass distributions before (left) and after (right) a selection on the Event Score, with a two-prong Z' signal contamination of 0.5\%.}
	\label{fig:m_JJ}
\end{figure}

Figure \ref{fig:m_JJ} shows the dijet invariant mass distributions before and after the Event Score selection, along with the local significance of the signal computed in each bin using the {\sc BinomialExpZ} function from {\sc RooStats} with a relative background uncertainty of 15\%~\cite{moneta2011roostats}.
Applying this selection dramatically increases the significance of the excess from $0.18\sigma$ to $2.2\sigma$ without significantly sculpting the shape of the background.

\begin{figure}[h!]
	\begin{center}
		\includegraphics[width=0.47\textwidth]{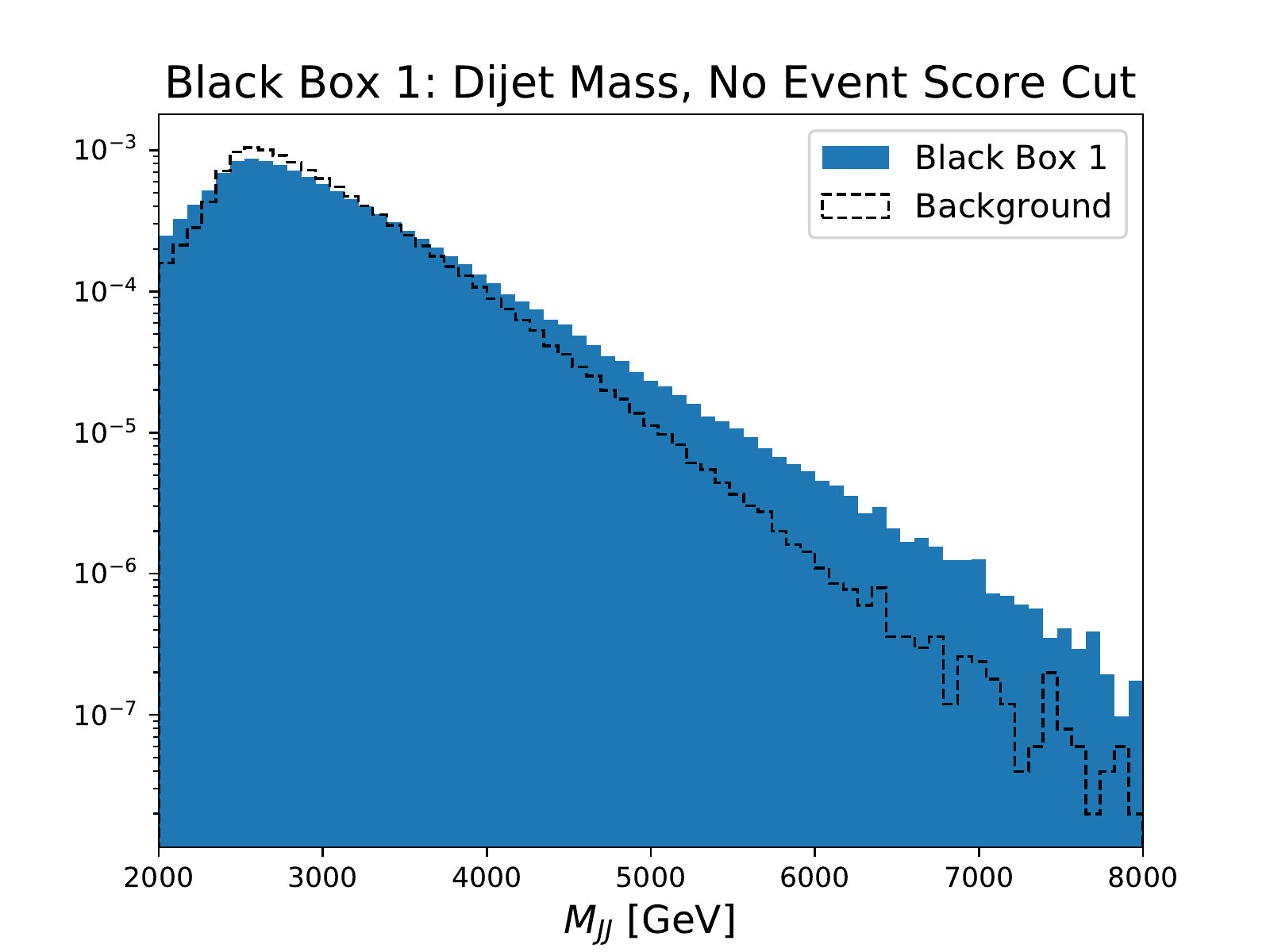}
		\includegraphics[width=0.47\textwidth]{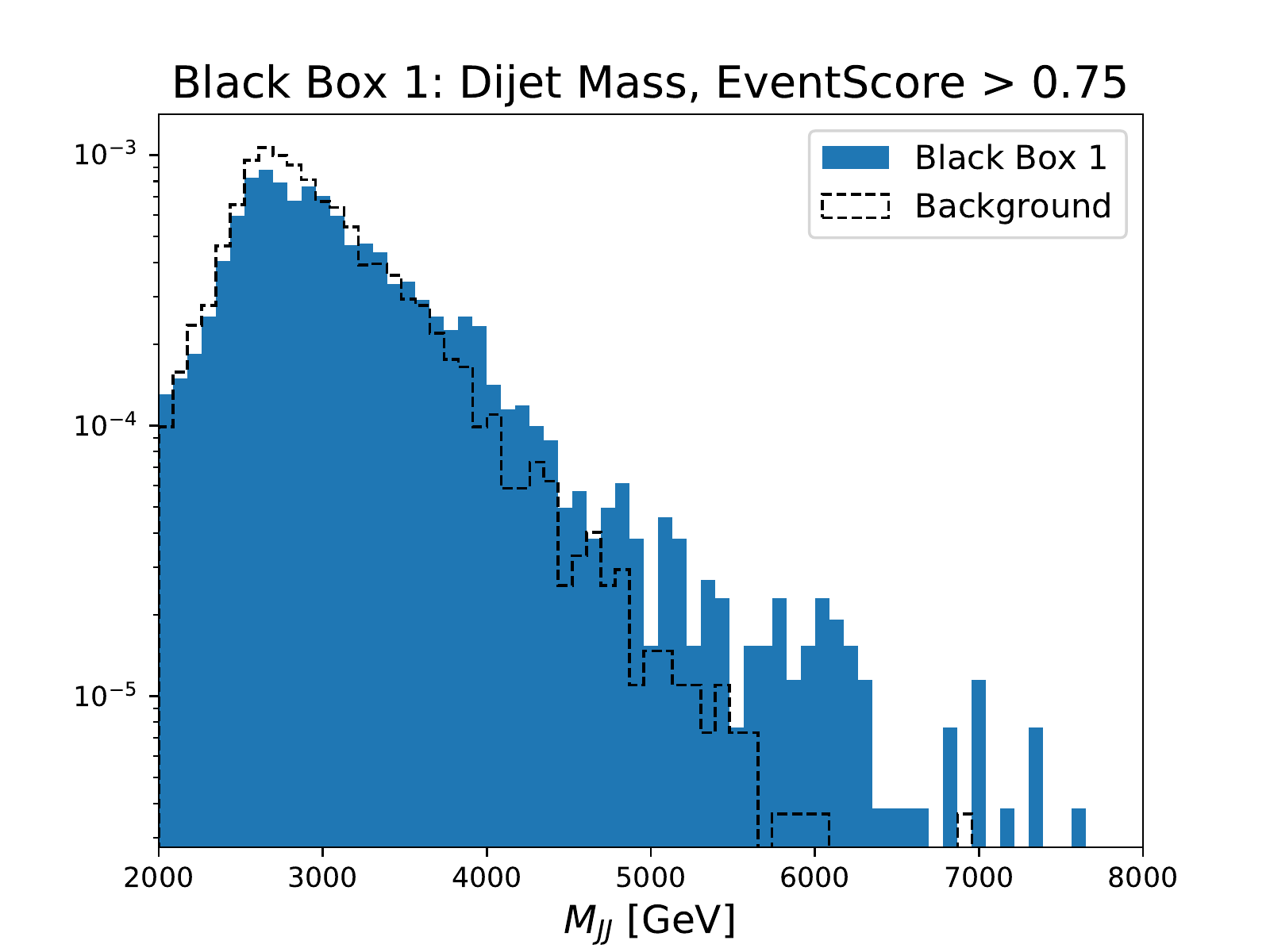}
	\end{center}
	\caption{Dijet invariant mass distributions before (left) and after (right) a selection on the Event Score from the Black Box 1 dataset. The signal present is a $Z'$ boson with a mass of 3800 GeV.}
	\label{fig:bb1}
\end{figure}

Once the method was validated in the R\&D dataset, it was applied to Black Box 1, with a re-optimized tighter selection on the Event Score of 0.75, as well as a requirement on the pseudorapidity of the leading and sub-leading jets to be less than 0.75, to ensure that central, high momentum transfer events are considered. Figure {\ref{fig:bb1}} shows the dijet invariant mass for both the Black Box 1 and Background datasets. The Event Score selection reveals an enhancement in $m_{JJ}$ just below 4000 GeV. This is consistent with the Black Box 1 signal, which is a new $Z'$ boson with a mass of 3800 GeV decaying to two new particles, each decaying hadronically.

\begin{figure}[h!]
	\begin{center}
		\includegraphics[width=0.47\textwidth]{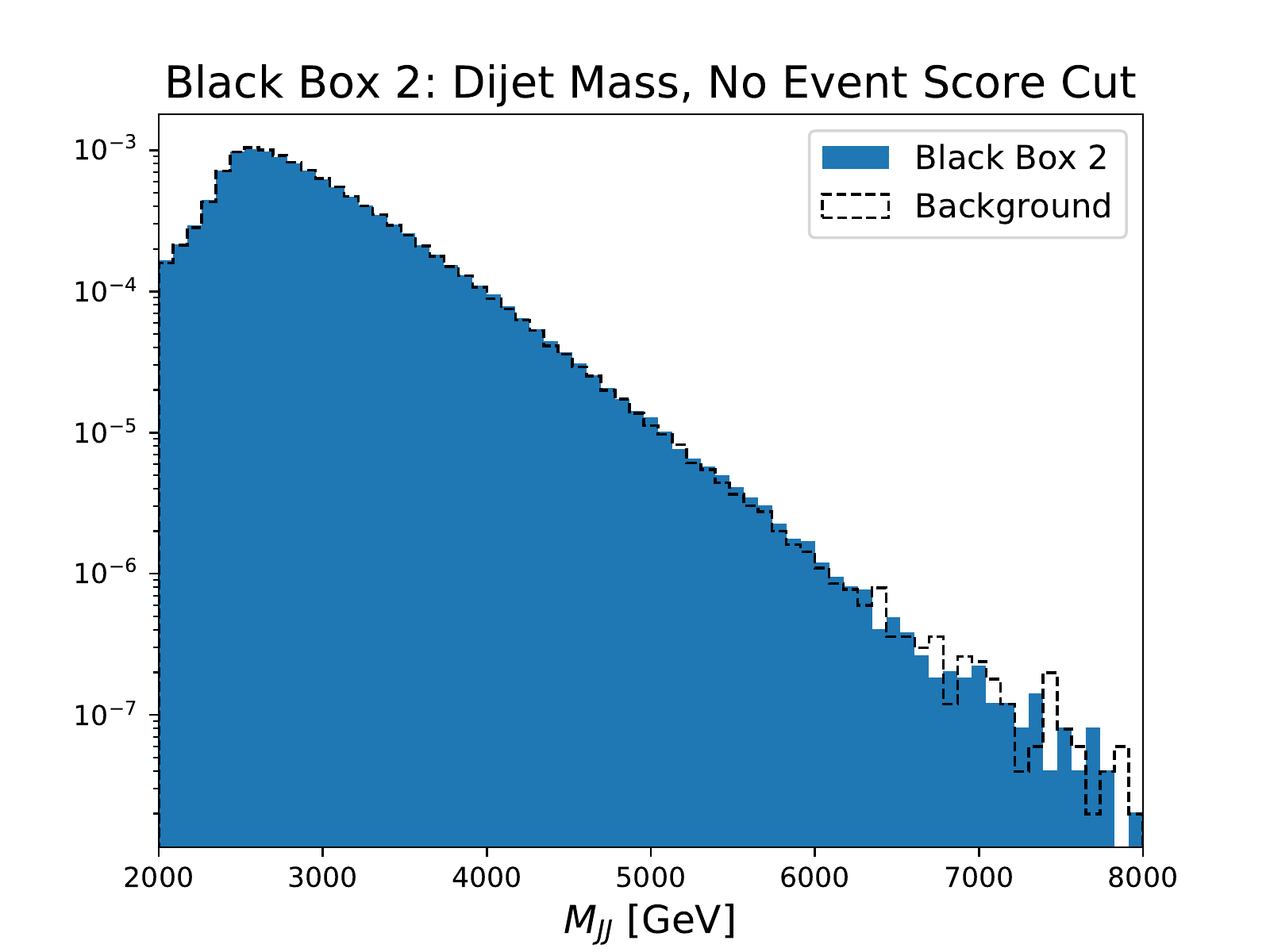}
		\includegraphics[width=0.47\textwidth]{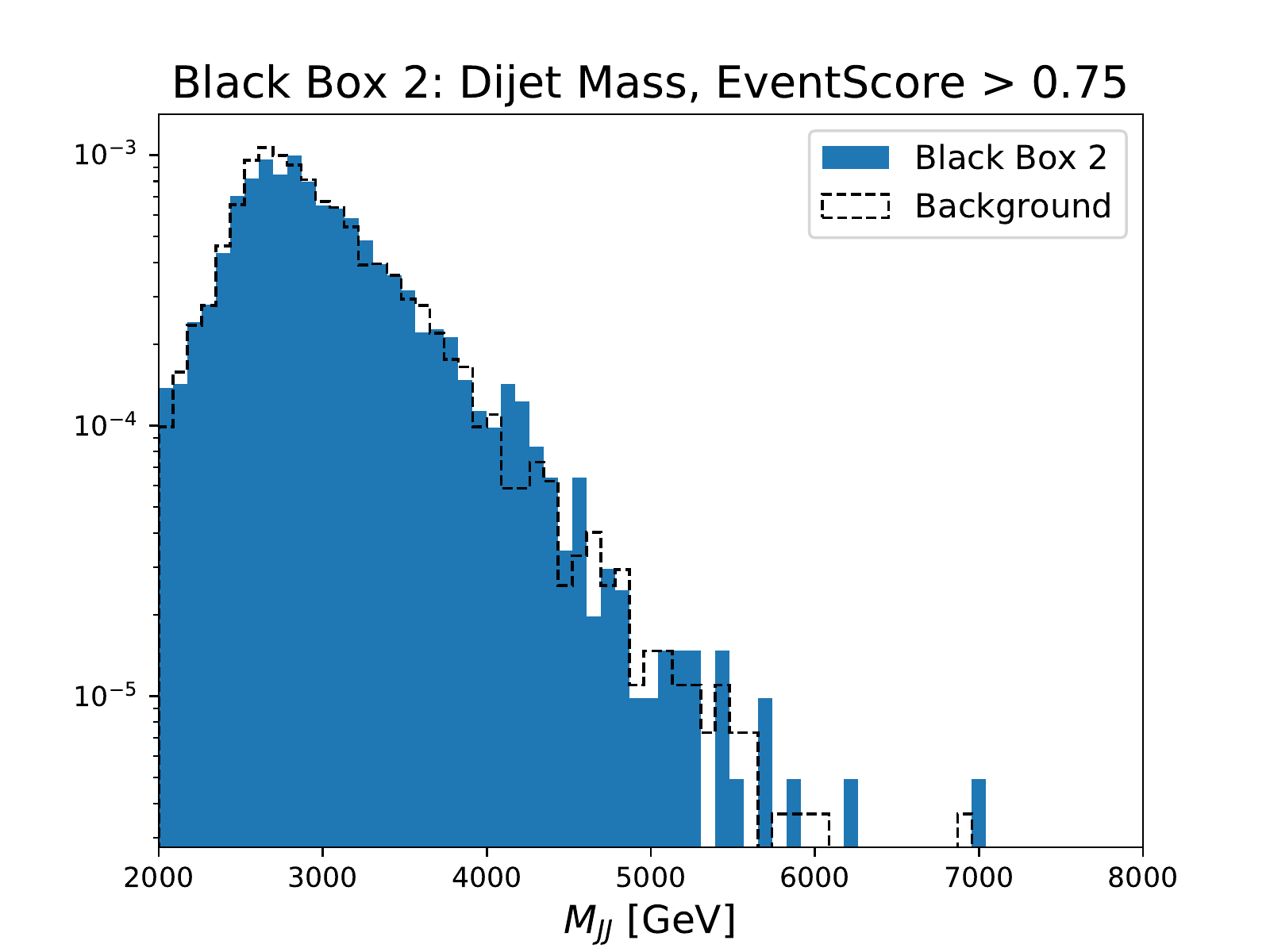}
	\end{center}
	\caption{Dijet invariant mass distributions before (left) and after (right) a selection on the Event Score from the Black Box 2 dataset. No signal is present, and the dataset shown consists entirely of multijet background events.}
	\label{fig:bb2}
\end{figure}

The same method applied to Black Box 2, shown in Fig.~\ref{fig:bb2}, results in no significant excesses in the invariant mass distribution. Additionally, the effect of the Event Score selection on the $m_{JJ}$ shapes is similar between the Black Box 2 and Background datasets. Black Box 2 does not contain any beyond-the-Standard-Model events, and therefore these results are consistent with a QCD-only sample. It is important to note that the model was trained independently on each dataset, and the resulting Event Scores are from entirely unique sets of network weights.

\begin{figure}[H]
	\begin{center}
		\includegraphics[width=0.47\textwidth]{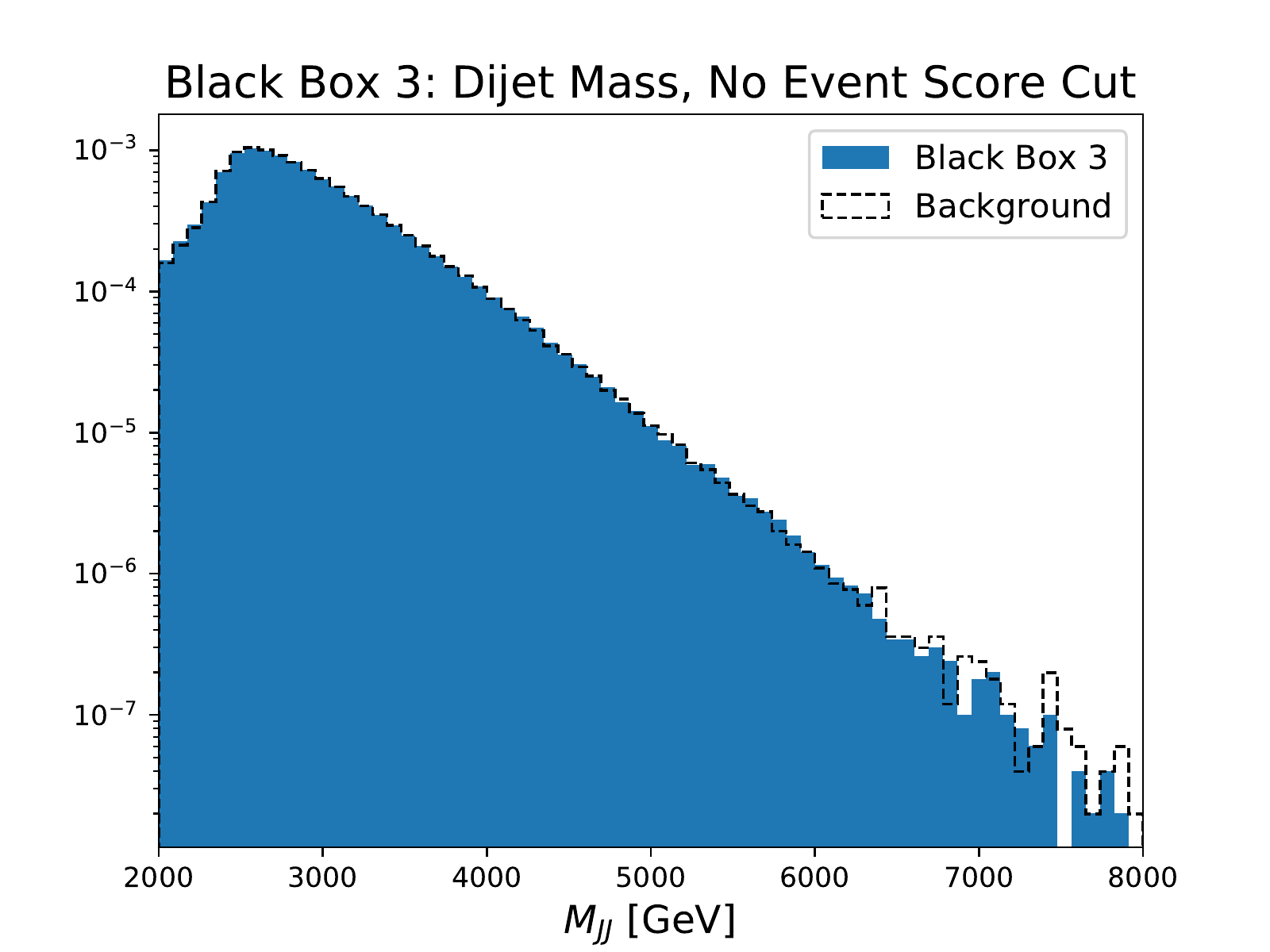}
		\includegraphics[width=0.47\textwidth]{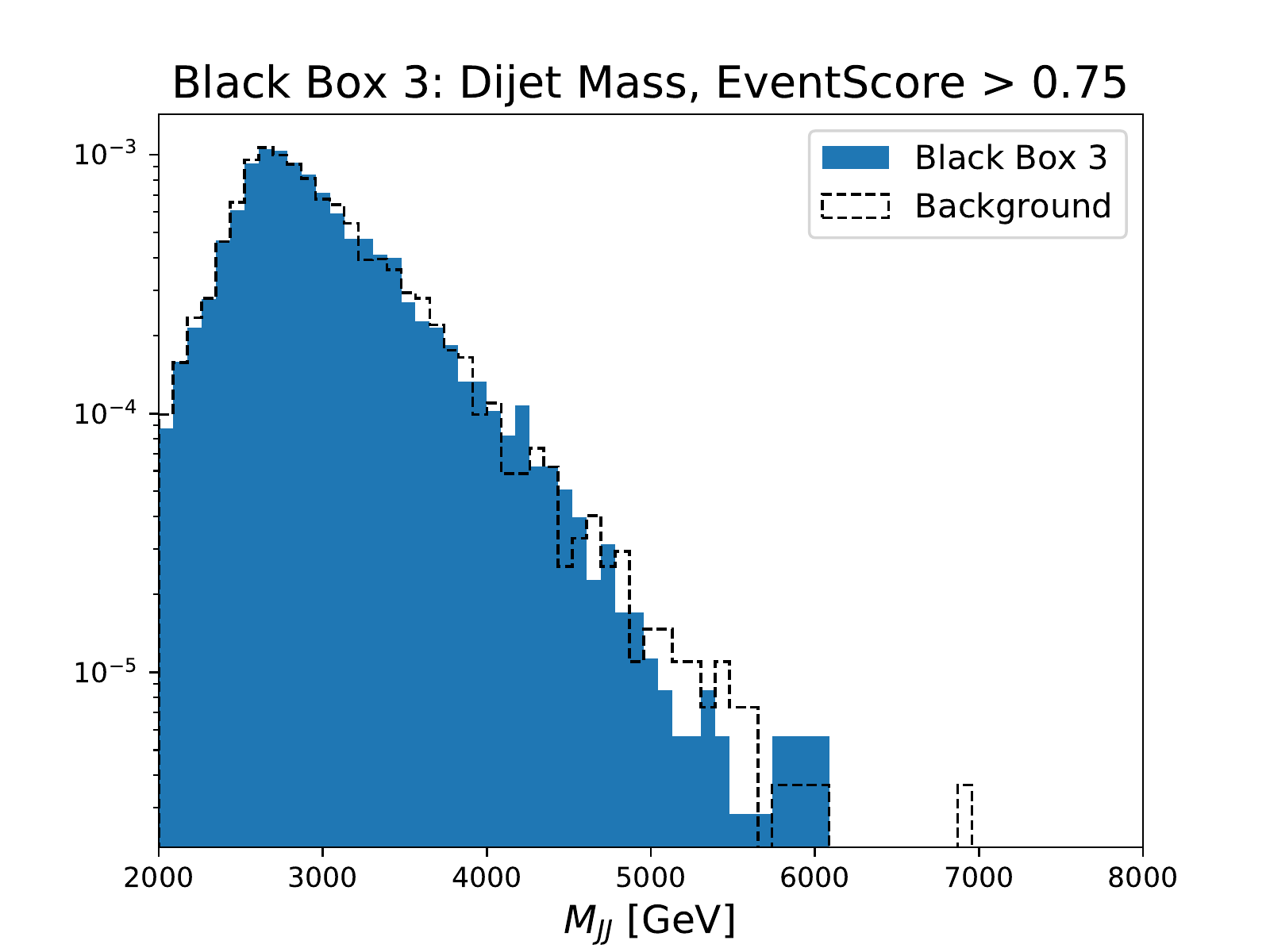}
	\end{center}
	\caption{Dijet invariant mass distributions before (left) and after (right) a selection on the Event Score from the Black Box 3 dataset. The signal present is a new boson with a mass of 4200 GeV.}
	\label{fig:bb3}
\end{figure}

Figure \ref{fig:bb3} shows results for Black Box 3. The signal in Black Box 3 consists of a new 4200 GeV particle, with varied final states beyond the two-prong large-$R$ jets described earlier. As the model described here is specifically sensitive to substructure within a large-$R$ jet, it is insensitive to the signal present in this Black Box. 

\subsubsection{Lessons Learned}
\label{sec:lessons}

This challenge presented a highly useful avenue for the development of our model. 
Results from the R\&D and Black Box dataset analyses indicate that the VRNN is capable of identifying anomalies via sequence modeling, as we have shown in the context of searching for anomalous substructure within boosted hadronically decaying objects.
We learned that the pre-processing method is hugely influential on the performance of the model, in particular the choice of $k_{t}$-ordered sequencing. 
We feel that this is a generalizable conclusion from our study which can be applied to the understanding and use of jet substructure in future analyses.

Given these lessons, there are a variety of future opportunities with this application of the VRNN architecture to jet-level anomaly detection. 
Since the VRNN takes constituent information as input and learns jet substructure without explicit reliance on high level variables, it is expected to have less correlation with jet mass than standard substructure variables such as $n$-subjettiness. Further characterization of this point could reveal a key advantage in using such an approach in an analysis context.
While we limited our scope in this study to be entirely unsupervised with no signal or background model information, the RNN and VAE elements of the VRNN give potential for accommodating more supervised training scenarios. 
Furthermore, a number of advancements to the architecture, such as a dedicated adversarial mass de-correlation network, or an additional input layer representing high-level features, are worthwhile avenues of exploration to enhance performance while minimizing unwanted correlations. 
 \FloatBarrier
\subsection[Anomaly Detection with Density Estimation]{Anomaly Detection with Density Estimation\footnote{Authors: Benjamin Nachman and David Shih.}}
\label{sec:ANODE}

This section introduces an approach called \textit{ANOmaly detection with Density Estimation} (ANODE) that is complementary to existing methods and aims to be largely background and signal model agnostic.  Density estimation, especially in high dimensions, has traditionally been a difficult problem in unsupervised machine learning. The objective of density estimation is to learn the underlying probability density from which a set of independent and identically distributed examples were drawn. In the past few years, there have been a number of breakthroughs in density estimation using neural networks and the performance of high dimensional density estimation has greatly improved. The idea of ANODE is to make use of these recent breakthroughs in order to directly estimate the probability density of the data. Assuming the signal is localized somewhere, one can attempt to use sideband methods and interpolation to estimate the probability density of the background. Then, one can use this to construct a likelihood ratio generally sensitive to new physics.

\subsubsection{Method}

This section will describe the ANODE proposal for an unsupervised method to search for resonant new physics using density estimation. 

Let $m$ be a feature in which a signal (if it exists) is known to be localized around some $m_0$.  The value of $m_0$ will be scanned for broad sensitivity and the following procedure will be repeated for each window in $m$.  It is often the case that the width of the signal in $m$ is fixed by detector properties and is signal model independent.  A region $m_0\pm\delta$ is called the signal region (SR) and $m\not\in[m_0-\delta,m_0+\delta]$ is defined as the sideband region (SB). A traditional, unsupervised, model-agnostic search is to perform a bump hunt in $m$, using the SB to interpolate into the SR in order to estimate the background.

Let $x\in\mathbb{R}^d$ be some additional discriminating features in which the signal density is different than the background density.  If we could find the region(s) where the signal differs from the background and then cut on $x$ to select these regions, we could improve the sensitivity of the original bump hunt in $m$. The goal of ANODE is to accomplish this in an unsupervised and model-agnostic way, via density estimation in the feature space $x$. 

More specifically, ANODE attempts to learn two densities: $p_\text{data}(x|m)$ and $p_\text{background}(x|m)$ for $m\in {\rm SR}$.  Then, classification is performed with the likelihood ratio
\begin{align}
\label{eq:Rx}
R(x|m)=\frac{p_\text{data}(x|m)}{p_\text{background}(x|m)}.
\end{align}
In the ideal case that $p_\text{data}(x|m)=\alpha\, p_\text{background}(x|m)+(1-\alpha)\,p_\text{signal}(x|m)$ for $0\leq\alpha\leq 1$ and $m\in\text{SR}$, Eq.~\ref{eq:Rx} is the optimal test statistic for identifying the presence of signal.  In the absence of signal, $R(x|m)=1$, so as long as $p_\text{signal}(x|m)\neq p_\text{background}(x|m)$, $R_\text{data}(x|m)$ has a non-zero density away from 1 in a region with no predicted background.

In practice, both $p_\text{data}(x|m)$ and $p_\text{background}(x|m)$ are approximations and so $R(x|m)$ is not unity in the absence of signal.  The densities $p(x|m)$ are estimated using conditional neural density estimation.  The function $p_\text{data}(x|m)$ is estimated in the signal region and the function $p_\text{background}(x|m)$ is estimated using the sideband region and then interpolated into the signal region.  The interpolation is done automatically by the neural conditional density estimator.  Effective density estimation will result in $R(x|m)$ in the SR that is localized near unity and then one can enhance the presence of signal by applying a threshold $R(x|m)>R_\text{cut}$, for $R_\text{cut}>1$.  The interpolated $p_\text{background}(x|m)$ can then also be used to estimate the background.

The ANODE procedure as described above is completely general with regards to the method of density estimation. In this work we will demonstrate a proof-of-concept using normalizing flow models for density estimation. Since normalizing flows were proposed in Ref.~\cite{pmlr-v37-rezende15}, they have generated much activity and excitement in the machine learning community, achieving state-of-the-art performance on a variety of benchmark density estimation tasks. 

\subsubsection{Results on LHC Olympics}

The conditional MAF is optimized\footnote{Based on code from \url{https://github.com/ikostrikov/pytorch-flows}.} using the log likelihood loss function, $\log(p(x|m))$.  All of the neural networks are written in PyTorch~\cite{NEURIPS2019_9015}.  For the hyperparameters, there are 15 MADE blocks (one layer each) with 128 hidden units per block.  Networks are optimized with Adam~\cite{adam} using a learning rate $10^{-4}$ and weight decay of $10^{-6}$.  The SR and SB density estimators are each trained for 50 epochs. No systematic attempt was made to optimize these hyperparameters and it is likely that better performance could be obtained with further optimization. For the SR density estimator, the last epoch is chosen for simplicity and it was verified that the results are robust against this choice.  The SB density estimator significantly varies from epoch to epoch.  Averaging the density estimates point-wise over 10 consecutive epochs results in a stable result.  Averaging over more epochs does not further improve the stability.  All results with ANODE present the SB density estimator with this averaging scheme for the last 10 epochs.

\begin{figure}[h!]
\centering
\includegraphics[scale=0.55]{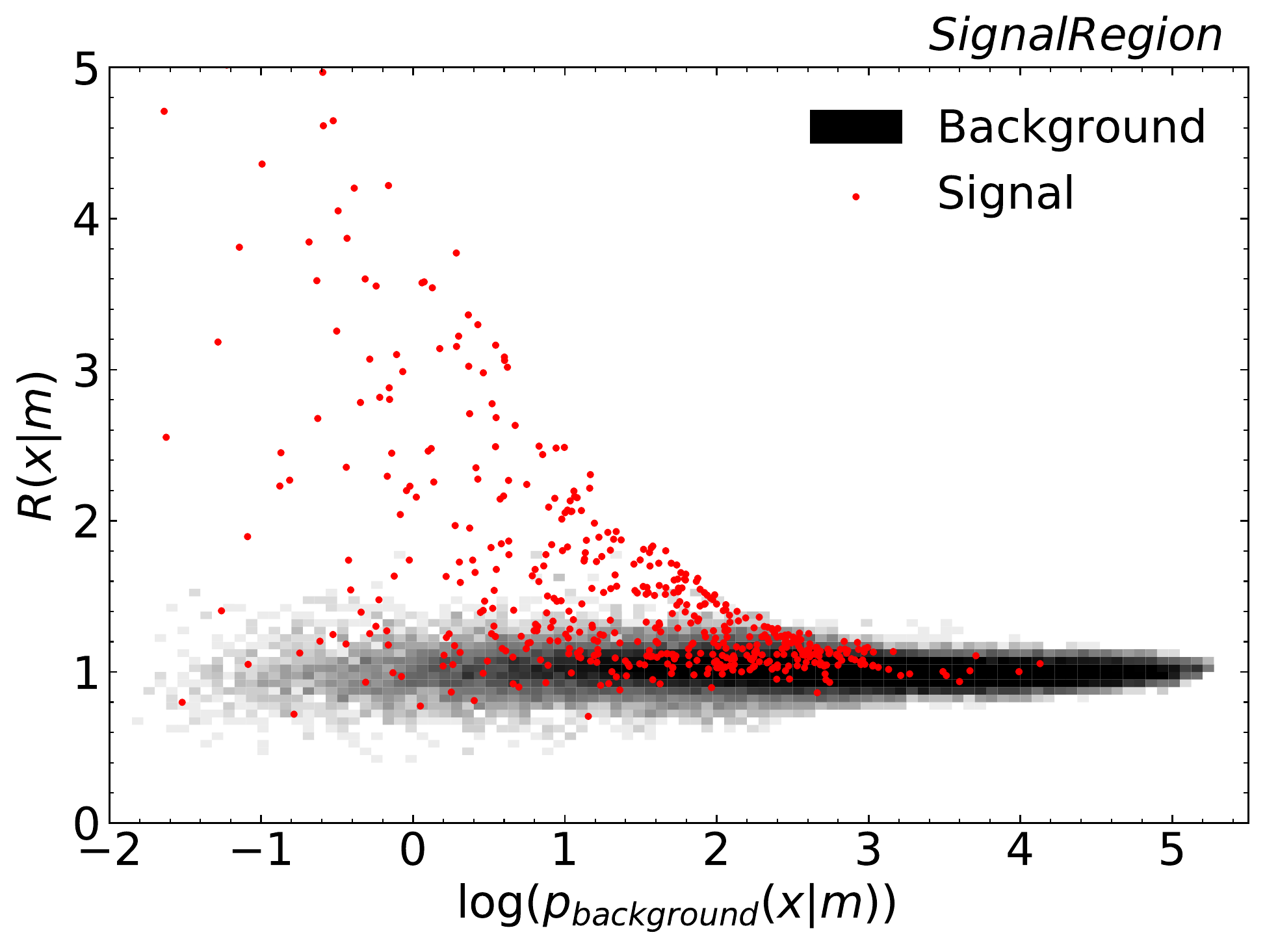}
\caption{Scatter plot of $R(x|m)$ versus $\log p_\text{background}(x|m)$ across the test set in the SR.  Background events are shown (as a two-dimensional histogram) in grayscale and individual signal events are shown in red.  Ref.~\cite{Nachman:2020lpy}.}
\label{fig:RvslogPbg}
\end{figure}

Figure~\ref{fig:RvslogPbg} shows a scatter plot of $R(x|m)$ versus $\log p_\text{background}(x|m)$ for the test set in the SR.  As desired, the background is mostly concentrated around $R(x|m)=1$, while there is a long tail for signal events at higher values of $R(x|m)$ and between $-2<\log p_\text{background}(x|m) <2$.   This is exactly what is expected for this signal: it is an over-density ($R>1$) in a region of phase space that is relatively rare for the background ($p_\text{background}(x|m)\ll 1$).

The background density in Fig.~\ref{fig:RvslogPbg} also shows that the $R(x|m)$ is narrower around $1$ when $p_\text{background}(x|m)$ is large and more spread out when $p_\text{background}(x|m)\ll 1$.    This is evidence that the density estimation is more accurate when the densities are high and worse when the densities are low. This is also to be expected: if there are many data points close to one another, it should be easier to estimate their density than if the data points are very sparse. 

\begin{figure}[h!]
\centering
\includegraphics[width=0.45\textwidth]{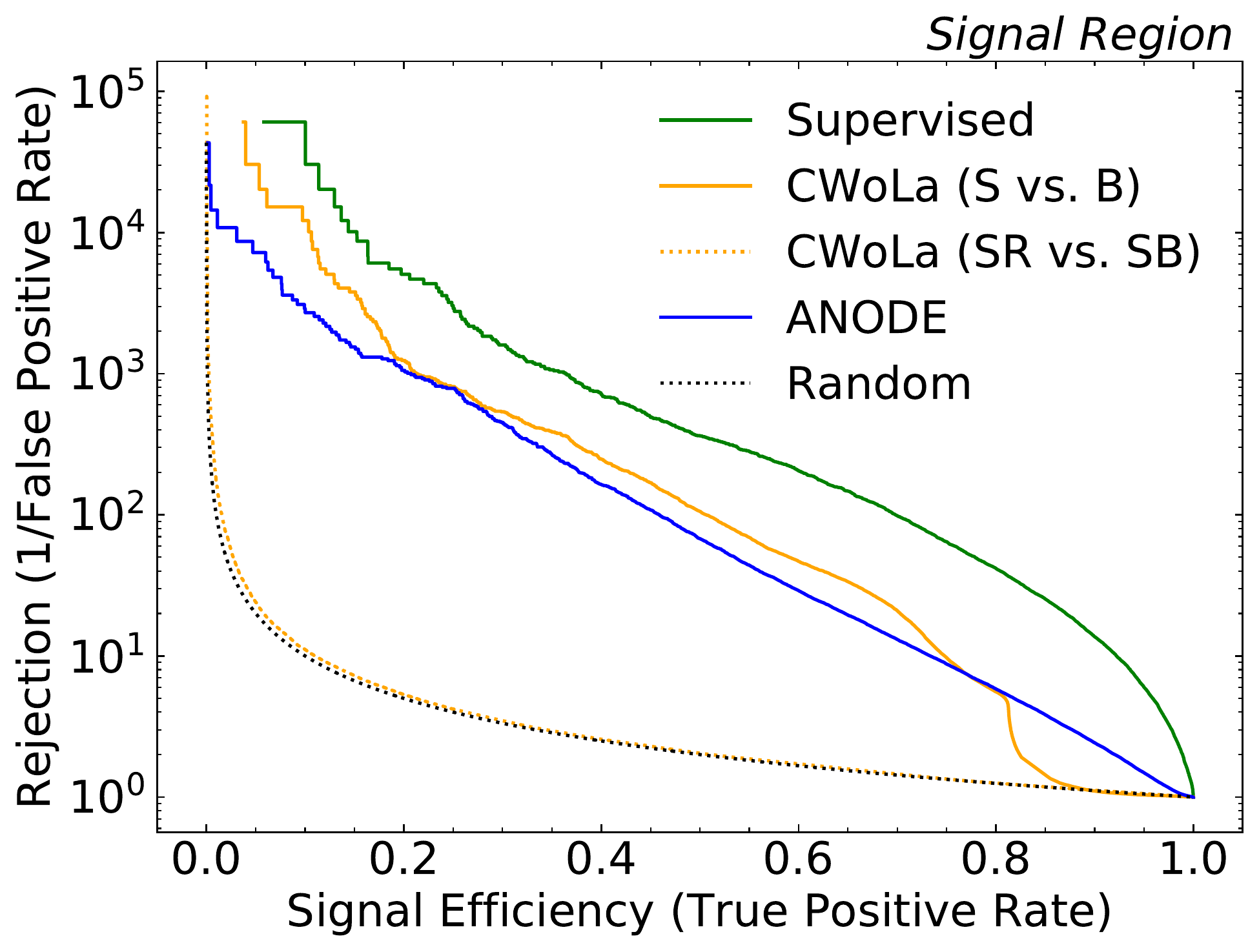}\hspace{5mm}
\includegraphics[width=0.45\textwidth]{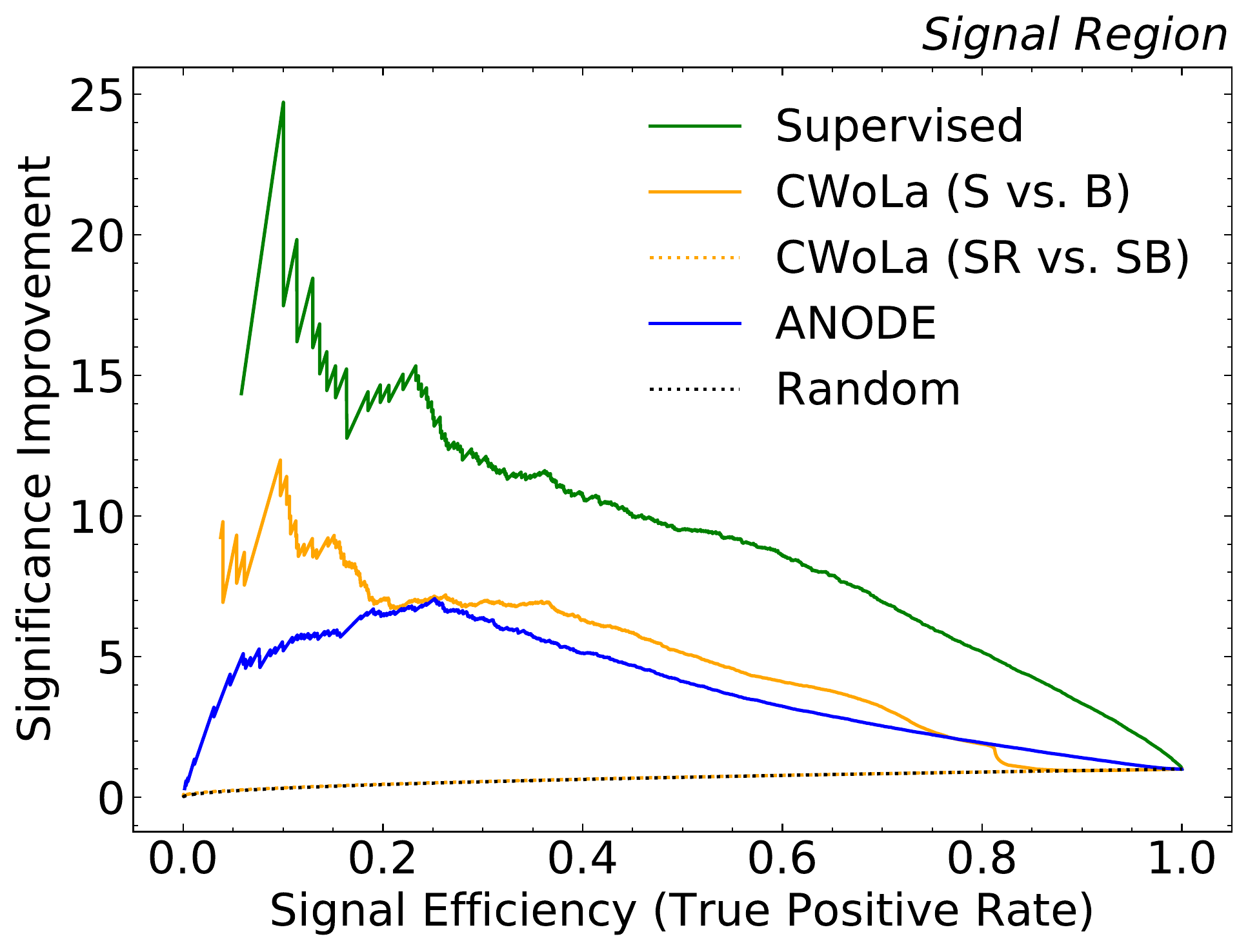}
\caption{Receiver Operating Characteristic (ROC) curve (left) and Significance Improvement Characteristic (SIC) curve (right).  Figure reproduced from Ref.~\cite{Nachman:2020lpy}.}
\label{fig:ROC}
\end{figure}

The performance of $R$ as an anomaly detector is further quantified by the Receiver Operating Characteristic (ROC) and Significance Improvement Characteristic (SIC) curves in Fig.~\ref{fig:ROC}.  These metrics are obtained by scanning $R$ and computing the signal efficiency (true positive rate) and background efficiency (false positive rate) after a threshold requirement on $R$.  The Area Under the Curve (AUC) for ANODE is 0.82. For comparison, the CWoLa hunting approach is also shown in the same plots.  The CWoLa classifier is trained using sideband regions that are 200 GeV wide on either side of the SR.  The sidebands are weighted to have the same number of events as each other and in total, the same as the SR.  A single NN with four hidden layers with 64 nodes each is trained using Keras~\cite{keras} and TensorFlow~\cite{tensorflow}.   Dropout~\cite{JMLR:v15:srivastava14a} of 10\% is used for each intermediate layer.  Intermediate layers use rectified linear unit activation functions and the last layer uses a sigmoid.  The classifier is optimized using binary cross entropy and is trained for 300 epochs.  As with ANODE, 10 epochs are averaged for the reported results\footnote{A different regularization procedure was used in Ref.~\cite{Collins:2018epr,Collins:2019jip} based on the validation loss and $k$-folding.  The averaging here is expected to serve a similar purpose.}.

The performance of ANODE is comparable to CWoLa hunting in Fig.~\ref{fig:ROC}, which does slightly better at higher signal efficiencies and much better at lower signal efficiencies.  This may be a reflection of the fact that CWoLa makes use of supervised learning and directly approaches the likelihood ratio, while ANODE is unsupervised and attempts to learn both the numerator and denominator of the likelihood ratio.  With this dataset, ANODE is able to enhance the signal significance by about a factor of 7 and would therefore be able to achieve a local significance above $5\sigma$ given that the starting value of $S/\sqrt{B}$ is 1.6.

\subsubsection{Lessons Learned}

While ANODE appears to be robust to correlations in the data (see Ref.~\cite{Nachman:2020lpy}), it is challenging to obtain precise estimates of the background density to very values of small $S/B$.  Another challenge is extending the density estimation to higher dimensions.  While the demonstrations here were based on the innovative MAF density estimation technique, the ANODE method can be used in conjunction with any density estimation algorithm.  Indeed, there are numerous other neural density estimation methods from the past few years that claim state-of-the-art performance, including Neural Autoregressive Flows \cite{DBLP:journals/corr/abs-1804-00779} and Neural Spline Flows \cite{durkan2019neural}; exploring these would be an obvious way to attempt to improve the results in this section. 

 \FloatBarrier
\subsection[BuHuLaSpa: Bump Hunting in Latent Space]{BuHuLaSpa: Bump Hunting in Latent Space\footnote{Authors: Blaz Bortolato, Barry M. Dillon, Andrej Matevc, Jernej F. Kamenik, Aleks Smolkovic.  The code used in this project can be found at \url{https://github.com/alekssmolkovic/BuHuLaSpa}.}}

\label{sec:buhula}

\subsubsection{Method}
\label{sec:method}

\noindent The BuHuLaSpa method assumes that the LHCO event data was generated through a stochastic process described by an underlying probabilistic generative model with continuous latent variables.
We use neural networks as approximators to the likelihood and posterior distributions of the model, and use the variational autoencoder (VAE) architecture as a means of optimising these neural networks.
For each event in the dataset we cluster the hadrons, select the resulting two leading $p_\text{T}$ jets, and order these by mass, $m_{j_1}>m_{j_2}$.
The data representation we use for the LHCO consists of the following observables for each jet: jet mass $m_{j}$, the $n$-subjettiness observables $\tau_2/\tau_1$ and $\tau_3/\tau_2$, and an observable similar to soft drop defined by clustering the jets with the C/A algorithm, then de-clustering them branch by branch, and summing the ratios of parent to daughter subjet masses along the way, stopping at some pre-defined mass scale which we have chosen to be $20$~GeV.
We denote these input measurements for the $i^{\text{th}}$ event in the sample by a vector $\vec{x}_i$.

The probablistic model underlying the VAE architecture can be viewed as a generative process through which the event data is generated from some underlying distributions.
The generative process for one event $\vec{x}_i$ starts with the sampling of a latent vector $\vec{z}_i$ from a prior distribution $p(\vec z)$.
Given this latent vector, the data for a single event is then sampled from the likelihood function $p(\vec{x}_i|\vec{z}_i)$.
The goal is then to approximate the posterior distribution, $p(\vec{z}_i|\vec{x}_i)$, i.e. perform posterior inference,  which maps a single event back to its representation in latent space.

The neural networks used as approximators to the posterior and likelihood functions are denoted by, $q_\phi(\vec{z}_i|\vec{x}_i)$ and $p_\theta(\vec{x}_i|\vec{z}_i)$, where $\phi$ and $\theta$ represent the weights and biases (i.e. the free parameters) of the encoder and decoder networks, respectively.
The sampling proceure is re-formulated using the re-parameterisation technique which allows the neural networks to be optimised through traditional back-propagation methods.
Specifically the encoder network consists of dim$(\vec{x})$ neurons in the input layer, followed by some number of hidden layers, and $2\times$dim$(\vec{z})$ neurons in the output layer.
Each element in $\vec{z}_i$ corresponds to two neurons in the output layer of the encoder network, one representing the mean and one representing the variance.
Elements of the latent vector $\vec{z}_i$ are then sampled from Gaussian distributions parameterised by these means and variances.
The resulting latent vector $\vec{z_i}$ is then fed to the decoder network which consists of dim$(\vec{z})$ neurons in the input layer, some number of hidden layers, and dim$(\vec{x})$ neurons in the output layer.

The VAE method is important because it allows us to frame the posterior inference task as an optimisation problem, and the loss function that is optimised is the Stochastic Gradient Variational Bayes (SGVB) estimator:
\begin{equation}
    \mathcal{L} = -D_\text{KL}(q_\phi(\vec{z}_i|\vec{x}_i)|p(\vec{z}_i)) + \beta_\text{reco}\log p_\theta(\vec{x}_i|\vec{z}_i)\,,
\end{equation}
where the first term is the KL divergence between the posterior approximation for event $i$ and the prior, and the second term is the reconstruction loss term.
We have added a re-scaling term $\beta_\text{reco}$ which alters how much influence the reconstruction loss has over the KL divergence term in the gradient updates.
We fix $\beta_\text{reco}=5000$ for this work, but our studies indicate that the results are insensitive to within order of magnitude changes to this number.

\paragraph{Invariant mass as latent dimension}

\noindent Once we have a fully trained VAE, the goal is then to use the latent representation of the data obtained from the posterior approximation to perform classification on the LHCO events.
To search for anomalies we typically look for excesses in the invariant mass distribution of the events. Thus it is important to understand any observed correlations between the latent vectors $\vec{z}_i$ and the invariant mass.
The latent space dimensions are each some non-linear function of the input observables.
In presence of correlations between the input observables and the invariant mass of the events, the latent dimensions are expected to encode some information on the invariant mass of the events. 
Crucially though, if signal events are localised in the invariant mass distribution and the VAE learns how to accurately encode and reconstruct the signal events, then part of the correlation the VAE networks learn must indeed correspond to the presence of the signal events in the data.

We then propose to make the invariant mass dependence of the VAE network explicit by imposing that one of the latent dimensions corresponds exactly to the invariant mass of the events.
We do this by modifying the generative process for a single event $\vec{x}_i$ with mass $m_i$ such that $\vec{z}_i$ is sampled from $p(\vec{z}_i)$, while $\tilde{m}_i$ is sampled from a gaussian prior, centered at $m_i$ and with a width $\sigma(m_i)$ reflecting a realistic uncertainty of the invariant mass reconstruction. In the LHCO case we take $\sigma(m_i) = 0.1 m_i$ for definiteness.
Both the latent vector $\vec{z}_i$ and the sampled mass variable $\tilde{m}_i$ are fed to the decoder which now has dim$(\vec{z})+1$ neurons in the input layer.
The encoder remains exactly the same as in the original VAE set-up and in particular can be made completely agnostic to invariant mass by decorrelating  the input variables $\vec{x}_i$ from $m_i$ using standard techniques. 
Now however the decoder is able to use the invariant mass information for each event to help in the reconstruction of the event data $\vec{x}_i$. At the same time the encoder network is no longer incentivized to learn the correlations between $\vec{x}_i$ and $m_i$ even if these are present in the data.
This has a number of potential benefits:
\begin{enumerate}
    \item The optimisation occurs locally in the invariant mass variable.
        Events with similar latent representations, i.e. similar $\vec{z}$, but very different invariant masses will now be treated differently by the decoder, therefore the network will no longer be forced to use the latent vector $\vec{z}$ to distinguish between events with different invariant masses.
    \item We can visualise the correlations between the latent space and the invariant mass explicitly without relying on data.
        By scanning over $\vec{z}_i$ and $\tilde{m}_i$ and feeding the values into the decoder we can visualise the latent space structure in terms of different observables at different invariant masses.
        This novel way of inferring on what the network has learned could lead to new approaches to bump hunting with machine learning at colliders, or even more broadly to machine learning applications in high-energy physics.
\end{enumerate}

\paragraph{Optimization and classification}

\noindent Using the R\&D dataset we investigated how best to train the VAE, and then applied what we learned here to the analysis on the black box datasets.
After an extensive scan over the hyper-parameters of the model, and monitoring the behaviour of the network throughout the training, we have have come to the following conclusions regarding optimization and classification:
\begin{itemize}
    \item The Adagrad and Adadelta optimizers consistently outperform momentum-based optimizers like Adam and Nadam, which we expect is due to the smoothing of gradients in the latter which in effect reduce the sensitivity of the gradient updates to outliers in the data.
    \item The norm of the latent vector $|\vec{z}_i|$ performs best as a classifier for the signal events.
    \item Classification performance does not converge throughout training, instead it peaks and then dies off at larger epochs.
        The epoch at which the peak performance occurs is correlated with a minimum in the reconstruction loss of the signal-only events, indicating that the network begins to ignore outliers in the data in order to reduce the overall reconstruction loss.
    \item It appears that the reason for this is that at some point during the training the network learns to reconstruct just one or two of the eight observables well, while mostly ignoring the others.  What we have found is that this can be avoided if we monitor the variance of the per-observable reconstruction losses through the training, and stop the training at the minima of this variance.  This is very strongly correlated with the peak in the classification performance.
\end{itemize}
For the training we used just one latent dimension, SeLU activation functions, two layers of 100 nodes each, the Adadelta optimizer with a learning rate of $0.001$, Mean-Squared-Error reconstruction loss, and batch sizes of $10$k.
The correlations used in the early-stopping procedure are more robust and precise when using larger batch sizes.

\subsubsection{Results on LHC Olympics}
\label{sec:results}

\noindent For the blackbox datasets and the R\&D dataset we trained the VAE networks on the whole event sample, without any cuts or binning in invariant mass, and followed the early stopping procedure outlined above.
In Fig.~\ref{fig:fig1} we show an example of a ROC curve obtained by training on the R\&D data with an S/B of $0.1\%$.
In Fig.~\ref{fig:fig2} we show a bump in the invariant mass spectrum in the Black Box 1 data after applying a classifier trained with this method.
The bump is at a mass of $\sim 3.8$ TeV and if we study the jet mass (Fig.~\ref{fig:fig3}) and $\tau_2/\tau_1$ distributions of the events that pass the cuts we clearly see that they correspond to events with jet masses $\sim 750$ GeV and $\sim400$ GeV, with $\tau_2/\tau_1$ values from the lower end of the spectrum.
Our analyses of the Black Box 2 and Black Box 3 data did not result in any clear signals in the data.

\begin{figure}[h!]
\centering
\includegraphics[width=0.5\textwidth]{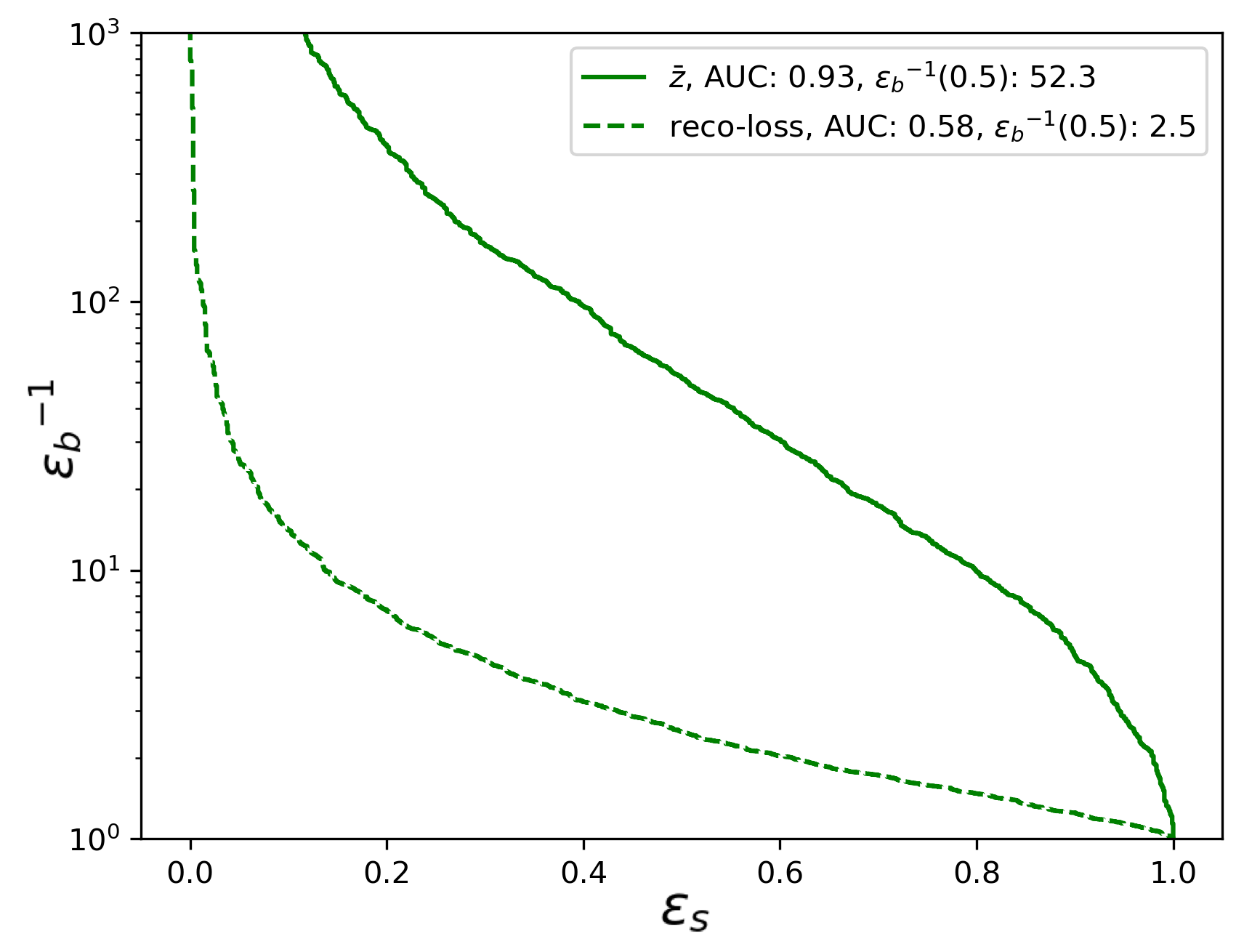}
    \caption{ROC curve obtained with the VAE classifier on the R\&D data.\label{fig:fig1}}
\label{fig:fig1}
\end{figure}

\begin{figure}[h!]
\centering
\includegraphics[width=0.5\textwidth]{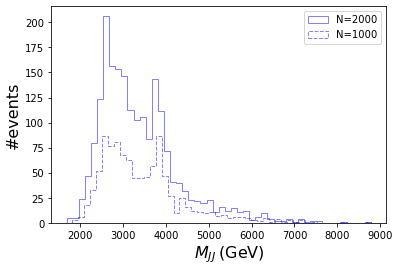}
    \caption{The invariant mass distribution for the blackbox 1 data after applying the VAE classifier.\label{fig:fig2}}
\label{fig:fig2}
\end{figure}

\begin{figure}[h!]
\centering
\includegraphics[width=0.45\textwidth]{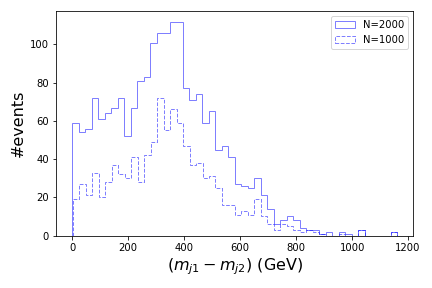}
\includegraphics[width=0.45\textwidth]{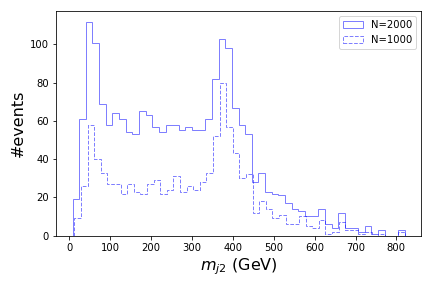}
    \caption{The jet mass distributions for the blackbox 1 data after applying the VAE classifier and restricting to the invariant mass range $[3.6,4.0]$ TeV.\label{fig:fig3}}
\end{figure}

\subsubsection{Lessons Learned}
\label{sec:lessons}

\noindent The first interesting lesson learned through this analysis was that the choice of the optimizer can play an important role in different machine-learning tasks.
While in standard classification tasks the momentum-based optimizers such as Adam perform very well, we found when using a VAE for anomaly detection this was not the case.
Instead, when the VAE is tasked with learning an effective latent representation of the dataset, including a small subset of anomalous signal events, it performs much better when using either the Adagrad or Adadelta optimizers.
The reason for this appears to be that the momentum updates in the Adam optimizer tend to smooth out the effects of anomalous events in the gradient updates, in turn ignoring the signal events in the data.
This may also be the case for other anomaly detection techniques, but has not been tested here.

The second lesson we learned was that after some number of epochs the VAE has a tendancy to `over-train' on just one or two of the eight inputs we used.
This results in an overall reduction in the loss function, but interestingly it results in an increase in the loss of signal-only events.
This increase in the reconstruction loss of signal-only events is inevitably correlated with a reduction in the peformance of the classifier.
We remedied this by introducing an early-stopping procedure in which we stop the training when the variance of the per-observable reconstruction losses reach a minimum.
This allowed us to achieve the optimal performance in an entirely unsupervised manner.

 \FloatBarrier

\subsection[GAN-AE and BumpHunter]{GAN-AE and BumpHunter\footnote{Authors: Louis Vaslin and Julien Donini.  All the scripts used to train and apply the GAN-AE algorithm are given at this link: \href{"https://github.com/lovaslin/GAN-AE\_LHCOlympics"}{"https://github.com/lovaslin/GAN-AE\_LHCOlympics"}.  The implementation of the BumpHunter algorithm used in this work can be found at this link: \href{https://github.com/lovaslin/pyBumpHunter}{https://github.com/lovaslin/pyBumpHunter}. In near future, it is planed that this implementation of BumpHunter becames a official package to be included in the scikit-HEP toolkit.}}
\label{sec:GAN-AE}

\subsubsection{Method}
\label{sec:method}

\noindent The methods presented in this section combine two independent anomaly detection algorithm.
The objective is to have a full analysis workflow that can give a global $p$-value and evaluate the number of signal events in any black-box dataset.

\paragraph{GAN-AE}

\noindent The GAN-AE method is an attempt at associating an Auto-Encoder architecture to a discriminant neural network in a GAN-like fashion.
The reason for this particular setting is to use information that does not come only from the ``reconstruction error" usually used to train AEs.
This could be seen as an alternative way to constrain the training of an AE.
As discriminant network, a simple feed-forward Multi-Layer Perceptron (MLP) is used. \\

\noindent This method been inspired by the GAN algorithm, the two participants (AE and MLP) are trained alternatively with opposite objectives :

\begin{itemize}
	\item The MLP is trained for a few epochs using the binary crossentropy (BC) loss function on a labeled mixture of original and reconstructed events, the objective being to expose the weaknesses of the AE.
	
	\item The AE is trained for a few epochs using a loss function combining the reconstruction error (here, the Mean Euclidean Distance between the input and output, or MED for short) and the BC loss of the MLP.
	In order to decorrelate as much as possible the reconstruction error and the invariant mass, the distance correlation (DisCo) term is used~\cite{DiscoFever}.
	The loss is then given by :
	$$\text{loss}_\text{AE} = \text{BC}+\varepsilon\times \text{MED}+\alpha\times \text{DisCo}$$
	\noindent With $\varepsilon$ and $\alpha$ two hyperparameters used to balance the weights of each terms. In this case, the BC term is evaluated by giving reconstructed events to the MLP, but this time with the ``wrong label", the objective being to mislead the MLP. 
	
	\item Then the AE is evaluated on a validation set using a Figure of Merit (FoM) that also combines the reconstruction error and some information from the MLP.
	The FoM used is given by :
	$$\text{FoM} = \text{MED}+(1-\text{Mean}~\text{MLP}~\text{output})$$
	\noindent This second term is preferred over the binary crossentropy because it seems to be more stable, which makes it more suitable to set a early stopping condition.
	As for the reconstruction error, $1-(\text{Mean}~\text{MLP}~\text{output})$ must be minimized.
	In fact, the closer to zero is this term, the better the AE is at misleading the MLP.
\end{itemize}

\noindent These three steps are repeated in a loop until the FoM fails to improve for five cycles. Once the AE has been trained, the MLP can be discarded since it is not needed anymore. Then, the AE can be used by taking the reconstruction error (Euclidean distance) as discriminative feature.\\

\noindent The GAN-AE hyperparameter used for the LHC Olympics are shown in Tab.~\ref{tab:param_GAE}
\begin{table}[h!]
	\centering
	\begin{tabular}{|c|c|c|}
		\hline
		 & AE & MLP \\
		\hline
		Neurons per hidden layer & 30/20/10/20/30 &  150/100/50 \\
		\hline
		Number of epochs per cycle & 4 & 10 \\
		\hline
		Activation function & ReLU (sigmoid for output) & LeakyReLU (sigmoid for output) \\
		\hline
		Dropout & \multicolumn{2}{|c|}{ 0.2 (hidden layers only)} \\
		\hline
		Early-stopping condition & \multicolumn{2}{|c|}{5 cycles without improvment} \\
		\hline
	\end{tabular}
	\caption{Hyperparameters used for the GAN-AE algorithm.}
	\label{tab:param_GAE}
\end{table}

\paragraph{BumpHunter}
\label{sec:BH}

The BumpHunter algorithm is a hypertest that compares a data distribution with a reference and evaluates the p-value and significance of any deviation.
To do so, BumpHunter will scan the two distributions with a sliding window of variable width.
For each position and width of the scan window, the local p-value is calculated. The window corresponding to the most significant deviations is then defined as the one with the smallest local p-value.

 In order to deal with the look elsewhere effect and evaluate a global p-value, BumpHunter generates pseudo-experiment by sampling from the reference histogram.
The scan is then repeated for each pesudo-data histogram by comparing with the original reference.
This gives a local p-value distribution that can be compared with the local p-value obtained for the real data.
Thus, a global $p$-value and significance is obtained. The BumpHunter hyperparameters used for the LHC Olympics are shown in Tab.~\ref{tab:param_BH}
\begin{table}[h!]
	\centering
	\begin{tabular}{|c|c|}
		\hline
		min/max window width & 2/7 bins \\
		\hline
		width step & 1 bins \\
		\hline
		scan step & 1 bin \\
		\hline
		number of bins & 40 \\
		\hline
		number of pseudo-experiments & 10000 \\
		\hline
	\end{tabular}
	\caption{Hyperparameters used for the BumpHunter algorithm.}
	\label{tab:param_BH}
\end{table}

\paragraph{Full analysis workflow}

The objective of this work is to use the Auto-Encoder trained withe the GAN-AE algorithm to reduce the background and then use the BumpHunter algorithm to evaluate the (global) $p$-value of a potential signal.  However, the use of this second algorithm requires the use of a "reference background" to be expected in the data.
Unfortunately, such reference is not always available, as it is the case for the LHC Olympics black-box dataset.
Thus, in order to use BumpHunter, one must first extract a background model for the data. Another point that has to be taken into account is the fact that, despite the use of the DisCo term, the dijet mass spectrum is not totally independent from the reconstruction error. Thus, simply rescaling the full dataset precut to fit the mass spectrum postcut will not work. 

One way to do this is to use a small subset of the data to compute a shaping function.
The objective of this function is to capture how the mass spectrum behaves when a cut on the reconstruction error is applied.
This function is computed bin per bin on the dijet mass histogram by doing the ratio of the bin yields postcut and precut.\\
Of course, the presence of signal in the subset used for this calculation might impact this shaping function.
In order to mitigate this effect, the shaping function can be fitted using the tools available in the scikit-learn toolkit.
This will minimize the effect of the signal on the shaping function.\\
Once the shaping function is defined, it can be used to reshape the mass spectum precut in order to reproduce the behaviour of the background postcut.\\

\noindent With this final step, the full analysis workflow is the following :
\begin{itemize}
	\item Data preprocessing (anti-$k_t$ clusturing, precut on dijet mass)
	\item Training of GAN-AE on the R\&D background
	\item Application of the trained AE on the black-box dataset
	\item Use 100k events for the black-box to compute a shaping function
	\item Use the shaping function to build a reference to use the BumpHunter algorithm
\end{itemize}

\subsubsection{Results on LHC Olympics}
\label{sec:results}

\noindent The results shown were obtained with an AE trained with the GAN-AE algorithm on 100k events from the R\&D background.
Note that before the training and application, cuts were applied on the dijet mass at 2700 GeV and 7000 GeV.

\paragraph{R\&D dataset}
\label{sec:RnD}

\begin{figure}[h!]
\centering
\includegraphics[width=0.49\textwidth]{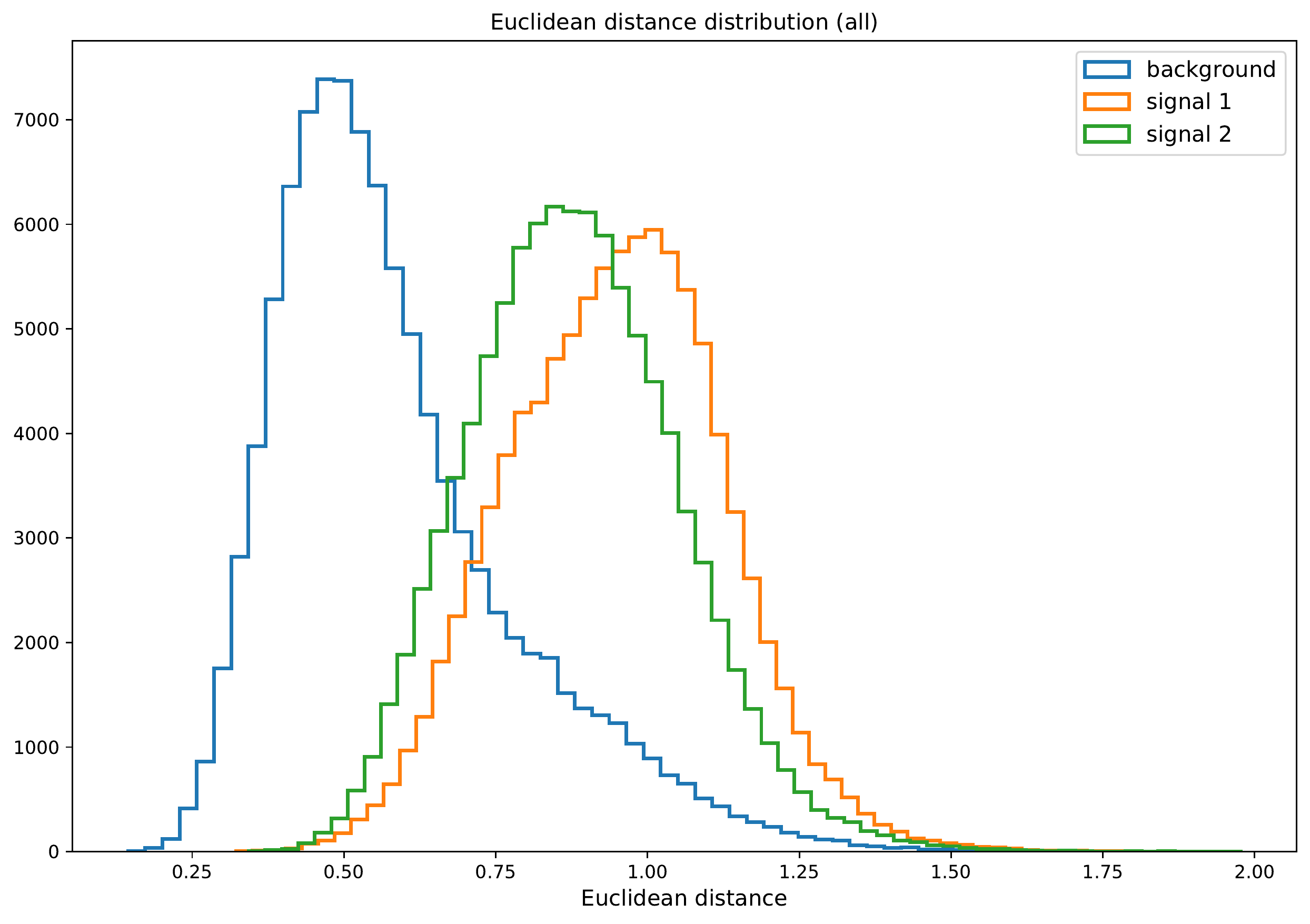}
\includegraphics[width=0.49\textwidth]{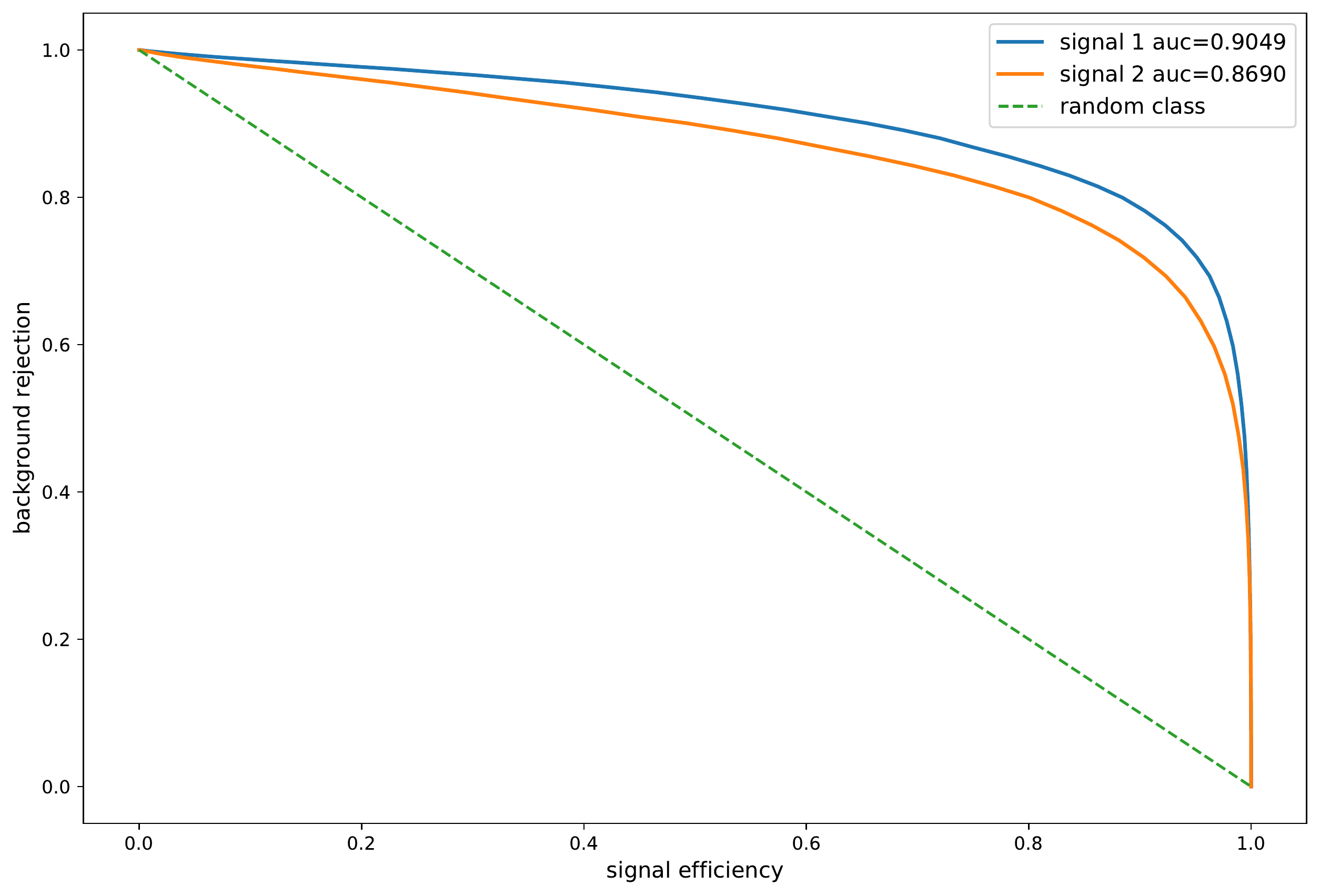}
\caption{Euclidean distance distributions and ROC curves obtained for the R\&D dataset.}
\label{fig:GAE_RnD}
\end{figure}

\noindent Here we discuss the result obtained on the R\&D dataset.
The trained AE have been tested on 100k background events (not used during the training), as well as on the two signals provided.
Fig.~\ref{fig:GAE_RnD} shows the Euclidean distance distributions (left) and the corresponding ROC curves (right).\\

\noindent This result illustrates the potential of the GAN-AE algorithm to obtain a good discrimination between the background and signals, event though only the background was used during the training.
However, if the obtained AUC is good, it also appears that the Euclidean distance is still very correlated with the dijet mass.
This might have a negative impact on the bump hunting algorithm performance.

\paragraph{Black Boxe datasets}
\label{sec:BB}

\noindent Here we discuss the results obtained for the black box dataset provided for the LHC Olympics challenge.\\

\begin{figure}[h!]
\centering
\includegraphics[width=0.5\textwidth]{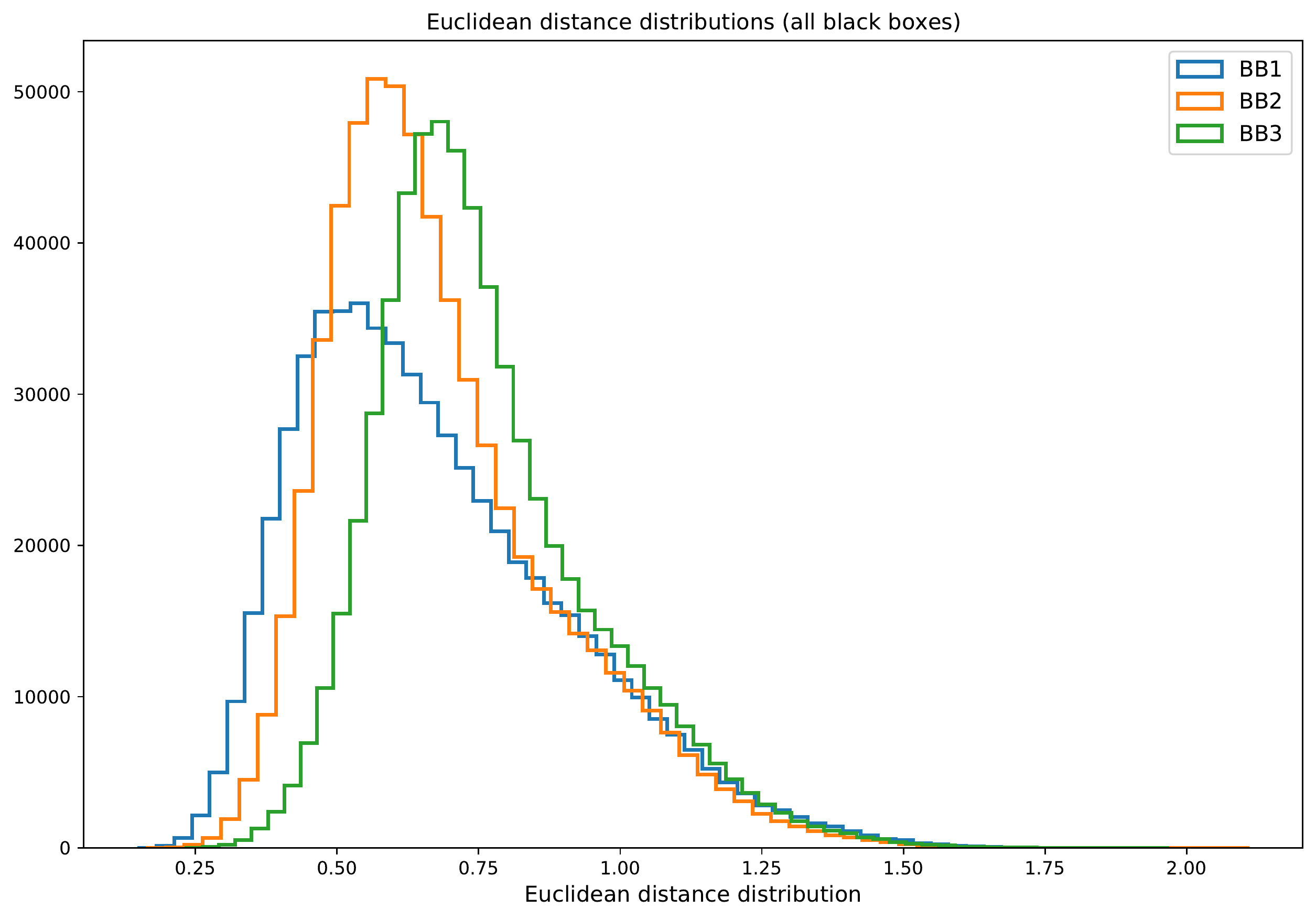}
\caption{Euclidean distance distributions and ROC curves obtained for the black boxes datasets.}
\label{fig:GAE_BB_dist}
\end{figure}

\noindent Figure ~\ref{fig:GAE_BB_dist} shows the Euclidean distance distribution obtained for each black box.
Compared to what was obtained with the R\&D background, the distributions seem larger and globally shifted to the right.
This is most likely due to the difference between the R\&D background and the background generated in the black boxes.
This fact shows that the method used is quite sensitive to the modeling of the background.

\begin{figure}[h!]
\centering
\includegraphics[width=0.33\textwidth]{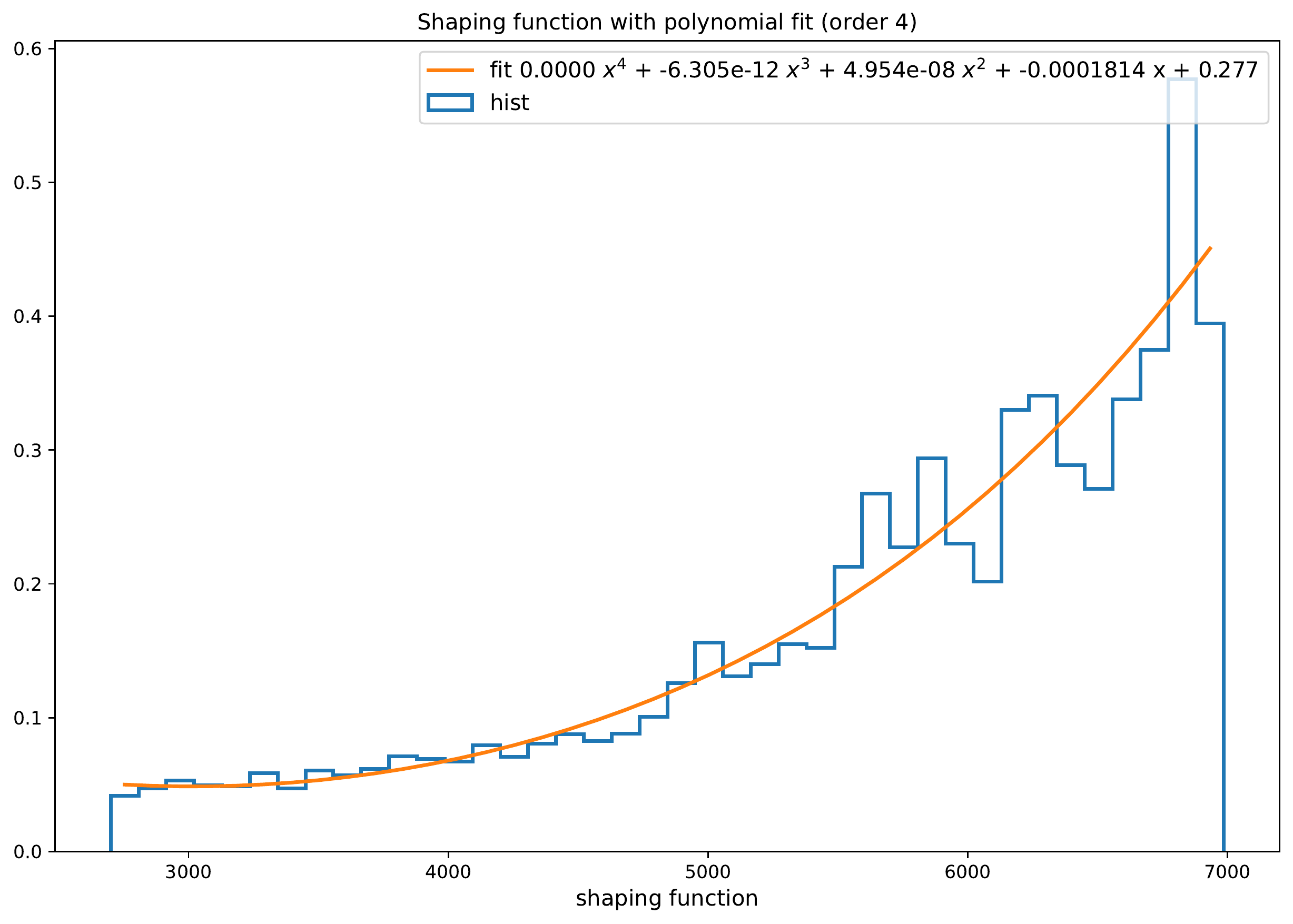}
\includegraphics[width=0.33\textwidth]{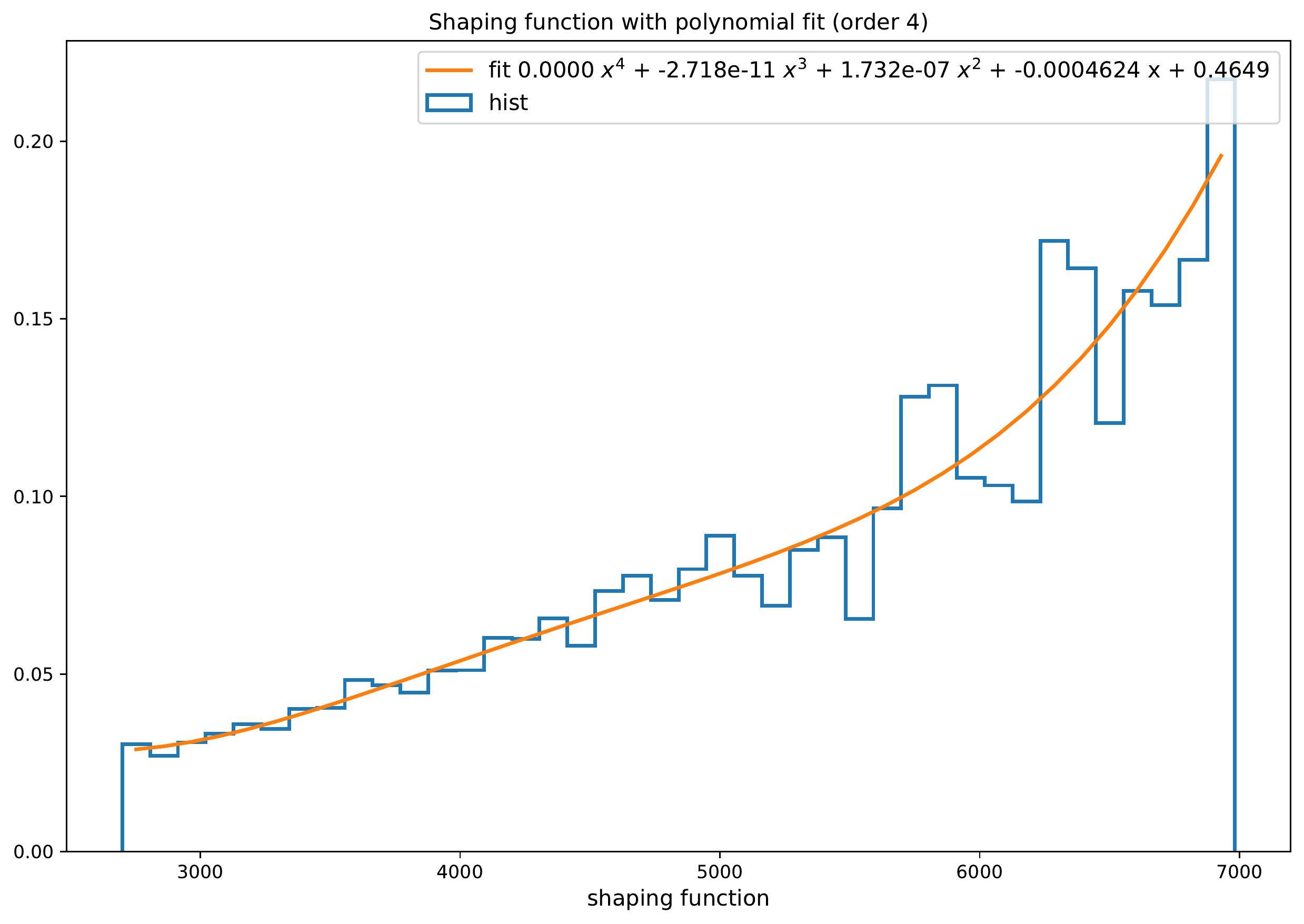}
\includegraphics[width=0.32\textwidth]{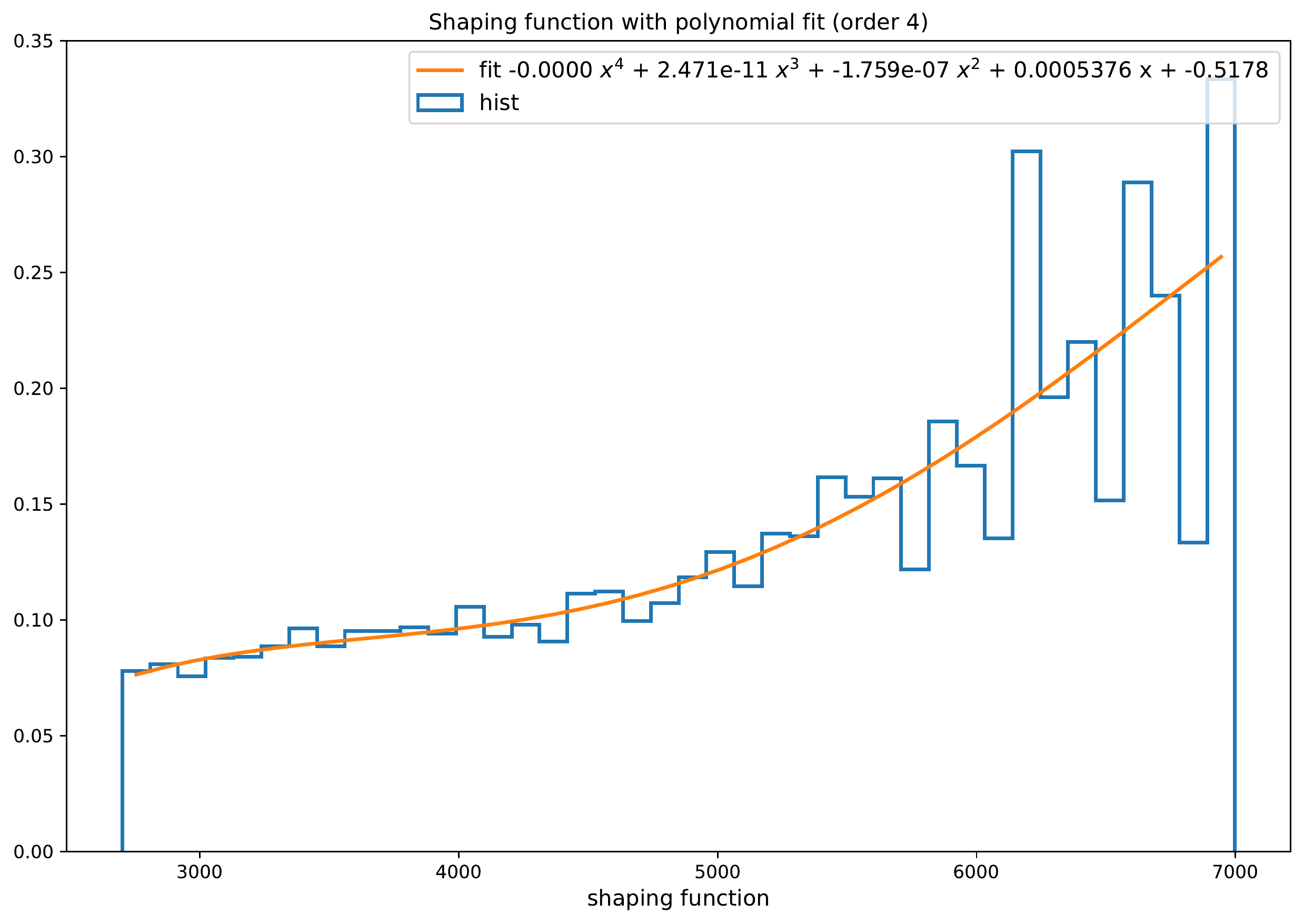}
\caption{Shaping function obtained for each black box.  From left to right, black box 1, 2 and 3.}
\label{fig:GAE_BB_shape}
\end{figure}

\noindent Figure~\ref{fig:GAE_BB_shape} shows the shaping function obtained using 100k events from each black box dataset. A preliminary fit was made to each of the distribution. Since the fit is suboptimal this might lead to the appearance of fake bump or fake deficit during the BumpHunter scan.

\begin{figure}[h!]
\centering
\includegraphics[width=0.33\textwidth]{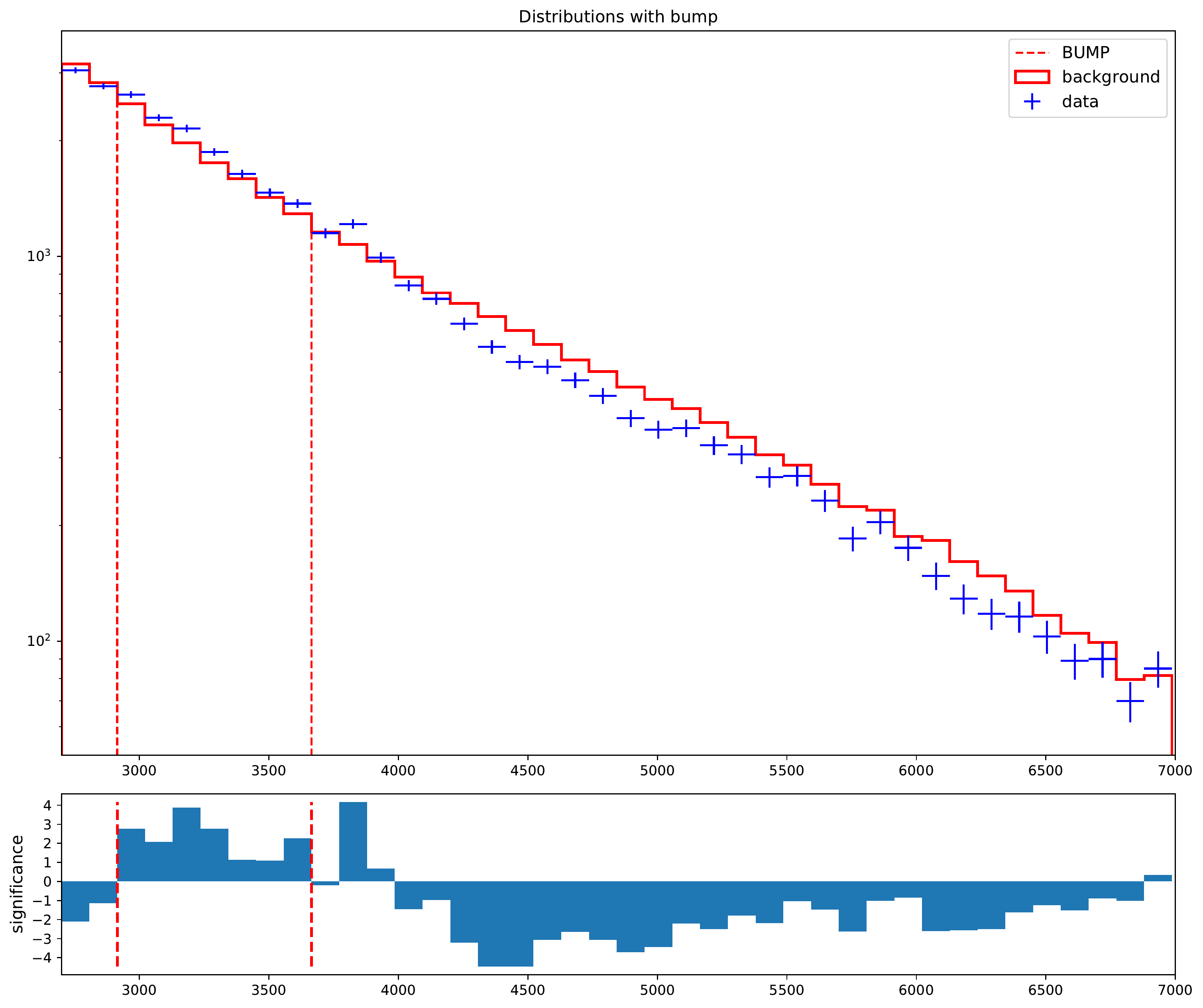}
\includegraphics[width=0.33\textwidth]{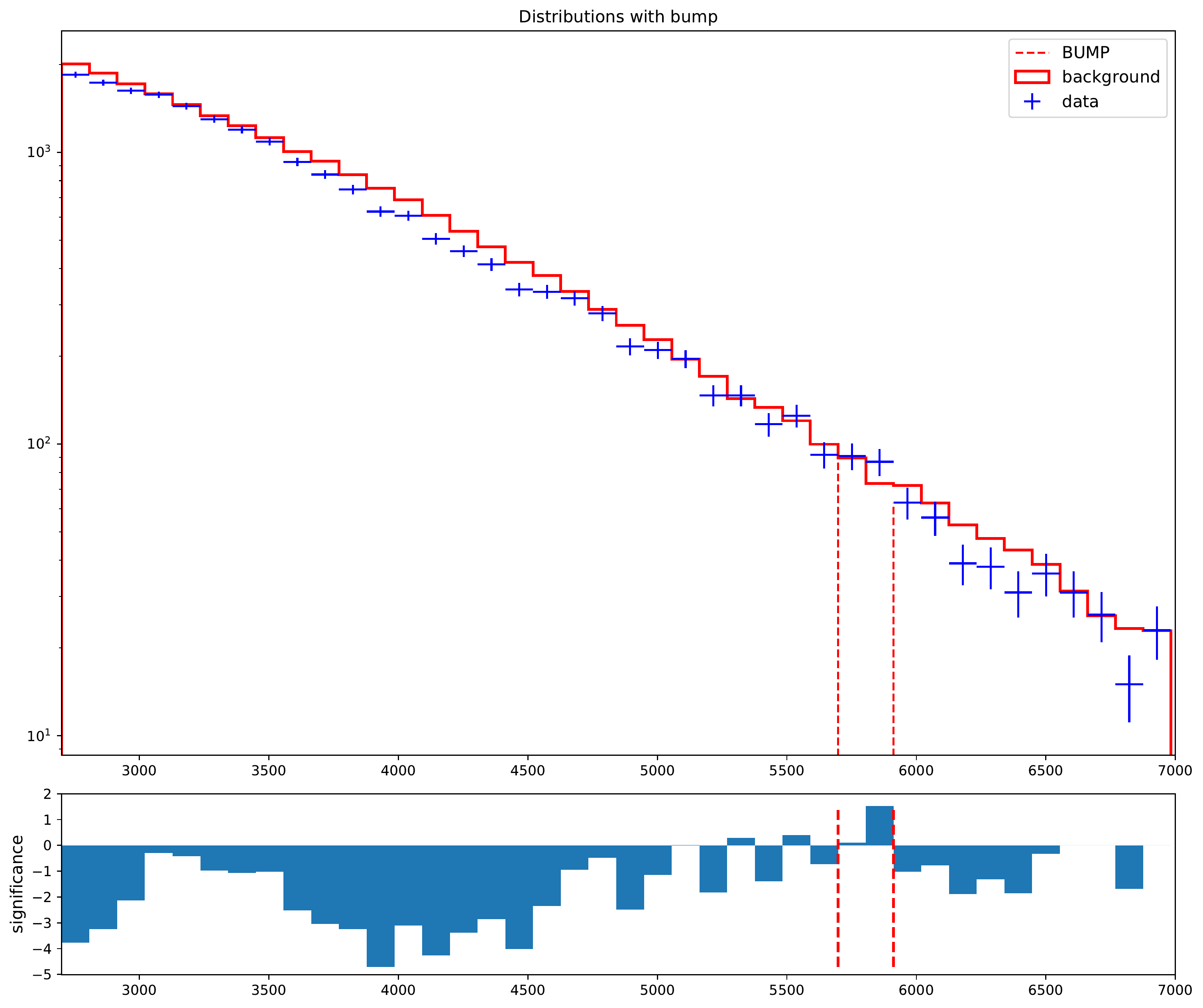}
\includegraphics[width=0.32\textwidth]{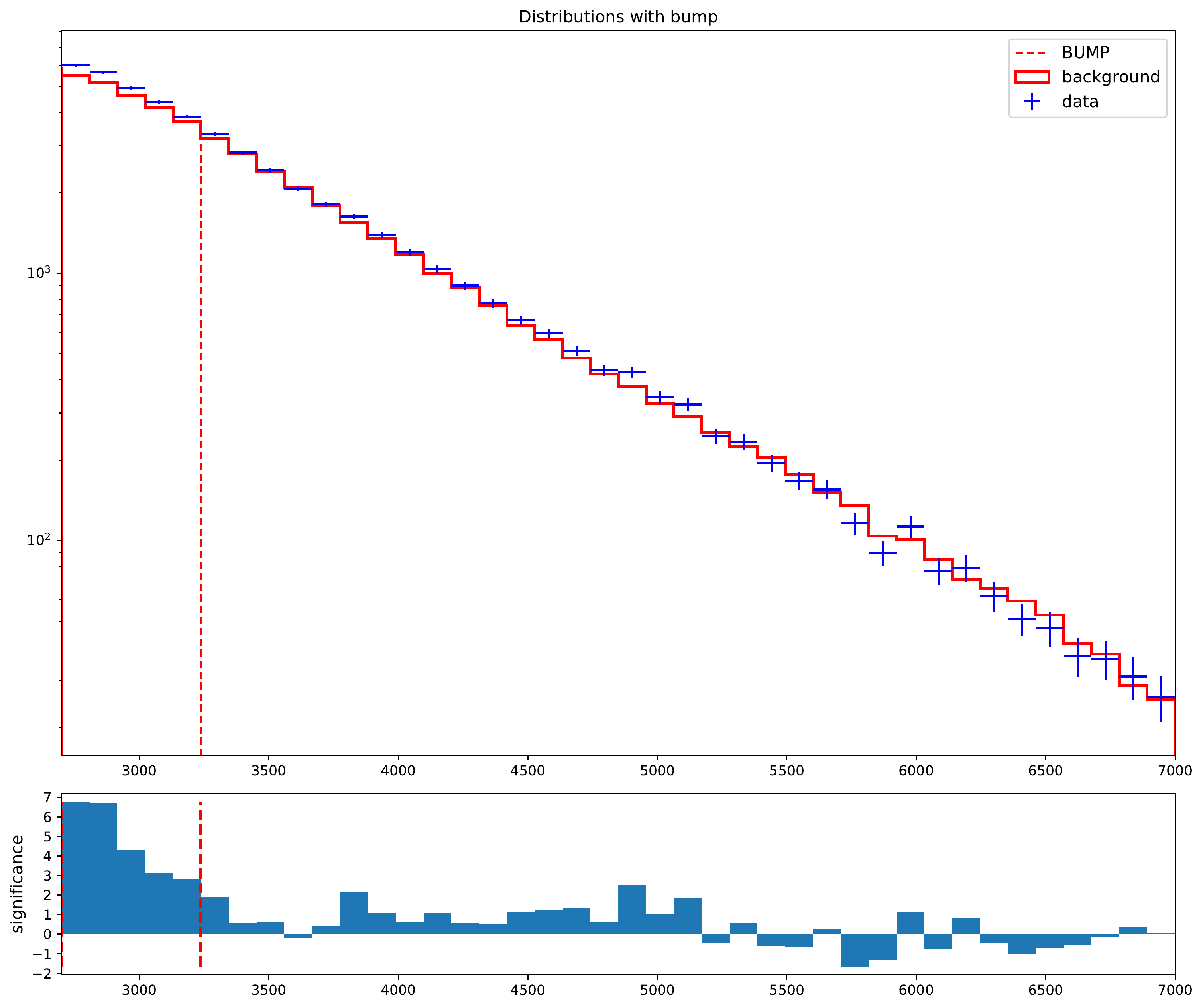}
\caption{Result of the BumpHunter scan obtained for each black box.  From left to right, Black Box 1, 2 and 3.}
\label{fig:GAE_BB_BH}
\end{figure}

\noindent Finally Fig.~\ref{fig:GAE_BB_BH} shows the results obtained with BumpHunter for all black boxes.
As foreseen with the poor fit of the shaping functions, the constructed reference backgrounds do not fit well the data after cut on the Euclidean distance.
In this condition and at the current stage of this work we can not really evaluate a meaningful $p$-value for a potential signal.
If the results were good on the R\&D dataset, it seems that the method is more challenging to apply without a good modeling of the background shape.

\subsubsection{Lessons Learned}
\label{sec:lessons}

The LHC Olympics challenge has been a good opportunity to test the potential of the GAN-AE algorithm that we have been developing.
This shows the potential of this method with the good results on the R\&D dataset, but also its limits.  

The results obtained revealed the sensibility of GAN-AE to the modeling of the background and to the correlation of the distance distribution with the dijet mass, despite the use of DisCo term. In addition, the fact that no background simulation that fits the black boxes data were available made the use of the BumpHunter algorithm difficult to apply.  
 \FloatBarrier
\subsection[Gaussianizing Iterative Slicing (GIS): Unsupervised In-distribution Anomaly Detection through Conditional Density Estimation]{Gaussianizing Iterative Slicing (GIS): Unsupervised In-distribution Anomaly Detection through Conditional Density Estimation\footnote{Authors: George Stein, Uro\u{s} Seljak, Biwei Dai.  The Gaussianizing Iterative Slicing (GIS) used in this work was an early form of what is now called Sliced Iterative Generation (SIG). More details on SIG can be found at \cite{sig}, and code will be made publicly available when ready. The results discussed in this section were also presented in Ref.~\cite{stein2020unsupervised}.}}

\label{sec:gis}

\noindent We approached the LHC signal detection challenge as an example of in-distribution anomaly detection. Rather than searching for samples near the tails of various data distributions as is typically done in out-of-distribution anomaly detection applications, 
the strategy we pursue is to look for excess density in a narrow region of a parameter of interest, such as 
the invariant mass. We term this {\textit{in-distribution anomaly detection}}. We perform conditional density estimation with Gaussianizing Iterative Slicing (GIS) \cite{sig}, and construct a local over-density based in-distribution anomaly score to reveal the signal in a completely blind manner. The results presented here are unchanged from our blind submission to the LHC Olympics in January 2020. 
Parallel and independent to our development and application of our conditional density estimation method, a similar one was applied in \cite{Nachman:2020lpy}, to great results on the R\&D dataset.

The R\&D dataset \cite{lhc_randd} was used for constructing and testing the method, while the first of the `black boxes' \cite{lhc_bb1} was the basis of our submission to the winter Olympics challenge. As the up to 700 particles given for each event are likely the result of hadronic decays we expect them to be spatially clustered in a number of jets. By focusing on the jet summary statistics rather than the particle data from an event we are able to vastly reduce the dimensionality of the data space. We note that this form of dimensionality reduction requires a small amount of prior knowledge and understanding of the data, and the assumption that the detected jets contain the anomaly, and other data-agnostic dimensionality reduction methods could instead be used. We used the python interface of FastJet \cite{Cacciari:2011ma,Cacciari:2005hq} - pyjet \cite{pyjet} - to perform jet clustering, setting $R=1.0$ as the jet radius and keeping all jets with $|\eta| < 2.5$. Each jet $J$ is described by a mass $m_J$, a linear momentum $p=(p_\text{T}, \eta, \phi)$, and n-subjettiness ratios $\tau^J_{n n-1}$ \cite{Thaler:2010tr,Thaler:2011gf}, which describe the number of sub-jets within each jet. A pair of jets has an invariant mass $M_{JJ}$. Additional parameters beyond these few may be necessary in certain scenarios, or at minimum useful, but our lack of familiarity with the field limited our search to use only these standard jet statistics. To construct images of the jets we binned each particles transverse momentum $p_\text{T}$ in $(\eta,\phi)$ and oriented using the moment of inertia. For the final black box 1 run we limited events to 2250 GeV $<$ $M_{JJ}$ $<$ 4750 GeV, resulting in 744,217 events remaining after all data cuts.

\subsubsection{Method}
\label{sec:method}

Our in-distribution anomaly detection method relies on a framework for conditional density estimation. Current state-of-the-art density estimation methods are those of flow-based models, popularized by \cite{realnvp} and comprehensively reviewed in \cite{normalizing_flows}. A conditional normalizing flow (NF) aims to model the conditional distribution $p(x|x_c)$ of input data $x$ with conditional parameter $x_c$  by introducing a sequence of $N$ differentiable and invertible transformations $f = f_1 \circ f_2 \circ \dots \circ f_N $ to a random variable $z$ with a simple probability density function $\pi (z)$, generally a unit Gaussian. Through the change of variables formula the probability density of the data can be evaluated as the product of the density of the transformed sample and the associated change in volume introduced by the sequence of transformations: 
\begin{equation}
    \label{eq:density}
    p(x|x_c) = \pi(f_{x_c}(x)) \left| \mathrm{det} \left(\frac{\partial f_{x_c}(x)}{\partial x} \right) \right| =  \pi(f_{x_c}(x)) \prod_{i=1}^{i=N} \left| \mathrm{det} \left(\frac{\partial f_{x_c,i}(x)}{\partial x} \right) \right| .
\end{equation}
While various NF implementations make different choices for the form of the transformations $f_i$ and their inverse $f_i^{-1}$, they are generally chosen such that the determinant of the Jacobian, $\mathrm{det} (\partial f_{x_c,i}(x)/\partial x)$, is easy to compute. Mainstream NF methods follow the deep learning paradigm: parametrize the transformations using neural networks, train by maximizing the likelihood, and optimize the large number of parameters in each layer through back-propagation.

In this work we use an alternative approach to the current deep learning methodology, a new type of normalizing flow - Gaussianizing Iterative Slicing (GIS) \cite{sig}. GIS works by iteratively matching the 1D marginalized distribution of the data to a Gaussian. 
At iteration $i$, the transformation of data $X_i$, $f_{x_c,i}$, can be written as
\begin{equation}
X_{i+1} = X_i - W_iW_i^TX_i + W_i \mathbf{\Psi}_{x_c,i}(W_i^T X_i),
\end{equation}

where $W_i$ is the weight matrix that satisfies $W_i^TW_i=I$, and $\mathbf{\Psi}_{x_c,i}$ is the 1D marginal Gaussianization of each dimension of $W_i^T X_i$.
To improve the efficiency, the directions of the 1D slices $W_i$ are chosen to maximize the PDF difference between the data and Gaussian using the Wasserstein distance at each iteration. The conditional dependence on $x_c$ is modelled by binning the data in $x_c$ and estimating a 1D mapping $\mathbf{\Psi}_i$ for each $x_c$ bin, then interpolating ($W_i$ is the same for different $x_c$ bins). 
GIS can perform an efficient parametrization and calculation of the transformations in Equation~\ref{eq:density}, with little hyperparameter tuning. We expect that standard conditional normalizing flow methods would also work well for this task, but did not perform any comparisons.

\begin{figure}[h]
  \centering
  \includegraphics[width=0.9\textwidth]{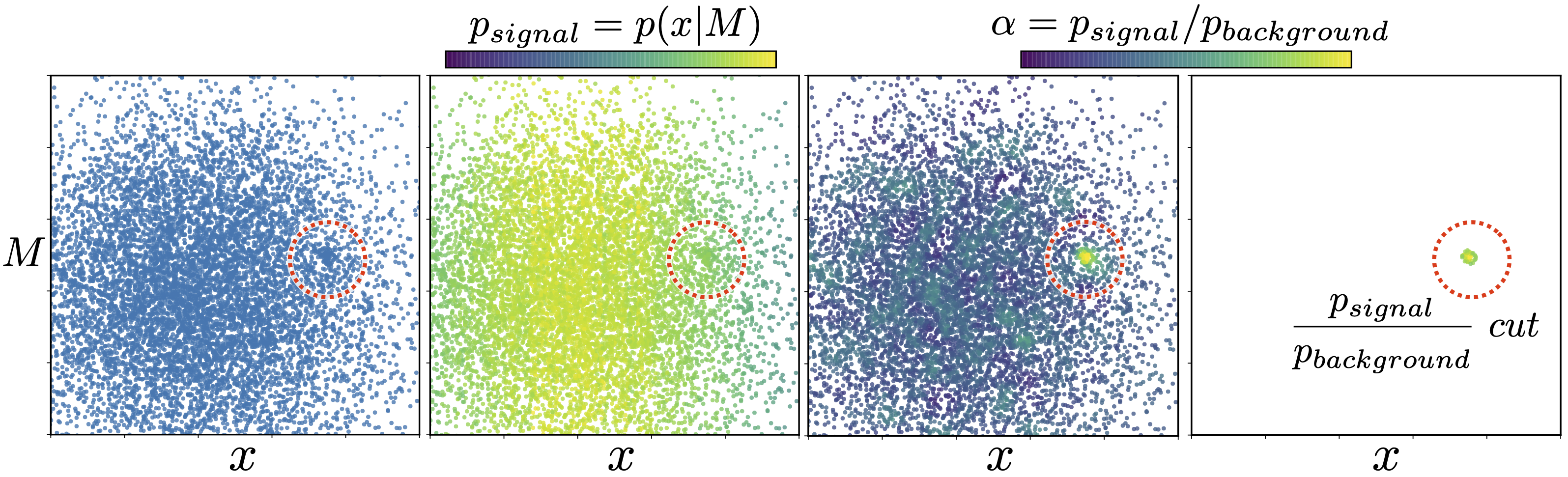}
   \caption{In-distribution anomaly detection through conditional density estimation. Consider samples of a 1D feature $x$ and a conditional parameter of interest $M$ (left panel), drawn from a smooth Gaussian `background' with a small number of anomalous `signal' events added (inside red circle for clarity). The conditional density values at each data point do not allow the anomaly to be distinguished from the background (center left panel), as they only identify the outliers. However, the local over-density anomaly ratio $\mathrm{\alpha}$ peaks at the anomalous data points (center right panel), and implementing a minimum cut on the anomaly ratio reveals the anomalous events (right panel).}
  \label{fig:method}
\end{figure}

With the GIS NF trained to calculate the conditional density, our in-distribution anomaly detection method, illustrated in Fig.~\ref{fig:method}, works as following: 
\begin{enumerate}
    \itemsep-0em 
    \item Calculate the conditional density at each data point $p(x|M_{JJ})$, denoting this $\mathrm{p_{signal}}$, using the jet masses and n-subjettiness ratios as the data $x$ and the invariant mass of a pair of jets $M_{JJ}$ as the conditional parameter.\item Calculate the density at neighbouring regions along the conditional dimension, $p(x | M_{JJ} \pm \Delta)$, and interpolate to get a density estimate in the absence of any anomaly. This is denoted $\mathrm{p_{background}}$. Explore various values of $\Delta$ and interpolation/smoothing methods. 
    \item The local over-density ratio (or anomaly score $\alpha$), $\mathrm{\alpha = p_{signal}/p_{background}}$, will be $\approx1$ in the presence of a smooth background with no anomaly. 
A sign of an anomalous event is $\alpha>1$. 
    Individual events can also be selected based on the desired $\alpha$ characteristic.
\end{enumerate}

\subsubsection{Results on LHC Olympics}
\label{sec:results}

We reasoned that if there is an anomalous particle decay in the data, its jet decay products would likely be located in a narrow range of masses corresponding to the mass of the particle itself. For this reason we chose the invariant mass $M_{JJ}$ of two jets as the conditional parameter to conduct the anomaly search along. We iterated on selections of jets $i$ and $k$, and selections of n-subjettiness ratios, and found the most significant anomaly when investigating the lead two jets and the first n-subjettiness ratio, so we used \{$M_{JJ}$, $m_{J_1}$, $m_{J_1}-m_{J_2}$, $\tau_{21}^{J_1}$,  $\tau_{21}^{J_2}$\} as the 5 parameters describing each event. 

We also experimented with training a convolutional autoencoder on the jet images, reasoning that rare events (anomalies) would have a higher reconstruction error and different latent space variables than more common ones, as seen in \cite{Farina:2018fyg}. While we found a larger than average reconstruction error for signal events, and latent space parameters to be noticeably different between background and signal events, on the R\&D dataset, these autoencoder-based variables introduced more noise in the  density estimation than the physics-based parameters, so they were not used in our final submission. 

Simple investigations of the dataset showed that it was smoothly distributed, and no anomalies were apparent by eye. We trained the conditional GIS on all events, and evaluated the anomaly score $\alpha$ for each datapoint. On the R\&D set we found that point estimates of the conditional densities resulted in a larger noise level than convolving the conditional density with a Gaussian PDF of width $\sigma=\Delta$ (1-PDF convolution for the background), discretely sampled at 10 points, so used the Gaussian-convolved probability estimates. $\mathrm{\sigma=250\ GeV}$ provided the most strongly peaked signal. 

\begin{figure}[h]
  \centering
  \includegraphics[width=0.8\textwidth]{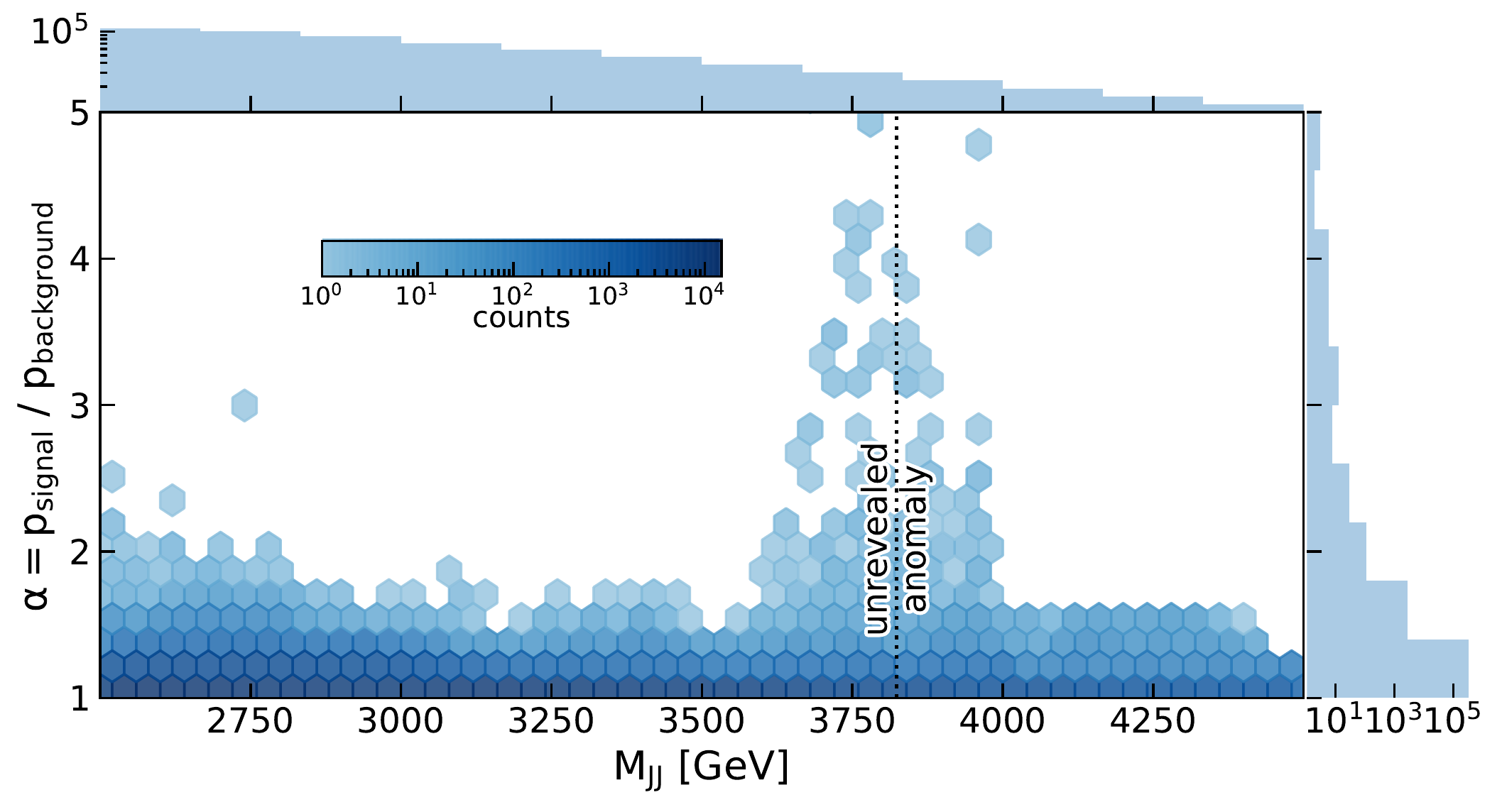}
    \caption{The anomaly score for each event as a function of the invariant mass of the leading two jets. A number of anomalous events are clearly seen near $\mathrm{M_{JJ}\approx 3750 GeV}$.}
  \label{fig:anomaly_score}
\end{figure}

As seen in Fig.~\ref{fig:anomaly_score}, the anomaly score strongly peaks around $\mathrm{M_{JJ}\approx 3750\ GeV}$. If these events are truly from a particle decay we expect that their resulting jet statistics will be clustered around some mean value, unlike if it is simply a result of noise in the model or background. To investigate the anomaly we remove data outside of $\mathrm{3600\ GeV < M_{JJ} < 3900\ GeV}$, and look at the events that remain after a series of cuts on the anomaly score $\alpha$. 

\begin{figure}[h]
  \centering

\includegraphics[width=0.8\textwidth]{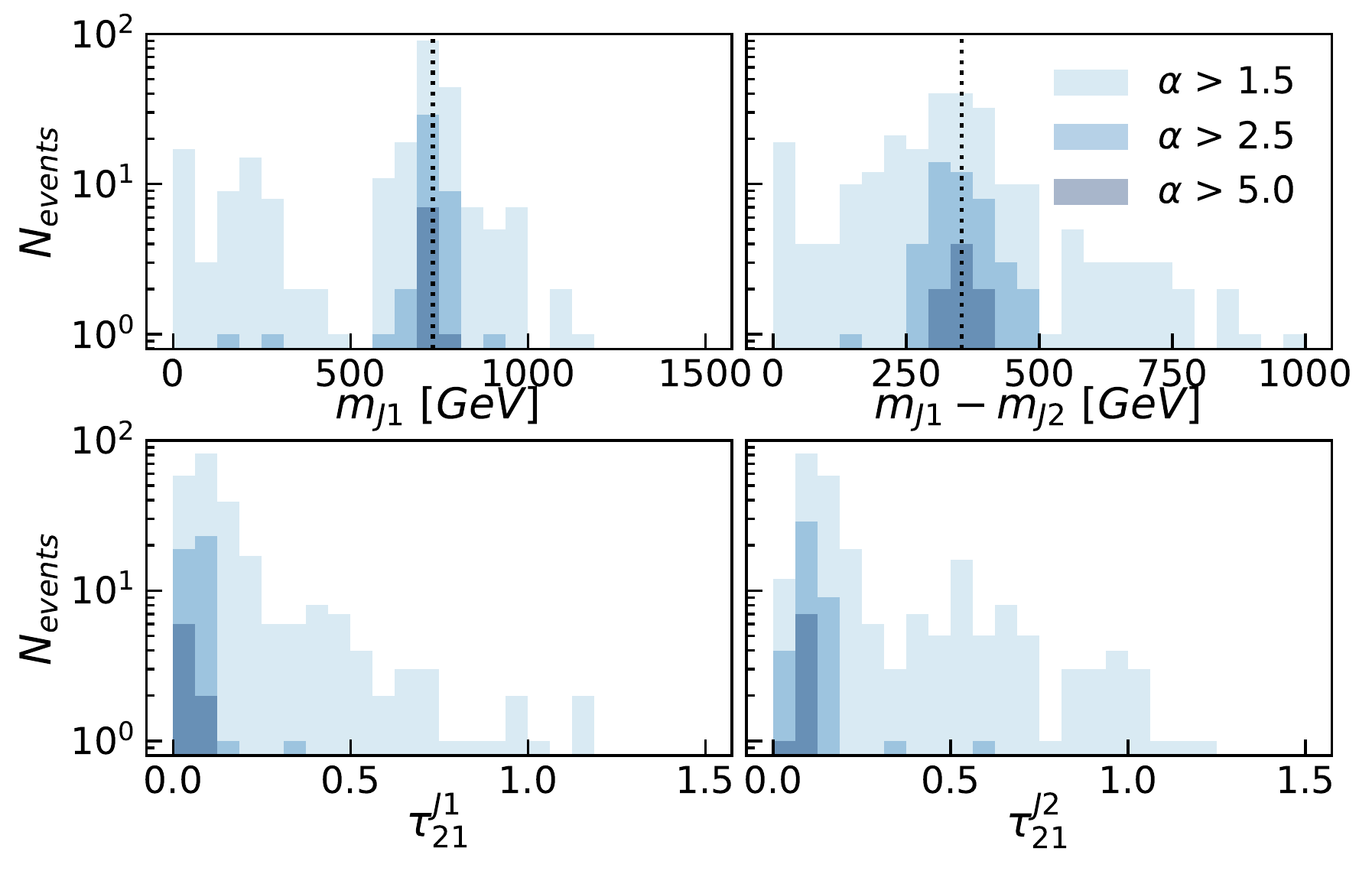}

  \caption{Parameter distributions of the events that remain after imposing cuts on the anomaly score $\alpha$, and limiting the mass range to  $\mathrm{3600\ GeV < M_{JJ} < 3900\ GeV}$. Vertical dashed lines are the true anomalous events that were unveiled after the close of the competition.}
  \label{fig:histograms}
\end{figure}

In Fig.~\ref{fig:histograms} we show the parameter distributions of the events that remain after imposing $\alpha > [1.5, 2.5, 5.0]$ cuts in the right four panels, and find that the most anomalous events are centered in $M_{J1}$ and $M_{J1}-M_{J2}$, and have small values of n-subjettiness $\tau_{21}$. This strongly indicates that we found a unique over-density of events that do not have similar counterparts at neighbouring $M_{JJ}$ values - i.e. an anomaly.

\begin{figure}[h]
  \centering
  \includegraphics[width=1.0\textwidth]{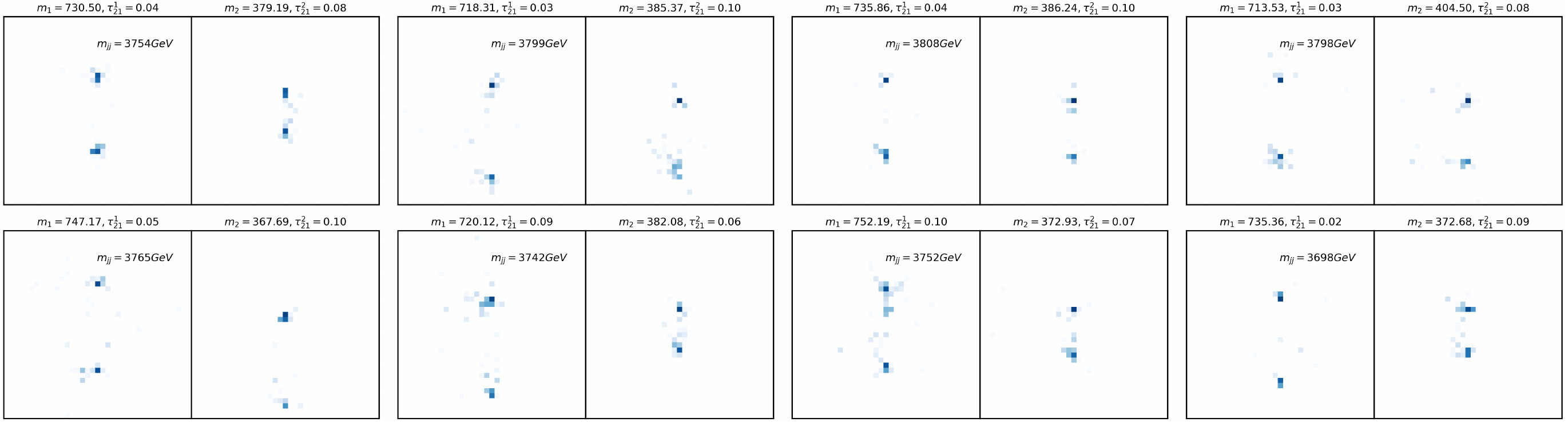}
    \caption{The eight most anomalous events in the black box. Each pair of images visualizes the particles belonging to the lead two jets. Images were constructed by binning the transverse momentum of each particle belonging to the jet in ($\eta$, $\phi$), and oriented along the y axis using using the $p_\text{T}$ weighted moment of inertia. Color is log scaled.}
  \label{fig:images}
\end{figure}

We visualized the events ranked by decreasing anomaly score in Fig.~\ref{fig:images}, and found that each of the leading two jets for events with a high anomaly score additionally have very similar visual appearances. Using the events that remain after an $\alpha > 2.0$ cut we can summarize the anomalous events as follows: a $\mathrm{3772.9\pm 8.3\ GeV}$ particle decays into 2 particles, one with $\mathrm{M_{1}=727.8 \pm 3.8\ GeV}$, and the other with $\mathrm{M_2 = 374.8 \pm 3.5\ GeV}$. Each of these decayed into  two-pronged jets. Based on the corresponding analysis of the R\&D data, by limiting the number of signal events until the results visually resembled Fig.~\ref{fig:anomaly_score}, we estimated that there were a total of $1000\pm 200$ of these events included in the black box of a million total events. While this is not a robust technique to estimate the number of events in all cases, as the anomaly characteristics may be much more broad or peaked in a black box than they were in the R\&D set, it nevertheless gave an accurate result here.

\subsubsection{Lessons Learned}
\label{sec:lessons}
The availability of a low-noise and robust density estimation method such as GIS was key throughout this work, as the lack of hyperparamater tuning allowed us to focus on the blind search rather than worrying that failing to detect an anomaly may purely stem from some parameters in the method. We also learned plenty of interesting particle physics along the way, and thank the organizers greatly for taking the time to design and implement this challenge.

 \FloatBarrier

\subsection[Latent Dirichlet Allocation]{Latent Dirichlet Allocation\footnote{Authors: B. M. Dillon, D. A. Faroughy, J. F. Kamenik, M. Szewc.  The implementation of LDA used here for the unsupervised jet-substructure algorithm is available at \url{http://github.com/barrydillon89/LDA-jet-substructure}. }}

\label{sec:lda}

\noindent Latent Dirichlet allocation (LDA) is a generative probabilistic model for discrete data first introduced to particle physics for unsupervised jet tagging and event classification in Refs.~\cite{Dillon:2019cqt, 1797846}. In general, a single collider event can be represented by a set of measurements $(o_1,o_2,\ldots)$. For example, the set of all particle four-momenta in the space $(p_\text{T},\eta,\phi)$, or any set of substructure observables extracted while declustering jets. The basic assumption of LDA is that individual events can be modelled by mixtures of a finite number of latent distributions, referred to as {\it themes} or {\it topics}. These themes are multinomial distributions over the binned space of observables where event measurements are generated from. Therefore, sampling a single measurement $o_i$ from a theme consists in drawing a bin from a discretized phase space containing the particular measurement. The simplest case is to assume two underlying themes, the two-theme LDA model. 
In this case the generative process for a single event goes as follows: 
(i) from a suitable prior distribution draw a random number $\omega$ between zero and one, 
(ii) select a theme by drawing from the binomial distribution with bias $\omega$, 
(iii) sample one measurement from the selected theme's multinomial space of observables. 
Repeat steps (ii-iii) until all measurements in the event are generated. 
Repeat the procedure above for each event in the event sample. 
The above setting can be generalized to more than two themes by replacing the two-theme mixing proportion  $\omega$ with a set of mixing proportions $(\omega_1,\ldots,\omega_T)$ living in a $(T-1)$-dimensional simplex\footnote{The simplex is the space of all $T$ dimensional vectors satisfying $0\le\omega_t\le1$ and $\sum_{t=0}^T\omega_t=1$.} where $T$ is the number of themes. The $\omega_t\,$'s reflect the preponderance of each theme within an individual event. The themes are then drawn from the multinomial distributions with biases $\omega_t$. In contrast to a mixture model\footnote{In a mixture model all measurements from an individual event are drawn from a single underlying distribution.}, in a {\it mixed membership model} like LDA different measurements within an event can originate from different themes, leading to a more flexible probabilistic model. LDA has a set of hyper-parameters $\alpha$ parametrizing the prior distribution from which the theme mixing proportions $\omega_t$ are to be drawn for each event (step (i) of the generative process described above). In particular, the prior is the Dirichlet distribution $\mathcal{D}(\alpha_0,\ldots,\alpha_T)$. Different choices of the concentration parameters $\alpha_t>0$ yield different shapes over the simplex. For the two-theme model, the Dirichlet reduces to a beta distribution $\mathcal{D}(\alpha_0,\alpha_1)$ over the unit interval.

Once the Dirichlet hyper-parameter $\alpha$ and the number of themes $T$ is fixed, we can train a LDA model by ``reversing" the generative process described above to infer from unlabelled collider data the latent parameters, namely the mixing proportions $\omega_t$ of each theme and the multinomial parameters $0\le \beta_{t,m}\le 1$ of the theme distributions $p(o|\beta)$, where $t$ labels the theme and $m$ labels the bins in the space of observables. To learn these parameters in this work we use the standard method of stochastic variational inference (SVI). Once these parameters are learned from the data, we can then use LDA to classify events in an unsupervised fashion. In the case of a two-theme LDA  model ($T=2$) we can conveniently use the likelihood ratio of the learned themes of an event $e=(o_1,\ldots o_N)$:
\[
L(o_1,\ldots, o_N|\alpha)=\prod_{i=1}^N\frac{p(o_{i}|\hat\beta_1(\alpha))}{p(o_{i}|\hat\beta_2(\alpha))}\,.
\] 
Here $\hat \beta_t$ are the estimators of the theme parameters extracted from SVI. 
Notice that the above expression is dependent on the Dirichlet hyper-parameter $\alpha$ leading to a landscape of classifiers.  In principle there are no hard criteria for choosing one set of hyper-parameters over the other. One way to guide the choice is by using the resulting model's {\it perplexity}, see Ref.~\cite{1797846} for details. After training LDA models for different points in the landscape, the LDA classifier with the lowest perplexity (corresponding to the LDA model that best fits the data) has been shown in examples to be correlated with truth-level performance measures like the AUC.

\subsubsection{Method}
\label{sec:method}

As shown in Refs.~\cite{Dillon:2019cqt, 1797846}, the two-theme LDA model can be used for anomaly detection in events with large radius jets. The jets are declustered, and at each splitting a set of substructure observables is extracted and binned. We refer to these binned measurements as $o_{j,i}$, with an added categorical variable that tags the jet to which the splitting belongs to. 
In the limit of exchangeable splittings, De Finetti's theorem allows us to derive, with the help of some additional assumptions, the latent substructure of such jets, characteristic of a mixed-membership model. In practice, exchangeability is a reasonable approximation since most of the interesting physical information contained in jet substructure is in the kinematical properties of the splittings, not in their ordering.

The choice of data representation and suitable binning are fundamental for LDA performance. Here we refer to data representation as both the kinematical information we use from each splitting as well as the kinematic cuts determining the splittings to be considered. As shown in Ref.~\cite{1797846}, the data representation and binning on one hand must allow for discrimination between signal and background, while at the same time produce co-occurrences of measurements within the same event. The former is obvious  considering the classification task at hand, while the latter is needed for the SVI procedure to be able to extract the latent distributions. This results in a trade-of of using relatively coarse binning in order to ensure co-occurrence of measurements without sacrificing too much discriminatory power. In a fully unsupervised setting, one does not know a priori which data representation is best for any given possible signal, and any data representation carries some assumptions on how the signal is imprinted in jet substructure. In this work we consider two fairly general bases of jet substructure observables, the so called mass-basis and the Lund-basis.
In the mass basis we only include splittings from subjets of mass above $30$ GeV.
In the Lund basis we only include splittings from subjets which lie in the primary Lund plane.
We emphasise that the resulting two data representations do not only differ in the observables included,
but also in the set of splittings kept for each jet due to the different declustering cuts. In our current setting, the number of considered jets in an event is fixed to two (of highest $p_\text{T}$).\footnote{When considering a variable number of jets, LDA tends to cluster together events based on jet multiplicity rather then jet substructure.}

After choosing a suitable data representation and binning, the procedure is as follows:  We first split the dataset into overlapping invariant mass bins. In each bin, we perform a hyper-parameter optimization using perplexity to find the best LDA model. Selecting the signal and background themes in the model by looking at the latent distributions of the themes over the vocabulary and the weight distributions of the events, we build a test statistic and define a threshold for data selection. Finally, we perform a bump hunt on the selected data invariant mass distribution.  In order to provide a background-only hypothesis, we consider the uncut invariant mass distribution as a background template and fix the total number of background events using the sideband regions. We can then produce a local p-value after also estimating the systematic errors due to possible classifier correlation with the invariant mass using the simulated background sample.

\subsubsection{Results on LHC Olympics}
\label{sec:results}

For Black Box 1 we assumed a di-jet resonance and consequently applied the LDA method to the two leading jets in each event using the mass-basis data representation. 
The invariant mass bin of 2.5-3.5 TeV yields themes shown in Fig.~\ref{fig:fig2}. We deem the signal theme to be the one with resonant substructre, uncharacteristic of QCD. 
\begin{figure}[h]
\begin{center}
\includegraphics[width=0.45\textwidth]{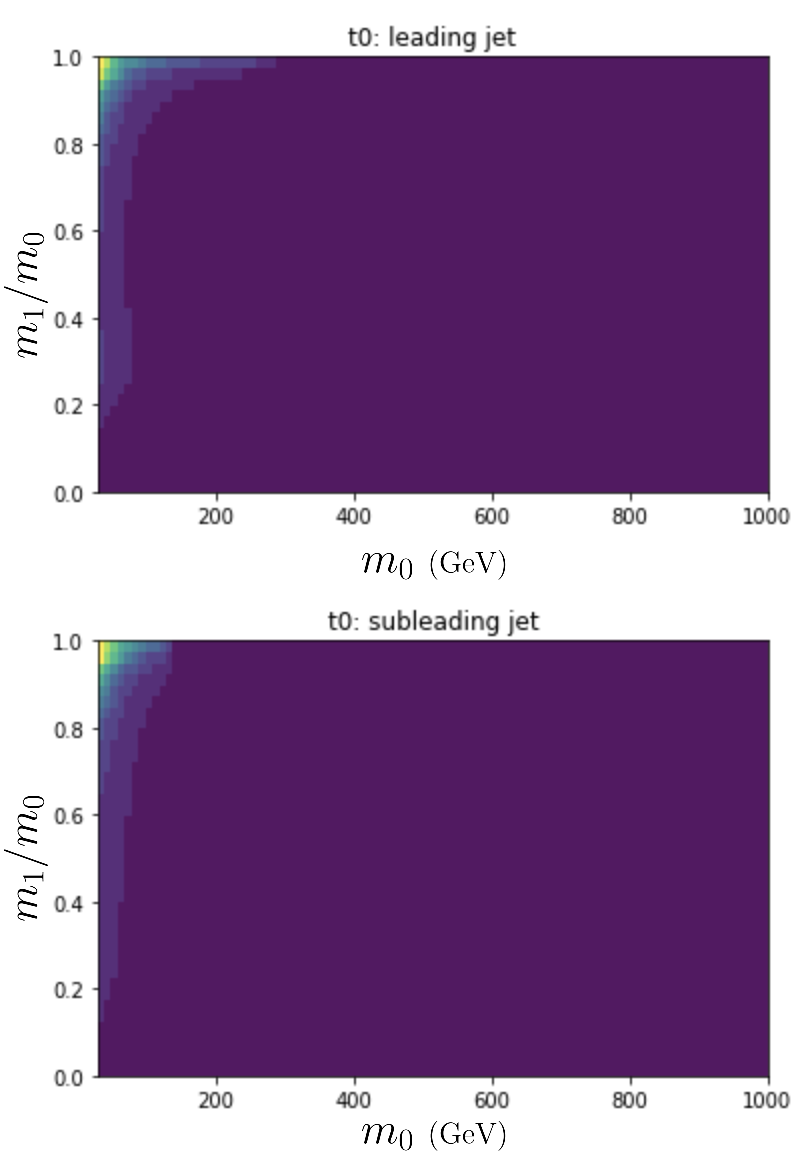}
\includegraphics[width=0.45\textwidth]{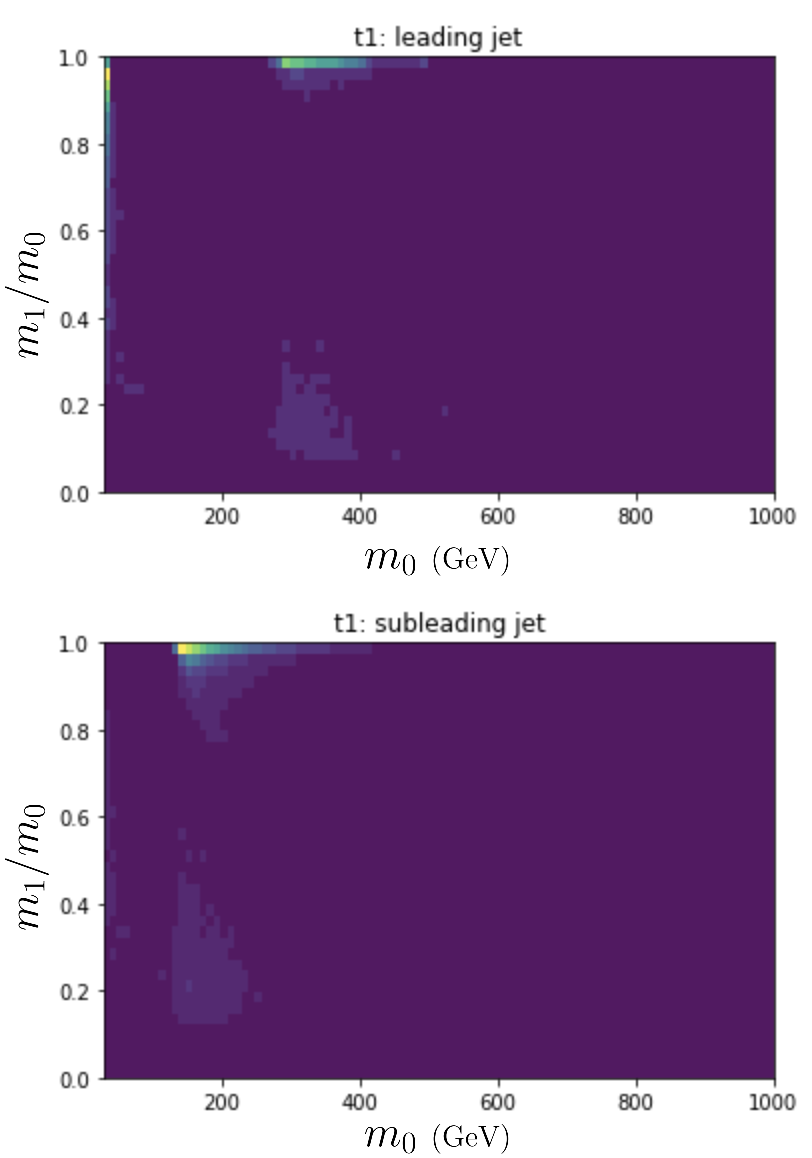}
\caption{ Best inferred latent distributions of the two themes (left and right column) for Black Box 1 with the LDA method. Shown is the $m_0, m_1/m_0$ plane of the mass-basis for the heavier (top row) and the lighter (bottom row) of the two jets.  \label{fig:fig2}}
\end{center}
\end{figure}
We perform a bump hunt with this model on Black Box 1 and on the simulated background sample. We show the invariant mass distribution after cutting using this LDA and the resulting BumpHunter excess in Fig.~\ref{fig:fig3}. In both cases we also show the background estimation used to compute the p-value. The reported significances are 1.8$\sigma$ and 3.8$\sigma$ for the background sample and the Black Box 1 sample respectively.
\begin{figure}[h]
\begin{center}
\includegraphics[width=0.35\textwidth]{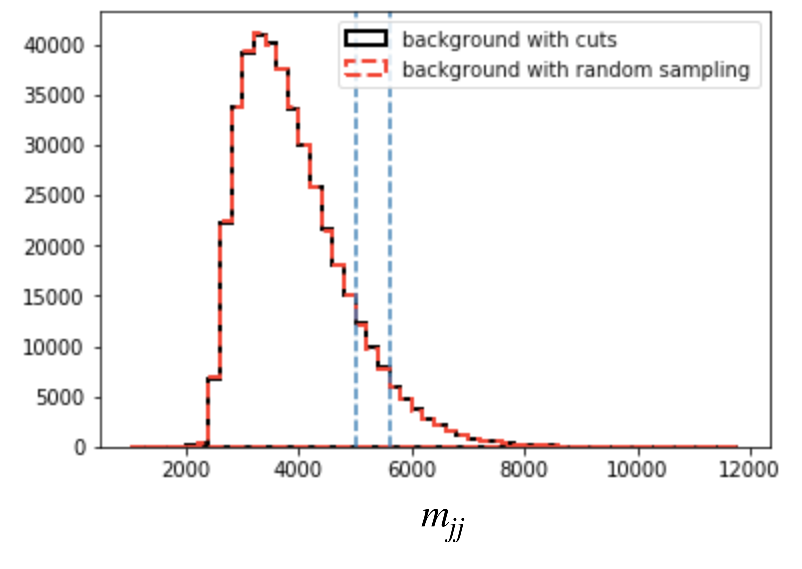}
\includegraphics[width=0.35\textwidth]{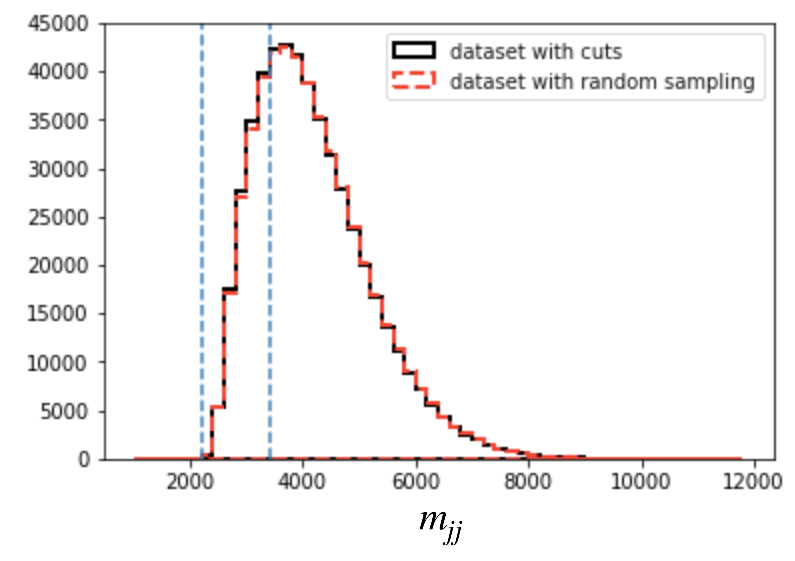}\\
\includegraphics[width=0.35\textwidth]{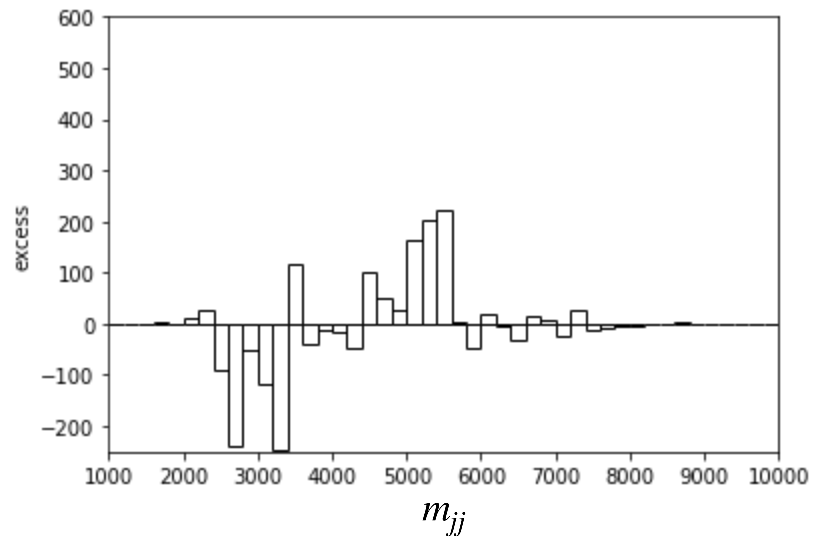}
\includegraphics[width=0.35\textwidth]{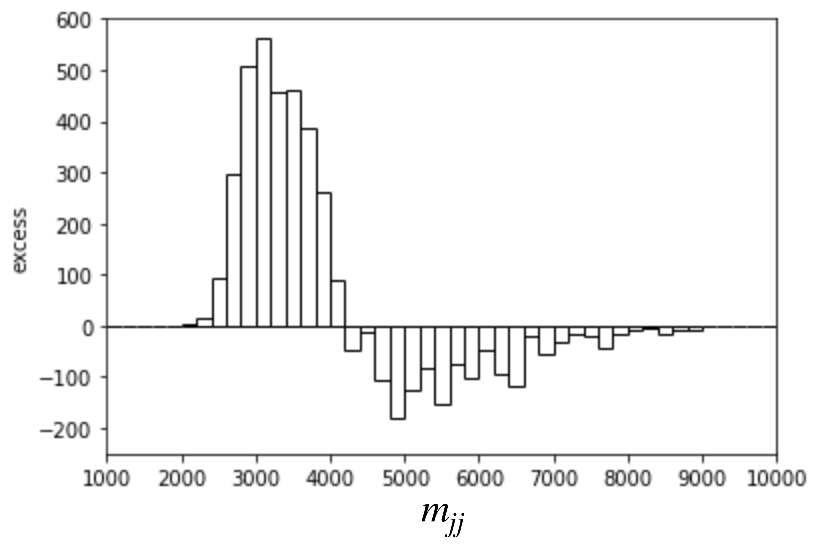}
\caption{ Invariant mass event distribution of the simulated background (left) and Black Box 1 (right) after performing an LDA-based cut along with the background estimation using the uncut invariant mass distribution. Bottom row displays the corresponding excess found by BumpHunter.
 \label{fig:fig3}}
\end{center}
\end{figure}
When comparing our estimates to the unveiled results, the LDA inferred di-jet resonance mass is not incompatible with the actual value of 3.8 TeV. 
However, the two decay products of this resonance have masses which are significantly above LDA estimates (732 and 378 GeV). 
The discrepancy is possibly due to an unfortunate choice of binning, since having bins narrow in $m_0$  may have reduced the strength of the co-occurrences, which in turn may have caused the signal features to be washed out by sculpting effects in the jet mass distribution coming from the $p_\text{T}$ cut.
On the other hand, we could not find compelling new physics candidates in neither Black Box 2, where no signal was present, nor Black Box 3.

\subsubsection{Lessons Learned}
\label{sec:lessons}

The main lesson we take from the LHCO challenges is that a realistic LDA implementation should consider several different data representations and binnings. As we limited ourselves to di-jet jet-substructure observables we missed the characteristics of a rare signal which does not produce a rich jet substructure in the two leading jets.  In the future, the search pipeline should allow to consider a larger number of jets but also include data representations which are not focused exclusively on jet substructure, e.g. by considering global jet or event variables.

 \FloatBarrier

\subsection[Particle Graph Autoencoders]{Particle Graph Autoencoders\footnote{Authors: Steven Tsan, Javier Duarte, Jean-Roch Vlimant, Maurizio Pierini.  All code is publicly available at \url{https://github.com/stsan9/AnomalyDetection4Jets}.}}
\label{sec:pga}

\subsubsection{Method}
\label{sec:method}

We propose particle graph autoencoders (PGAEs) based on graph neural networks~\cite{1808887} for unsupervised detection of new physics in multijet final states at the LHC. 
By embedding particle jet showers as a graph, GNNs are able to exploit particle-particle relationships to efficiently encode and reconstruct particle-level information within jets.
We posit that this can improve the capacity of autoencoders to learn a compressed representation of a jet and consequently help identify anomalous beyond-the-standard-model (BSM) multijet signal events from LHC data.

\begin{figure}[htpb]
    \centering
    \includegraphics[width=\textwidth]{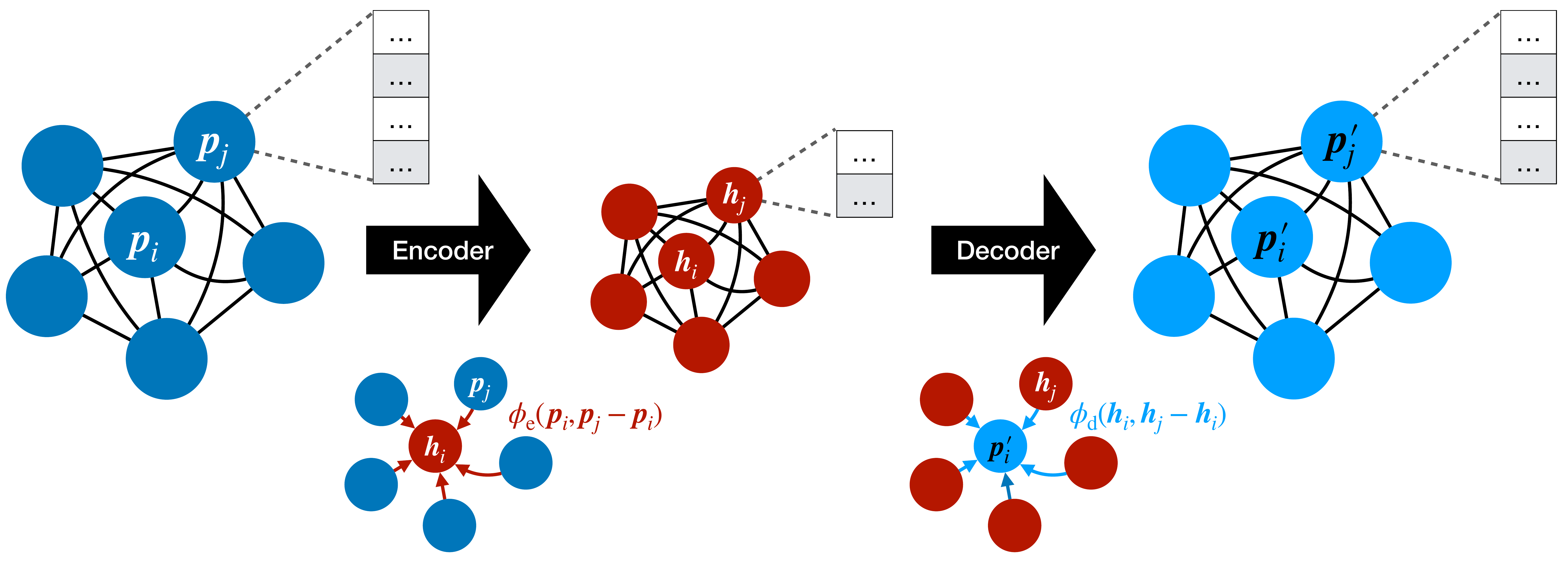}
    \caption{Schematic of the particle graph autoencoder model proposed. 
    Each input jet is represented as a graph in which each particle of the jet is a node, and each node has an edge connecting it to every other particle in the jet.
    After an edge convolution layer~\cite{DGCNN}, each particle is encoded in a reduced two-dimensional latent space, before another edge convolution layer reconstructs each particle's four-momentum $(E, p_x, p_y, p_z)$.
}
    \label{fig:gae}
\end{figure}

In our PGAE model, we represent each input jet as a graph in which each particle of the jet is a node, and each node has an edge connecting it to every other particle in the jet (i.e. a fully-connected particle graph).
When encoding and decoding, the graph structure of the data remains the same, but the nodes' features, initially the particle's four-momentum $(E, p_x, p_y, p_z)$, have their dimensionality reduced during the encoding phase.
We note the model can be expanded to consider additional particle-level information, such as particle type, electromagnetic charge, and pileup probability weight~\cite{Bertolini:2014bba}.
For the encoder and decoder, we use the edge convolution layer from Ref.~\cite{DGCNN}, which performs message passing along the edges and aggregation of messages at the nodes of the graphs.
A schematic of this is shown in Fig.~\ref{fig:gae}.

The PGAE model is constructed using the PyTorch Geometric library~\cite{PyTorchGeometric}.
In this model, the input node features are first processed by a batch normalization layer~\cite{batchnorm}.
The encoder is an edge convolution layer~\cite{DGCNN}, built from a fully connected neural network $\phi_\mathrm{e}$ with layers of sizes $(8, 32, 32, 2)$ and rectified linear activation unit (ReLU) activation functions~\cite{relu}.
The first layer of dimension $8$ represents the input, which is given by $(\boldsymbol{p}_i, \boldsymbol{p}_j-\boldsymbol{p}_i)$, where $\boldsymbol{p}_i$ ($\boldsymbol{p}_j$) is the four-momentum for particle $i$ ($j$) and $i\neq j$.
The final layer produces a two-dimensional message vector from each pair of distinct particles.
These two-dimensional message vectors are aggregated (using a mean function) for each receiving particle
\begin{equation}
\boldsymbol{h}_i = \frac{1}{|\mathcal N(i)|}\sum_{j\in \mathcal N(i)} \phi_\mathrm{e}(\boldsymbol{p}_i, \boldsymbol{p}_j-\boldsymbol{p}_i)\,,
\end{equation}
where $\mathcal N(i)$ is the neighborhood of particles connected to the $i$th particle, which corresponds to all other particles in this case.
This summed message $\vec h_i$ is the bottleneck or encoded representation for the $i$th particle.
The decoder is also an edge convolution layer, containing a network $\phi_\mathrm{d}$ with layers of sizes $(4, 32, 32, 4)$ and ReLU activation functions, except for the final layer, which reconstructs each particle's momentum.
We note that the architecture itself is insensitive to the ordering of the input particles. 
PyTorch Geometric supports variable-size input graphs so there is no need for zero-padding.

The model is trained on the QCD background dataset with two different loss functions. 
The first is the mean squared error (MSE) between the input and output particles. 
This choice of loss function violates the permutation invariance of the algorithm because the particles must be reconstructed in the same order as they are input to achieve a small value of the loss function.
For this reason, we also investigate a second, alternative loss function, the Chamfer distance loss, whose value does not depend on either the order of the input particles or the reconstructed particles~\cite{10.5555/1622943.1622971,Fan_2017_CVPR,Zhang2020FSPool}.
Given two input sets of particles $\mathcal{M}$ and $\mathcal{N}$, expressed in terms of the momentum vectors $\boldsymbol{p}_i$ and $\boldsymbol{p}_j$ (with $i \in \mathcal{M}$ and $j \in \mathcal{N}$), the loss function is defined as
\begin{equation}
D^\mathrm{NN}(\mathcal{M}, \mathcal{N}) =  \frac{1}{|\mathcal{M}|}\sum_{i \in \mathcal{M}} \min_{j \in \mathcal{N}} \left(||\boldsymbol{p}_i - \boldsymbol{p}_j||\right)^2 + \frac{1}{|\mathcal{N}|}\sum_{j \in \mathcal{N}} \min_{i \in \mathcal{M}} \left(||\boldsymbol{p}_i - \boldsymbol{p}_j||\right)^2\,,
\label{eq:L_NND}
\end{equation}
where $||\boldsymbol{p}_i-\boldsymbol{p}_j||$ is the Euclidean distance.

\begin{figure}[htpb]
\centering
\includegraphics[width=0.24\textwidth]{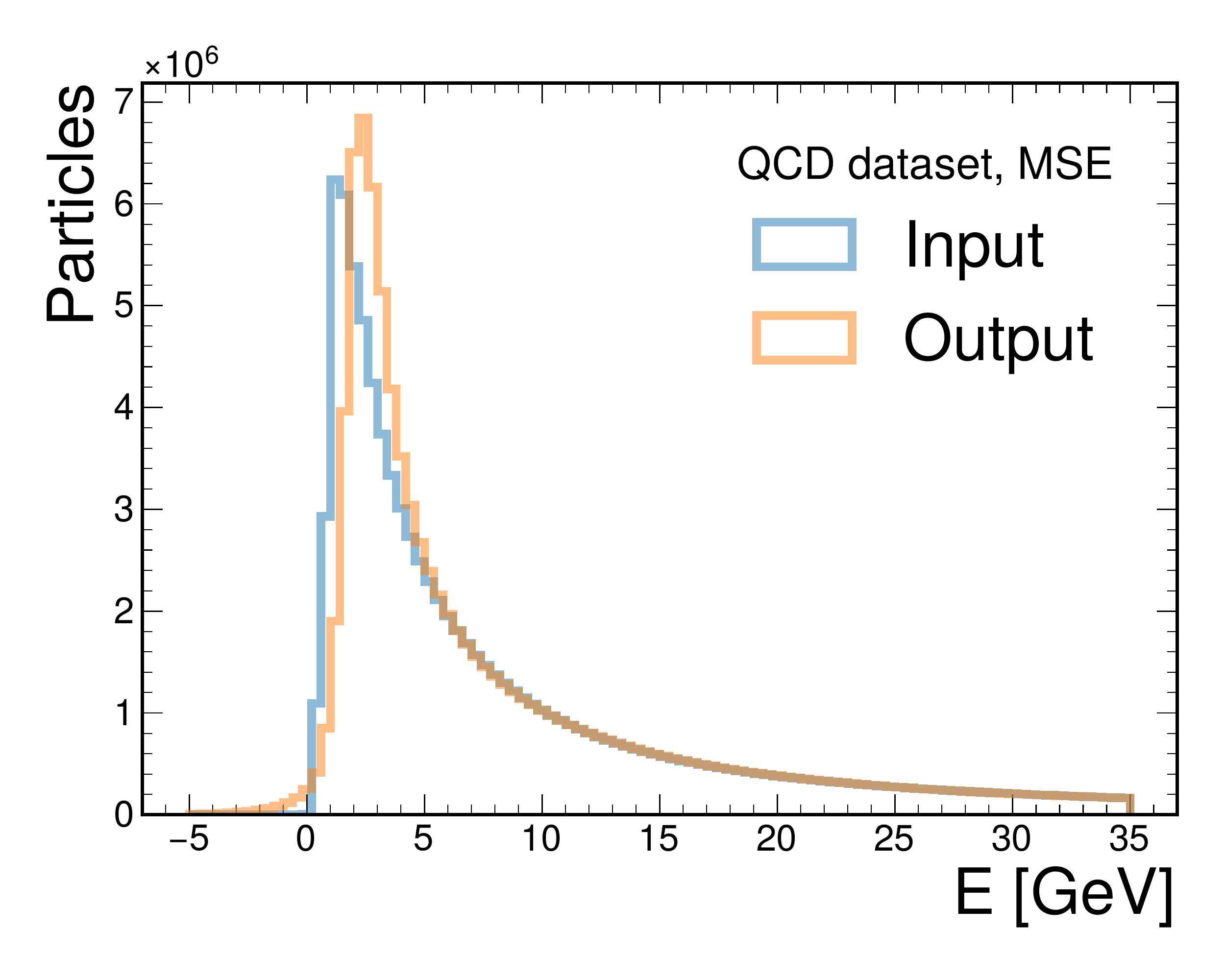}
\includegraphics[width=0.24\textwidth]{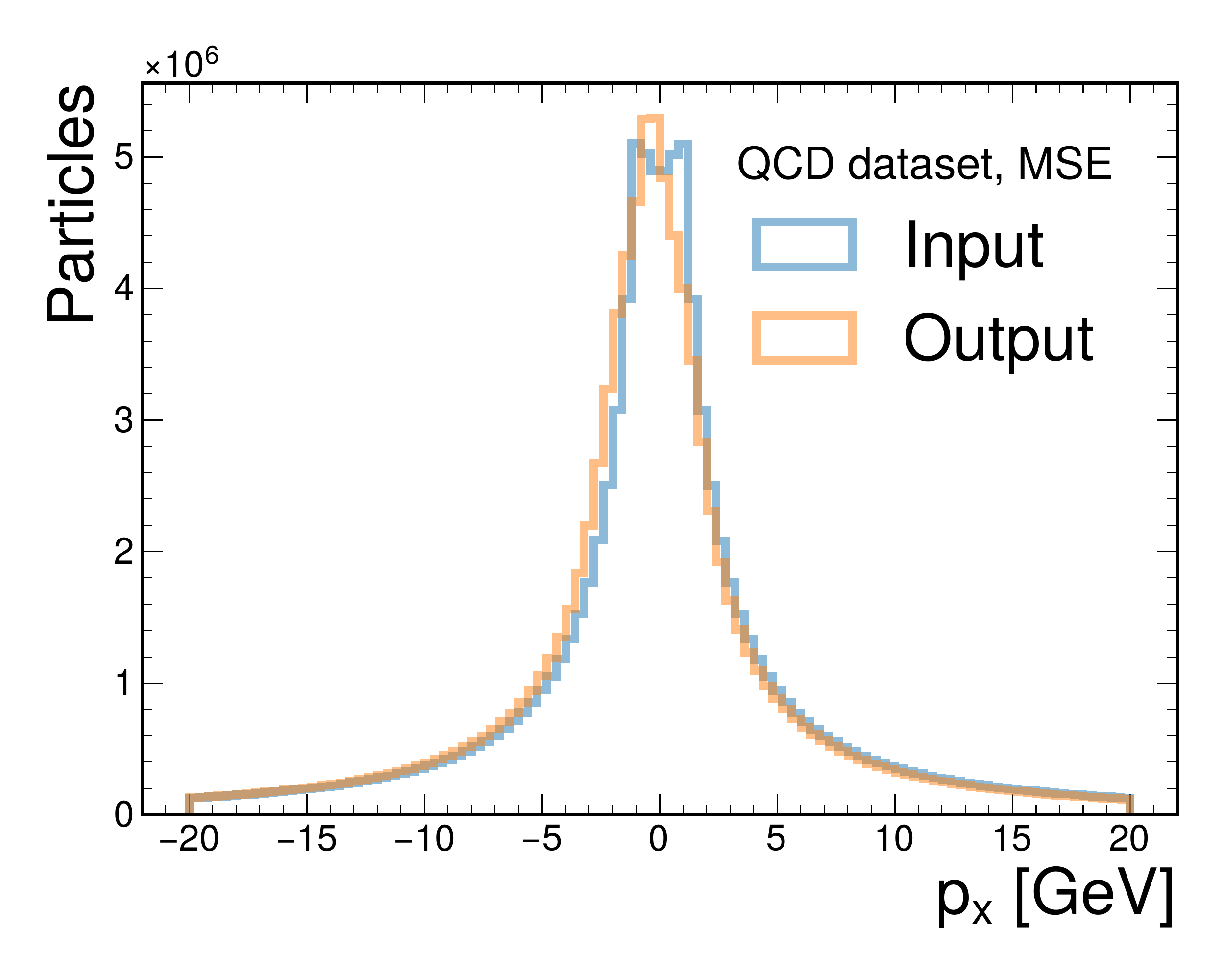}
\includegraphics[width=0.24\textwidth]{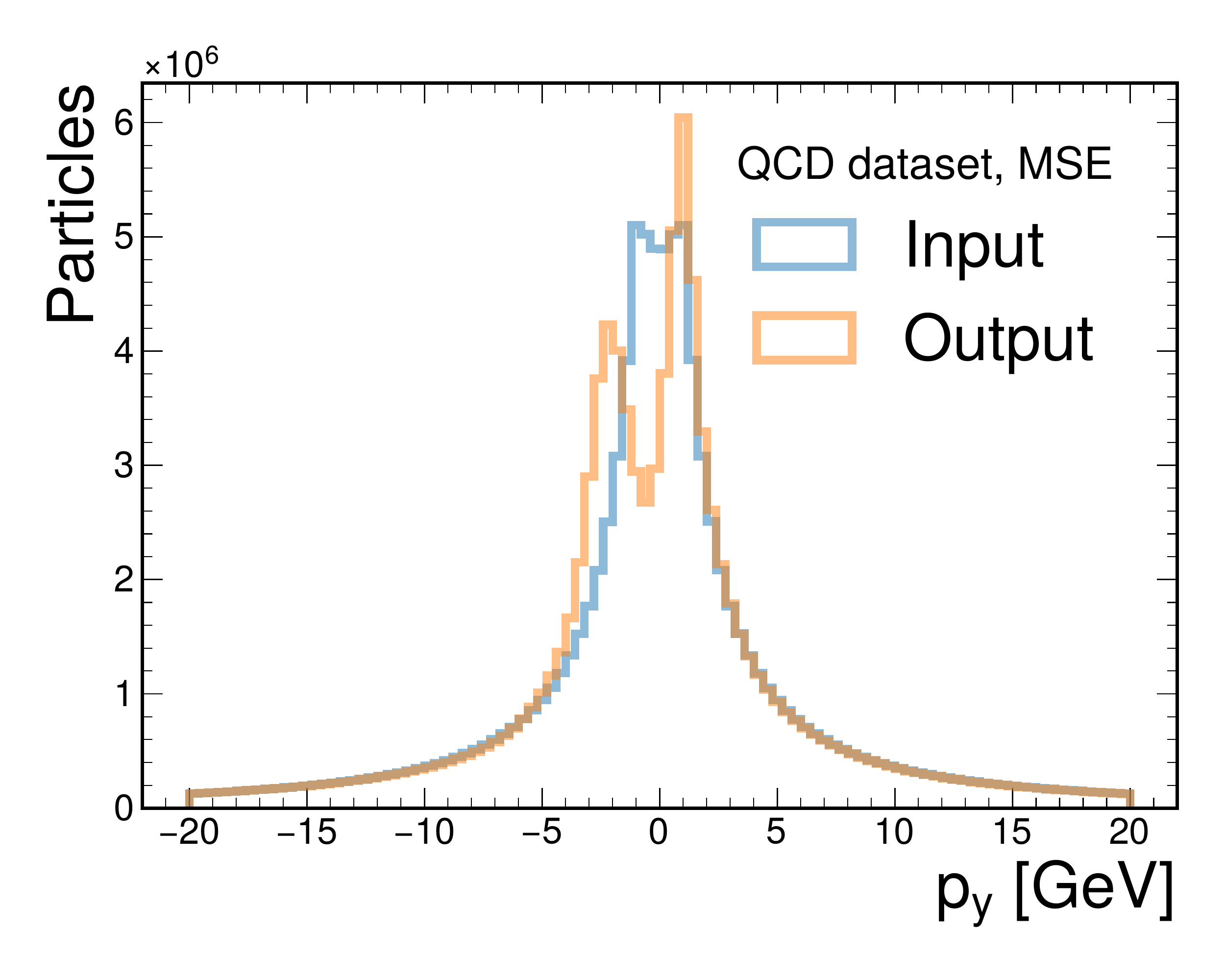}
\includegraphics[width=0.24\textwidth]{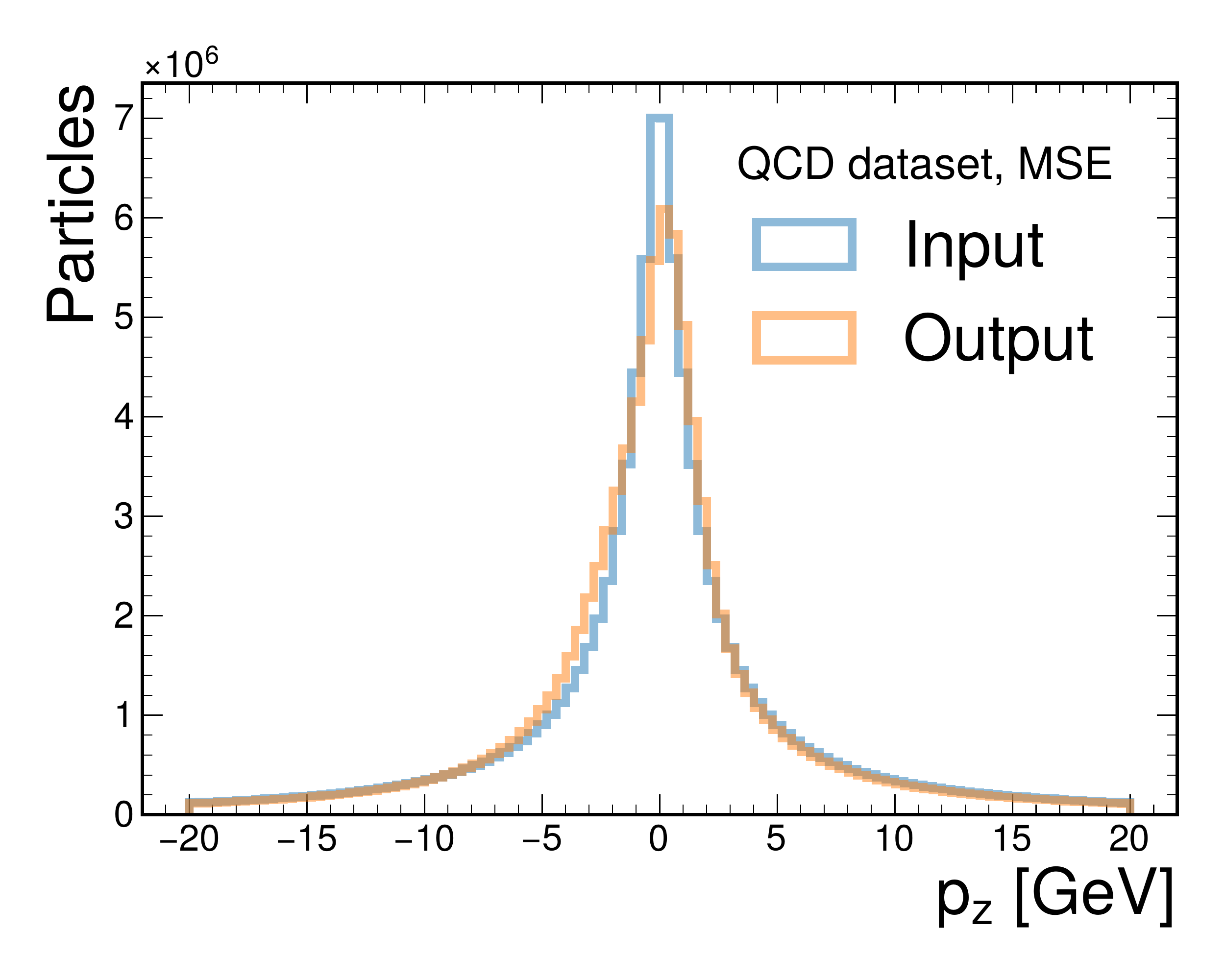}\\
\includegraphics[width=0.24\textwidth]{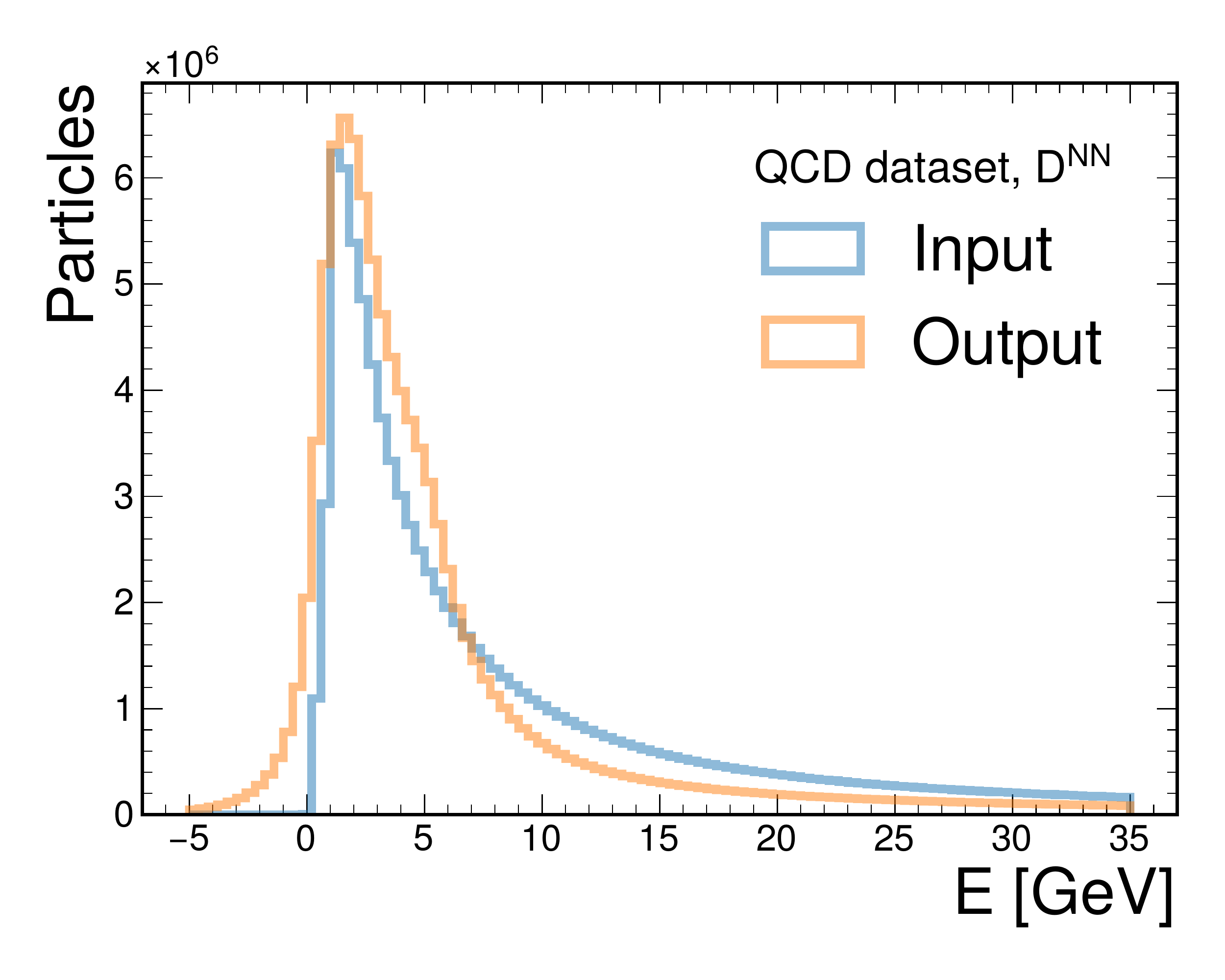}
\includegraphics[width=0.24\textwidth]{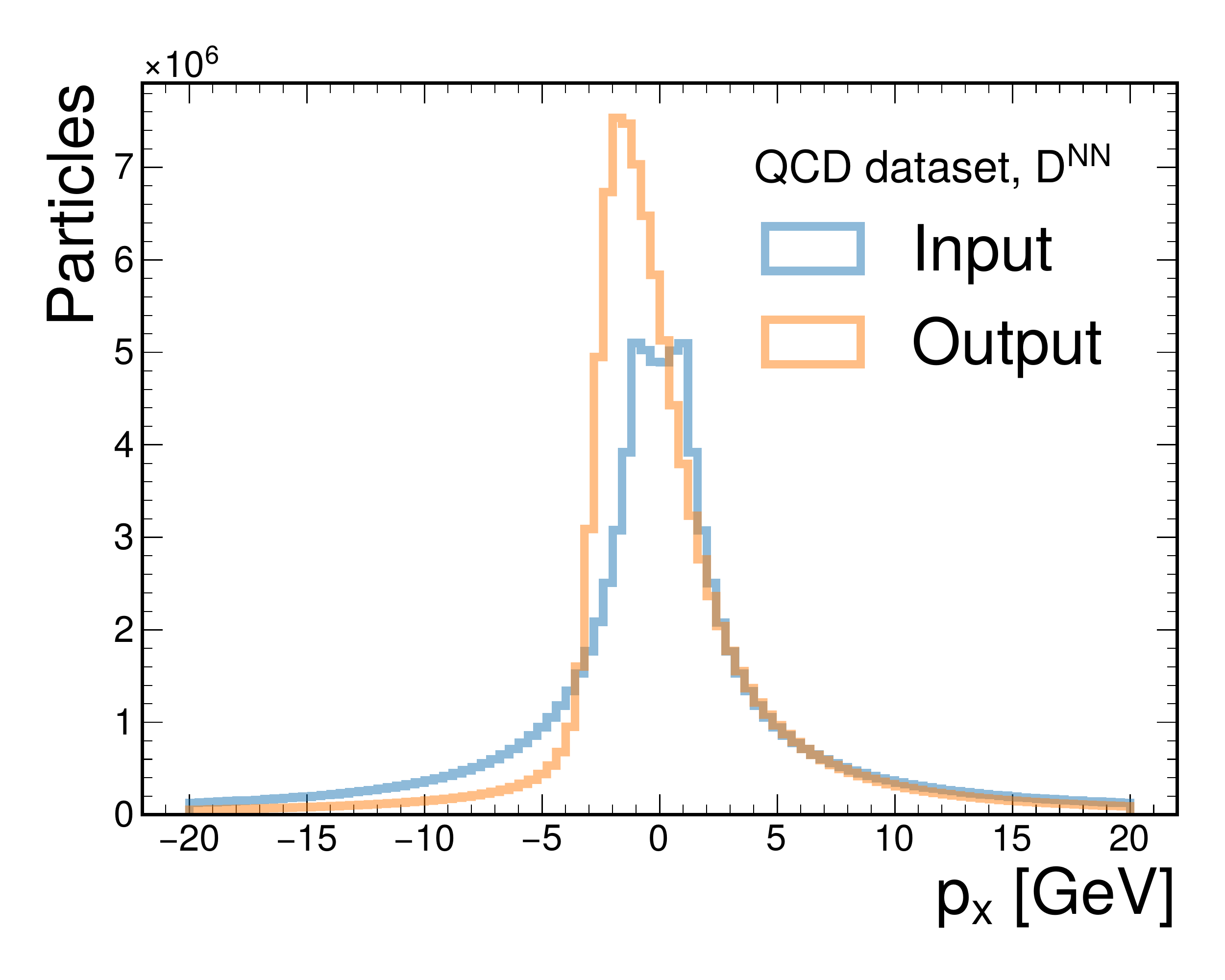}
\includegraphics[width=0.24\textwidth]{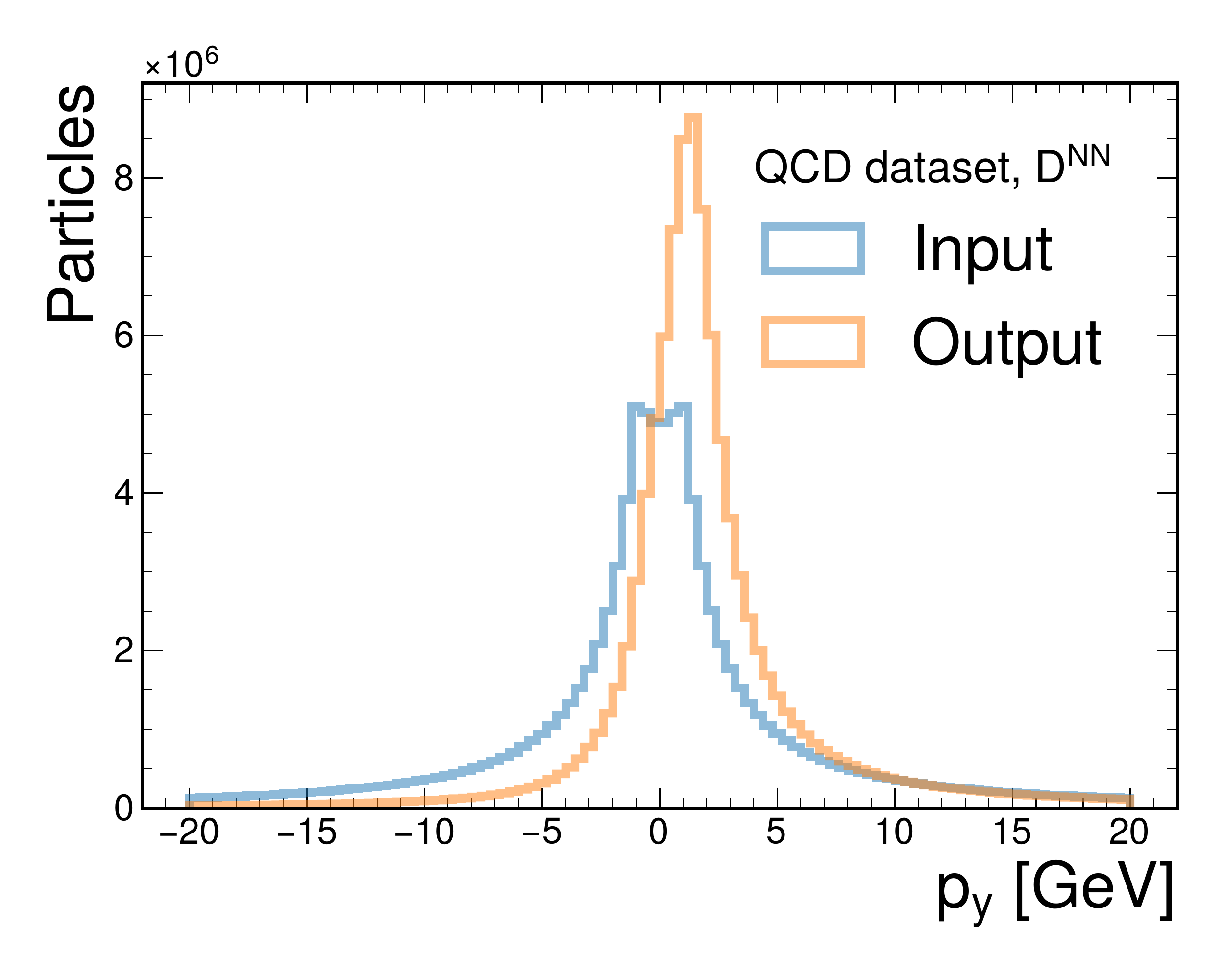}
\includegraphics[width=0.24\textwidth]{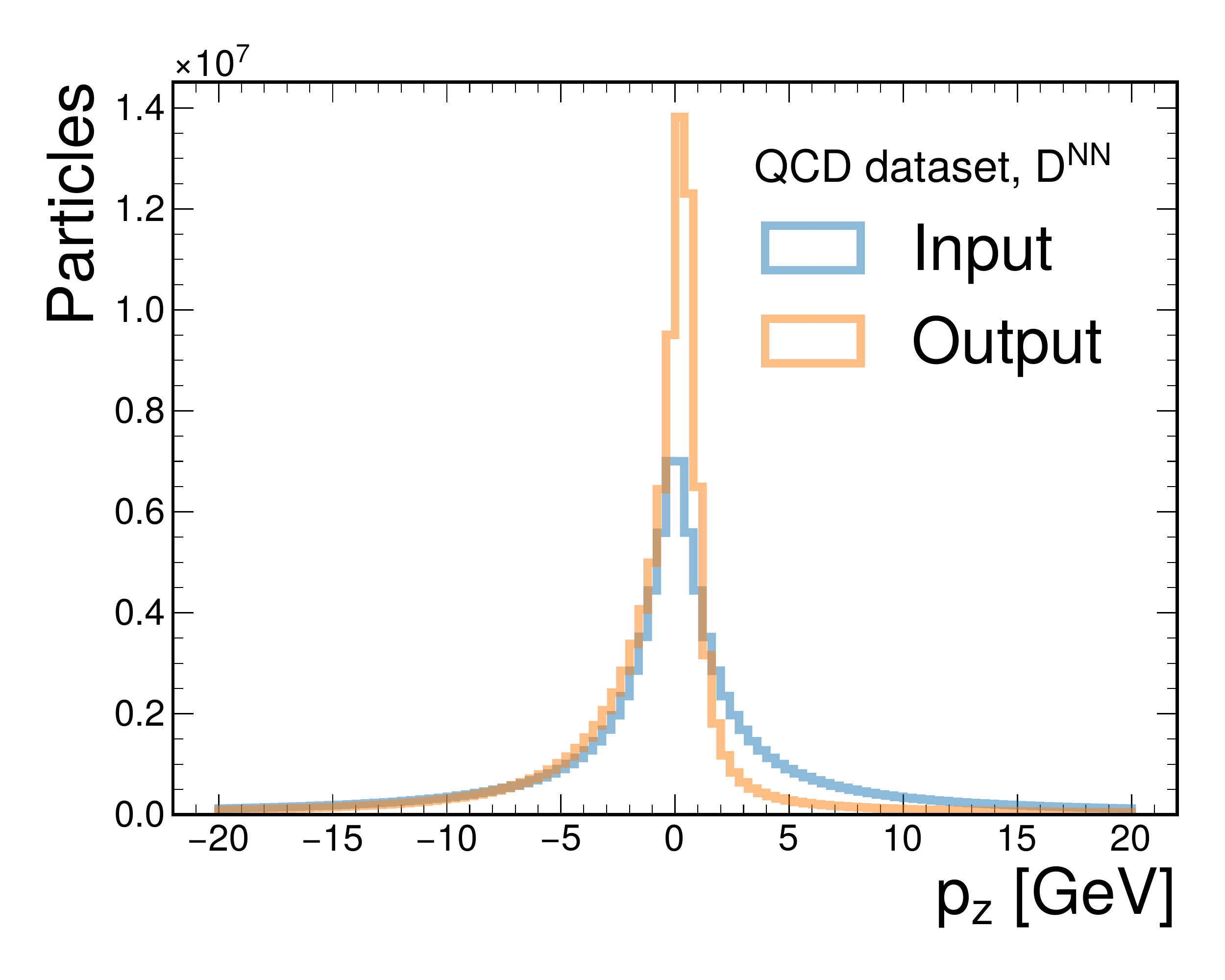}
\caption{Comparison of input and reconstructed features $E$ (far left), $p_x$ (center left), $p_y$ (center right), and $p_z$ (far right) for the models trained with MSE (top) and Chamfer (bottom) loss functions on the QCD testing dataset.
}
\label{fig:reconstruction}
\end{figure}

\subsubsection{Results on LHC Olympics}

\begin{figure}
    \centering
    \includegraphics[width=0.45\textwidth]{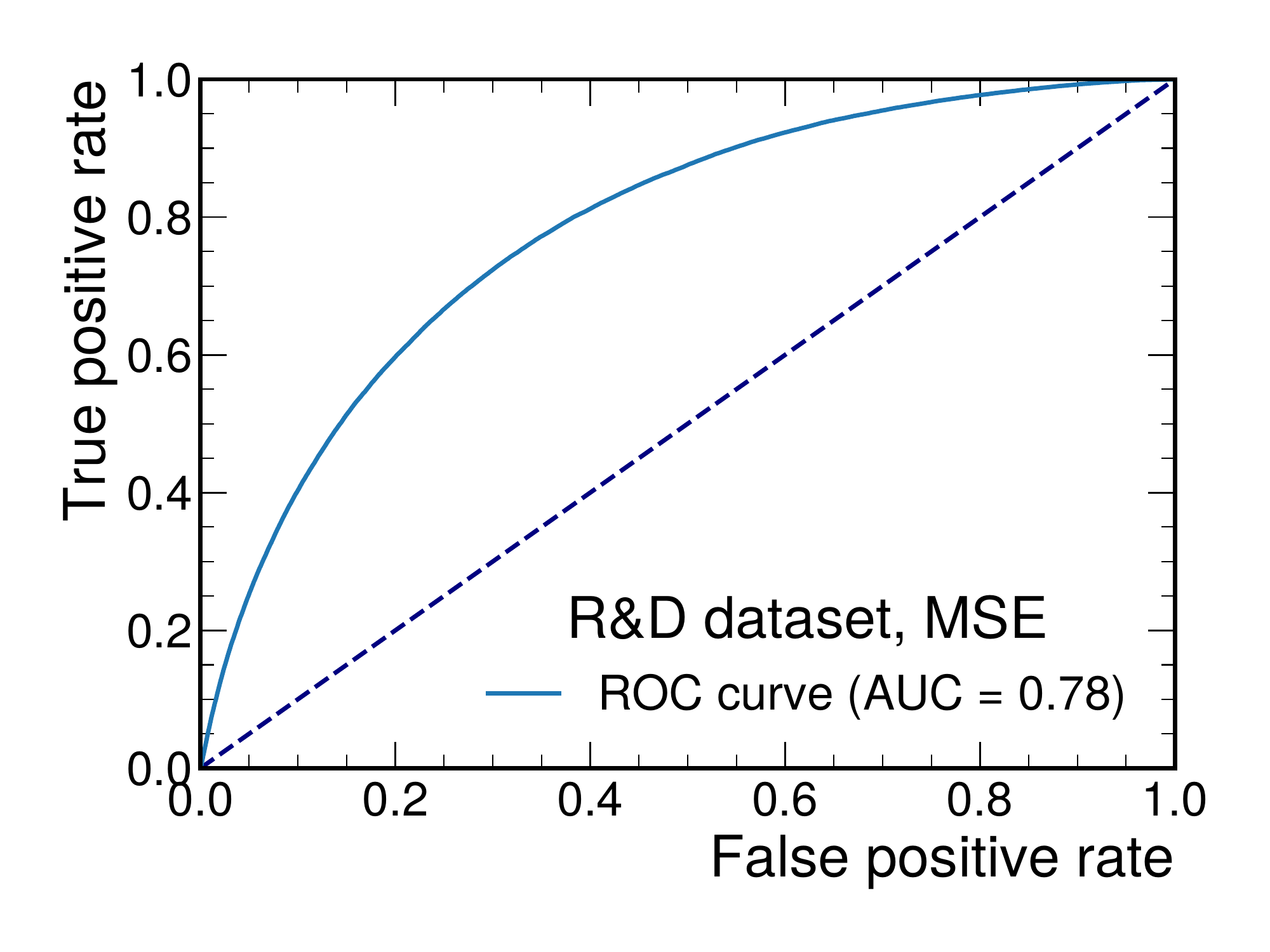}
    \includegraphics[width=0.45\textwidth]{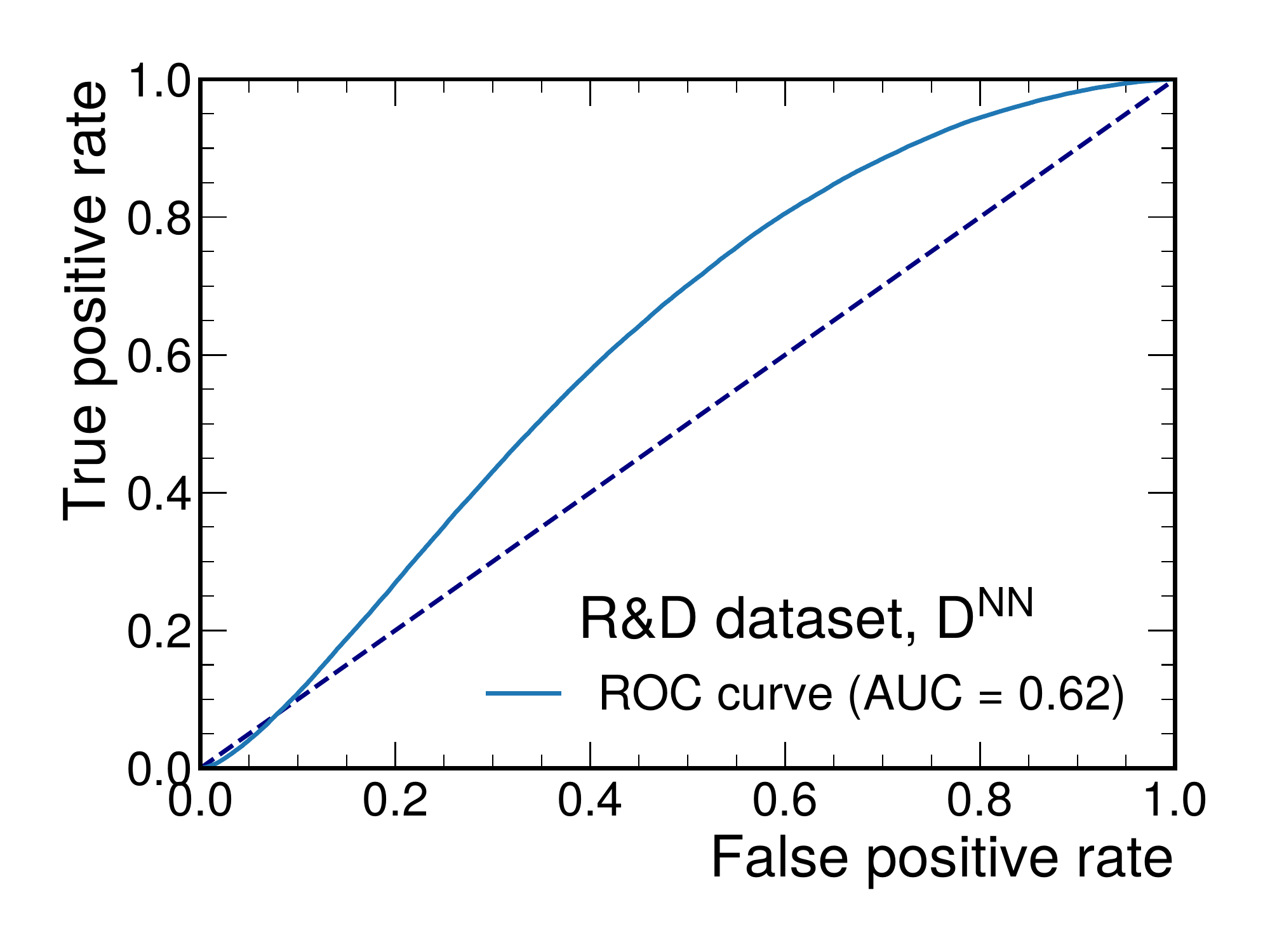}
    \caption{ROC curves for the PGAE trained with the MSE (left) and Chamfer loss (right).}
    \label{fig:rocs}
\end{figure}

First, we studied our algorithm on the R\&D dataset.
As the truth information is provided, we can create a receiver operating characteristic (ROC) curve to determine the effectiveness of the PGAE to identify a signal ($\PWpr\to\PX\PY$, $\PX\to\Pq\Pq$, and $\PY\to\Pq\Pq$ with $m_{\PWpr} = 3.5$~TeV, $m_\PX = 500$~GeV, and $m_\PY= 100$~GeV) that it did not observe during training.
The ROC curves for both the MSE and Chamfer loss functions are shown in Fig.~\ref{fig:rocs}.
Although the MSE loss is not permutation invariant, we find it provides better discrimination for a new unseen signal.

To evaluate our model's performance for anomaly detection, we perform a resonance search (or ``bump hunt'') in the dijet invariant mass $m_\mathrm{jj}$, computed from the two jets with highest $\pt$ in the event. 
We perform this dijet search in black box (BB) 1, which contains a resonant dijet signal at $m_\mathrm{jj}\sim 3.8$~TeV, and BB 2, which contains no signal.
We require both of the jets to be ``outliers,'' which we define as jets with a reconstruction loss exceeding a threshold corresponding to the 90\% quantile of the loss distribution for the leading two jets in the corresponding evaluation dataset.
We note that because our algorithm is jet-focused, it is straightforward to generalize this search to multijet events.

\begin{figure}[htpb]
    \centering
    \includegraphics[width=0.9\textwidth]{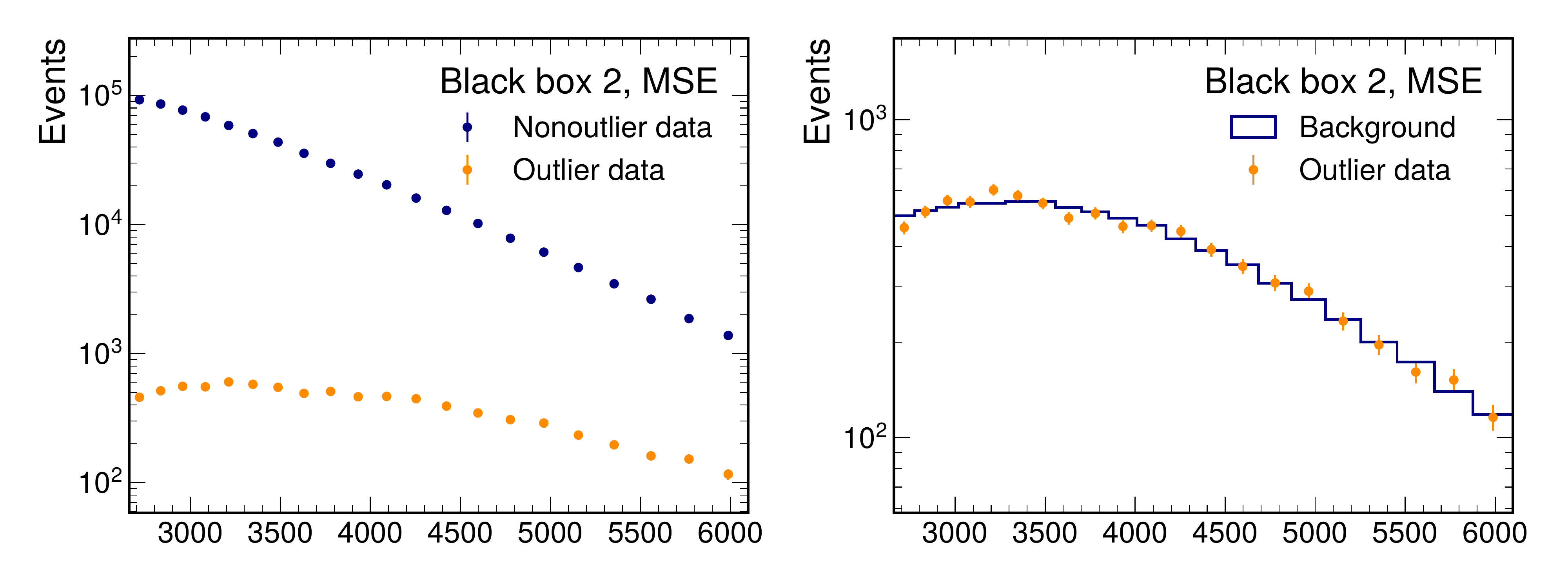}\\
    \includegraphics[width=0.9\textwidth]{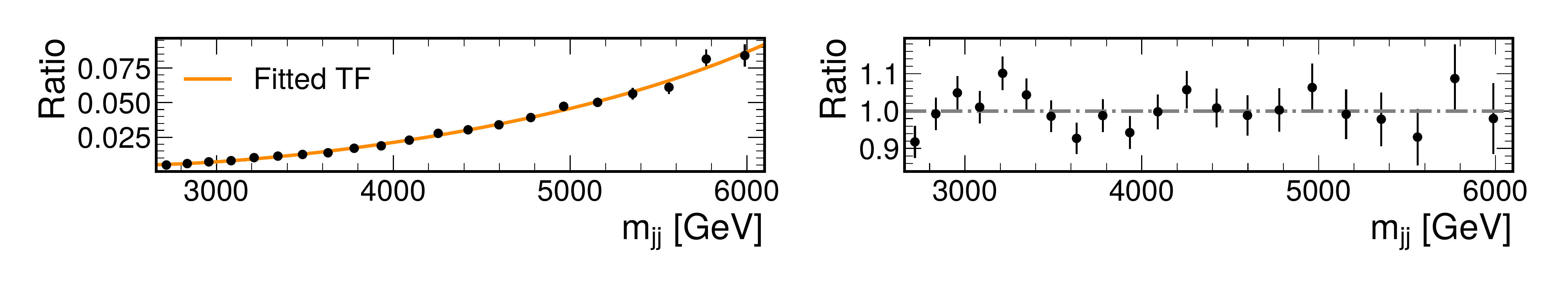}
    \caption{Illustration of the simplified background estimation procedure in BB 2 for the GAE trained with MSE loss. 
    A comparison between the nonoutlier and outlier jet mass distribution is shown (upper left). 
    The ratio of the two distributions is fit with a fourth-order polynomial to derive a transfer factor (lower left). 
    The corresponding postfit prediction is also shown (upper right). 
    The postfit ratio is randomly scattered around one as expected for BB 2, which contains no signal.}
    \label{fig:fit}
\end{figure}

For the background prediction in the signal-enriched outlier region, we perform a simplified analysis using the shape of the data in the background-enriched nonoutlier region.
Specifically, we fit the ratio of the nonoutlier-to-outlier dijet mass distribution with a fourth-order polynomial to derive a transfer factor (TF).
We take nonoutlier data distribution weighted by the TF as an estimate of the expected background in the outlier region. 
We do not consider systematic uncertainties associated to the TF although these could be taken into account in a more complete analysis in the future.
The procedure is illustrated in Fig.~\ref{fig:fit} for BB 2.

\begin{figure}[htpb]
    \centering
        \includegraphics[width=0.45\textwidth]{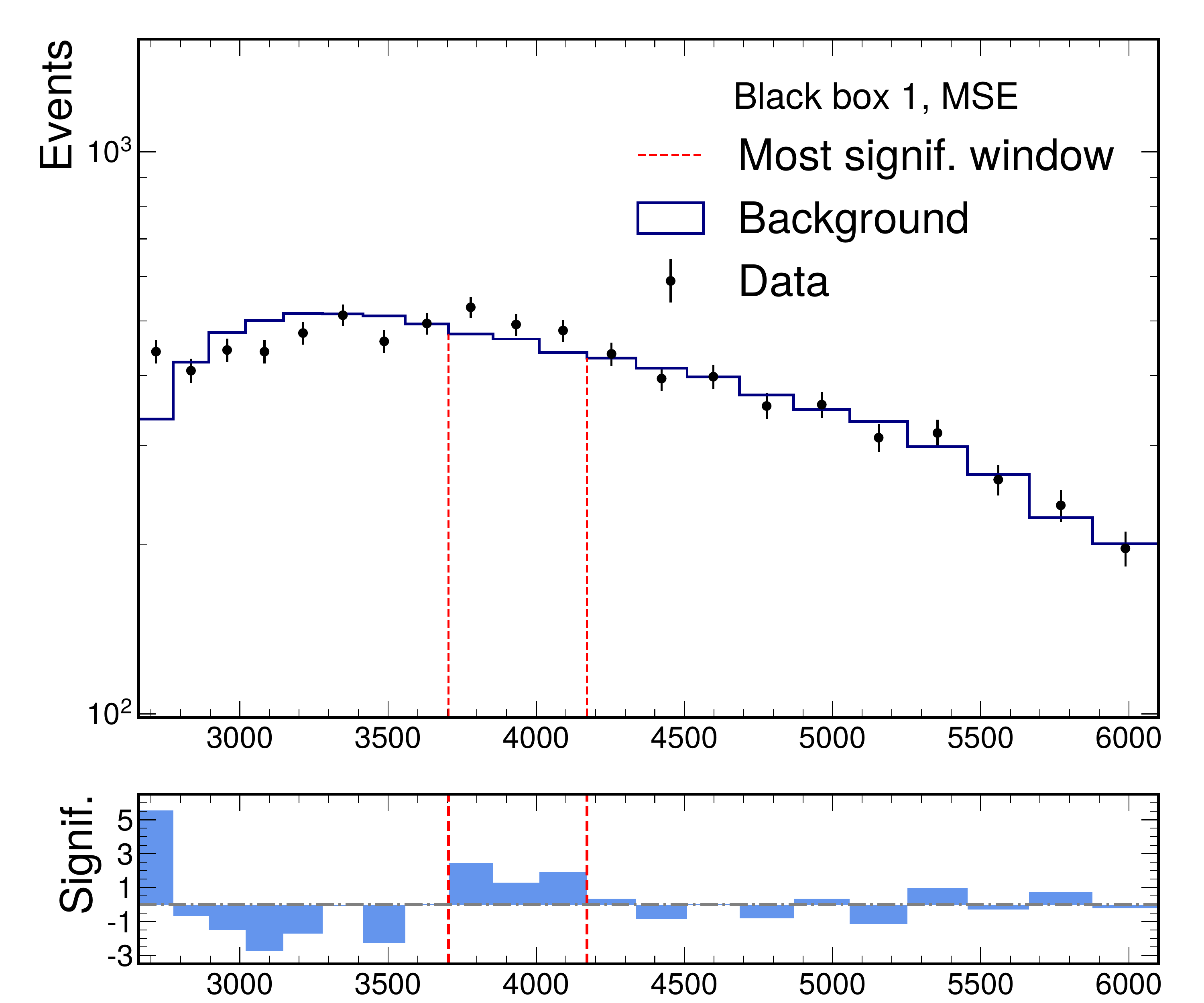}
        \includegraphics[width=0.45\textwidth]{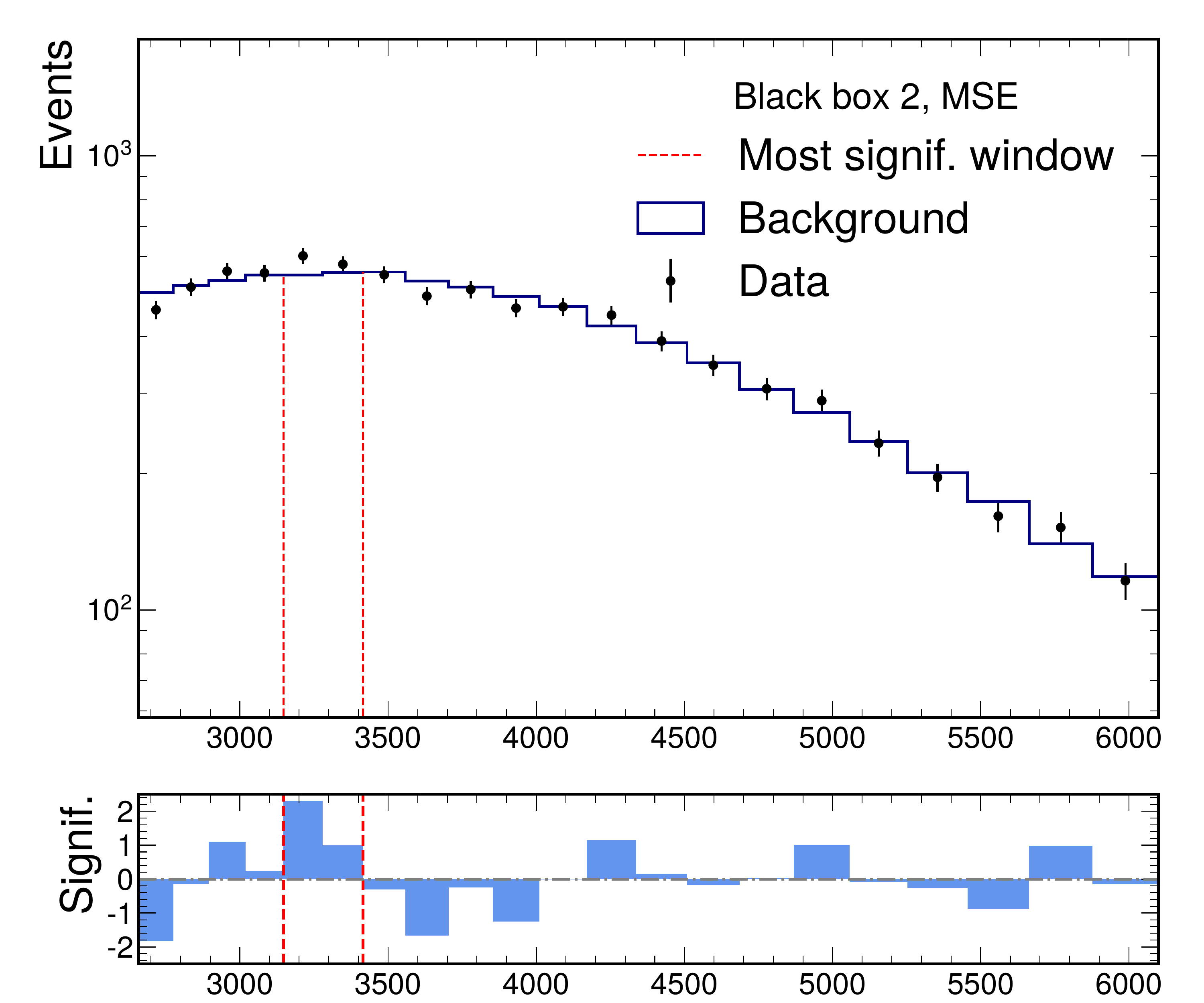}
        \includegraphics[width=0.45\textwidth]{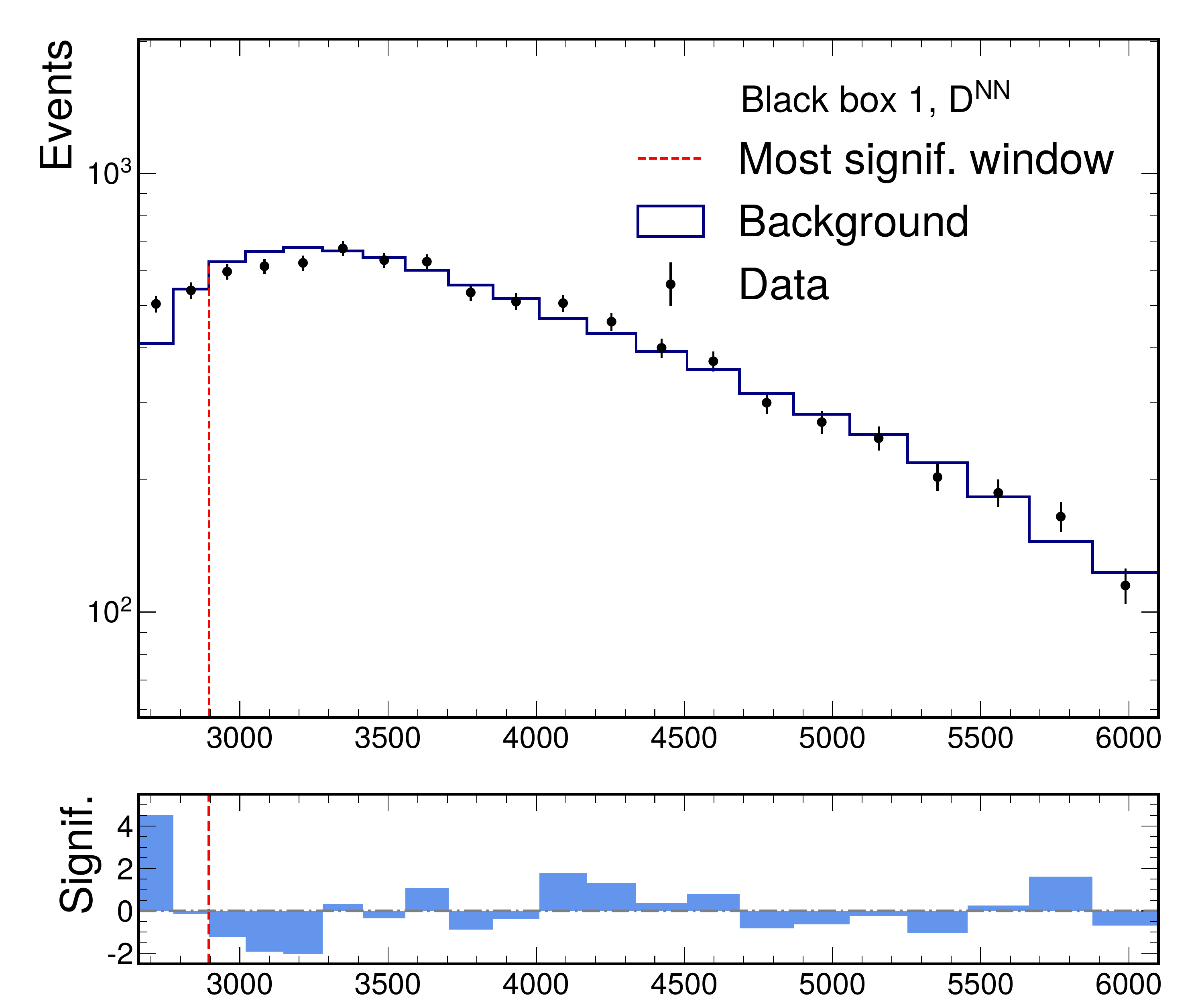}
        \includegraphics[width=0.45\textwidth]{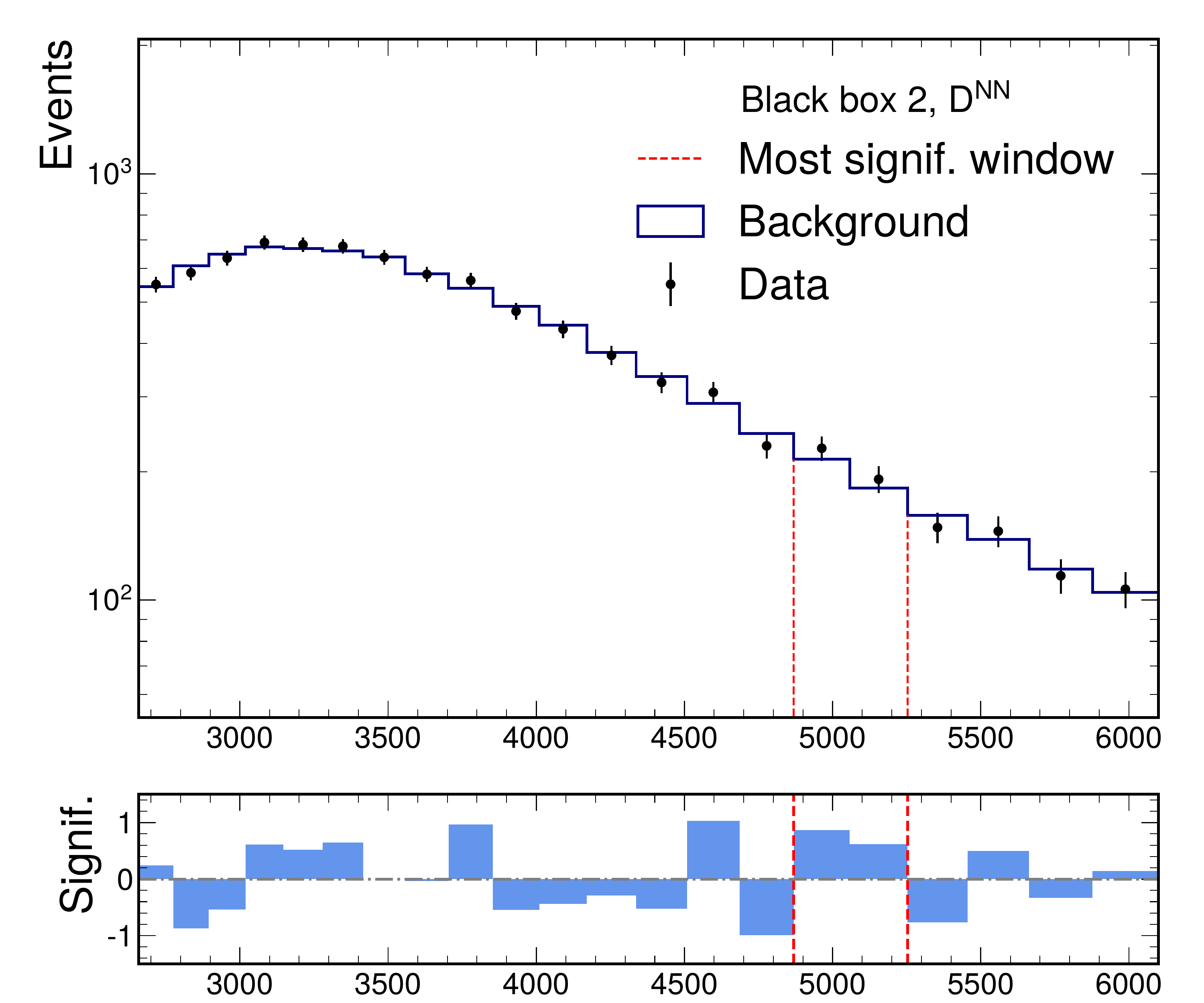}
\caption{BB 1, MSE, $2.1\,\sigma$ at $3.9$~TeV, BB 2, MSE, $0.8\,\sigma$ at $3.3$~TeV, BB 1, Chamfer, $1.5\,\sigma$ at $2.8$~TeV, BB 2, Chamfer, $-1.4\,\sigma$ at $5.1$~TeV.  Bump hunt in the dijet invariant mass in BB 1 (left) and 2 (right) using MSE (top) and Chamfer (bottom) as the loss functions.
Outlier jets have a reconstruction loss in the top 10\% with respect to the corresponding BB. 
Outlier events are required to have both jets be outliers.
BB 1 has an anomalous large-radius dijet signal $\PZpr \to \PX\PY \to (\Pq\Pq)(\Pq\Pq)$ injected at $m_\PZpr=3823$~GeV (with $m_\PX = 732$~GeV and $m_\PY= 378$~GeV), while BB 2 has no injected anomalies.
}
\label{fig:gnnaebumps}
\end{figure}

To derive the observed significance with the simplified background prediction, we use the bump hunter (BH) algorithm~\cite{Choudalakis:2011qn}, recently implemented in Python~\cite{pybumphunter}.
We choose the variable-width mass binning from the CMS dijet searches~\cite{Sirunyan:2018xlo} in the range from 2659~GeV to 6099~GeV.
We look for resonances in windows spanning two to five bins.
With the MSE model in BB 1, we identify a possible resonance around $3.9$~TeV with a local significance of $2.1\,\sigma$, which is close to the region of the injected dijet resonance with $m_\PZpr=3823$~GeV.
In BB 2 using the same model, the most discernable bump lies around $3.3$~TeV with a small local significance of $0.8\,\sigma$, which agrees with the fact that BB 2 has no injected signal. 
For the model trained with the Chamfer loss, a $1.5\,\sigma$ excess is seen at $2.8$~TeV in BB 1 and a $-1.4\,\sigma$ excess at $5.1$~TeV in BB 2. 
Neither is significant.
As noted previously, the permutation invariant Chamfer loss performs worse at the unsupervised anomaly detection task.
This may be due to the minimization, which will often return a smaller loss value than MSE even for poorly reconstructed, anomalous jets.
Fig.~\ref{fig:gnnaebumps} shows the BH results for BBs 1 and 2 using the models trained with both losses.

\subsubsection{Lessons Learned}
\label{sec:lessons}

Graph neural networks, like our proposed particle graph autoencoder, are promising methods for anomaly detection.
However, further work is needed to define a permutation-invariant loss function for use with such architectures that is more performant for anomaly detection.
In addition, a more generic resonance search procedure, such a multimensional fit in the trijet, dijet, trijet, and single-jet mass distributions possibly using methods like Gaussian process fitting~\cite{Frate:2017mai}, would be appropriate to use in combination with this algorithm.
In our experience, the R\&D dataset was extremely helpful in preparing our anomaly detection algorithms and gauging whether the algorithm we were developing was on the right track. 
In the future, more extensive R\&D datasets, together with additional black boxes with different signals, may be useful.
Finally, it may be productive to host a future competition on a well-known platform, such as Kaggle, to increase engagement with the broader machine learning community.

 \FloatBarrier

\subsection[Regularized Likelihoods]{Regularized Likelihoods\footnote{Authors: Ioan-Mihail Dinu.  Most of the machine learning heavy lifting was done with the help of the existing code base from the original $\mathcal{M}$-flow model introduced in Ref.~\cite{Brehmer:2020vwc} by Johann Brehmer and Kyle Cranmer. \href{https://github.com/johannbrehmer/manifold-flow}{https://github.com/johannbrehmer/manifold-flow}.}}

\label{sec:regualrizedlikes}

\subsubsection{Method}
\label{sec:method}
The method presented in this section attempts to use the power of generative models for the downstream task of Anomaly Detection. We have mainly explored the possible applications of flow-based methods, since they have the advantage of providing an explicit likelihood.

Normalizing Flows (NF) are one of the best methods available at the moment for density estimation in high-dimensional data (Ref. \cite{pmlr-v37-rezende15}). Those types of models work by learning a bijective mapping between the data distribution and a multivariate gaussian (with the same number of dimensions). Experience shows that, unfortunately, the likelihood that NF models provide is not sufficient as a stand-alone anomaly detection metric. 

In an attempt to \textit{regularize} the likelihood obtained with such density estimation techniques we have explored several alternatives to the vanilla NF models. One particularly interesting approach is the $\mathcal{M}$-flow model introduced originally in Ref. \cite{Brehmer:2020vwc}. 

\paragraph{$\mathcal{M}$-flows}

The $\mathcal{M}$-flow model combines the idea of reconstruction error from autoencoders with the tractable density of NF. If there exists a lower-dimensional  data manifold embedded in the data space, this method attempts to learn both the shape of this data manifold $\mathcal{M}$ and the density over that manifold.

In order to create a $\mathcal{M}$-flow we start with a bijective mapping $\mathrm{f}$ between the latent space $\mathrm{U} \times \mathrm{V}$ to the data space $\mathrm{X}$, as in Eq.~\ref{eq:eq1}. The latent space is split in two components: $\mathbf{u}$, which is the latent space representation that maps to the learned manifold, and $\mathbf{v}$, which represents the remaining latent variables that are “off the manifold”. 
\begin{equation}
    \begin{split}
        \mathrm{f}: \mathrm{U} \times \mathrm{V} & \rightarrow \mathrm{X} \\ u,v &\rightarrow \mathrm{f}(u,v)
    \end{split}
    \label{eq:eq1}
\end{equation}

The transition from the space $\mathrm{U} \times \mathrm{V}$ space to the space $\mathrm{U}$ is implemented as a projection operation, the $\mathbf{v}$ component being basically discarded. The inverse of this transition is implemented with zero-padding, $\mathbf{u}$ remains unchanged and $\mathbf{v}$ is filled with zeros. We notate the previous operations with the function $\mathrm{g}$, characterizing the transformation of a latent representation $\mathbf{u}$ to a data point $\mathbf{x}$ (shown in Eq.~\ref{eq:eq2}). 

\begin{equation}
    \begin{split}
        \mathrm{g}: \mathrm{U} & \rightarrow \mathcal{M} \subset \mathrm{X} \\ 
        u &\rightarrow \mathrm{g}(u) = \mathrm{f}(u,0) 
    \end{split}
    \label{eq:eq2}
\end{equation}

Finally the density in the space $\mathrm{U}$ is learned using a regular NF model denoted as $\mathrm{h}$. A schematic representation of those operations is presented in Fig.~\ref{fig:fig1}.

\begin{figure}[h!]
    \centering
    \includegraphics[width=0.7\textwidth]{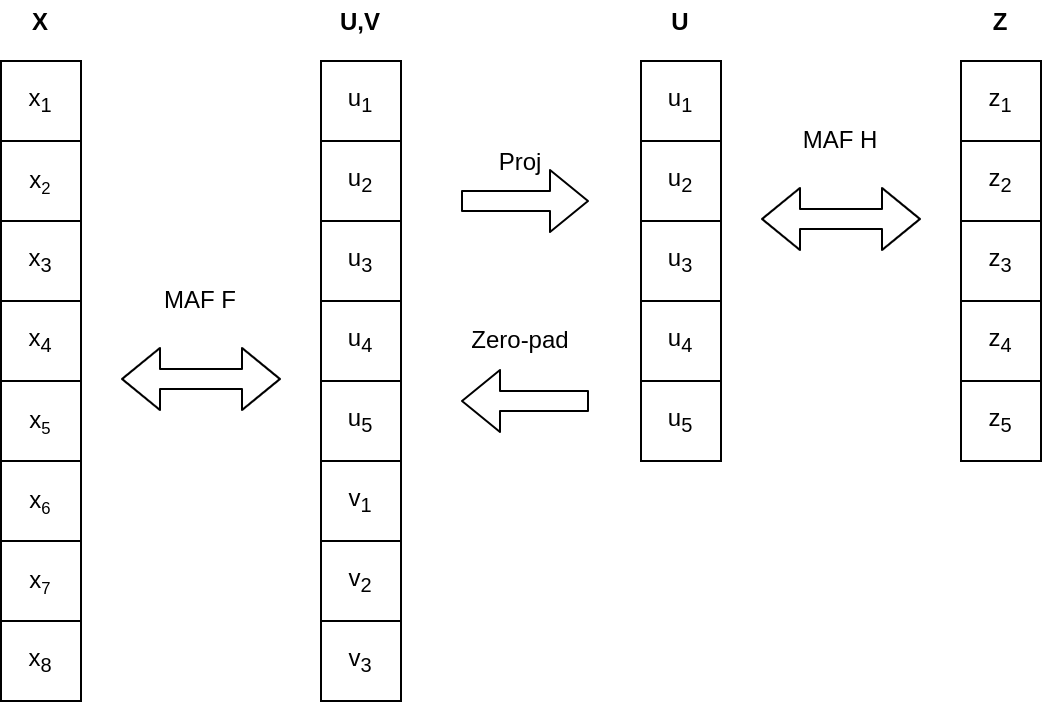}
    \caption{An  example representation of dependencies between the data $\mathbf{x}$, latent variables $\mathbf{u}$, $\mathbf{v}$ and the normally distributed variable $\mathbf{z}$. Here the example data has 8 dimensions and the latent space has 5 dimensions. The bijective transformations are learned with Masked Autoregressive Flows (MAFs).}
    \label{fig:fig1}
\end{figure}

The training of this model is split in two phases completed sequentially for every batch. Firstly, the parameters of $ \mathrm{f}$ are updated by minimizing reconstruction error from the projection onto the manifold (loss function in Eq.~\ref{eq:eq3}). The second phase of training consists in updating the parameters of $\mathrm{h}$ by minimizing the negative log likelihood from Eq.~\ref{eq:eq4}. 

\begin{equation}
    \mathcal{L}_{manifold} = \lvert \lvert x-g(g^{-1}(x)) \lvert \lvert ^2
    \label{eq:eq3}
\end{equation}

\begin{equation}
    \mathcal{L}_{density} = \log p_{u}(g^{-1}(x))
    \label{eq:eq4}
\end{equation}

Regarding the preprocessing steps, the LHC Olympics datasets have been clustered and the following features have been selected for each of the two leading jets: $p_T$, $\eta$, $E$, $m$, $\tau_3/\tau_2$, $\tau_2/\tau_1$, where $\tau_n$ is the n-subjettiness. For these 12 features, the best performing manifold size was 8.

This model offers the possibility to calculate both the density on the manifold and the reconstruction error from the projection on the manifold. We tried to use both of those metrics in order construct a robust anomaly score as in Eq.~\ref{eq:eq5}. This metric performs the anomaly detection task better on the R\&D dataset than its components and better than a basic normalizing flow model trained on the same data, judging by the ROC curves in Fig.~\ref{fig:fig2}.

\begin{equation}
    \mathcal{R}_{exp}(x) = \frac{\lvert \lvert x-g(g^{-1}(x)) \lvert \lvert ^2}{1+p_{u}(g^{-1}(x))}
    \label{eq:eq5}
\end{equation}

While experimenting with this anomaly score, it became apparent that it generates a bias towards events with high dijet mass ($m_{jj}$). In order to decouple $ \mathcal{R}_{exp}$ from $m_{jj}$ we included the marginal likelihood of $m_{jj}$, that was modeled using Kernel Density Estimation (KDE), as a term into the anomaly score. The resulting metric, denoted ${R}_{m_{jj}}$, uses the ratio between the likelihood on the manifold and marginal $m_{jj}$ likelihood as in Eq.~\ref{eq:eq6}.

\begin{equation}
    \mathcal{R}_{m_{jj}}(x) = \frac{\lvert \lvert x-g(g^{-1}(x)) \lvert \lvert ^2}{1+\frac{p_{u}(g^{-1}(x))}{p_{KDE}(m^{x}_{jj})}}
    \label{eq:eq6}
\end{equation}

Translating the performance obtained on the R\&D data to the black boxes proved to be a big challenge. The small differences in modeling from a black box to another are often enough to introduce significant biases. The only apparent solution seems to be training and applying the method on the same dataset.

\subsubsection{Results on LHC Olympics}
\label{sec:results}

The R\&D dataset was heavily used for benchmarking different approaches, Fig.~\ref{fig:fig2} shows the anomaly detection performance of different metrics on the R\&D dataset. 

\begin{figure}[h!]
    \centering
    \includegraphics[width=0.5\textwidth]{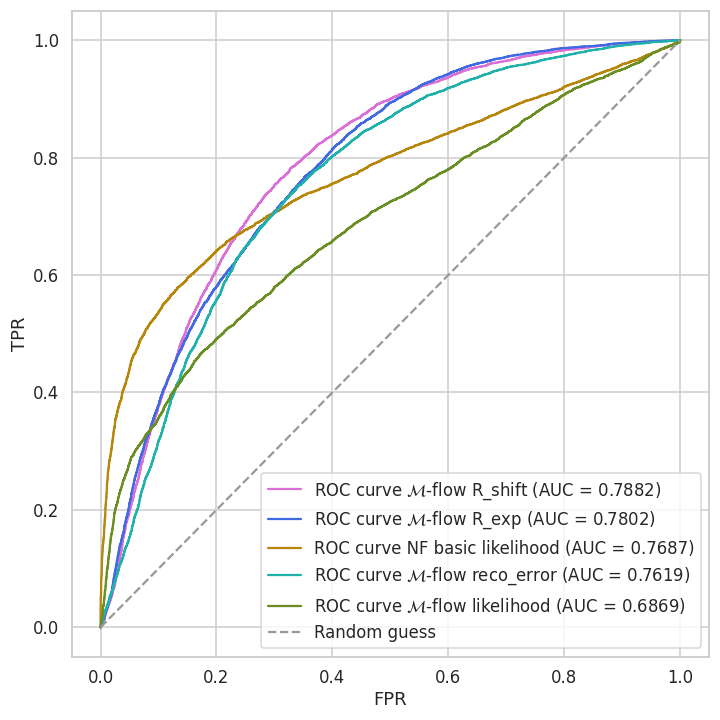}
    \caption{Signal detection ROC curves in the R\&D dataset for different anomaly scores}
    \label{fig:fig2}
\end{figure}

In order to evaluate the performance of this method in the absence of pure background training data, a small fraction ($\sim 1\%$) of signal was introduced into a subsample from the R\&D dataset. The resulting data sample was used both for training and evaluation of the model.

Several cuts have been applied on $\mathcal{R}_{m_{jj}}$ while trying to find any indication of a resonance in the $m_{jj}$ spectrum. Although less apparent, there is still a bias towards identifying higher $m_{jj}$ events as being anomalous. The right plot in Fig.~\ref{fig:fig3} shows the $m_{jj}$ distribution for events above the $50^{th}$ percentile of $\mathcal{R}_{m_{jj}}$ vs events above the $70^{th}$ percentile of $\mathcal{R}_{m_{jj}}$. If we were to take the $50^{th}$ cut as a baseline, it is clear that increasing the threshold has the effect of selecting events with slightly higher $m_{jj}$. Unfortunately there is no sharp peak in the $m_{jj}$ distribution that would indicate a possible resonance, but rather the tail of the distribution seems to get bigger.

\begin{figure}[h!]
    \centering
    \begin{minipage}[b]{.5\textwidth}
        \centering
        \includegraphics[width=\textwidth]{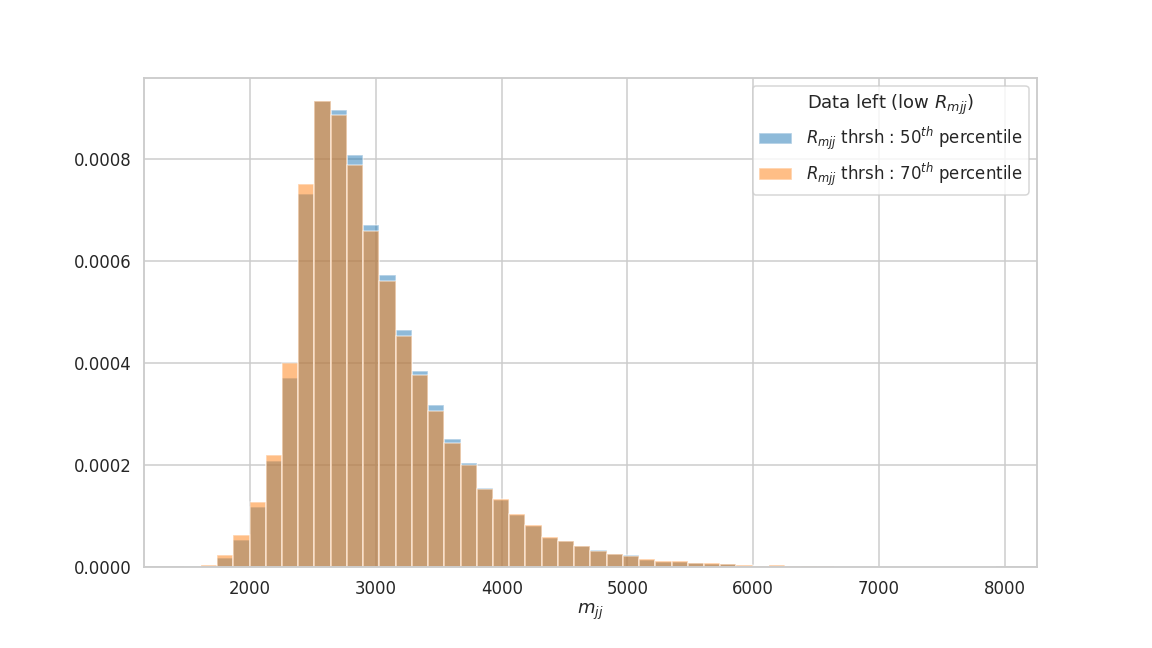}
\end{minipage}\begin{minipage}[b]{.5\textwidth}
        \centering
        \includegraphics[width=\textwidth]{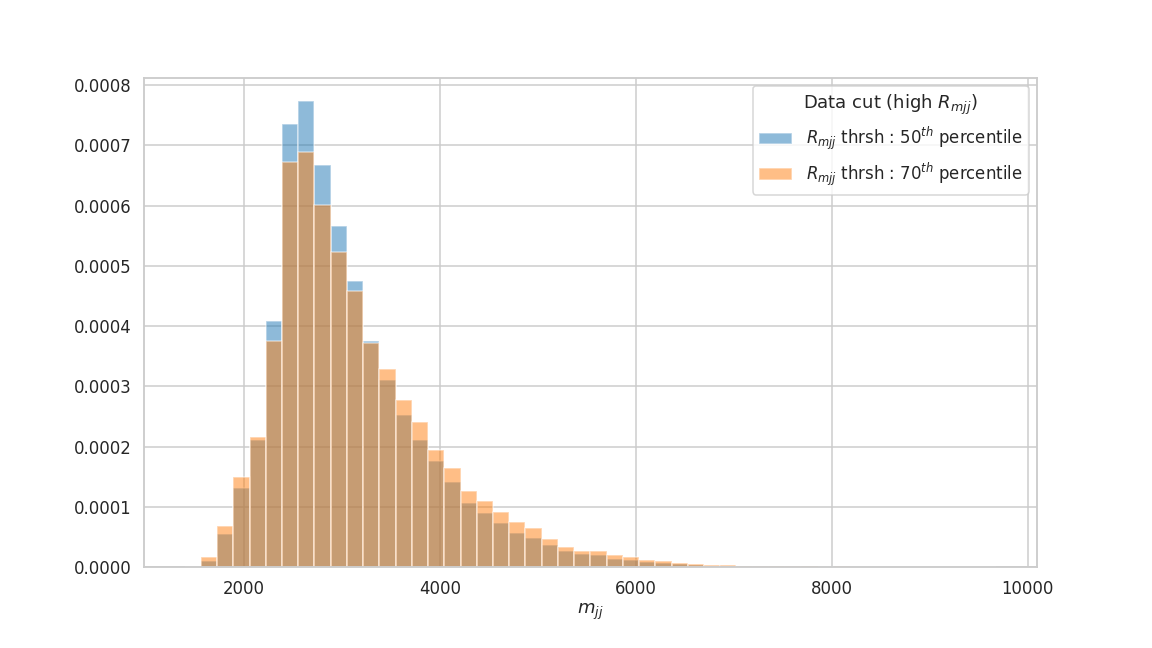}
\end{minipage}
    \label{fig:fig3}
    \caption{Overlapping $m_{jj}$ distributions below (left) and above (right) two threshold cuts on $\mathcal{R}_{m_{jj}}$. Distributions for a $50^{th}$ percentile cut are in blue, while  distributions for a $70^{th}$ percentile cut are in orange. The $x$ axis is in $GeV/c^2$.}
\end{figure}

The results so far suggest that this method can not be used reliably to find the hidden signal within the black-boxes. This behavior is consistent regardless of the choice of $\mathcal{R}_{m_{jj}}$ thresholds.

\subsubsection{Lessons Learned}
\label{sec:lessons}

One of the main lessons learned during this challenge is that: in absence of a good background model, the neural networks by themselves can not achieve good anomaly detection performance. 

For the winter LHC Olympics, we approached the problem with a simple autoencoder that was trained on the full background black box. Applying that model on Black Box 1 (BB1) introduced a lot of bias that ended up acting like a fake signal. Special precautions should always be taken in order to avoid this scenario.

With the experience gained from studying BB1 we were a lot more careful to avoid creating fake signal. The subsequent problem proved to be the lack of a good background model. Since we could not rely on the full background black box, the alternative was to train on data, but this comes with its own issues. 

All of the attempts so far came short of providing a good background modeling and therefore the current anomaly detection performance leaves a lot to be desired. Those trials taught us that a good machine learning anomaly detection algorithm is not just about the neural network itself, but many other analysis details should be treated with the same amount of attention.

 \FloatBarrier
\subsection[UCluster: Unsupervised Clustering]{UCluster: Unsupervised Clustering\footnote{Authors: Vinicius Mikuni and Florencia Canelli.  UCluster is available at: \href{https://github.com/ViniciusMikuni/UCluster}{https://github.com/ViniciusMikuni/UCluster}.}}

\label{sec:ucluster}

\subsubsection{Method}
\label{sec:method}

The properties of physics beyond the Standard Model (BSM) are not yet known. However,
we can expect anomalous events, from the same physics processes, to carry similar event signatures. In this section, we introduce a method for Unsupervised Clustering (UCluster). The goal of UCluster is to reduce the data dimensionality using a neural network that retains the main properties of the event collision. In this reduced representation, a clustering objective is added to the training to encourage points embedded in this space to be close together when they share similar properties and far apart otherwise. 

To create meaningful event embeddings, a per-particle jet mass classification is chosen.
We first start clustering particles into jets with the Fastjet implementation of the anti-$k_t$ algorithm with $R = 1.0$ for the jet radius. Each particle associated to a clustered jet receives a label, proportional to the mass of the associated jet. For this task, we require the model to learn the mass of the associated jet the particle belongs to, and which particles should belong to the same jet. This approach is motivated by the fact that the invariant mass of a jet is correlated with jet substructure observables, which often contains useful information for distinguishing different physics processes. The mass labels are then created by defining 20 equidistant intervals from 10 to 1000 GeV. For simplicity, the first 100 particles associated to the two heaviest jets in the event are considered. If a smaller number of particles are found, events are zero-padded up to 100, otherwise truncated. 

The classification task is achieved by means of a classification loss ($L_{\mathrm{focal}}$), defined by the focal loss \cite{DBLP:journals/corr/abs-1708-02002}. The focal loss is usually applied to classification problems with unbalanced labels. This choice was made since a different number of events is expected for different mass intervals. The focal loss expression for a multiclass classification is defined as:
\begin{equation}
    L_{\mathrm{focal}}=  -\frac{1}{N}\sum_j^N \sum_m^M  y_{j,m} (1 - p_{\theta,m}(x_j))^{\gamma}\log(p_{\theta,m}(x_j))
    \label{eq:class_loss}
\end{equation}
where $p_{\theta,m}(x_j)$ is the network’s confidence, for event $x_j$ with trainable parameters $\theta$, to be classified as class $m$. The term $y_{j,m}$ is 1 if class $m$ is the correct assignment for event $x_j$ and 0 otherwise.
In this implementation, the parameter $\gamma=2$ is used. Different values of $\gamma$ were tested resulting in no significant changes in performance.

To cluster events with similar properties, a clustering loss ($L_{\mathrm{cluster}}$) is added to the overall loss function. $L_{\mathrm{cluster}}$ was introduced in \cite{DBLP:journals/corr/abs-1806-10069}, defined as:
\begin{equation}
    L_{\mathrm{cluster}} = \frac{1}{N}\sum_k ^K \sum_j^N \left \| f_{\theta}(x_j) - \mu_k \right \|^{2}\pi_{jk}.
    \label{eq:cluster_loss}
\end{equation}
The distance between each event $x_j$ and cluster centroid $\mu_k$ is calculated in the embedding space $f_{\theta}$, created by the classification task. 
The function $\pi_{jk}$ weighs the importance of each event to the clustering objective of the form:
\begin{equation}
    \pi_{jk} = \frac{e^{-\alpha \left \| f_{\theta}(x_j) - \mu_k \right \|}}{\sum_{k'} e^{-\alpha \left \| f_{\theta}(x_j) - \mu_k \right \|}},
    \label{eq:pi}
\end{equation}
with hyperparameter $\alpha$. Since $L_{\mathrm{cluster}}$ is differentiable, stochastic gradient descent can be used to optimize jointly the trainable parameters $\theta$ and the centroid positions $\mu_k$. 

The combined loss to be minimized is then:
\begin{equation}
    L =  L_{\mathrm{focal}} + \beta L_{\mathrm{cluster}}.
    \label{eq:loss}
\end{equation}
The hyperparameter $\beta$ controls the relative importance between the two losses.
The value of $\beta$=10 is used to give the two components the same relative order of magnitude.

As defined in Eq.~\ref{eq:cluster_loss}, $L_{\mathrm{cluster}}$ requires an initial value for the cluster centers. While the initial value can be corrected during training, a more stable performance is observed when the model is first pre-trained with only $L_{\mathrm{focal}}$ for 10 epochs. After the pre-training, the centroids are initialized by applying the K-Means algorithm \cite{10.2307/2346830} to the object embeddings. The full training is then carried out with the combined loss defined in \ref{eq:loss} for 100 epochs. The $\alpha$ parameter controls the importance of the initial cluster assignment and is set to a starting value of 1,  increasing by a factor 2 for every following epoch. 

UCluster was designed to be independent of the ML architecture. For these studies,  ABCNet \cite{Mikuni:2020wpr} is used as the backbone network. ABCNet is a graph-based implementation where each reconstructed particle is taken as a node in a graph. The importance of each node is then learned by the addition of attention mechanisms described in  \cite{velikovi2017graph}.

\begin{figure}[h!]
    \centering
    \includegraphics[width=0.8\textwidth]{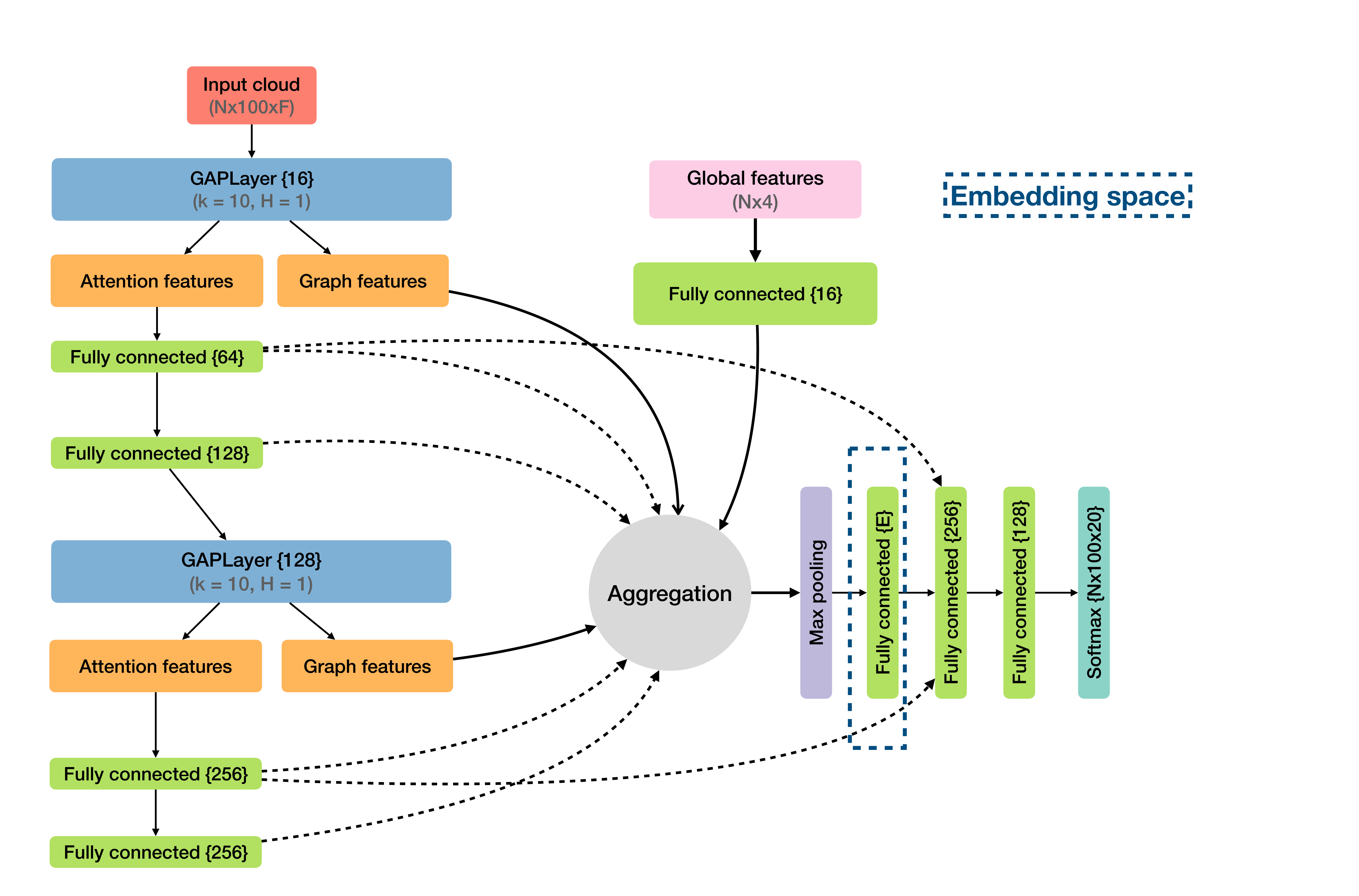}
    \caption{ABCNet architecture used in UCluster for a batch size N, F input features, and embedding space of size E. Fully connected layers and encoding node sizes are denoted inside ``\{\}''. For each GAPLayer, the number of k-nearest neighbors (k) and heads (H) are given. Full lines represent direct connections while dotted lines denote skip connections.}
    \label{fig:abc_model}
\end{figure}

The 10 nearest neighbors from each particle are used to calculate the GAPLayers \cite{2019arXiv190508705C}. The initial distances are calculated in the pseudorapidity-azimuth ($\eta-\phi$) space of the form $\Delta R = \sqrt{\Delta\eta^2 + \Delta\phi^2}$. The second GAPLayer uses the Euclidean distances in the space created by subsequent fully connected layers. The architecture used and the layer where the embedding space is define are depicted in Fig.~\ref{fig:abc_model}. No significant changes were observed when varying the number of neighbors and maximum number of training epochs. Additional hyperparameters of ABCNet were kept as is to avoid fine tuning. 

UCluster and ABCNet are implemented in v1.14 of Tensorflow~\cite{tensorflow}. The loss is optimized by Adam \cite{adam} and back-propagation to compute gradients. The learning rate starts from 0.001 and decreases by a factor 2 every three epochs, until reaching a minimum of 1e-5. The batch size is fixed to 1024.

\subsubsection{Results on LHC Olympics}
\label{sec:results}

Results are presented on the R\&D data set created for the LHC Olympics 2020. From this data set, 300k events are used for training, 150k for testing and 300k events used to evaluate the performance. The signal fraction in each of these samples is fixed at 1\% of the total amount of events.

The distributions used as input features for ABCNet are described in Tab.~\ref{tab:anomaly_vars}. 

\begin{table}[ht]
    \centering
    \caption{Descriptions of each feature used to define a point in the point cloud implementation for multiclass classification. The last two lines are the global information added to parameterize the network.}
    \label{tab:anomaly_vars}
	\begin{tabular}{ll}
    \hline\noalign{\smallskip} 
             Variable & Description  \\
            \hline
            $\Delta\eta$       &  \small{Pseudorapidity difference between the constituent and the associated jet}\\  
            $\Delta\phi$       &  \small{Azimuthal angle difference between the constituent and the  associated jet}\\  
            $\log(p_\text{T})$       &  \small{Logarithm of the constituent's $p_\text{T}$ }\\  
            $\log\mathrm{E}$       &  \small{Logarithm of the constituent's E }\\  
            $\log\frac{p_\text{T}}{p_\text{T}(\mathrm{jet})}$       &  \small{Logarithm of the ratio between the constituent's $p_\text{T}$ and the associated jet $p_\text{T}$}\\  
            $\log\frac{\mathrm{E}}{\mathrm{E}(\mathrm{jet})}$       &  \small{Logarithm of the ratio between the constituent's E and the associated jet E}\\  
            $\Delta\mathrm{R}$       &  \small{Distance in the $\eta-\phi$ space between the constituent and the associated jet}\\  
            \hline
            $\log m_{J\{1,2\}}$ & \small{Logarithm of the masses of the two heaviest jets in the event} \\
            $\tau_{21}^{\{1,2\}}$ & \small{Ratio of $\tau_1$ to $\tau_2$ for the two heaviest jets in the event}  \\
    \noalign{\smallskip}\hline  
	\end{tabular}
\end{table}

\begin{figure}[h!]
    \centering
    \includegraphics[width=0.4\textwidth]{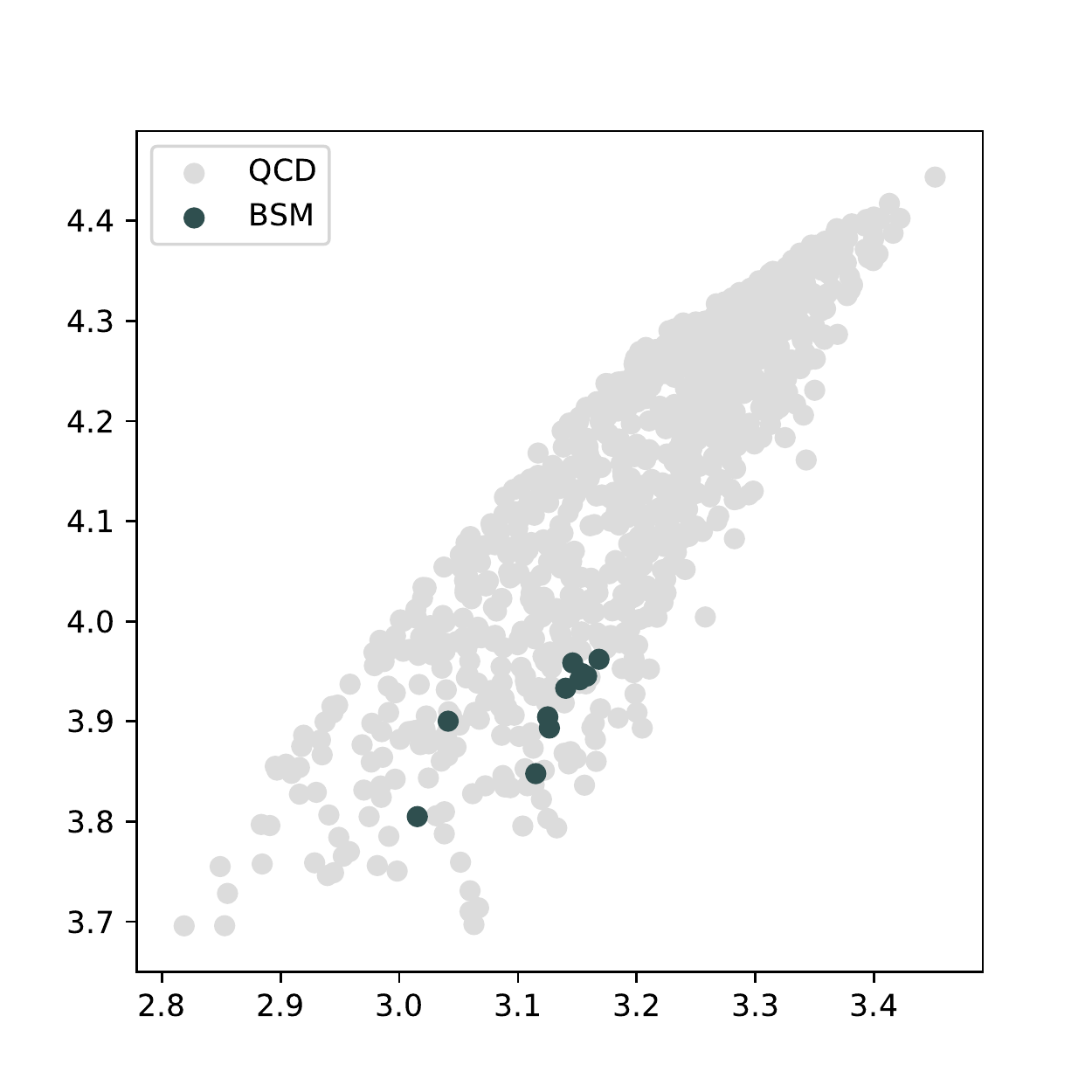}
    \includegraphics[width=0.4\textwidth]{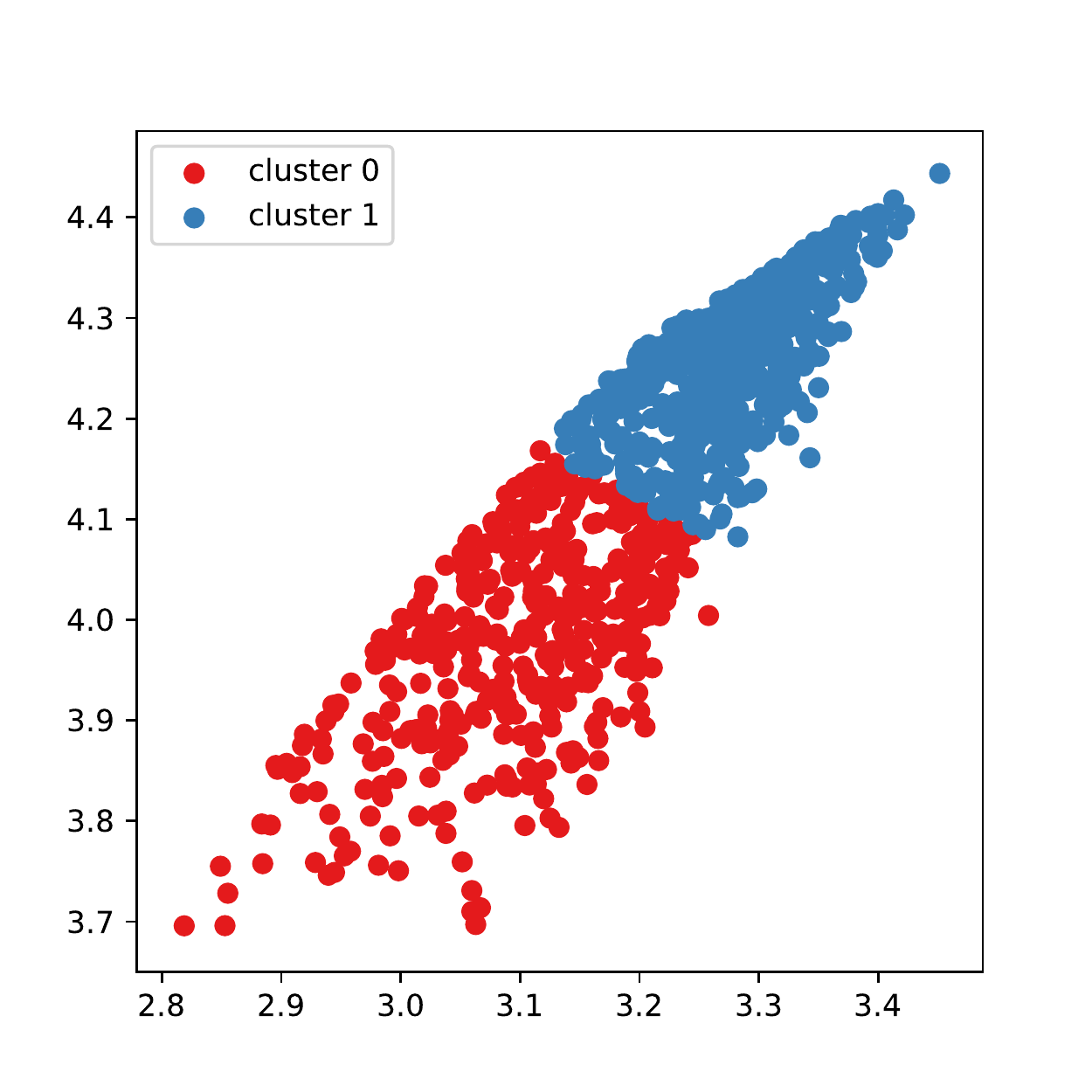}
    \caption{Visualisation of the embedding space created for anomaly detection for 1000 events. The true labels are show in the left, while the cluster labels created by UCluster are shown in the right. Figure from Ref. \cite{Mikuni:2020qds}.}
    \label{fig:embedding_space}
\end{figure}

We first evaluate the performance of UCluster by requiring the presence of two clusters in an embedding space of same size. Fig.~\ref{fig:embedding_space} shows the result of the event embeddings, superimposed for 1000 events of the evaluation sample. 

\begin{figure}[h!]
    \centering
    \includegraphics[width=0.4\textwidth]{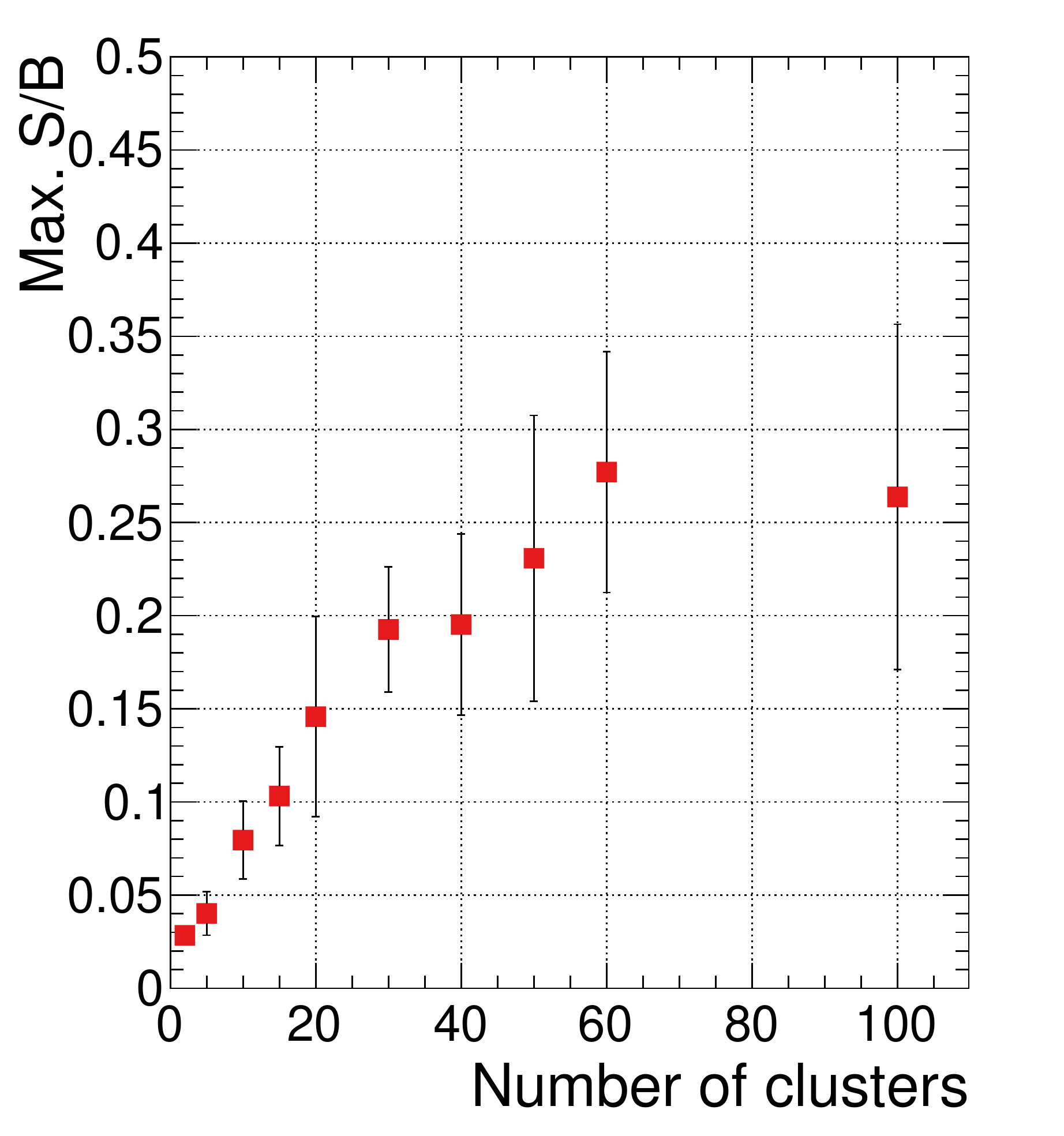}
    \includegraphics[width=0.4\textwidth]{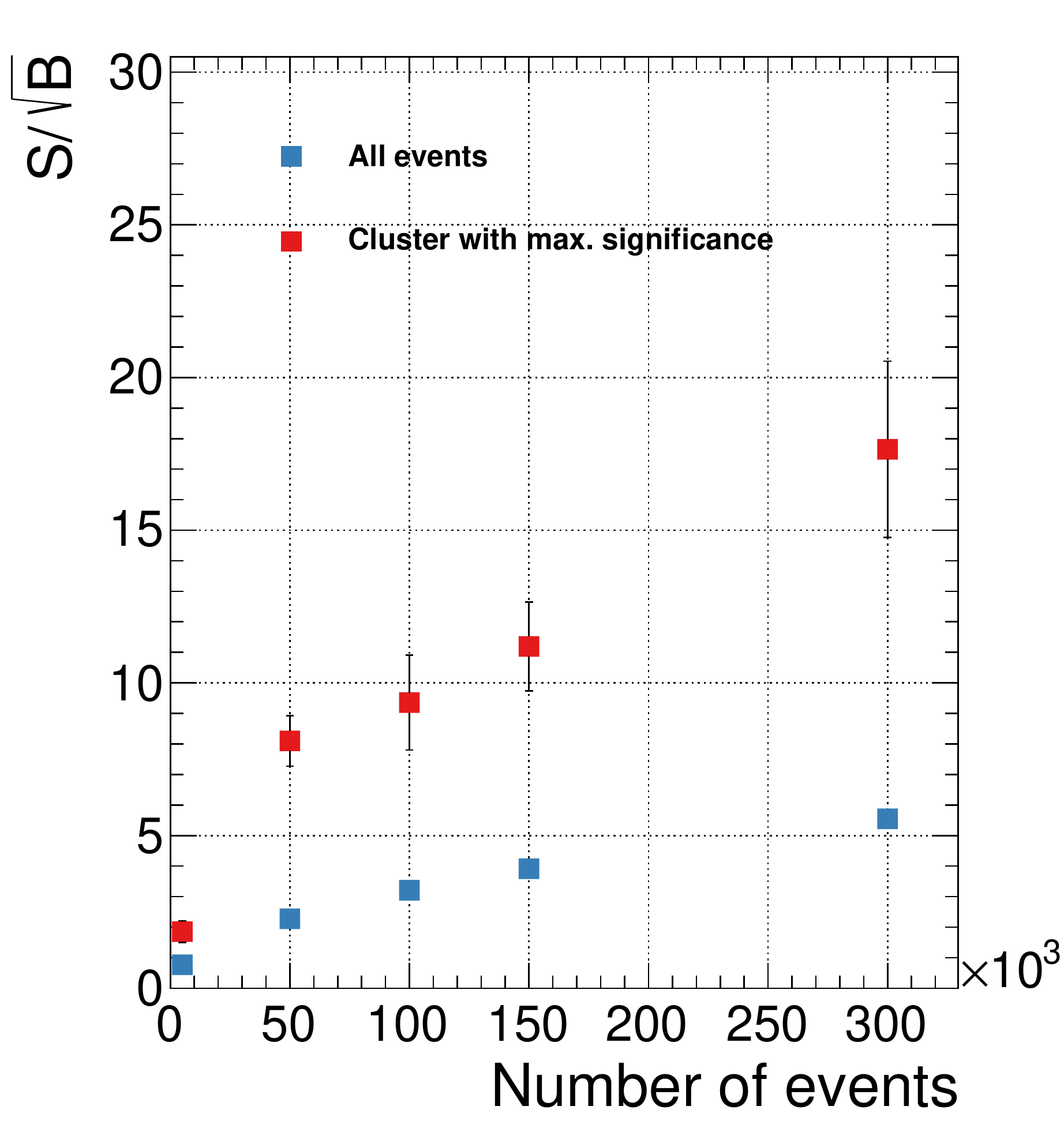}
    \caption{Maximum signal-to-background ratio found for different clustering sizes (left) and maximum approximate significance found for UCluster trained and evaluated on different number of events with cluster size fixed to 30 (right). The uncertainty corresponds to the standard deviation of five trainings with different random weight initialization. Figure from Ref. \cite{Mikuni:2020qds}.}
    \label{fig:anomaly_sig}
\end{figure}

A large fraction of BSM events are found in the same cluster, confirming the earlier assumption that anomalous events would end up close together in the embedding space. However, the QCD background contamination in the same cluster only leads to a signal-to-background ratio (S/B) increase from 1\% to 2.5\%. The S/B can be further enhanced by partitioning the events into more clusters. This assumption is correct if the properties of the anomalous events are different than the QCD signatures. To verify this behavior, the cluster size is varied while keeping all other network parameters fixed. In Fig.~\ref{fig:anomaly_sig} (left), the maximum S/B found in a single cluster is shown as a function of the cluster multiplicity. The S/B increases up to around 28\% as the number of clusters increases. The effect of the sample size used for training was also checked by varying the amount of training and evaluation examples while keeping the initial S/B fixed. In Fig.~\ref{fig:anomaly_sig} (right), the approximate significance (S/$\sqrt{\mathrm{B}}$) is shown as a function of the different sample sizes for UCluster  trained with a fixed cluster size of 30. The red markers show the maximum significance found in a single cluster, compared to the initial significance of the sample shown in blue. For initial significances in the range 2-6, we observe enhancements by factors 3-4. The training stability is tested by retraining each model five times.  The standard deviation of the independent trainings is shown by the error bars in Fig.~\ref{fig:anomaly_sig}. When many clusters are used, the clustering stability starts to decrease, as evidenced by larger error bars.

\subsubsection{Lessons Learned}
\label{sec:lessons}
The development of UCluster was carried based on the  R\&D data set. While the conceptual implementation to cluster events with similar properties was achieved in this data set, an additional step to identify interesting clusters for further inspection was also required. The latter step, while important, was not fully investigated by the time the results were announced, leading to no conclusive results when the method was applied to the black boxes. For future endeavors, an automatic procedure to evaluate the cluster importance will be necessary. The classification task, paired with the clustering objective, is paramount to the ability of UCluster to reduce the data dimensionality while providing meaningful event embeddings. During the development of the method, the substructure observables of each jet in the dijet event carried information to characterize the anomaly. Because of that, a classification task that took advantage of this property was defined. However, for different decay topologies, like the one presented in BB3, this approach would not necessarily be optimal. The reason is that only one of the decay modes presented jets with substructure properties that would differ from the main QCD background. To alleviate this issue, a different classification task could be adapted. However, a more general approach to create the embedding space could be used. In particular, auto-encoders applied to particle physics are suitable candidates for a summary statistic that can encapsulate the event information in a lower dimensional representation.

 \FloatBarrier

\section{Weakly Supervised}
\label{sec:weaklysupervised}

\subsection[CWoLa Hunting]{CWoLa Hunting\footnote{Authors: Jack H Collins and Benjamin Nachman.  The code can be found at \url{https://github.com/Jackadsa/CWoLa-Hunting/tree/tf2/LHCO-code}.}}

\label{sec:cwola}

\subsubsection{Method}
\label{sec:method}

CWoLa (Classification Without Labels) Hunting is a strategy for searching for resonant anomalies, in which the signal is hypothesized to be localized in one chosen resonant variable (e.g. some invariant mass, $m_{\text{res}}$) and the background is known to follow some smooth and simple distribution in that variable. Given a hypothesis resonance mass $m_{\text{hyp}}$ and width, a signal region is constructed by selecting events in a window around the resonance mass hypothesis, and upper and lower sideband regions are constructed by selecting events in windows adjacent to the signal region. Additional features $\{y\}$ orthogonal to the resonance mass (e.g. jet substructures) are used to distinguish a potential signal from the background. A binary classifier is trained to distinguish signal region events from sideband events using these additional features. If the features are chosen such that the distribution of background events in the signal region is indistinguishable from those in the sideband, then in the absence of a signal the classifier will be driven by statistical fluctuations between the two event samples and will have poor performance on test data. If, however, the signal region contains an additional population of signal events that is not present or is very rare in the sideband, then the classifier may learn the distribution of the signal events in $\{y\}$.

Given that the black-box data is simulated with a di-jet invariant mass trigger, we use as our resonant mass variable the invariant mass between the two highest $p_\text{T}$ $R=1$ anti-$k_t$ jets in the event, and the orthogonal features will be the jet substructure variables
\begin{equation}
    \mathrm{Features:} ~~ m_{J, A}, \; m_{J, B}, \; \tau_{21, A}^{(1)}, \; \tau_{21, B}^{(1)}, \; \tau_{32, A}^{(1)}, \; \tau_{32, B}^{(1)},
\end{equation}
where $A, B$ refer to the two $p_\text{T}$-ordered jets. In order to remove some amount of correlation between the jet masses and $m_{JJ}$ in background QCD events, we rescale the jet masses before they are input into the classifiers
\begin{equation}
    m_J \rightarrow m_J' = \frac{m_J-30\;\text{GeV}}{m_{JJ}} + \frac{30 \; \text{GeV}}{3000 \; \text{GeV}}.
\end{equation}
The key part is the rescaling by dividing by $m_{JJ}$, since the $m_J$ distributions have a strong scaling with $m_{JJ}$. The additional offset by $30\;\text{GeV}$ is not important, but was judged by eye to result in smaller correlation between $m_J$ and $m_{JJ}$.

By construction, this strategy is sensitive only to signals that result from the decay of a heavy resonance into two significantly lighter particles which each decay into largely hadronic boosted final states. This still covers a broad range of phenomenological possibilities, as the space of possible jet substructures is large. This has the potential to be sensitive to the the signal in the R\&D dataset and BB1, but not to that in BB3. We attempted to apply a modified form to the signal in BB3 without success, as briefly described at the end.

The statistical independence of training and test sets is critical, and in order to retain as much statistical power as possible we perform a nested cross-validation procedure to select signal-like events. A detailed explanation follows. There are four training loops including the scan over $m_{\text{hyp}}$, the additional ones running over loops indexed by the labels $k, l, i$.

\begin{enumerate}
    \item[$k$] We split the entire dataset (including events outside the signal and sideband regions) randomly into five $k$-folds, and when searching for a signal in the $k$th fold we train a classifier using the remaining $k-1$ folds. Given a pre-determined threshold efficiency $\epsilon_\text{sel}$, that fraction of highest scoring events is chosen from the $k$th fold as judged by the classifier trained on the other folds. The selected events from each fold are then combined into a single histogram in $m_{\text{res}}$. A bump hunt is then performed at $m_{\text{hyp}}$ using a fit of a simple function to data outside the signal region to predict the expected background in the signal region. A simple Poisson hypothesis test is performed on the observed event rate in the signal region compared to the background expectation, with uncertainties in the fit parameters assumed to follow a Gaussian distribution.

    \item[$l$] The training of the classifier for a single $k$-fold involves another layer of cross-validation. This is due to the difficulty of a single classifier learning a small difference between two distributions that are otherwise identical besides statistical fluctuations, and overfitting to these fluctuations is unavoidable. Multiple classifiers are liable to overfit in different ways, and an ensemble model consisting of an average of multiple individually-trained neural networks tends to be more robust, due to destructive interference of overfitting and constructive interference of a true signal. For each $k$ four classifiers are trained labelled by $1 \leq l \leq 5$, $l \neq k$. The $l$th classifier uses the $l$th fold of data as a validation set and the remaining three folds as training data. The ensemble model consists of the mean of the outputs of the individual neural networks.

    \item[$i$]For each $l$, multiple networks (in this work, three) are trained on the same data and the best performing one is chosen as the corresponding input to the ensemble model. The performance metric (evaluated on validation data) is the selection efficiency on the signal region events of a selection cut on the neural network output above a threshold determined to have a given efficiency $\epsilon_\text{cut}$ on sideband events. In the present study $\epsilon_\text{cut}$ is chosen to be 0.01, in order to be as small as possible while avoiding being dominated by statistical fluctuations when the number of validation events is small.

    The neural networks are coded in Keras~\cite{keras} with Tensorflow~\cite{tensorflow} backend. The architecture consists of four hidden layers each with 128 nodes. The activation function of the first hidden layer is Leaky ReLU with inactive gradient of 0.1, while the other hidden layers have elu activation functions. Dropout layers with probability 0.1 are added between each pair of hidden layers. Adam~\cite{adam} is used for optimization with hyperparameters: $\text{lr} = 0.001$, $\beta_1 = 0.8$, $\beta_2 = 0.99$, $\text{decay} = 5\times10^{-4}$. The model is trained with batch size of 5000, the large number being chosen to increase the chance of true signal events being in each batch. The metric $\epsilon_\text{cut}$ is monitored on validation data; the model is saved at the maximum value and training is halted if 250 epochs pass without improvement. Training and validation events are reweighted so that the lower and upper sidebands each have equal weight (which ensures that one is not favoured over the other in training), and together they have the same total weight as the signal region.
\end{enumerate}

No scan or systematic optimization of hyperparameters was performed and many of these choices are likely to be suboptimal. 

Data is selected in the window $2632 \; \text{GeV} \leq m_{JJ} \leq 6000 \; \text{GeV}$, and split into 16 equally log-spaced bins. A signal region is defined as three adjacent bins, which corresponds to a width of around 15\%. The two bins adjacent above and below the signal region are defined as the upper and lower sidebands. There are therefore ten overlapping signal regions, starting centered at the fourth bin and ending centered at the 13th bin. This strategy was chosen so that a signal cannot hide by being centered at a bin boundary, split equally between signal region and sideband. The signal region background is determined by a fit of the following function to the $m_{JJ}$ distribution in bins outside the signal region
\begin{equation}
    \frac{dN}{d m_{JJ}} = p_0 \frac{(1-y)^{p_1}}{y^{p_2 + p_3 \log(y)}}, ~~~ y = \frac{m_{JJ}}{13 \; \text{TeV}}
\end{equation}
where $p_i$ are four free fit parameters. This function is used in ATLAS and CMS diboson searches~\cite{Aaboud:2017eta,Sirunyan:2016cao}.

\subsubsection{Results on LHC Olympics}
\label{sec:results}

\begin{figure}[h!]
    \centering
    \includegraphics[width=0.7\textwidth]{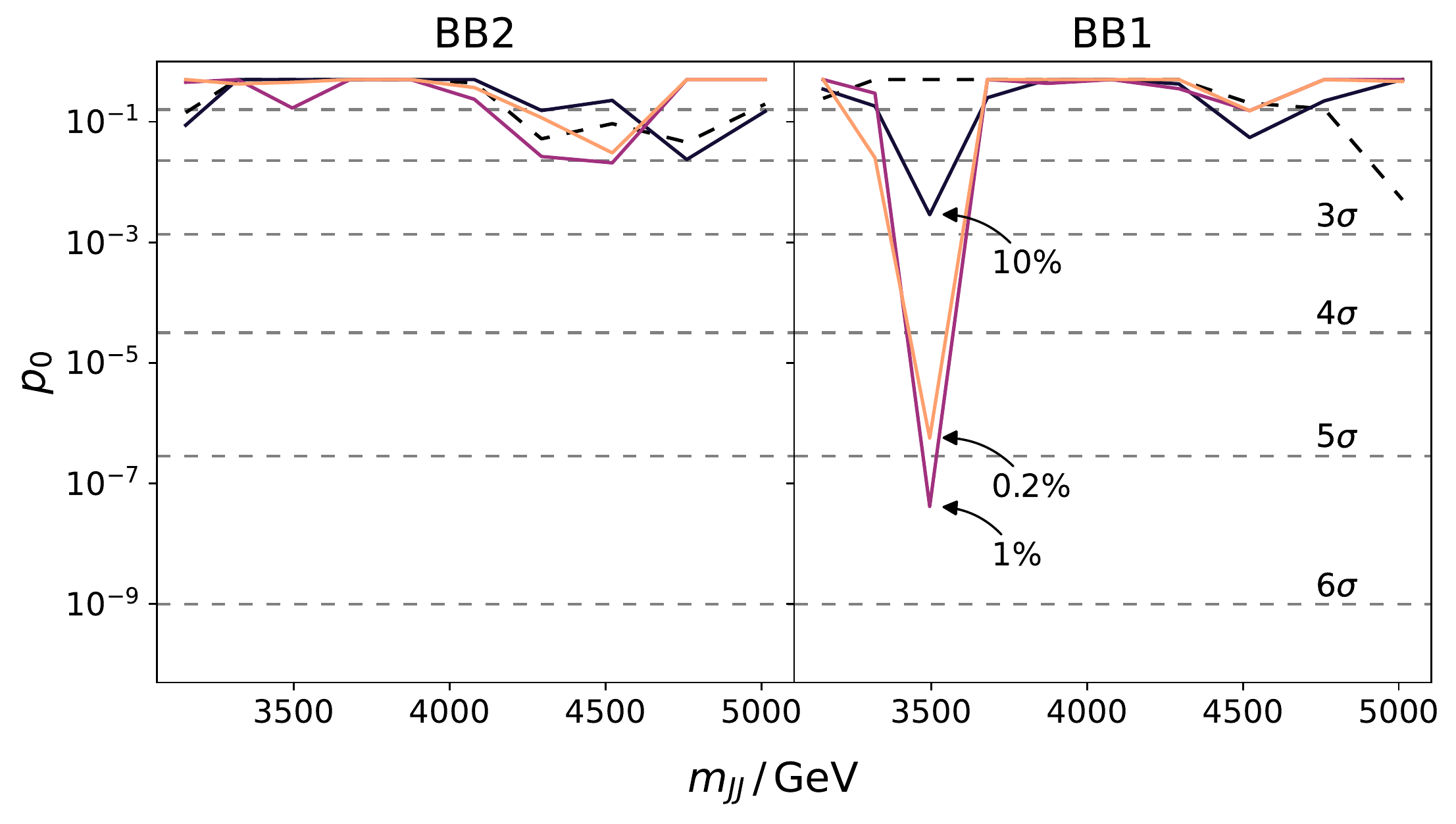}
    \caption{$p$-values obtained from the analysis in the resonance mass scan for BB2 (left) and BB1 (right) at selection efficiencies 10\%, 1\%, 0.2\%. The dashed black line is the result with no selection cut.}
    \label{fig:pval}
\end{figure}

\begin{figure}[h!]
    \centering
    \includegraphics[width=0.49\textwidth]{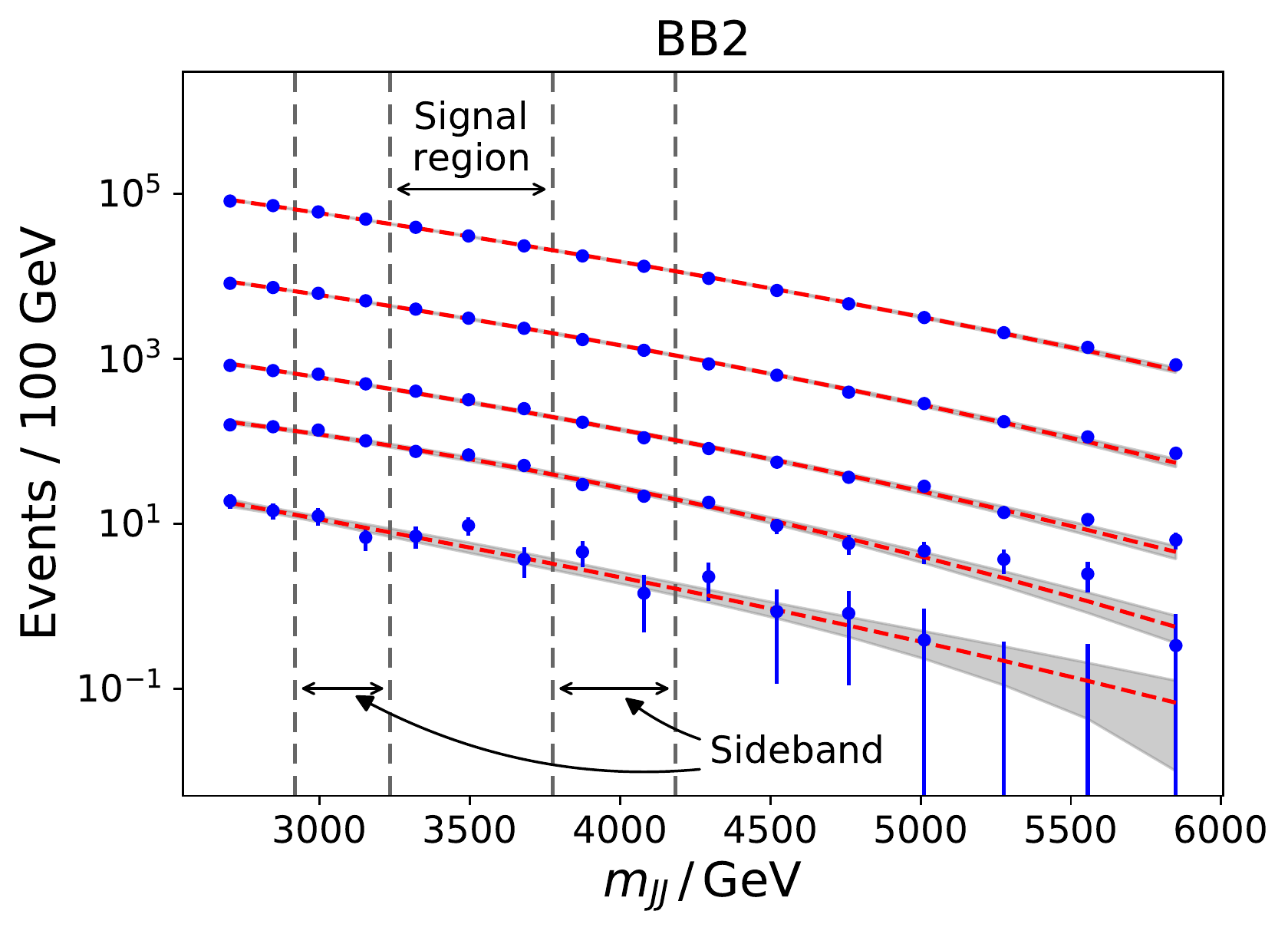}
    \includegraphics[width=0.49\textwidth]{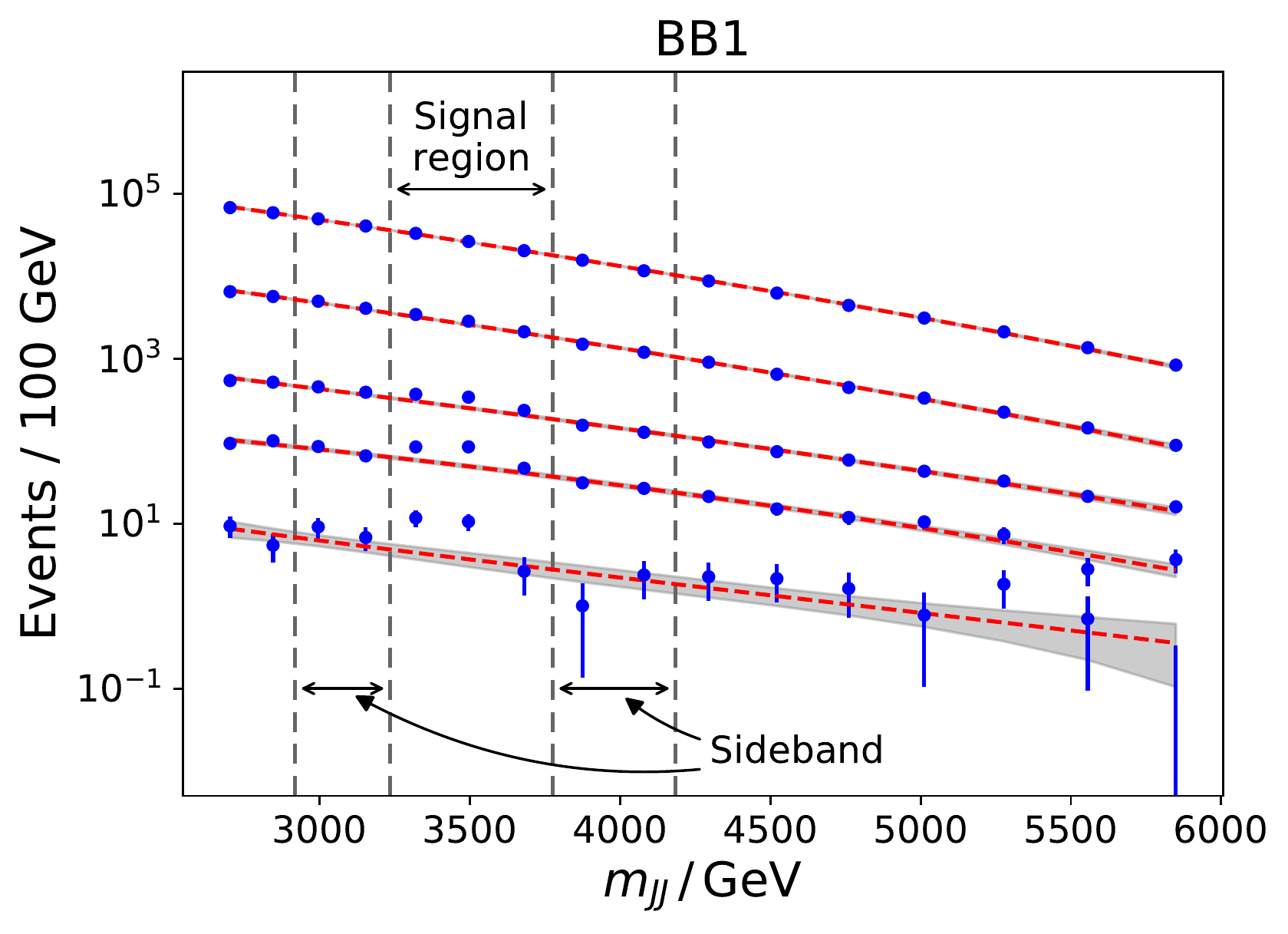}
    \caption{$m_{JJ}$ distributions obtained for BB2 (left) and BB1 (right) for the signal region centered around $3500\,\mathrm{GeV}$ after a series of selection cuts. The top line and data points corresponds to no selection cut.}
    \label{fig:mjj}
\end{figure}

This study was performed on BB1 and BB2 after the signal was unblinded. However, no changes were made in the algorithm compared to the original study~\cite{Collins:2018epr,Collins:2019jip} that were chosen on the basis of knowledge of the signal. The $p$-values obtained are shown in Fig.~\ref{fig:pval}, for cuts at efficiency 10\%, 1\%, and 0.2\% (the solid black line is the result before any selection). We find no significant excess in BB2, but a large $5\sigma$ excess in BB1 at a resonance mass of $3500\;\mathrm{GeV}$. Fig.~\ref{fig:mjj} shows the distributions in $m_{JJ}$ obtained for the signal region centered around $3500\,\mathrm{GeV}$ for BB2 (left) and BB1 (right) after a series of cuts.

\begin{figure}[h!]
    \centering
    \includegraphics[width=\textwidth]{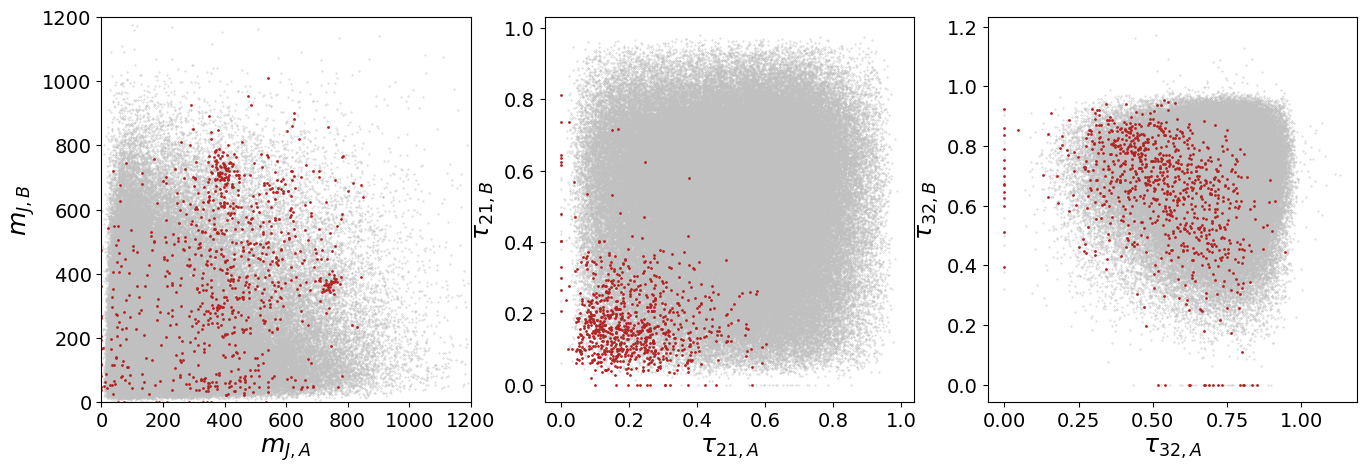}
    \caption{Substructure distributions in the anomalous BB1 signal region for signal-like (red), and background-like (grey) events. For this figure, signal-like is defined by a selection on the classifier output with efficiency 0.5\%}
    \label{fig:substructure}
\end{figure}

We can study the signal observed in BB1 in more detail by plotting substructure distributions of selected events in the anomalous signal region, Fig.~\ref{fig:substructure}. Grey points are the distribution of all events in the signal region sample, while red points are the events in that sample that have been selected by a cut on the classifier output with efficiency 0.5\%. In the leftmost plot, we see two clusters of events with jet masses of around $400,750\;\text{GeV}$ and the reverse, indicating that the two fat jets are produced from the decay of boosted particles of these masses. The middle plot indicates that the signal-like events all have small $\tau_{21}$ for both jets, indicating that they have a two-pronged structure. No strong clustering is observed in $\tau_{32}$ (right plot). 

\subsubsection{Lessons Learned}
\label{sec:lessons}

Compared to the original study~\cite{Collins:2018epr,Collins:2019jip}, we found that rescaling $m_J$ by $m_{JJ}$ is effective in sufficiently eliminating the correlation between these variables. In the original study we instead removed events with high jet mass over $500 \; \text{GeV}$, since this is where the neural networks focussed on finding these correlations and a cut on high jet masses severely distorts the QCD background shape by rejecting a very high fraction of events at low $m_{JJ}$. The same strategy applied to BB1 would have missed the signal.

Of course, the method stricty defined is clearly limited in finding signals that do not look like two fat jets with substructure, and would therefore fail in identifying the signal in BB3. We attempted to apply a modificatied version of the strategy, called `Tossed CWoLa SALAD' (a variation on CWoLa SALAD (Simulation Assisted Likelihood-free Anomaly Detection)~\cite{1815227}). In this attempt, the top ten jets in each event have recorded their 4-momenta which act as the inputs to the classifiers (with zero-padding in the case of fewer jets in an event), and the total invariant mass of the system acts as the resonant variable. The jet-momenta are rescaled by this mass in an attempt to avoid correlations. The classifiers are trained simultaneously on BB3 data and also on QCD simulation from the R\&D dataset, but in this second dataset the sideband and signal region labels are reversed (`tossed'). If the simulated background is similar to the true background, then this training strategy penalizes attempts to learn background correlations. Nonetheless, all attempts to apply this strategy and the original CWoLa SALAD strategy on BB3 led to heavy sculpting of the background $m_{JJ}$ distribution. A more global decorrelation strategy is apparently needed.

 \FloatBarrier

\subsection[CWoLa and Autoencoders: Comparing Weak- and Unsupervised methods for Resonant Anomaly Detection]{CWoLa and Autoencoders: Comparing Weak- and Unsupervised methods for Resonant Anomaly Detection\footnote{Authors: Jack H. Collins, Pablo Mart\'in-Ramiro, Benjamin Nachman, and David Shih.}}

\label{sec:cwolaaecompare}

\subsubsection{Machine Learning Setup}
\label{sec:method}

There are two techniques that show great potential at model-independent anomaly detection: Classification Without Labels (CWoLa)~\cite{Metodiev:2017vrx, Collins:2018epr, Collins:2019jip} and deep autoencoders~\cite{Farina:2018fyg, Heimel:2018mkt, Cerri:2018anq, Roy:2019jae, Blance:2019ibf}. These techniques have two important advantages over supervised methods. First, they are model independent and therefore allow to extend the sensitivity of current new physics searches to model-agnostic BSM scenarios. Second, they can learn directly from real data and thus do not rely on simulations that may suffer from potentially large mismodeling effects. In this section, we provide a comparative study between CWoLa and an autoencoder (AE) using a similar signal than the one released in the R\&D dataset: two jets with masses $m_{j_{1}} = m_{j_{2}} = 500 \; \mathrm{GeV}$. We examine the ability of the two methods to identify the new physics signal at different cross sections and to increase the significance of the signal excess. CWoLa is expected to reach an excellent performance for large amounts of signal, while the AE should show a robust performance in the limit of low signal statistics. Therefore, these two approaches may have an intersection in performance at different cross sections that would be of great interest for real experimental searches.

The R\&D dataset represents a general new physics scenario where a signal is localized in one known dimension of phase space (in this case, the dijet invariant mass $m_{JJ}$) on top of a smooth background. In this scenario, CWoLa and the AE can be trained to exploit the information in the substructure of the two jets to gain discriminating power between the signal and background events. From the full dataset, we select all of the events in the range $m_{JJ} \in [2800, 5200] \; \mathrm{GeV}$ and split them uniformly in $\log(m_{JJ})$ in $15$ bins. After selecting this range, $537304$ background events remain in our sample. We consider the following set of input features for each jet:
\begin{equation}
Y_{i} = \left\{ m_{J}, \, \sqrt{\tau_{1}^{(2)}} / \tau_{1}^{(1)}, \, \tau_{21}, \, \tau_{32}, \, \tau_{43}, \, n_{\text{trk}} \right\} \, ,
\label{eq:PMinput}
\end{equation}
where $\tau_{ij}^{(\beta)}$ represent fractions of $N$-subjettiness variables (with angular exponent $\beta = 1$ unless otherwise specified in the superscript), $n_{\text{trk}}$ denotes the number of tracks of a given jet, and jets are ordered by mass in descending order. For the autoencoder we add two extra input features for each jet: $\left\{p_{\text{T}_{1}}, \, p_{\text{T}_{2}}, \eta_{1}, \, \eta_{2}\right\}$, which lead to a significant performance improvement. For CWoLa, using these extra input features produces an undesirable correlation between the jets $p_\text{T}$ and $m_{JJ}$, which may help CWoLa learn $m_{JJ}$ and sculpt artifical bumps on this distribution in the absence of signal.

\paragraph{Classification Without Labels (CWoLa)}
\label{sec:PMCWoLa}

The strategy that we follow to implement CWoLa is similar to the approach described in Ref.~\cite{Collins:2019jip}. First, we build a signal region and a sideband region to test for a signal hypothesis with mass $m_{JJ} = m_{\text{peak}}$, where $m_{\text{peak}}$ is the mean mass of the injected signal. The signal region contains all of the events in the three bins centered around $m_{\text{peak}}$, while the sideband region includes all of the events in the two bins below and above the signal region. The width of the signal region is $435 \; \mathrm{GeV}$, and the lower and upper sidebands have a width of $262 \; \mathrm{GeV}$ and $322 \; \mathrm{GeV}$, respectively. Note that in a real search the location of the mass peak of any potential signal would be unknown, and therefore the mass hypothesis must be scanned as described in Ref.~\cite{Collins:2019jip}.

After defining these two regions, the CWoLa approach is used to train a fully supervised classifier to distinguish the events of the signal region from the events of the sideband using the set of twelve input features that describe the jet substructure of each event, presented in Eq.~\eqref{eq:PMinput}. If a signal is present in the signal region with anomalous jet substructure, CWoLa should learn the information that is useful to distinguish the signal and sideband regions. This classifier can then be used to select signal-like events, producing a new distribution in the dijet mass that may enhance the significance of the signal excess. Note that the CWoLa performance should be poor when no signal is present in the signal region; in this case, the signal and sideband regions will be statistically identical and thus the classifier should not be able to distinguish between the two regions.

The classifier that we use is a dense neural network with four hidden layers. The first layer has $64$ nodes with ReLU activation, and the second through fourth layers have $32$, $16$ and $4$ nodes respectively, with ELU activation. The output layer has a sigmoid activation. The first three hidden layers are followed by dropout layers with a $20 \, \%$ dropout rate. We use the binary cross-entropy loss function and the Adam optimizer with learning rate of $0.001$ and learning rate decay of $5 \cdot 10^{-4}$, and a batch size of $20480$. The training data is reweighted such that the two sidebands have the same total weight, the signal region has the same total weight as the sum of the sidebands, and the sum of all events weights in the training data is equal to the total number of training events. Although the two sideband regions have different event rates, this reweighting procedure ensures that they contribute equally to the training process and that the classifier output peaks around $0.5$ if no signal is present in data.

In order to reduce any potential overfitting, a $5$-fold cross-validation procedure is implemented. After standardizing all the input features, we divide each bin of the full dataset in five parts to build five samples of events of equal size. First, four of these samples are used to perform four rounds of training and validation, using three different subsets for training and one for validation each time, and the other sample is saved for testing. For each cross-validation round, ten neural networks are trained for $200$ epochs on the same training and validation data using different initializations. The performance of each classifier is measured on validation data according to the metric $\epsilon_{\text{val}}$. This metric is defined as the true positive rate for classifying signal region events as such, calculated at a threshold with a false positive rate $z = 0.5 \, \%$ for incorrectly classifying sideband region events. The best of the ten models is saved at the end of each round, and the four selected models are used to build an ensemble model, which is used to classify the events in the test set. The output of this classifier can then be used to select the $x \, \%$ most signal-like events in the test set. We repeat the same procedure for the five choices of test set and combine the signal-like event subsamples into a final signal-like sample. If a signal is present in data and CWoLa is able to find it, the selected sample of signal-like events will show an enhanced excess in the signal region on the $m_{JJ}$ plane.

\paragraph{Autoencoder}
\label{sec:PMAE}

In order to use all the available information from the events, we build two different autoencoders, Autoencoder I and II, which are trained on Jet 1 and Jet 2, respectively. Both autoencoders are trained and tested on a mixed sample of signal and background events. The reason for this is that the signal contamination ratio in the full sample for the $S/B$ benchmarks that we consider is small enough for the AE to learn the potentially anomalous feature distribution of the signal events. For each jet, we build an autoencoder ensemble that is trained on a randomly selected sample of $50000$ events for only $1$ epoch. We train twenty different models (i.e. the ensemble components) and compute the reconstruction error for each event. The final reconstruction error of an event is obtained by computing the mean over the twenty different ensemble components. The autoencoder ensembles are then used to classify events in the test set, by selecting the $x \, \%$ most signal-like events applying a simultaneous cut in the reconstruction loss of the autoencoders trained on Jet $1$ and Jet $2$. Since the autoencoders are trained mostly on background events, signal events are expected to yield larger reconstruction losses if the signal is sufficiently different to the background.

In this work, the two autoencoders that we consider are dense neural networks with seven hidden layers. The autoencoders have an input layer with $8$ nodes. The encoder has three hidden layers of $64$, $32$ and $16$ nodes, and is followed by a bottleneck layer with $1$ node and linear activation. Finally, the decoder has 3 hidden layers of $16$, $32$ and $64$ nodes. All of the hidden layers have ReLU activation. The output layer is made of $8$ nodes and has linear activation. We use the Minimum Squared Error (MSE) loss function, the Adam optimizer with learning rate of $0.001$ and a batch size of $5120$. We standardize all the input features from the training and test sets using training information only.

\subsubsection{Results on LHC Olympics}
\label{sec:results}

The goal of this work is to compare the performance of CWoLa and the AE at different cross sections. For this purpose, we define a set of eight benchmarks with the same number of background events and different amounts of injected signal events. In particular, we consider a set of benchmarks distributed over the range $S/B \in [1.2 \cdot 10^{-3}, 6 \cdot 10^{-3}]$ in the signal region. To test the consistency of both models in the absence of signal, we consider a final benchmark with no signal events. For each $S/B$ benchmark, we present results averaged over five independent runs using a random subset of signal events each time. The $S/B$ range that we consider is key to observe the complementarity of the two methods for different amounts of signal, and the observed behaviors continue beyond these limits.

We analyze the performance of CWoLa and the AE according to two different metrics. First, we measure the performance of the two methods according to the AUC metric. The AUC score is computed using all the available signal events to reduce any potential overfitting. Second, we compare the performance of CWoLa and the AE at increasing the significance of the signal region excess. For this purpose, we use the following $4$-parameter function~\cite{Aad:2019hjw, Sirunyan:2018xlo} to fit the smooth background distribution: $d\sigma / dm_{JJ} = (p_{0}(1-x)^{p_{1}}) / (x^{p_{2}+p_{3}\ln(x)})$. This function is used to estimate the background density outside of the signal region and then the fit result is interpolated into the signal region. The number of expected and observed events in the signal region are compared and p-value is calculated to evaluate the significance of any potential excess.

\begin{figure}[h!]
\centering
\includegraphics[scale=0.45]{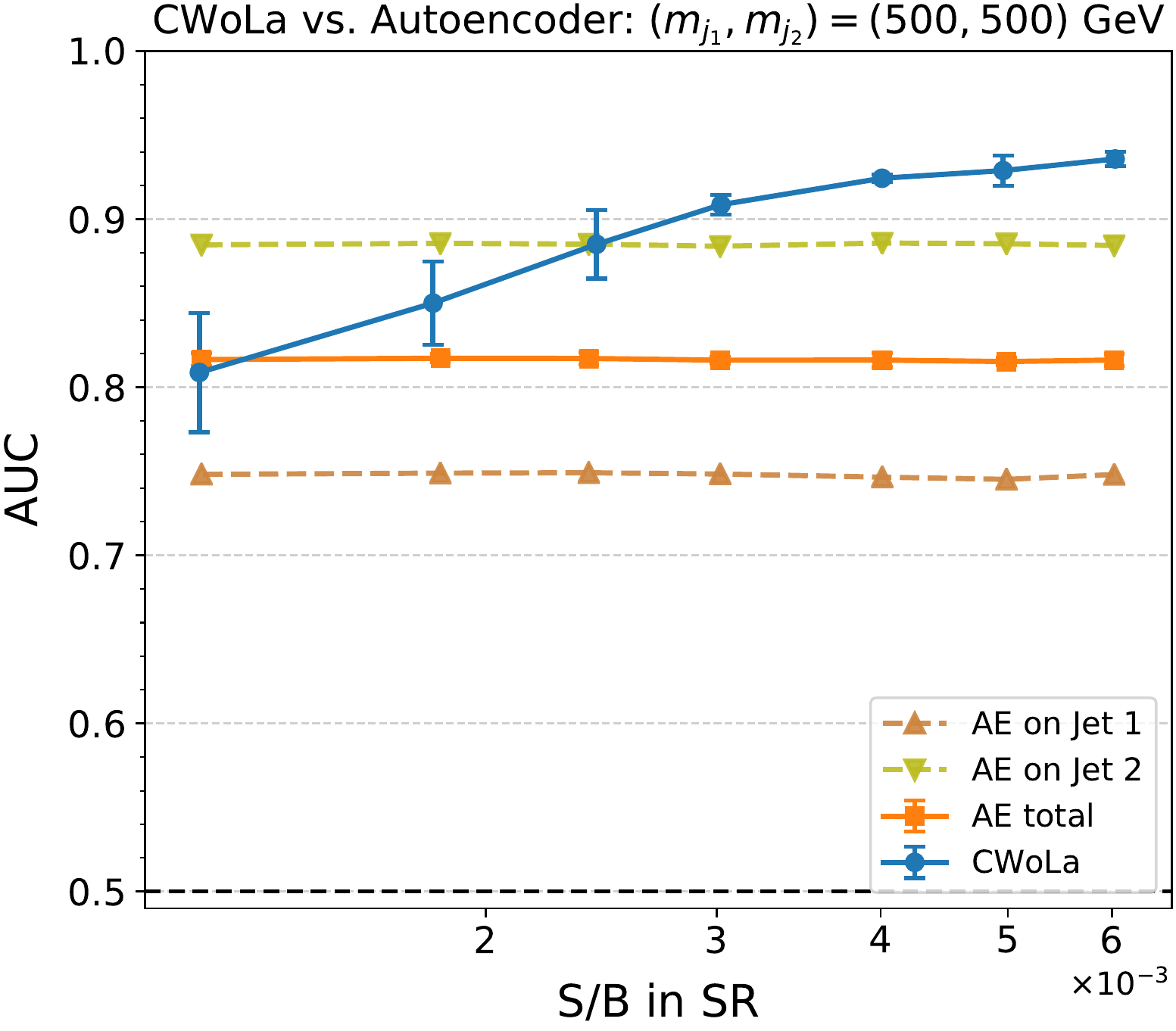}
\hspace{5pt}
\includegraphics[scale=0.45]{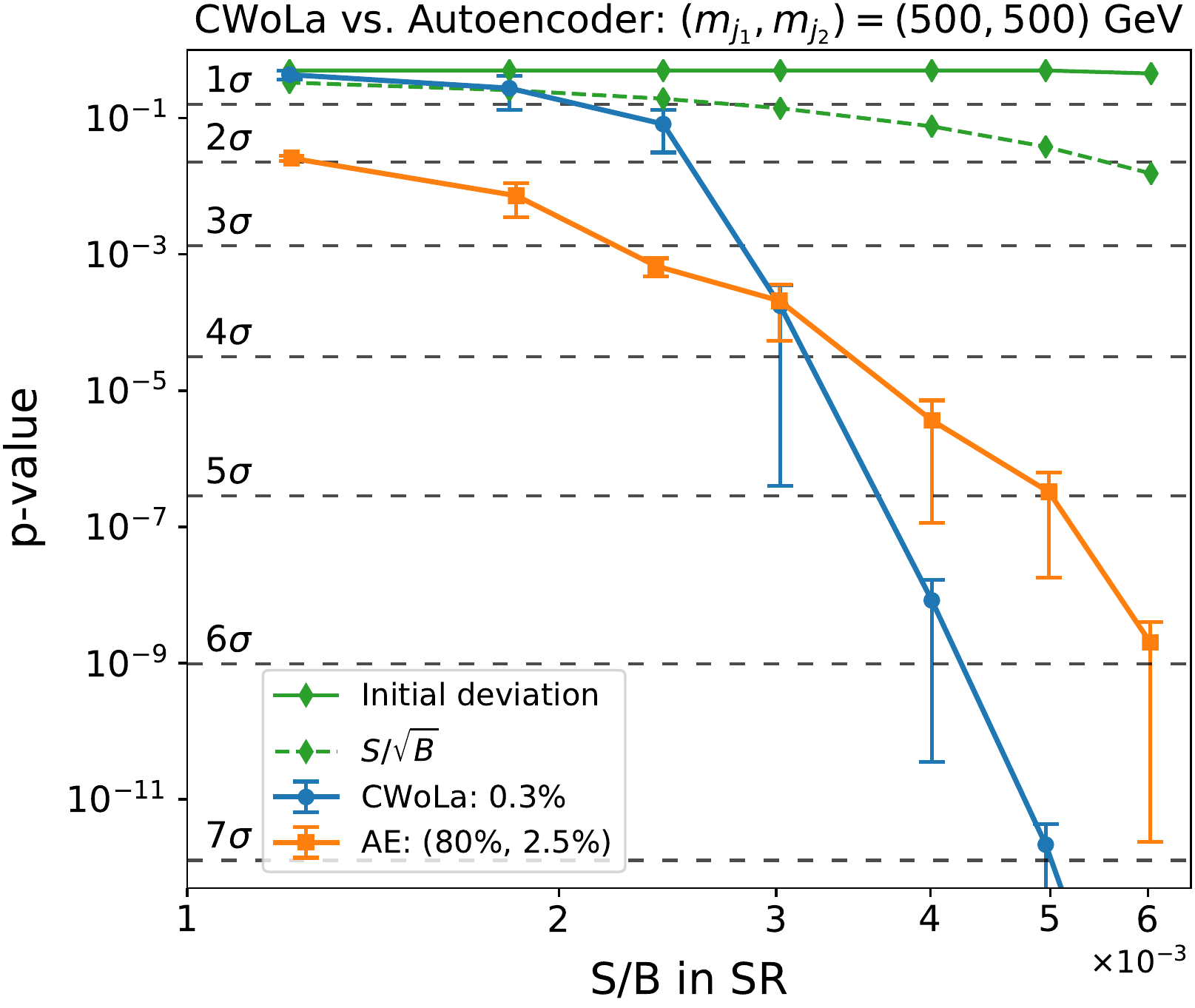}
\caption{\textbf{Left plot:} Performance of CWoLa (blue), the Autoencoder trained on Jet 1 (brown) and Jet 2 (green), and their average (orange), as measured by the AUC metric. The error bars denote the standard deviation on the AUC metric. \textbf{Right plot:} Significance of the signal region excess after applying different cuts for CWoLa (blue) and the Autoencoder (orange). The best cuts for CWoLa and the AE ensemble correspond to the $0.3 \%$ and the (Jet 1, Jet 2) = $(80 \, \%, 2.5 \, \%)$ event selections, respectively. The initial significance of the excess ($100 \, \%$ selection) is shown in green. Note that the fit to the raw distribution (i.e. no cut applied) is lower than the naive expected significance $S/\sqrt B$ due to a downward fluctuation in the number of background events in the signal region.}
\label{fig:PMCWoLa_vs_AE}
\end{figure}

We present results showing the performance of CWoLa and the AE for different $S/B$ ratios according to the two previously defined metrics in Fig.~\ref{fig:PMCWoLa_vs_AE}. The left plot shows results for the AUC metric, while the right plot shows the models performance at increasing the significance of the signal region excess. First, the AUC metric shows that CWoLa achieves very good discrimination power between signal and background events for large $S/B$ ratios, reaching AUC values above $0.90$ and approaching the $0.98$ score from a fully supervised classifier. As the amount of injected signal in the signal region is decreased, the amount of useful information that allows CWoLa to discriminate between the signal and sideband regions during training is reduced. As a consequence, the classifier struggles to learn the signal features and its performance drops in testing. By contrast, the AE shows a solid performance in the full $S/B$ range. This is caused by the fact that once the AE learns to reconstruct the event sample, its performance remains independent of the amount of signal present in this sample as long as the contamination ratio is sufficiently small. Interestingly, the performance of the AE trained on Jet 2 is superior to the one trained on Jet 1, which suggests that using full event information can be very important. Note that the AUC scores from CWoLa and the AE cross at $S/B \sim 3 \cdot 10^{-3}$.

The p-values analysis shows two interesting patterns. First, CWoLa is able to enhance the significance of the signal regions excess by $3 \sigma - 8 \sigma$ for $S/B$ ratios above $\sim 3 \cdot 10^{-3}$, even when the fit to the full event sample shows no deviation from the background-only hypothesis. Second, the AE shows a superior performance below this range, increasing the significance of the excess by at least $2 \sigma - 3 \sigma$ in the low $S/B$ region where CWoLa is not sensitive to the signal. Crucially, there is again an intersection in the performance of CWoLa and the AE as measured by their ability to enhance the significance of the signal region excess. Therefore, our results show that the two methods are complementary for resonant anomaly detection depending on the amount of signal.

\subsubsection{Lessons Learned}
\label{sec:lessons}

We have compared weakly-supervised and unsupervised anomaly detection methods in a fully hadronic dijet resonance search in the context of the LHC Olympics $2020$. We used CWoLa and deep autoencoders as representative models of the two classes, and examined their ability to identify the signal and enhance the sensitivity of the signal excess at different cross sections. Our results demonstrate that CWoLa is very effective for sizable amounts of signal, increasing the significance of a negligible excess above the $5\sigma$ discovery limit. The AE showed a solid performance at low signal rates, raising the significance of the excess by up to $3\sigma$ in a region where CWoLa was not sensitive to the signal. Therefore, both techniques are complementary and can be used together for anomaly detection at the LHC and beyond.

We feel the LHC Olympics $2020$ has been a very enriching experience that allowed us to deepen our understanding of machine learning methods for LHC physics and learn more about the work that has been done in this field. We really hope to repeat this experience next year.

 \FloatBarrier
\subsection[Tag N' Train]{Tag N' Train\footnote{Authors: Oz Amram and Cristina Mantilla Suarez.  Code to reproduce all of our results can be found on \href{https://github.com/OzAmram/TagNTrain}{https://github.com/OzAmram/TagNTrain}.}}
\label{sec:tnt}
\subsubsection{Method}

\noindent 
Tag N' Train \cite{Amram:2020ykb} is a technique to train classifiers on unlabeled events that is naturally employed in an anomaly search.
The Tag N' Train (TNT) approach  is based on the premise that signal events contain two or more anomalous objects (hereafter called Object-1 and Object-2)
in them that can be used independently for classification.
If this is the case, one can use the Object-1's in each event to \textit{tag} examples as signal-like or background-like. 
These signal-rich and background-rich samples can then be used to \textit{train} a classifier for Object-2. 
This training step uses the Classification Without Labels (CWoLa) method \cite{Metodiev:2017vrx}, in which a classifier is trained by using mixed samples of signal and background rather than fully labeled events.
One can then repeat the procedure to train a classifier for Object-1 as well. 

\begin{figure}[h!]
\centering
\includegraphics[width=0.8\textwidth]{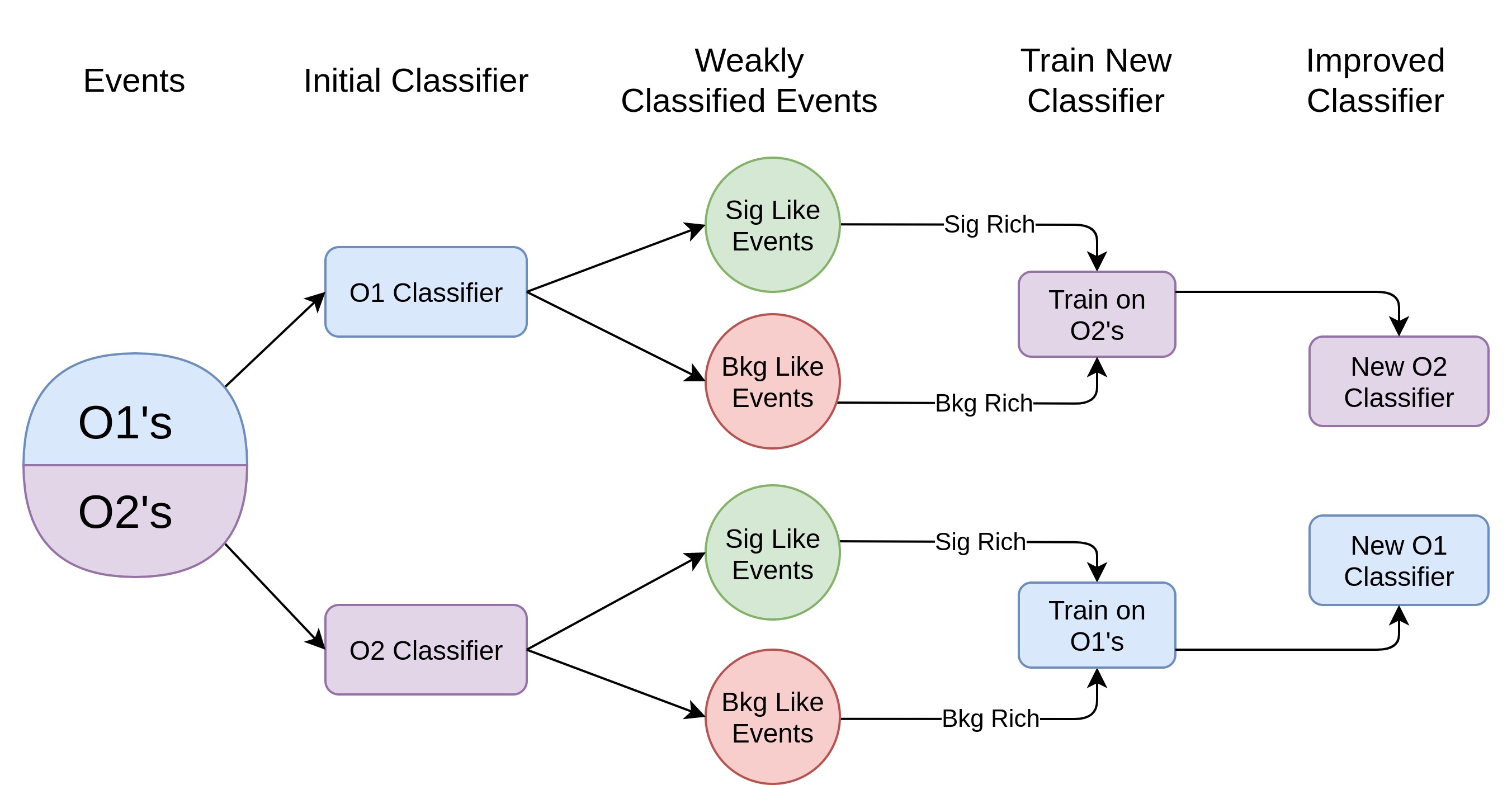}
\caption{An illustration of the Tag N’ Train technique. Here O1 and O2 represent Object-1 and
Object-2, the two components of the data one wishes to train classifiers for.}
\label{fig:TNT_algo}
\end{figure}

In order to perform the initial tagging, one must be able to at least weakly classify the anomalous objects to begin with, and so the technique must be seeded by initial classifiers.
In a jet-based anomaly search, autoencoders can be used as the initial classifiers because 
they were previously shown to be effective unsupervised classifiers of anomalous jets \cite{Farina:2018fyg,Heimel:2018mkt} . 
Overall, TNT takes as input a set of unlabeled data events and two initial classifiers, and outputs two new classifiers designed to have improved performance.
Because the technique works better if the initial classifier can create a larger separation between signal and background in the mixed samples, multiple iterations of this technique (where the output classifiers are used with a new data sample to train new classifiers) can further improve classification performance until a plateau is reached.
The technique is summarized graphically in Fig.~\ref{fig:TNT_algo}.

The usage of TNT in an anomaly search requires data events to be partitioned into three subsets. 
The first subset is used to train the autoencoders, the second subset is used to perform the TNT technique that trains improved classifiers, which are then used on the third subset to select anomalous events and search for a signal. 
A nested cross validation approach, where the different subsets are swapped from being used for training or searching, can be used in order to achieve maximum sensitivity. 

Our search only targeted dijet resonances, where we took the two highest p$_T$ jets as the dijet candidate.
In order to apply the Tag N' Train technique, we treat our Object-1 as the more massive jet and Object-2 as the less massive jet of these two jets.
We found that one can incorporate the assumption of a resonant signal by requiring signal-like events fall in particular dijet mass window and scanning this window over the full range during a search (as in \cite{Collins:2018epr,Collins:2019jip}). 
This requirement helps to better isolate resonant signals and improves the performance of the resulting classifier. 
Our implementation of the TNT based anomaly search used jet images as the inputs for both the autoencoders and the TNT classifiers and CNN based model architectures trained with the Adam optimizer. 
We chose a latent size of 6 for the autoencoder based on results of the previous studies in the literature \cite{Farina:2018fyg,Heimel:2018mkt}.
Based on results on the R\&D Dataset we found that the second iteration of the Tag N' Train technique generally reached the plateau performance and so we used 2 iterations in our search.
No optimization of the model architectures and optimizer hyperparameters was attempted.
A rough optimization of the selection of signal-like and background-like samples in the TNT technique was performed using the R\&D dataset.
In the first iteration, we used the 40\% of events with the lowest autoencoder reconstruction losses as the background-like sample and the 20\% with the highest as signal-like sample during the first iteration.
In the second iteration, we once again used the 40\% of events with the lowest scores as the background-rich sample, but tightened the signal-like cut to the top 10\% of events. 
On the R\&D dataset we found the performance was quite insensitive to the exact background-like cut used (as the resulting sample was always nearly pure background) and moderately sensitive to the signal-like cut used. 

On the Blackboxes we used 200k events to train the autoencoders, 400k to run Tag N' Train (200k for each iteration) and searched for a signal in remaining 400k events. 
Due to limited computational resources, we did not run the full cross validation, but rather switched the 400k events used for training and searching and kept the same autoencoders. 
Thus only 800k out of the 1M events were actually used to determine the significance of the anomaly. 
We used the alteration of TNT that assumes a resonance by requiring signal events fall in a dijet mass window and scanned over the dijet mass range of 3000 to 5000 with window sizes of 500 GeV.
In searching for a signal, we selected events where both jets scores were in the top 3\% most signal-like (for an overall efficiency of roughly 0.1\%) and then did a bump hunt.
We generally found that cutting as tightly as possible while still having enough statistics for a stable fit maximized our sensitivity.
We did not mix events from the two sub-samples but rather fit each simultaneously to produce a single p-value.

\subsubsection{Results on LHC Olympics}
\label{sec:results}

\noindent On the R\&D dataset we compared the performance of the Tag N' Train classifiers to autoencoders and the CWoLa hunting \cite{Collins:2018epr,Collins:2019jip} method for various amounts of signal in the dataset 
(9\%, 1\%, 0.3\% and 0.1\% of the total dataset respectively).  
We generally found the Tag N' Train approach to be competitive with these other methods.
For the 1\% signal test, TNT produced a classifier that is somewhat worse than the one produced with TNT with an additional dijet mass cut (TNT + M$_{jj}$), 
but still had significantly improved performance with respect to the autoencoder.
For the 0.3\% and 0.1\% signal tests, there was too little signal for the TNT classifier to learn from, and TNT performs significantly worse than the autoencoder. 
The TNT + M$_{jj}$ classifier performs similarly to the one trained using CWoLa hunting for the 3 tests with larger signal.
For the 0.1\% test the TNT + M$_{jj}$ classifier is able to achieve better performance better than that of the CWoLa hunting method, but does not improve with respect to the autoencoders approach.
More details along with ROC curves are in the TNT paper \cite{Amram:2020ykb}.

\begin{figure}[h!]
\centering
\includegraphics[width=0.5\textwidth]{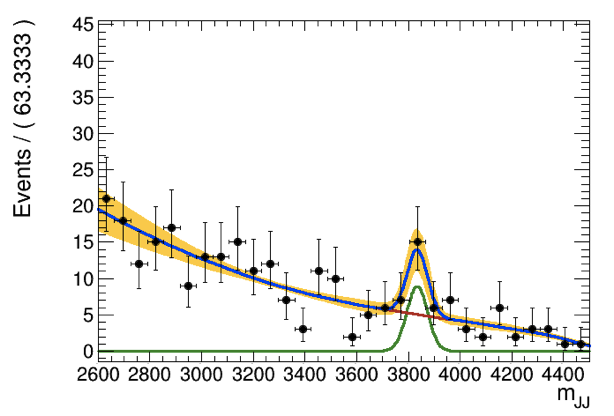}
\caption{Events in the first data subset after final selection for Blackbox 1. The signal peak can be seen slightly above 3800 GeV. 
The local p-value for just this subset of the data was around 3$\sigma$. }
\label{fig:BB1_bump}
\end{figure}

When applying the Tag N' Train search to Blackbox 1 we found a resonance at around 3800 $\pm$ 50 GeV with a local significance of 4$\sigma$. 
The bump-hunt plot for one of the subset is shown in Fig.~\ref{fig:BB1_bump}.

We had difficulty characterizing the nature of the signal as that was not extensively tested on the R\&D dataset. 
We reported that one of the resonance's daughters had a mass of 270 $\pm$ 40 GeV (this was meant to be the lighter daughter) and did a very rough guess of the total number of signal events present. 

When doing an initial run over Black boxes 2 and 3 we did not see any significant evidence of a signal
and we did not revisit Blackboxes 2 and 3 once the results of Blackbox 1 was revealed.  

\subsubsection{Lessons Learned}
\label{sec:lessons}

\noindent
It is not surprising our technique was not able to find the signal in Blackbox 3 because our implementation of TNT focused only dijet resonances where both jets had anomalous substructure
while Blackbox 3 had 3-jet decays and its dijet decays had gluon-jets.

We were happy to find the correct resonance on Blackbox 1 but had difficulty characterizing the signal. 
Because we were using jet images and CNN's it was not straightforward to interpret what our models had learned was signal-like. 
For future studies it may be interesting to explore using more sophisticated techniques to attempt to understand what a model like a CNN has learned, or use models with higher level features
that are more interpretable. 
Additionally, we tried plotting the distribution of jet masses of signal-like events, but we knew that our technique distorted the jet mass distribution (selecting higher jet masses as preferentially signal-like).
However we had not extensively studied this effect, making it difficult to extract the signal jet masses from the distributions of most signal-like events.
We think with more deliberate study of these effects and/or using more interpretable model architectures, 
characterizing basic aspects of the signal (number of prongs, jet masses, etc.) should be possible.
What poses a more significant challenge is trying to estimate the signal cross section (total amount of signal present in the dataset) with an anomaly detection search that features
a cut meant to isolate signal events. 
One can always set a lower bound based on the estimated number of signal events in the final fit, however because these events are selected with quite a low selection efficiency it will usually be a
poor lower bound. 
Without specifying a particular model, one cannot know the signal efficiency of the selection imposed so it is difficult to estimate how far this lower bound is from the true amount of signal. 
An approach that could be taken would be to try to calibrate the sensitivity of a technique in mock experiments on simulated datasets where the amount of signal is known. 
However it is likely that such a calibration, a mapping from observed p-value to total amount of signal present, depends on the nature of the signal and will not be universal. 
Some signals (e.g. those containing more exotic substructures) may be easier to find than others. 
Thus such a procedure would face difficult to estimate modeling uncertainties even if performed after signal characterization has been attempted.

 \FloatBarrier
\subsection[Simulation Assisted Likelihood-free Anomaly Detection]{Simulation Assisted Likelihood-free Anomaly Detection\footnote{Authors: Anders Andreassen, Benjamin Nachman, and David Shih.  The code can be found at \url{https://github.com/bnachman/DCTRHunting}.}}

\label{sec:salad}

While learning directly from data can mitigate model biases, it is also useful to incorporate information from background simulations.  These simulations are only an approximation to the Standard Model, but they include a wealth of physics knowledge at all energy scales relevant for collider reactions.  This section describes an approach that uses a background simulation in a way that depends as little as possible on the simulations.  In particular, a neural network based on the \textit{Deep neural networks using Classification for Tuning and Reweighting} (\textsc{Dctr}) protocal~\cite{Andreassen:2019nnm} is trained in a region of phase space that is largely devoid of signals.  In a resonance search, this region can be isolated using sidebands in the resonant feature.  The reweighting function morphs the simulation into the data and is parameterized in the resonant feature(s).  The model is then interpolated to the signal region region and the reweighted background simulation can be used for both enhancing signal sensitivity and estimating the background.  As deep learning classifiers can naturally probe high dimensional spaces, this reweighting model can in principle exploit the full phase space for both enhancing signal sensitivity and estimating the Standard Model background.

\subsubsection{Method}

Let $m$ be a feature (or set of features) that can be used to localize a potential signal in a signal region (SR).  Furthermore, let $x$ be another set of features which are useful for isolating a potential signal.  For the LHC Olympics, $m$ will be the invariant mass of two jets and $x$ includes information about the substructure of the two jets.   The Simulation Assisted Likelihood-free Anomaly Detection (\textsc{Salad})~method then proceeds as follows:

\begin{enumerate}
\item Train a classifier $f$ to distinguish data and simulation for $m\not\in\text{SR}$.  This classifier is parameterized in $m$ by simply augmenting $x$ with $m$, $f=f(x,m)$~\cite{Cranmer:2015bka,Baldi:2016fzo}.  If $f$ is trained using the binary cross entropy or the mean squared error loss, then asymptotically, a weight function $w(x|m)$ is defined by 

\begin{align}
\label{eq:weightfunction}
w(x|m)\equiv \frac{f(x)}{1-f(x)} = \frac{p(x|\text{data})}{p(x|\text{simulation})}\times\frac{p(\text{data})}{p(\text{simulation})},
\end{align}

\noindent where the last factor in Eq.~\ref{eq:weightfunction} is an overall constant that is the ratio of the total amount of data to the total amount of simulation.  This property of neural networks to learn likelihood ratios has been exploited for a variety of full phase space reweighting and parameter estimation proposals in high energy physics (see e.g.~\cite{Andreassen:2019nnm,Brehmer:2018hga,Brehmer:2018eca,Brehmer:2018kdj,Cranmer:2015bka,Andreassen:2019cjw}).

\item Simulated events in the SR are reweighted using $w(x|m)$.  The function $w(x|m)$ is interpolated automatically by the neural network.  A second classifier $g(x)$ is used to distinguish the reweighted simulation from the data.  This can be achieved in the usual way with a weighted loss function such as the binary cross-entropy:

\begin{align}
\text{loss}(g(x))=-\sum_{m_i\in\text{SR}_\text{data}} \log g(x_i)-\sum_{m_i\in\text{SR}_\text{simulation}} w(x_i|m_i)\log (1-g(x_i)).
\end{align}

\noindent Events are then selected with large values of $g(x)$.  Asymptotically\footnote{Sufficiently flexible neural network architecture, enough training data, and an effective optimization procedure.}, $g(x)$ will be monotonically related with the optimal classifier:

\begin{align}
\frac{g(x)}{1-g(x)}\propto \frac{p(x|\text{signal+background})}{p(x|\text{background})}.
\end{align}

\noindent It is important that the same data are not used for training and testing.  The easiest way to achieve this is using different partitions of the data for these two tasks.  One can make use of more data with a cross-validation procedure~\cite{Collins:2018epr,Collins:2019jip}.

\item One could combine the previous step with a standard data-driven background estimation technique like a sideband fit or the ABCD method.  However, one can also directly use the weighted simulation to predict the number of events that should pass a threshold requirement on $g(x)$:

\begin{align}
\label{eq:backgroundprediction}
N_\text{predicted}(c)=\sum_{m_i\in\text{SR}_\text{simulation}} w(x_i|m_i)\mathbb{I}[g(x_i)>c],
\end{align}

for some threshold value $c$ and where $\mathbb{I}[\cdot]$ is the indicator function that is one when its argument is true and zero otherwise.  The advantage of Eq.~\ref{eq:backgroundprediction} over other data-based methods is that $g(x)$ could be correlated with $m$; for sideband fits, thresholds requirements on $g$ cannot sculpt local features in the $m$ spectrum.

\end{enumerate}

\subsubsection{Results on LHC Olympics}

\begin{figure}[h!]
\centering
\includegraphics[width=0.45\textwidth]{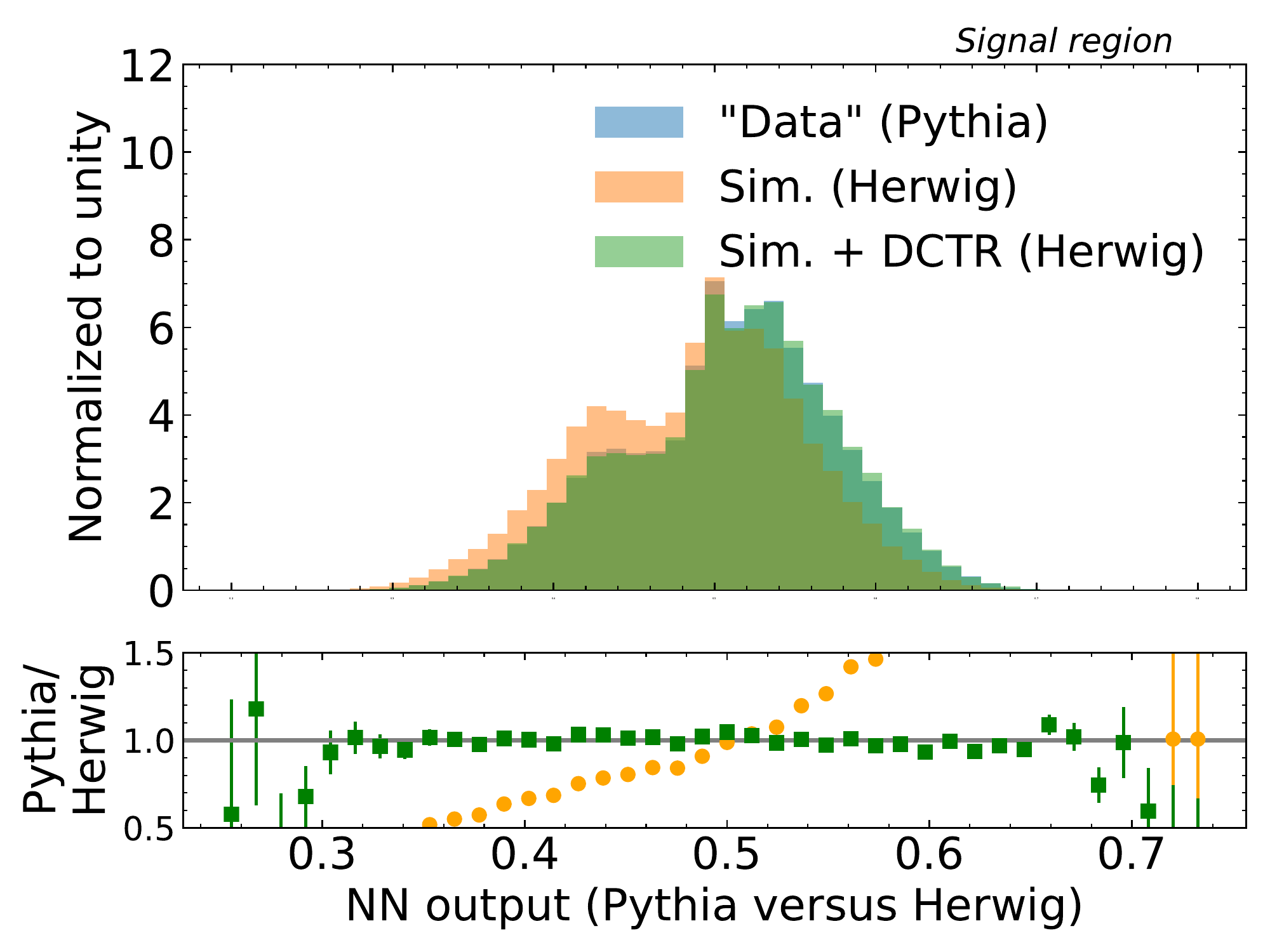}
\includegraphics[width=0.45\textwidth]{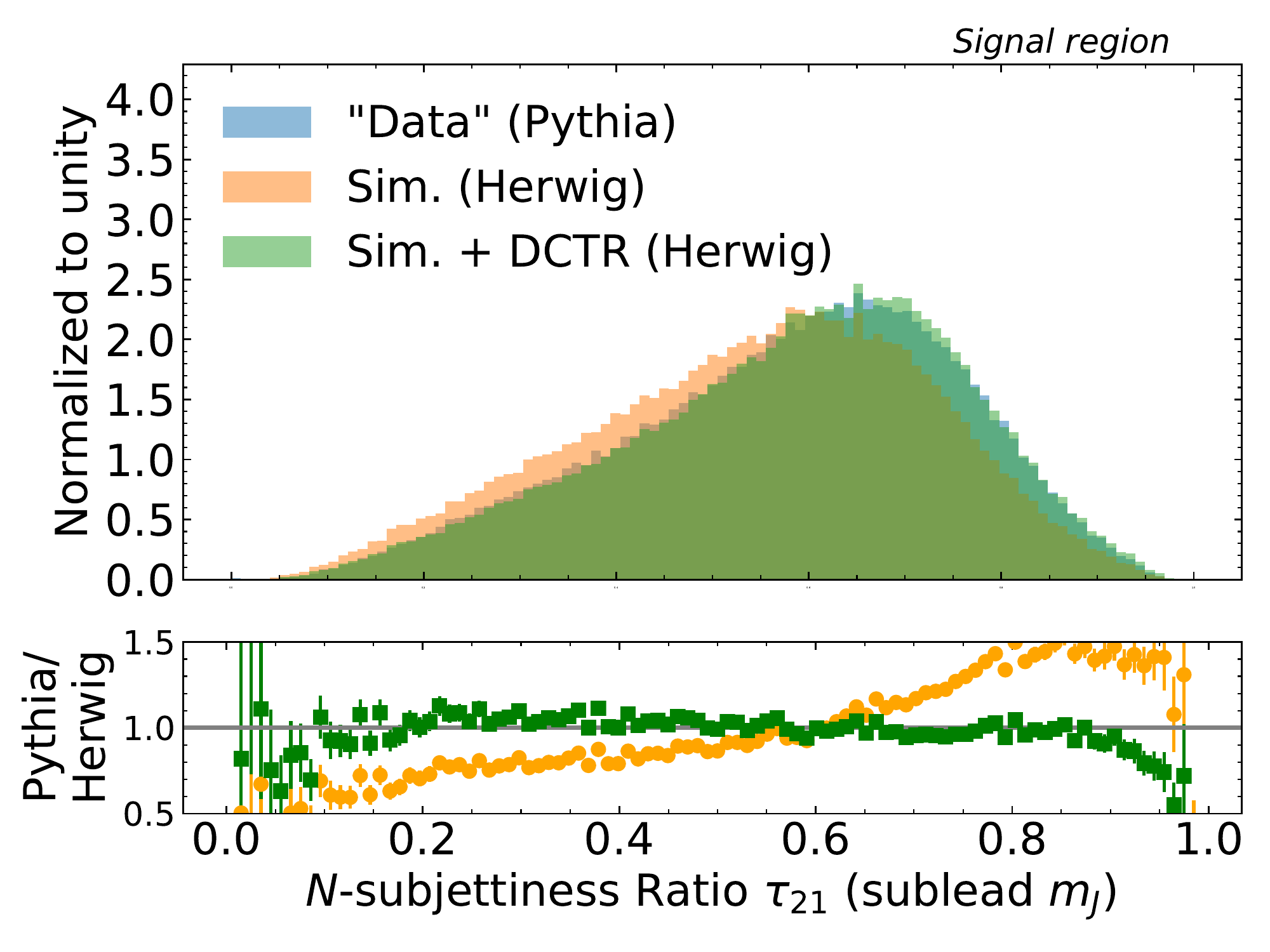}
\caption{A histogram of the classifier output (left) and the subleading $\tau_{21}$ (right) for a neural network trained to distinguish `data' (Pythia) and `simulation' (Herwig) in the signal region. The ratio between the `simulation' (Herwig) or `simulation + \textsc{Dctr}' and `data' (Pythia) is depicted by orange circles (green squares) in the lower panels.  Figure from Ref.~\cite{Andreassen:2020nkr}.}
\label{fig:dctrmodel_SR}
\end{figure}

The R\&D dataset was used for the results presented in this section. The first step of the \textsc{Dctr}~reweighting procedure is to train a classifier to distinguish the `data' (Pythia) from the `simulation' (Herwig) in a sideband region.  The next step for \textsc{Salad}~is to interpolate the reweighting function.  The neural network is trained conditional on $m_{jj}$ and so it can be evaluated in the SR for values of the invariant mass that were not available during the network training.  Note that the signal region must be chosen large enough so that the signal contamination in the sideband does not bias the reweighting function.  Figure~\ref{fig:dctrmodel_SR} shows a classifier trained to distinguish `data' and 'simulation' in the signal region before and after the application of the interpolated \textsc{Dctr}~model as well as $\tau_{21}$.  As expected, the neural network is a linear function of the likelihood ratio (as seen in the ratio), but this closure is excellent after the interpolated reweighting.

After reweighting the signal region to match the data, the next step of the search is to train a classifier to distinguish the reweighted simulation from the data in the signal region.  If the reweighting works exactly, then this new classifier will asymptotically learn $p(\text{signal}+\text{background})/p(\text{background})$.  If the reweighting is suboptimal, then some of the classifier capacity will be diverted to learning the residual difference between the simulation and background data.  If the reweighted simulation is nothing like the data, then all of the capacity will go towards this task and it will not be able to identify the signal.  There is therefore a tradeoff between how different the (reweighted) simulation is from the data and how different the signal is from the background.  If the signal is much more different from the background than the simulation is from the background data, it is possible that a sub-optimally reweighted simulation will still be able to identify the signal.

\begin{figure}[h!]
\centering
\includegraphics[width=0.45\textwidth]{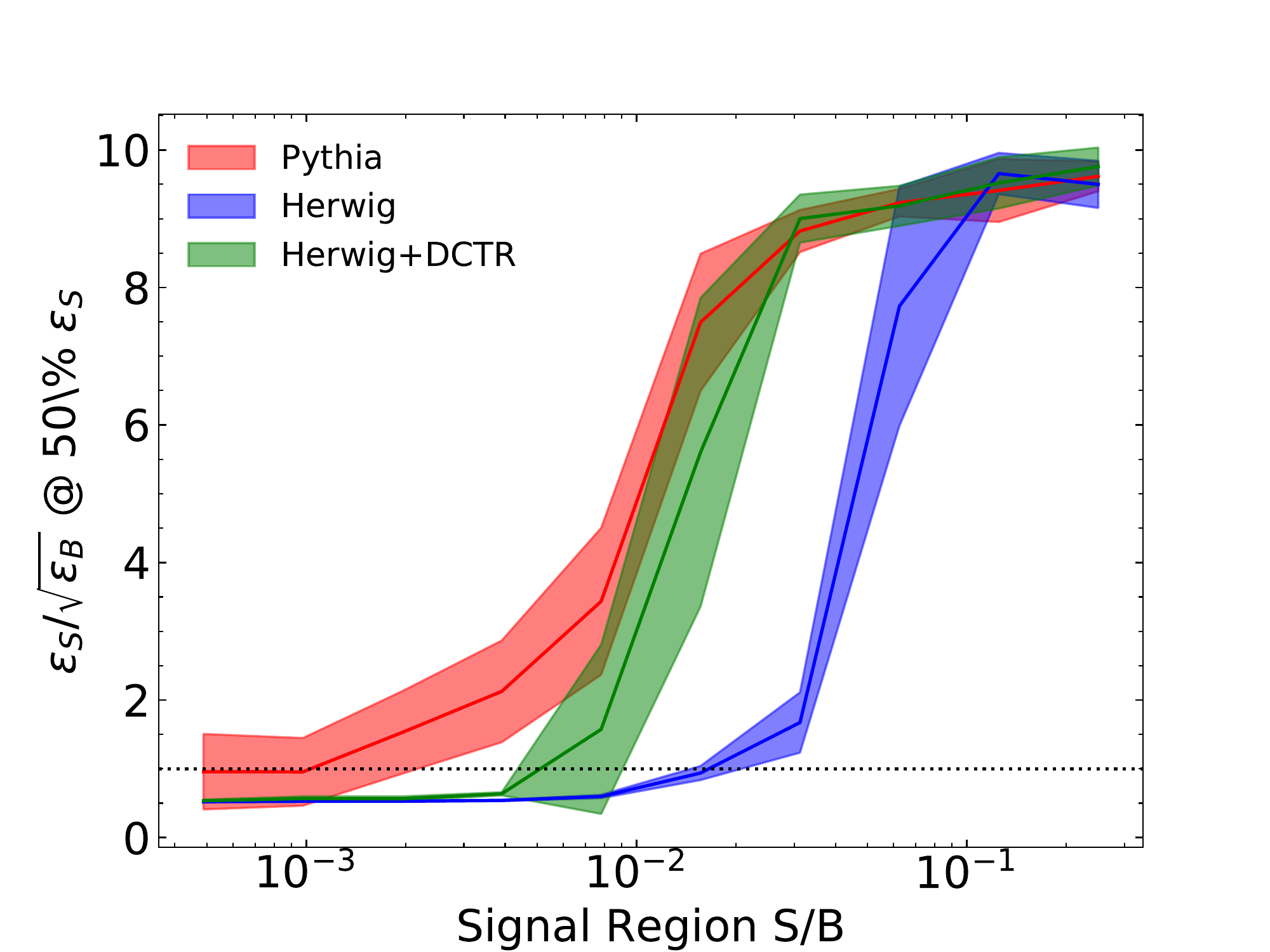}
\includegraphics[width=0.45\textwidth]{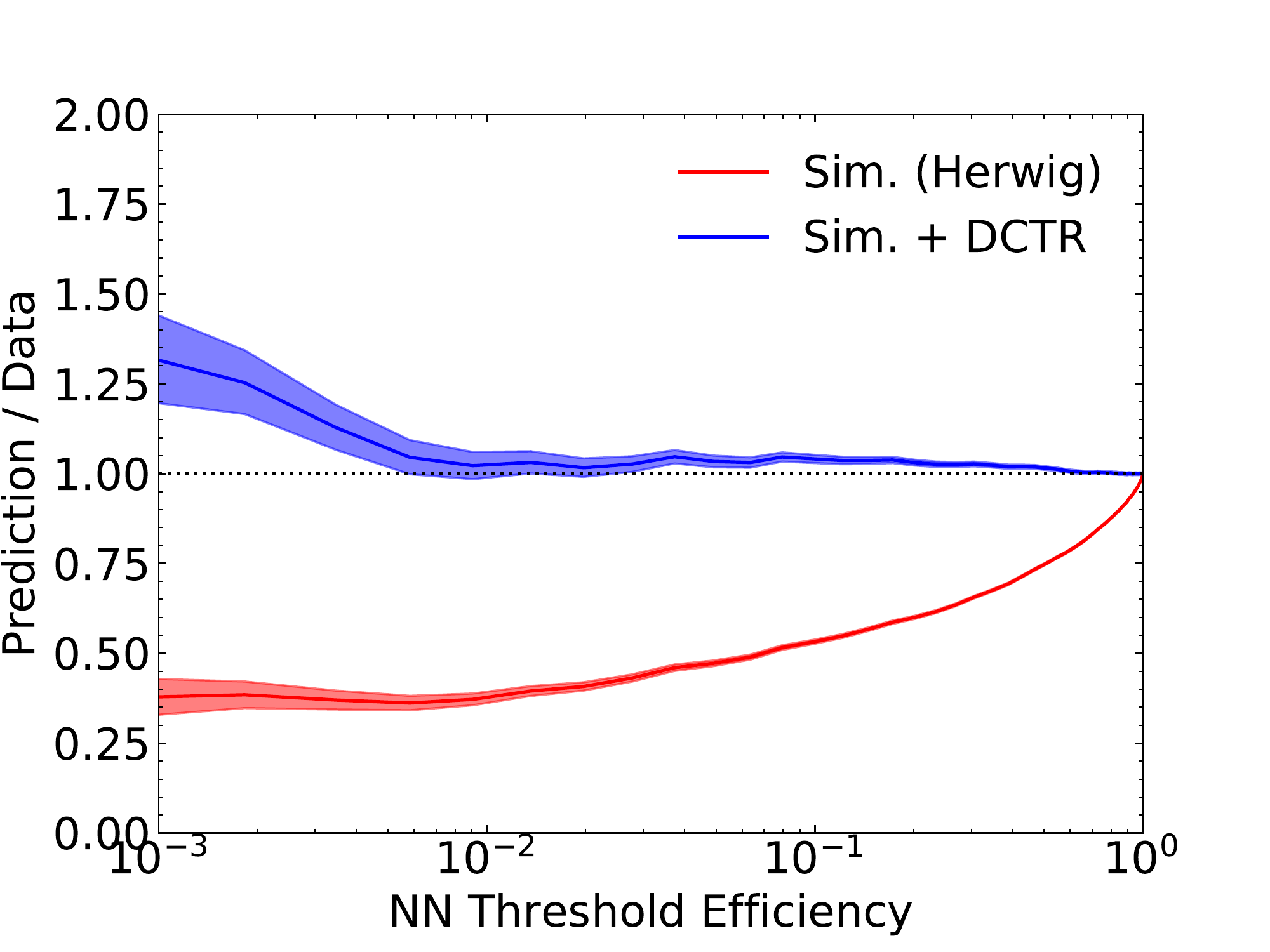}
\caption{Left: the significance improvement at the a fixed 50\% signal efficiency as a function of the signal-to-background ratio ($S/B$) in the signal region.  The evaluation of these metrics requires signal labels, even though the training of the classifiers themselves do not have signal labels.  Error bars correspond to the standard deviation from training five different classifiers.  Each classifier is itself the truncated mean over ten random initializations.  Right: The predicted efficiency normalized to the true data efficiency in the signal region for various threshold requirements on the NN.  The $x$-axis is the data efficiency from the threshold.  The error bars are due to statistical uncertainties.  Figure from Ref.~\cite{Andreassen:2020nkr}.}
\label{fig:sensitivity}
\end{figure}

Figure~\ref{fig:sensitivity} shows the sensitivity of the \textsc{Salad}~tagger to signal as a function of the signal-to-background ratio ($S/B$) in the signal region.  In all cases, the background is the QCD simulation using Pythia.  The Pythia lines correspond to the case where the simulation follows the same statistics as the data ($=$ Pythia).  When the $S/B\sim\mathcal{O}(1)$, then the performance in Fig.~\ref{fig:sensitivity} is similar to a fully supervised classifier.  As $S/B\rightarrow 0$, the Pythia curves approach the random classifier, with a max significance improvement of unity.  The significance improvement quickly drops to unity for Herwig when $S/B\lesssim 1\%$, indicating the the network is spending more capacity on differentiating Pythia from Herwig than finding signal.   \textsc{Salad}~significantly improves the performance of the Herwig-only approach.  In particular, the \textsc{Salad}~tagger is effective to about $S/B\lesssim 0.5\%$, whereas the Herwig-only tagger is only able to provide useful discrimination power down to about $S/B\sim 1\%$.   

The performance gains can be combined with a sideband background estimation strategy, as long as threshold requirements on the classifier do not sculpt bumps in the $m_{jj}$ spectrum.  However, there is also an opportunity to use \textsc{Salad}~to directly estimate the background from the interpolated simulation.  The right plot of Fig.~\ref{fig:sensitivity} illustrates the efficacy of the background estimation for a single classifier trained in the absence of signal.  Without the \textsc{Dctr}~reweighting, the predicted background rate is too low by a factor of two or more below 10\% data efficiency.  With the interpolated reweighting function, the background prediction is accurate within a few percent down to about 1\% data efficiency.

\subsubsection{Lessons Learned}

In practice, the difficulty in using \textsc{Salad}~to directly estimate the background is the estimation of the residual bias.  One may be able to use validation regions between the signal region and sideband region, but it will never require as much interpolation as the signal region itself.  One can rely on simulation variations and auxiliary measurements to estimate the systematic uncertainty from the direct \textsc{Salad}~background estimation, but estimating high-dimensional uncertainties is challenging~\cite{Nachman:2019dol,Nachman:2019yfl}.   With a low-dimensional reweighting or with a proper high-dimensional systematic uncertainty estimate, the parameterized reweighting used in \textsc{Salad}~should result in a lower uncertainty than directly estimating the uncertainty from simulation.  In particular, any nuisance parameters that affect the sideband region and the signal region in the same way will cancel when reweighting and interpolating.

While the numerical \textsc{Salad}~results presented here did not fully achieve the performance of a fully supervised classifier trained directly with inside knowledge about the data, there is room for improvement.  In particular, a detailed hyperparameter scan could improve the quality of the reweighting.  Additionally, calibration techniques could be used to further increase the accuracy~\cite{Cranmer:2015bka}.  Future work will investigate the potential of \textsc{Salad}~to analyze higher-dimensional feature spaces as well as classifier features that are strongly correlated with the resonant feature.  It will also be interesting to compare \textsc{Salad}~with other recently proposed model independent methods.  When the nominal background simulation is an excellent model of nature, \textsc{Salad}~should perform similarly to the methods presented in Ref.~\cite{DAgnolo:2018cun,DAgnolo:2019vbw} and provide a strong sensitivity to new particles.  In other regimes where the background simulation is biased, \textsc{Salad}~should continue to provide a physics-informed but still mostly background/signal model-independent approach to extend the search program for new particles at the LHC and beyond.

 \FloatBarrier
\subsection[Simulation-Assisted Decorrelation for Resonant Anomaly Detection]{Simulation-Assisted Decorrelation for Resonant Anomaly Detection\footnote{Authors: Kees Benkendorfer, Luc Le Pottier, Benjamin Nachman.  The code can be found at \url{https://github.com/bnachman/DCTRHunting}.}}

\label{sec:sacwola}

In this section, two weakly supervised approaches are studied: Classification without Labels (\textsc{CWoLa})~\cite{Metodiev:2017vrx,Collins:2018epr,Collins:2019jip,collaboration2020dijet} and Simulation Assisted Likelihood-free Anomaly Detection (\textsc{Salad})~\cite{Andreassen:2020nkr}. \textsc{CWoLa} is a method that does not depend on simulation and achieves signal sensitivity by comparing a signal region with nearby sideband regions in the resonance feature.  As a result, \textsc{CWoLa} is particularly sensitive to dependencies between the classification features and the resonant feature.  \textsc{Salad} uses a reweighted simulation to achieve signal sensitivity.  Since it never directly uses the sideband region, \textsc{Salad} is expected to be more robust than \textsc{CWoLa} to dependencies.  In order to recover the performance of \textsc{CWoLa} in the presence of significant dependence between the classification features and the resonant feature, a new method called simulation augmented CWoLa (SA-\textsc{CWoLa}) is introduced.  The SA-\textsc{CWoLa} approach augments the \textsc{CWoLa} loss function to penalize the classifier for learning differences between the signal region and the sideband region in simulation, which is signal-free by construction.  All of these methods will be investigated using the correlation test proposed in Ref.~\cite{Nachman:2020lpy}.

\subsubsection{Method}

For a set of features $(m,x)\in\mathbb{R}^{n+1}$, let $f:\mathbb{R}^n\rightarrow [0,1]$ be parameterized by a neural network. The observable $m$ is special, for it is the resonance feature that should be relatively independent from $f(x)$.  The signal region (SR) is defined by an interval in $m$ and the sidebands (SB) are neighboring intervals.

All neural networks were implemented in \textsc{Keras}~\cite{keras} with the \textsc{Tensorflow} backend~\cite{tensorflow} and optimized with \textsc{Adam}~\cite{adam}.   Each network is composed of three hidden layers with 64 nodes each and use the rectified linear unit (ReLU) activation function.  The sigmoid function is used after the last layer.  Training proceeds for 10 epochs with a batch size of 200.   None of these parameters were optimized; it is likely that improved performance could be achieved with an in-situ optimization based on a validation set.

\paragraph{Simulation Assisted Likelihood-free Anomaly Detection (SALAD)}

The \textsc{Salad} network~\cite{Andreassen:2020nkr} is optimized using the following loss:

\begin{align}
\mathcal{L}_\text{SALAD}[f]&=-\sum_{i\in\text{SR,data}}\log(f(x_i))-\sum_{i\in\text{SR,sim.}}w(x_i,m)\log(1-f(x_i))\,
\end{align}
where $w(x_i,m)=g(x_i,m)/(1-g(x_i,m))$ are a set of weights using the Classification for Tuning and Reweighting (\textsc{Dctr})~\cite{Andreassen:2019nnm} method.  The function $g$ is a parameterized classifier~\cite{Cranmer:2015bka,Baldi:2016fzo} trained to distinguish data and simulation in the sideband:

\begin{align}
\mathcal{L}[g]&=-\sum_{i\in\text{SB,data}}\log(g(x_i,m))-\sum_{i\in\text{SB,sim.}}\log(1-g(x_i,m))\,.
\end{align}
The above neural networks are optimized with binary cross entropy, but one could use other functions as well, such as the mean-squared error.  Intuitively, the idea of \textsc{Salad} is to train a classifier to distinguish data and simulation in the SR.  However, there may be significant differences between the background in data and the background simulation, so a reweighting function is learned in the sidebands that makes the simulation look more like the background in data.  

\paragraph{Simulation Augmented Classification without Labels (SA-CWoLa)}

The idea of \textsc{CWoLa}~\cite{Metodiev:2017vrx} is to construct two mixed samples of data that are each composed of two classes.  Using \textsc{CWoLa} for resonant anomaly detection~\cite{Collins:2018epr,Collins:2019jip}, one can construct the mixed samples using the SR and SB.  In the absence of signal, the SR and SB should be statistically identical and therefore the \textsc{CWoLa} classifier does not learn anything useful.  However, if there is a signal, then it can detect the presence of a difference between the SR and SB.  In practice, there are small differences between the SR and SB because there are dependencies between $m$ and $x$ and so \textsc{CWoLa} will only be able to find signals that introduce a bigger difference than already present in the background.  The \textsc{CWoLa} anomaly detection strategy was recently used in a low-dimensional application by the ATLAS experiment~\cite{collaboration2020dijet}.

We propose a modification of the usual \textsc{CWoLa} loss function in order to construct a simulation-augmented (SA) \textsc{CWoLa} classifier:

\begin{align}\nonumber
\mathcal{L}_\text{SA-CWola}[f]&=-\sum_{i\in\text{SR,data}}\log(f(x_i))-\sum_{i\in\text{SB,data}}\log(1-f(x_i))\\
\label{eq:sacwola}
&\hspace{25mm}+\lambda\left(\sum_{i\in\text{SR,sim.}}\log(f(x_i))+\sum_{i\in\text{SB,sim.}}\log(1-f(x_i))\right)\,,
\end{align}
where $\lambda > 0$ is a hyper-parameter.  The limit $\lambda\rightarrow 0$ is the usual \textsc{CWoLa} approach and for $\lambda > 0$, the classifier is penalized if it can distinguish the SR from the SB in the (background-only) simulation.  In order to help the learning process, the upper and lower sidebands are given the same total weight as each other and together, the same weight as the SR.

\subsubsection{Results on LHC Olympics}

The R\&D dataset is used for the results presented here.  For anomaly detection, the dijet invariant mass $m_{JJ}$ is used as the resonant feature. The classification features used are the invariant mass of the lighter jet, the mass difference between the two leading jets, and the $N$-subjettiness $\tau_{21}$ of the two leading jets.  As a benchmark, 1500 signal events corresponding to a fitted significance of about $2\sigma$ are injected into the data for training.  For evaluation, the entire signal sample (except for the small number of injected events) is used.  In order to demonstrate the breakdown of CWoLa in the presence of dependencies between the classification features and the resonant feature, we strengthen the dependence between the jet masses $m_J$ and invariant dijet mass $m_{JJ}$ by setting $m_J \mapsto m_J + 0.1 m_{JJ}$, as in \cite{Nachman:2020lpy}.

\begin{figure}[h!]
\centering
\includegraphics[height=0.36\textwidth]{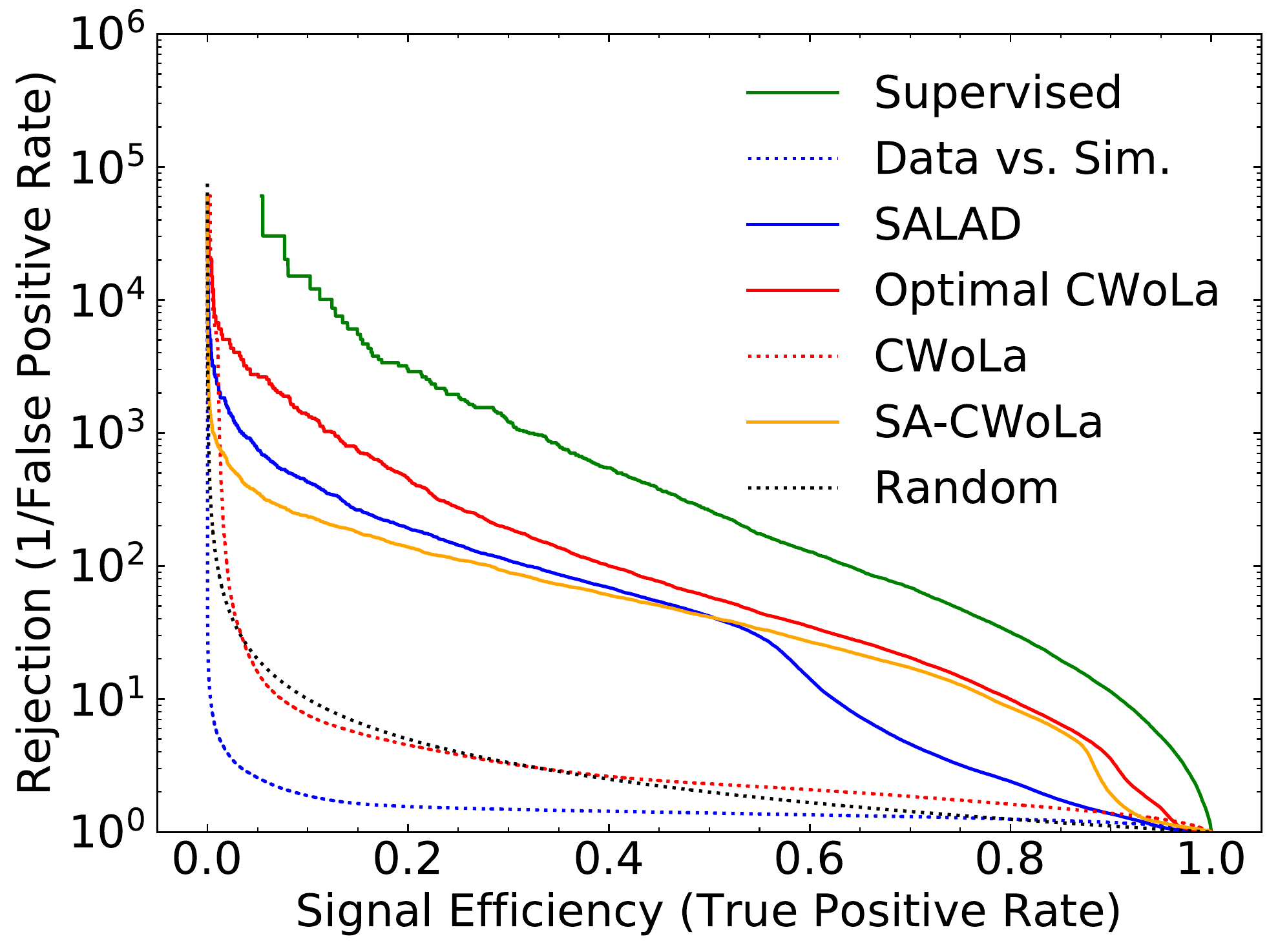}
\includegraphics[height=0.36\textwidth]{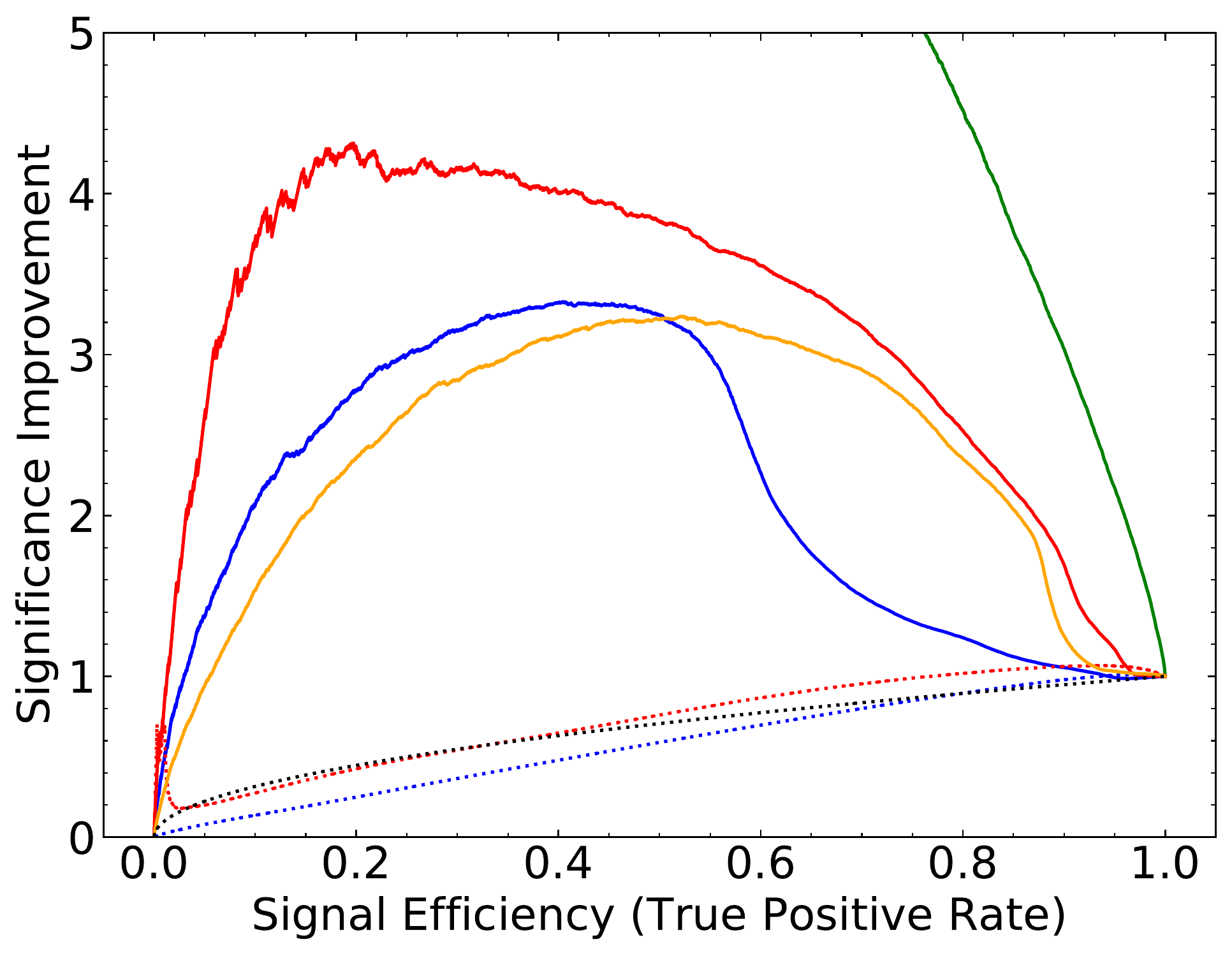}
\caption{A Receiver Operating Characteristic (ROC) curve (left) and significance improvement curve (right) for various anomaly detection methods described in the text.   The significance improvement is defined as the ratio of the signal efficiency to the square root of the background efficiency.  A significance improvement of 2 means that the initial significance would be amplified by about a factor of two after employing the anomaly detection strategy.  The supervised line is unachievable unless there is no mismodeling and one designed a search for the specific $W'$ signal used in this paper.  The curve labeled `Random' corresponds to equal efficiency for signal and background.  Figure from Ref.~\cite{1815227}.}
\label{fig:roc}
\end{figure}

Figure~\ref{fig:roc} shows the performance of various configurations.    The fully supervised classifier uses high statistics signal and background samples in the SR with full label information.  Since the data are not labeled, this is not achievable in practice.  A solid red line labeled `Optimal \textsc{CWoLa}' corresponds to a classifier trained using two mixed samples, one composed of pure background in the single region and the other composed of mostly background (independent from the first sample) in the SR with the 1500 signal events.  This is optimal in the sense that it removes the effect from phase space differences between the SR and SB for the background.  The Optimal \textsc{CWoLa} line is far below the fully supervised classifier because the neural network needs to identify a small difference between the mixed samples over the natural statistical fluctuations in both sets.  The actual \textsc{CWoLa} method is shown with a dotted red line.  By construction, there is a significant difference between the phase space of the SR and SB and so the classifier is unable to identify the signal.  At low efficiency, the \textsc{CWoLa} classifier actually anti-tags because the SR-SB differences are such that the signal is more SB-like then SR-like.  Despite this drop in performance, the simulation augmenting modification (solid orange) with $\lambda=0.5$ nearly recovers the full performance of \textsc{CWoLa}.  

For comparison, a classifier trained using simulation directly is also presented in Figure~\ref{fig:roc}.  The line labeled `Data vs. Sim.' directly trains a classifier to distinguish the data and simulation in the SR without reweighting.  Due to the differences between the background in data and the simulated background, this classifier is not effective.  In fact, the signal is more like the background simulation than the data background and so the classifier is worse than random (preferentially removes signal).  The performance is significantly improved by adding in the parameterized reweighting, as advocated by Ref.~\cite{Andreassen:2020nkr}.  With this reweighting, the \textsc{Salad} classifier is significantly better than random and is comparable to SA-\textsc{CWoLa}.  The Optimal \textsc{CWoLa} line also serves as the upper bound in performance for \textsc{Salad} because it corresponds to the case where the background simulation is statistical identical to the background in data.

\subsubsection{Lessons Learned}

This section has investigated the impact of dependencies between $m_{jj}$ and classification features for the resonant anomaly detection methods \textsc{Salad} and \textsc{CWoLa}. A new simulation-augmented approach has been proposed to remedy challenges with the \textsc{CWoLa} method.  This modification is shown to completely recover the performance of \textsc{CWoLa} from the ideal case where dependences are ignored in the training.  In both the \textsc{Salad} and SA-\textsc{CWoLa} methods, background-only simulations provide a critical tool for mitigating the sensitivity of the classifiers on dependences between the resonant feature and the classifier features.  
    
Each of the methods examined here have advantages and weaknesses, and it is likely that multiple approaches will be required to achieve broad sensitivity to BSM physics.  Therefore, it is critical to study the sensitivity of each technique to dependencies and propose modifications where possible to build robustness.  This paper is an important step in the decorrelation program for automated anomaly detection with machine learning.  Tools like the ones proposed here may empower higher-dimensional versions of the existing ATLAS search~\cite{collaboration2020dijet} as well as other related searches by other experiments in the near future.

 \FloatBarrier

\section{(Semi)-Supervised}
\label{sec:supervised}

\subsection[Deep Ensemble Anomaly Detection]{Deep Ensemble Anomaly Detection\footnote{Authors: Felipe F. De Freitas, Charanjit K. Khosa, Veronica Sanz.  The codes can be found at the following link \url{https://github.com/FFFreitas/Deep-Ensemble-Anomaly-Detection}.}}

\label{sec:CNNBDT}

\subsubsection{Method}
\label{sec:method}
For the LHC Olympics challenge we opted for a semi-supervised approach. This was partly motivated by lack of time, and partly by the way the challenge itself was set up. Indeed, previous to the releasing of the blackboxes, the organisers had provided warm-up examples including signal and background labels. 
At the end we focused on a mixture of neural networks, with convolutional layers, and Boosted Decision Trees (BDTs). This hybrid approach was based on previous studies by one of the authors~\cite{Alves:2019ppy}, which proposes to use a two step ``pipeline" to assign event-by-event probabilities in categories {\it signal} or {\it background}.  This model uses event input in two forms; raw data as an image as well as high level features (kinematic variables). We train the model for the labelled background and signal data sets (R $\&$ D data set). ResNet is used as a pre-classifier for the $\eta$-$\phi$ 2d images (weighted by pt) of the un-clustered particles of the event. Along with ResNet predictions of signal/background (event-by-event), we used the kinematics of fat jets (zero-padded in case of only one) for the BDTs.  This two-step approach provides an AUC increase of about  5$\%$  over the BDT trained only on kinematic observables. 
\paragraph{Data sets}
We start by describing the data preparation procedure
\begin{enumerate}
\item  We first create event images from the data. The images are generated from the uncluttered data display in a $224\times 224$ 2-D grid with the $x$ and $y$ positions given by the $\eta$ and $\phi$ information from the particles in an event. The 2-D grid is further converted into a RGB image with $224\times 224$ pixels, the pixels color values are normalized according to $W_{p_\text{T}} = \frac{p_\text{T}(i)}{max(p_\text{T}(i))}$, where $i$ runs over all the particles found in an event. 

\item The tabular data is a comma-separated values (CSV) file, where each row corresponds to an event and the columns are the kinematic and angular observables from the final state particles and jets from the event. In our analysis, we cluster  inclusive fat jets with $p_\text{T}^{min}>$ 500 GeV and R=1 with the anti-$k_T$ algorithm. For this analysis, we consider the first two leading jets. If there were only one jet, then all the kinematics of the second jet were set to zero values. We cluster again the constituents of the fat jets with R=0.4  and $k_T$ algorithm with minimum $p_T$ condition of 20 GeV. With these jets we construct the following observables :
$p_\text{T}^{j_1}$, $m_{j_1}$, $\eta_{j_1}$, $\phi_{j_1}$, $E_{j_1}$, 
$p_\text{T}^{j_2}$, $m_{j_2}$, $\eta_{j_2}$, $\phi_{j_2}$, $E_{j_2}$,
$\delta \eta$, $\delta \phi$, ${m/E}_{j_1}$, ${m/E}_{j_2}$,
$m_{jj}$, $P_T^A (j_1j_2)$, $\delta R^{j_1}_{12}$, $\delta R^{j_1}_{13}$,
$\delta R^{j_1}_{14}$, $\delta R^{j_1}_{23}$, 
$\delta R^{j_1}_{24}$, $\delta R^{j_1}_{34}$,
$\delta R^{j_2}_{12}$, $\delta R^{j_2}_{13}$,
$\delta R^{j_2}_{14}$, $\delta R^{j_2}_{23}$, 
$\delta R^{j_2}_{24}$, $\delta R^{j_2}_{34}$,
$n_{subjets1}$,$n_{subjets2}$. Some of the observables are constructed from the fat-jets kinematics and some of them are from their sub-jets.
\end{enumerate}

After we have a trained CNN model for the classification of the image dataset, we include in the tabular data the predicted scores from the CNN model for a given event. This additional information helps to improve further the classification power of the BDT model, allowing our framework to predict with fairly good confidence if an event is background or an anomaly.
\paragraph{The CNN architecture and training methodology}
We use a modified pre-trained ResNet-34 as a pre-classifier, the ResNet-34 consists of 34 convolutional (Conv2D) layers. In between each Conv2D layers, one has a series of batch normalizations, average pooling and rectified activations (ReLU). For our task, we replace the last fully connected layers of the ResNet-34, responsible for the classification, with the following sequence of layers:
\begin{itemize}
\item  An adaptive concatenated pooling layer ({\it AdaptiveConcatPool2d}), 
\item  A flatten layer,
\item  A block with batch normalization, dropout, linear, and ReLU layers,
\item  A dense linear layer with 2 units as outputs, corresponding to a $signal$ and $background$, and a softmax activation function.
\end{itemize}
The {\it AdaptiveConcatPool2d} layer uses adaptive average pooling and adaptive max pooling and concatenates them both. Such procedure provides the model with the information of both methods and improves the overall performance of the CNN. We also make use of the label smoothing methodology~\cite{DBLP:journals/corr/SzegedyVISW15} as well as the MixUp~\cite{2017arXiv171009412Z} training method.

Due to the high imbalance between the number of signal and background events in the full R \& D data set, we generate a smaller sample with 93390 images, with the same proportion of images for the signal and background events. We further separate these images into a training set containing 74712 (80\%) images and 18678 (20\%) for the validation and test sets.

\begin{figure}[h!]
\centering
\includegraphics[width=0.45\textwidth]{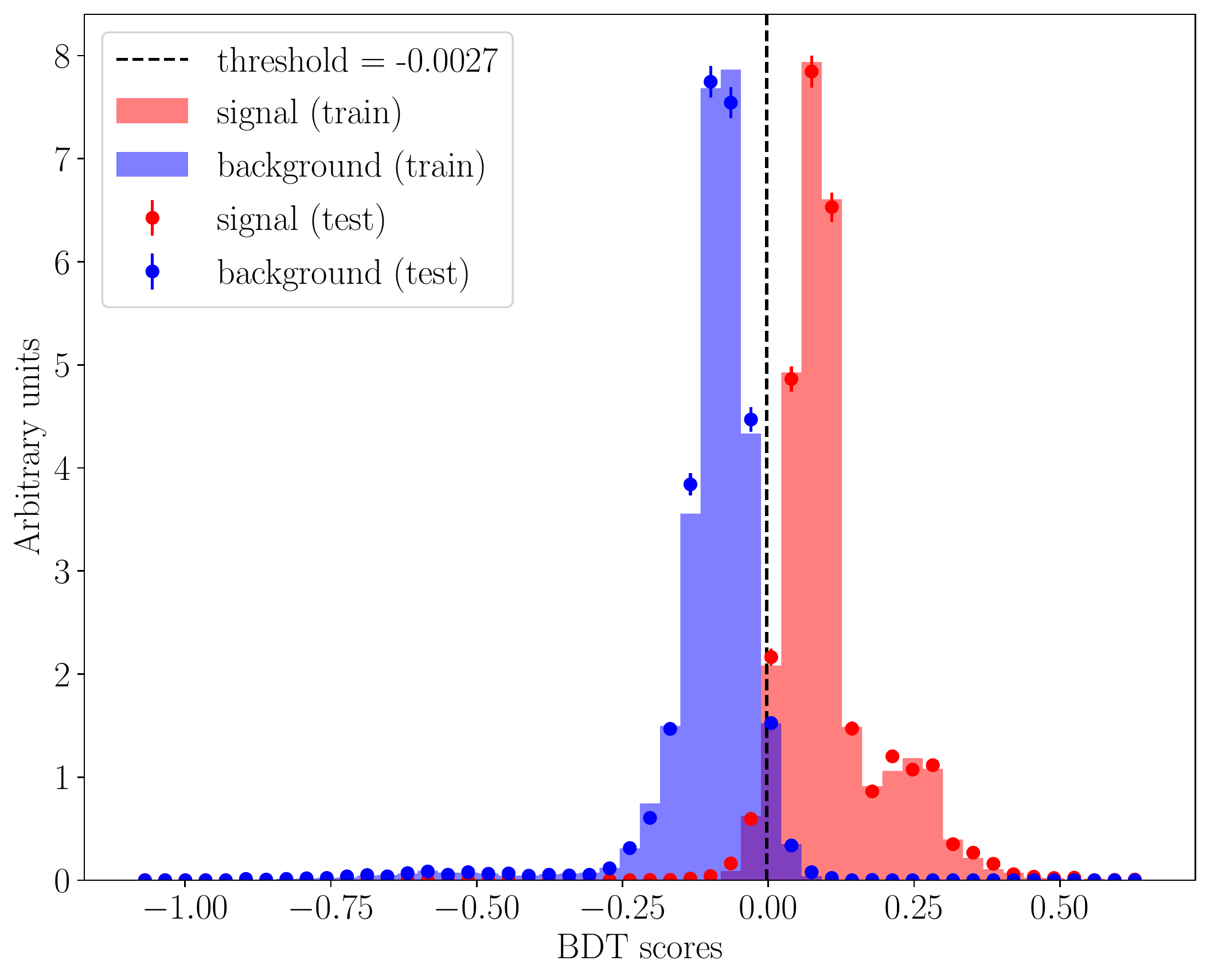}
\includegraphics[width=0.45\textwidth]{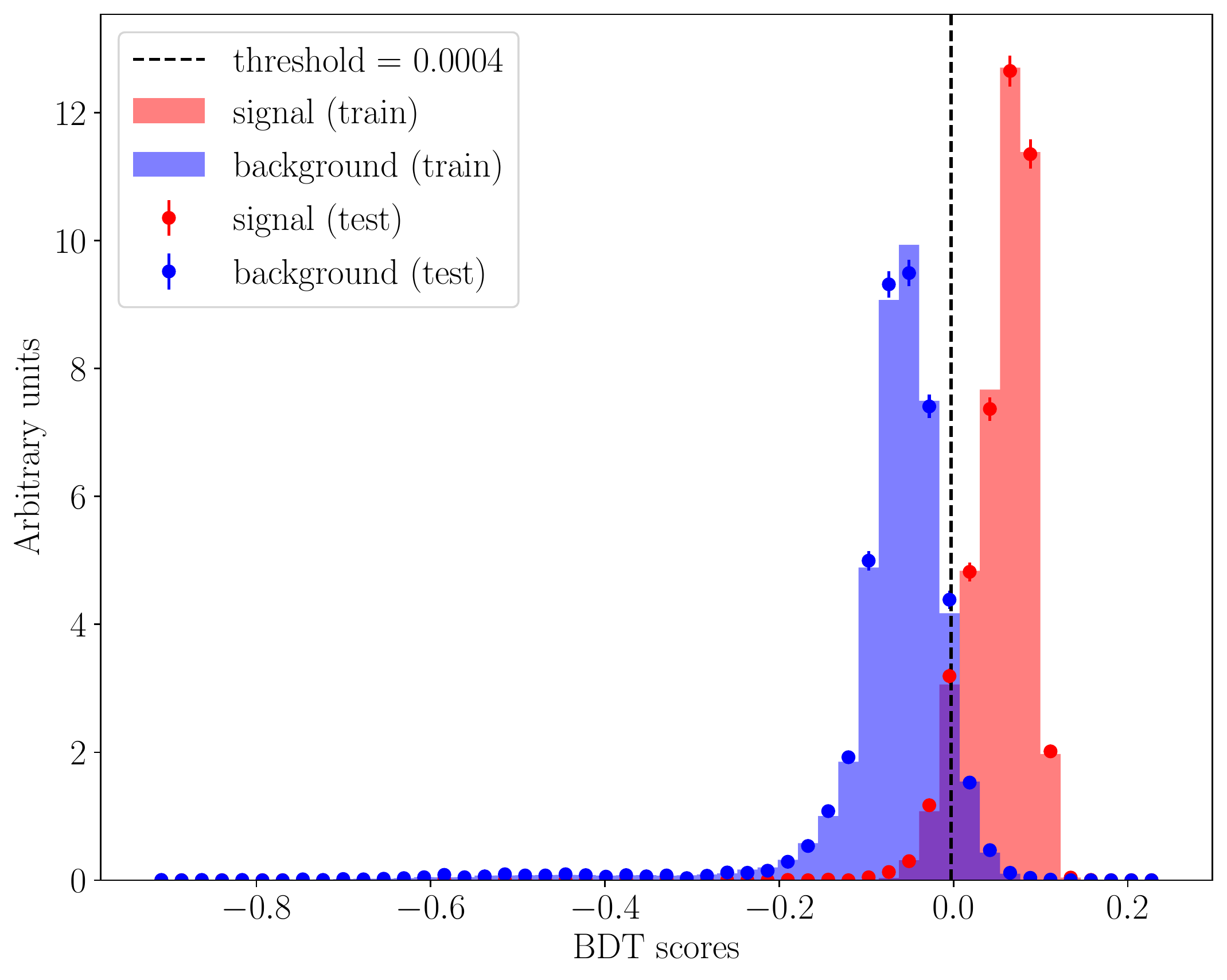}
\caption{Left: BDT scores using the kinematic observables and the scores from ResNet-34. Right: BDT scores using the kinematic observables only.} 
\label{fig:BDT_CNN}
\end{figure}

\begin{figure}[ht!]
\centering
\includegraphics[width=0.45\textwidth]{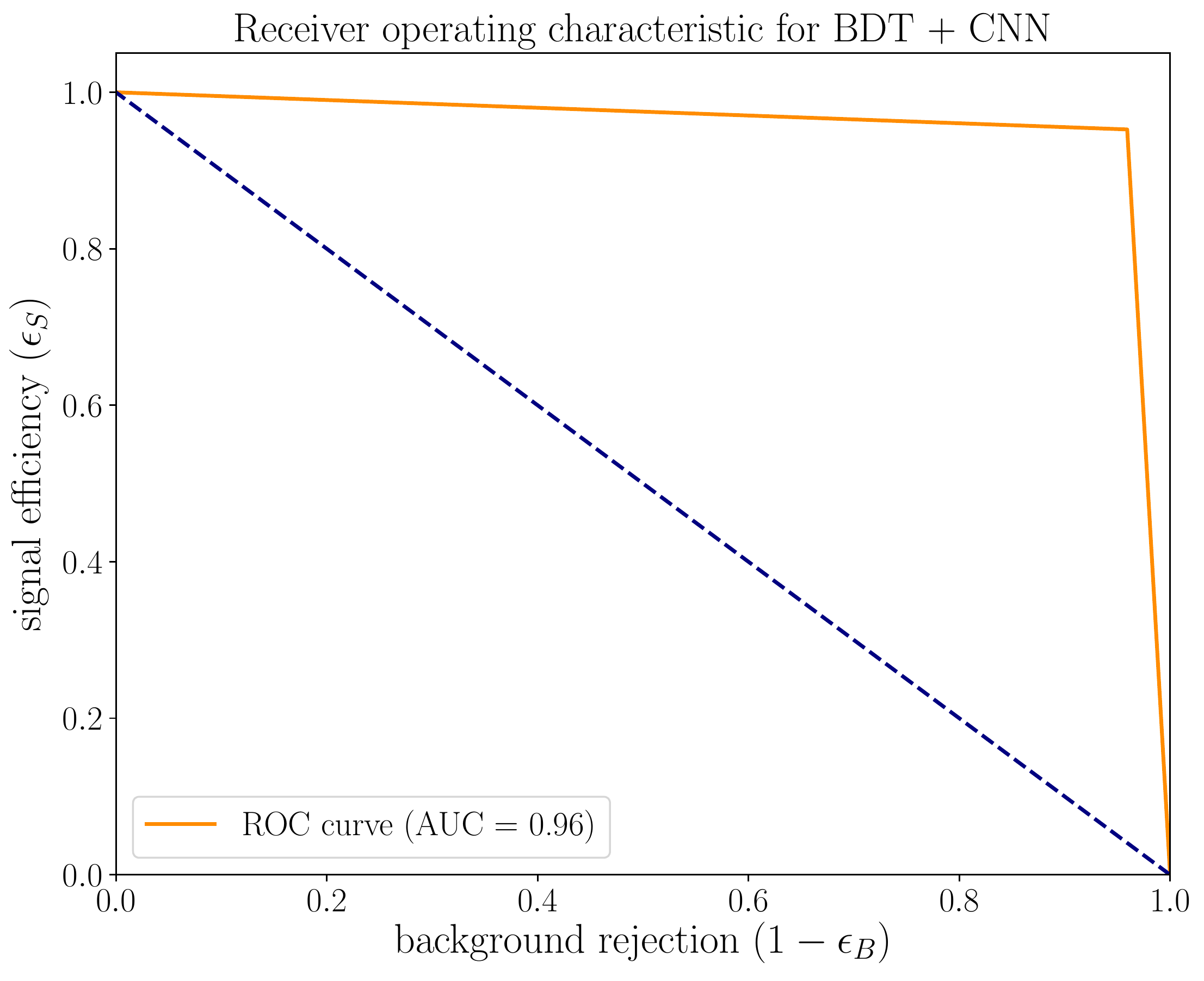}
\includegraphics[width=0.45\textwidth]{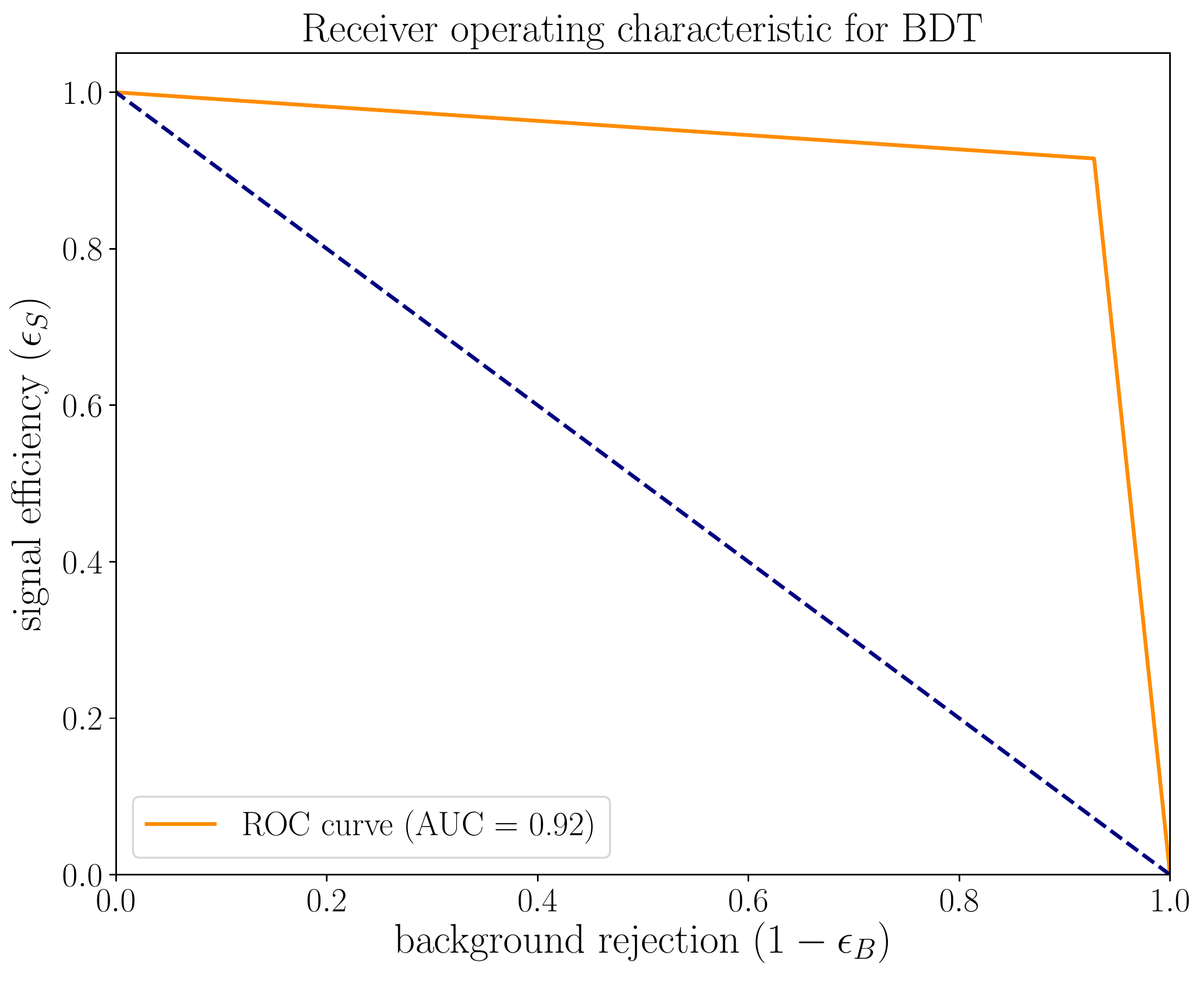}
\caption{Left: ROC curve for a BDT using the kinematic observables and the scores from ResNet-34.
Right: ROC curve for a BDT using the kinematic observables only.} 
\label{fig:ROC_BDT_CNN}
\end{figure}

We train the ResNet-34 for 25 epochs using fit one-cycle method~\cite{2018arXiv180309820S}. Our modified ResNet-34 achieves an accuracy of  92\%. We then use this CNN to get the predictions for each image and append these values to the tabular data, so we have both the kinematic information for a given event and also the score from the CNN of the same event to belong to $signal$ or $background$.
\paragraph{The BDT model}
After we gather the predictions from our modified ResNet-34 and the kinematic information, described in Sec.~\ref{sec:method}, with the appended tabular dataset we make use of scikit-learn~\cite{scikit-learn} and DEAP~\cite{DEAP_JMLR2012} to build an evolutionary search algorithm in order to find the best hyper-parameters of the BDT which maximize the accuracy. After we perform a evolutionary search over the space of hyper-parameters, we found a BDT with 95.38\% accuracy were achieved by the following configuration:
\begin{itemize}
    \item Estimators : 700 
    \item Learning rate : 0.1 
    \item Algorithm : SAMME.R 
    \item Max depth : 3 
    \item Criterion : entropy 
    \item Splitter : random
\end{itemize}
The BDT model gives us the final classification of events where we estimate all metrics presented in the section \ref{sec:results}. In Fig. \ref{fig:BDT_CNN} and \ref{fig:ROC_BDT_CNN} we show the BDT score and ROC curves for the test data sets. 
\subsubsection{Results on LHC Olympics}
\label{sec:results}

\begin{figure}
\centering
\includegraphics[width=0.45\textwidth]{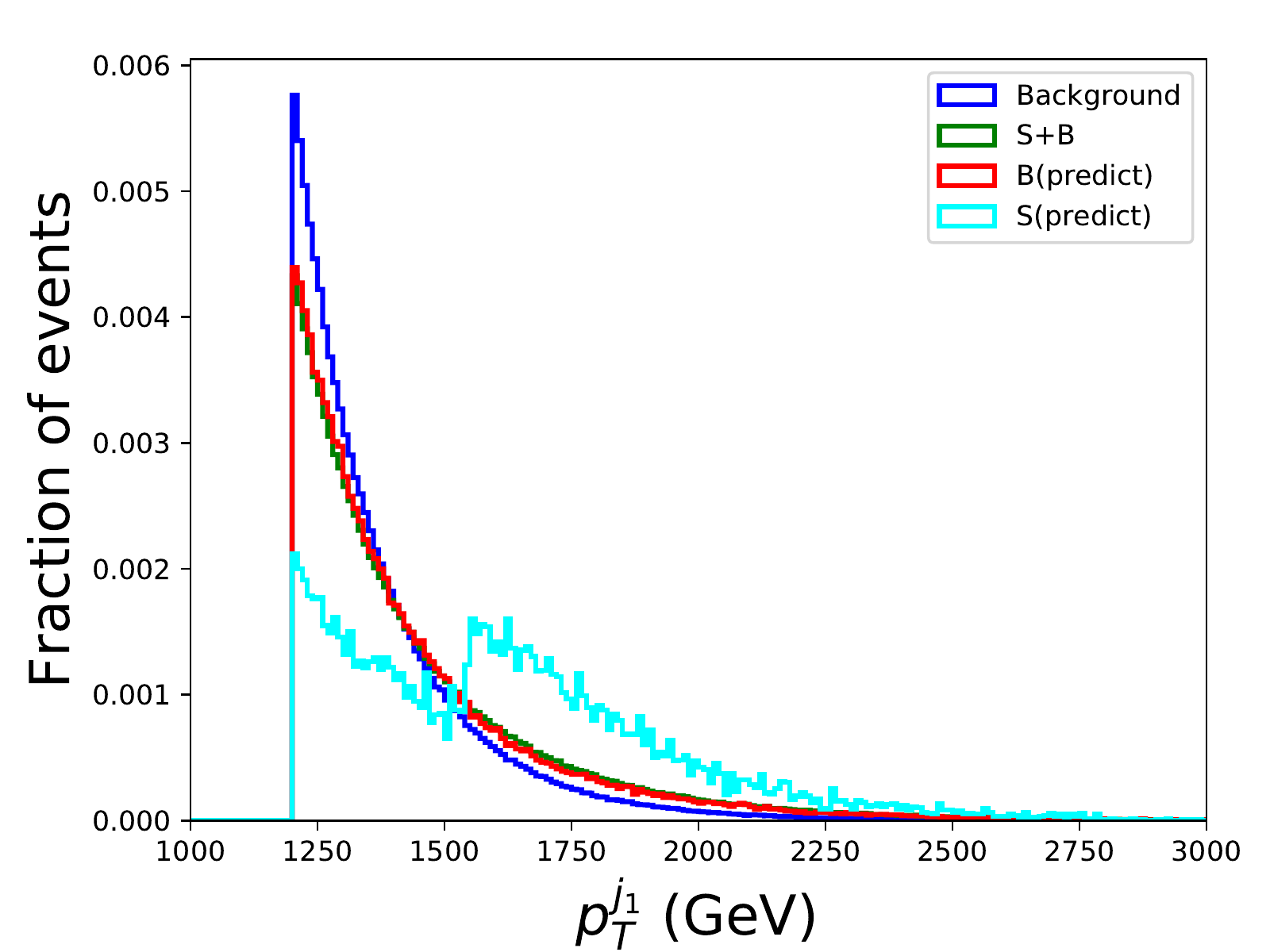}
\includegraphics[width=0.45\textwidth]{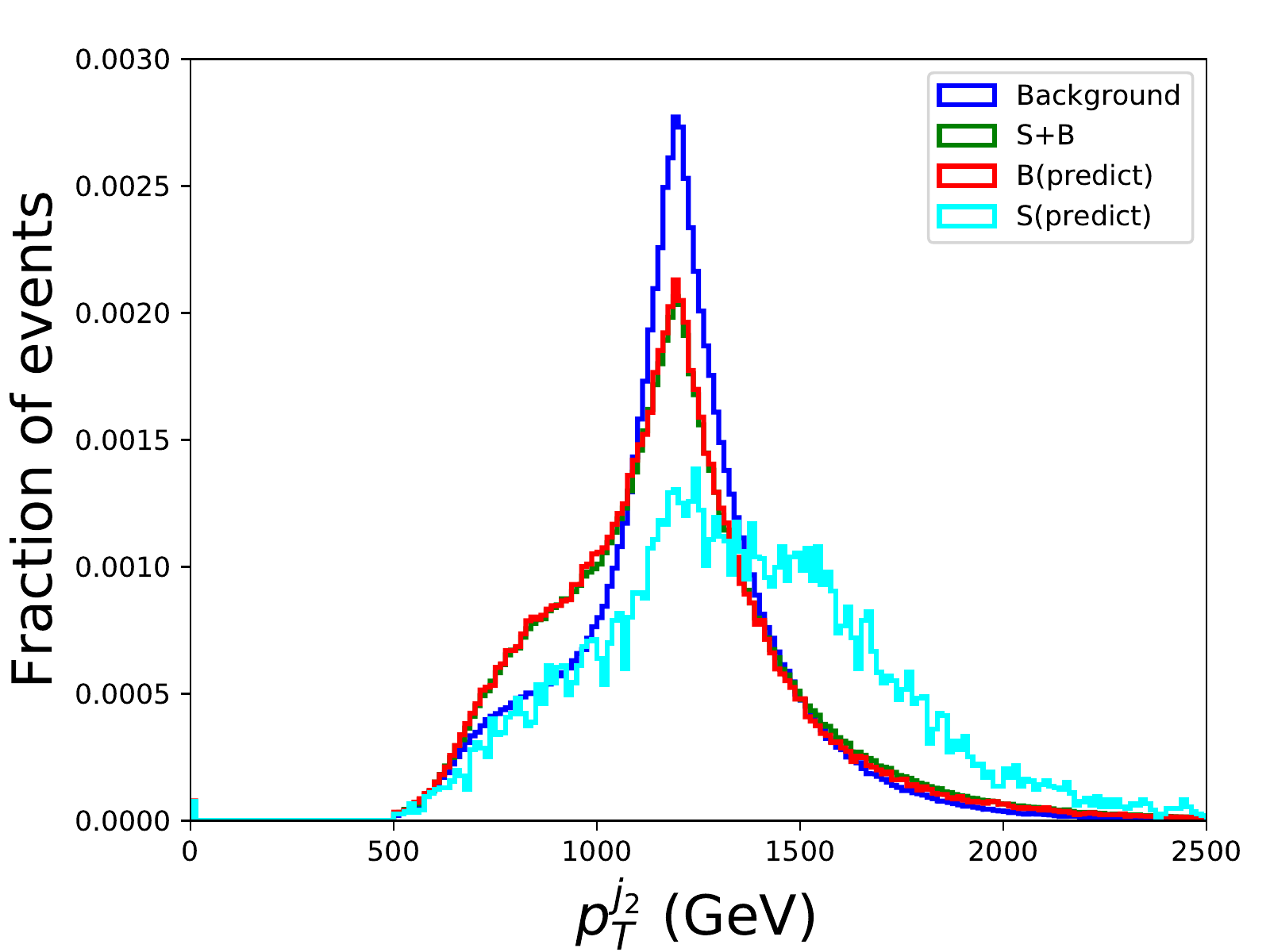}
\includegraphics[width=0.45\textwidth]{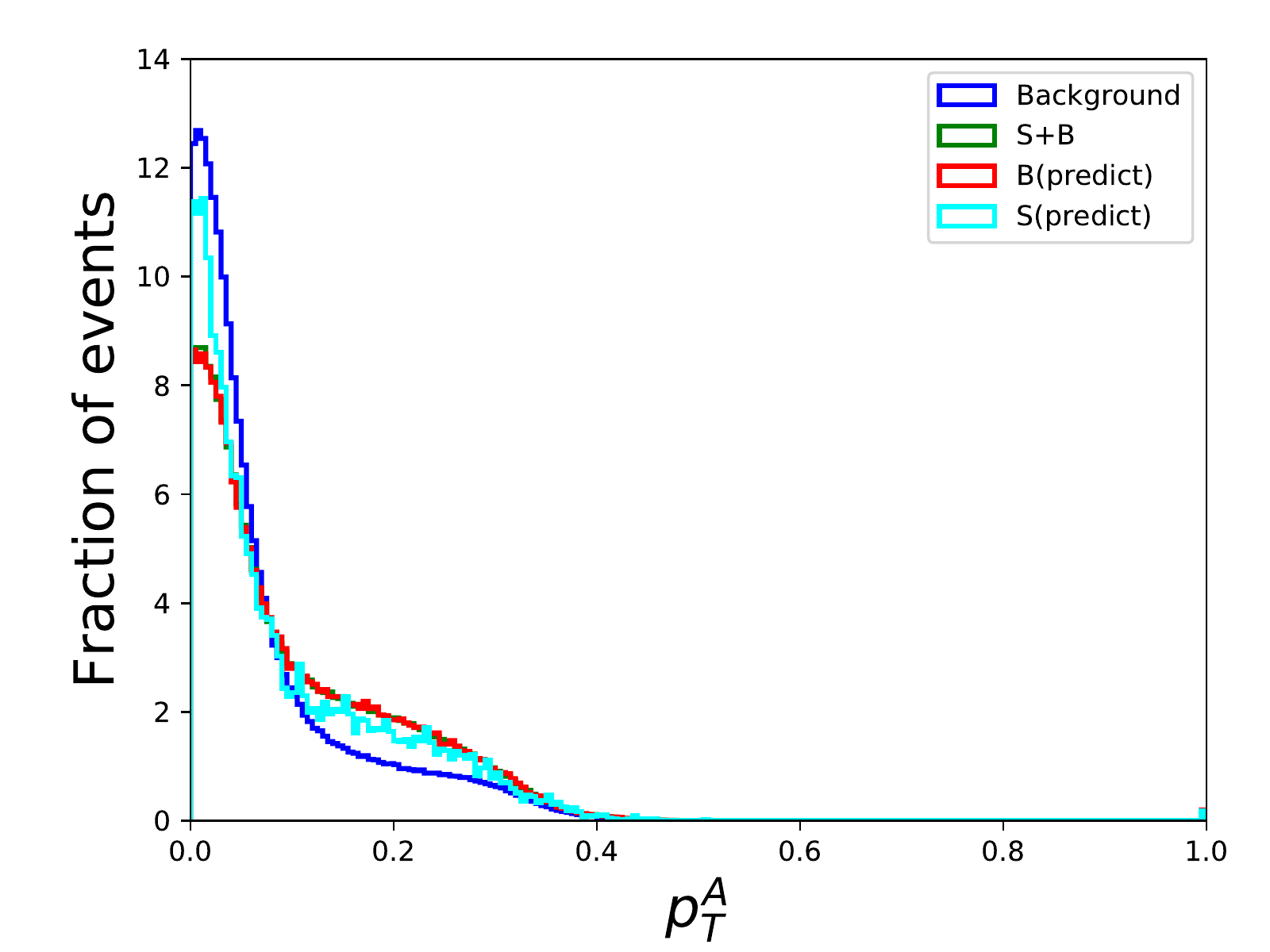}
\caption{Data Features for the Blackbox data 1. The dark blue line (background) refers to the labeled dataset, whereas the other three lines are distributions from the blackbox.  \label{fig:np1}}
\end{figure}

\begin{figure}
\centering
\includegraphics[width=0.45\textwidth]{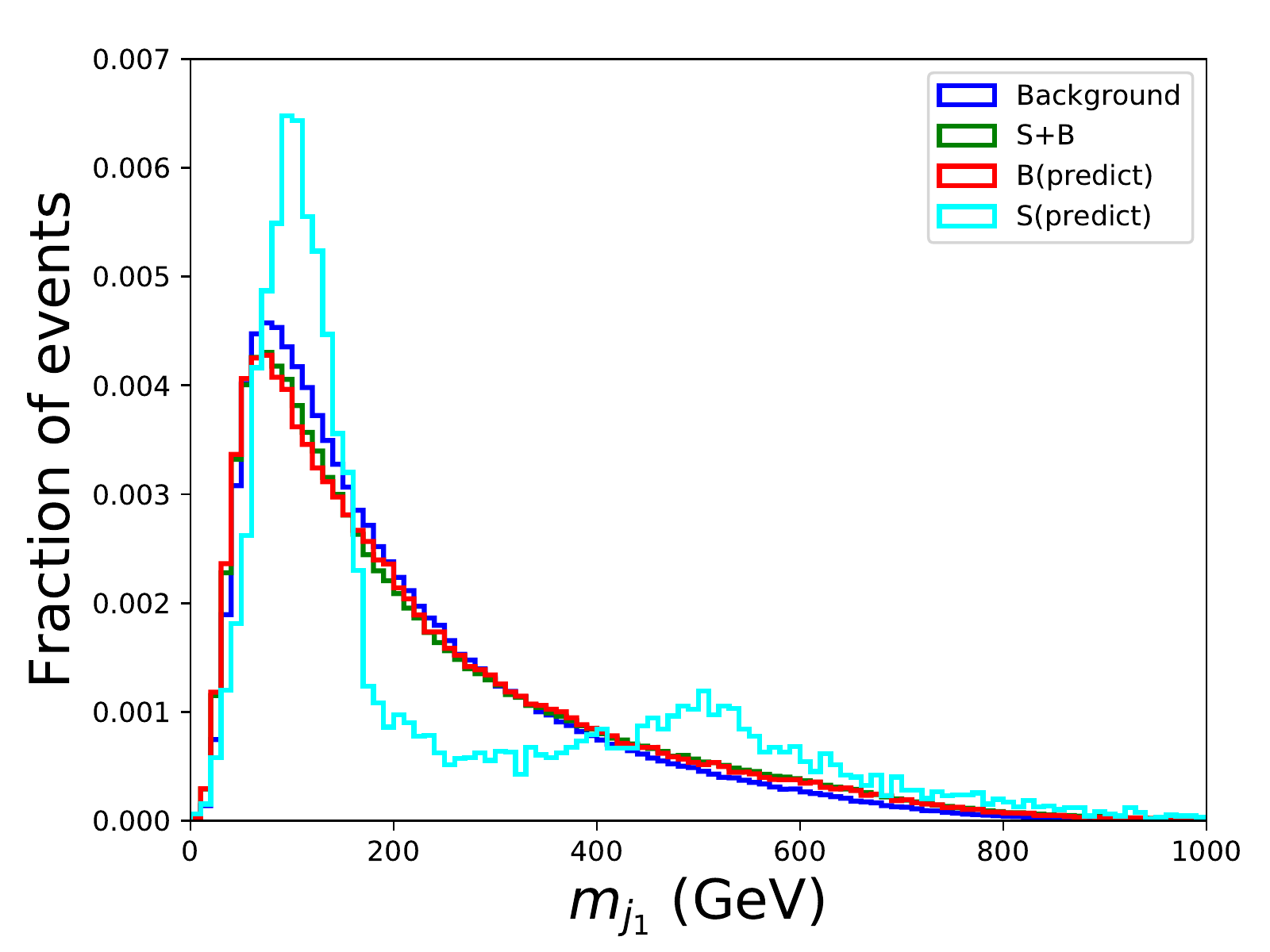}
\includegraphics[width=0.45\textwidth]{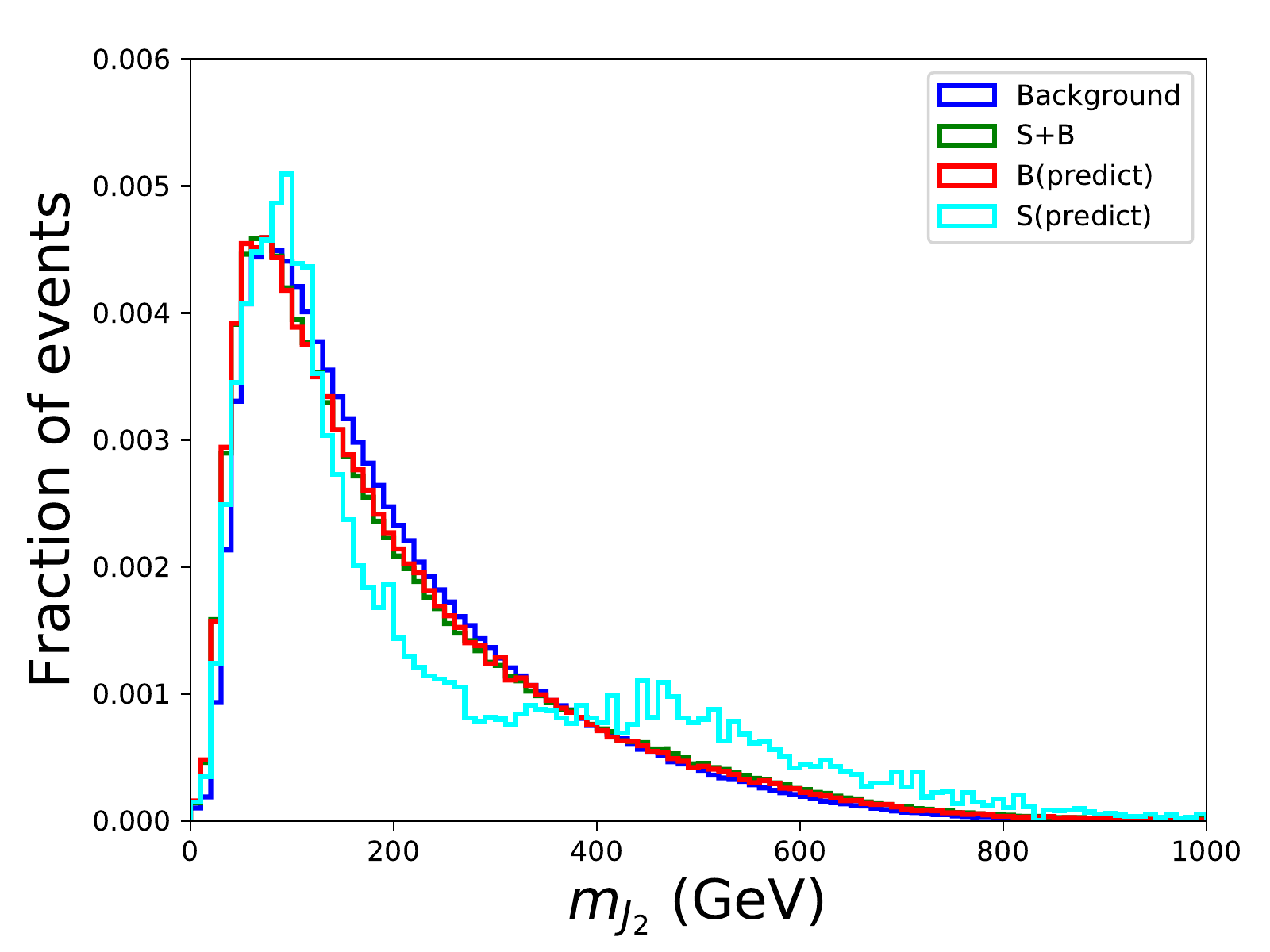}
\includegraphics[width=0.45\textwidth]{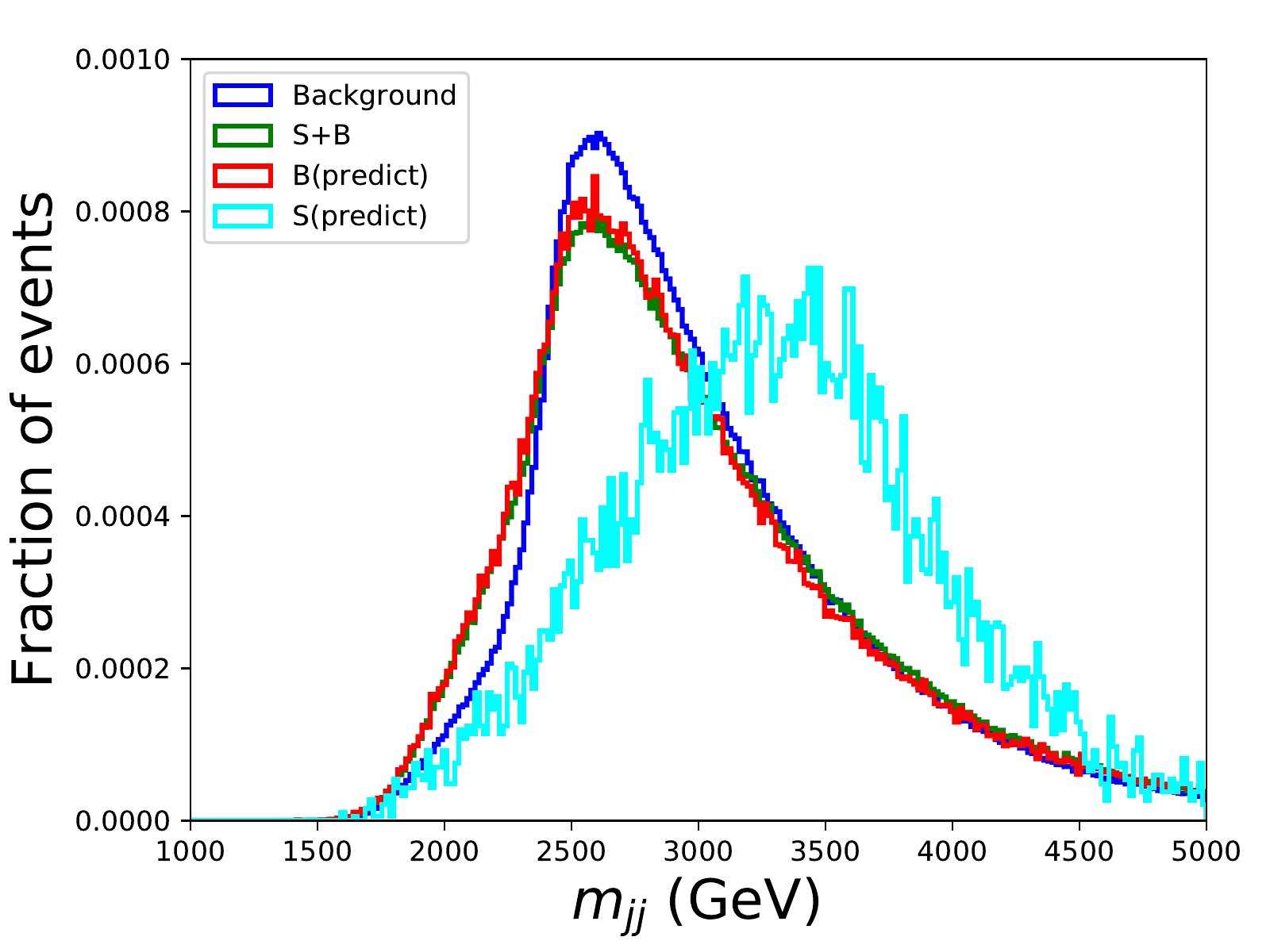}
\includegraphics[width=0.45\textwidth]{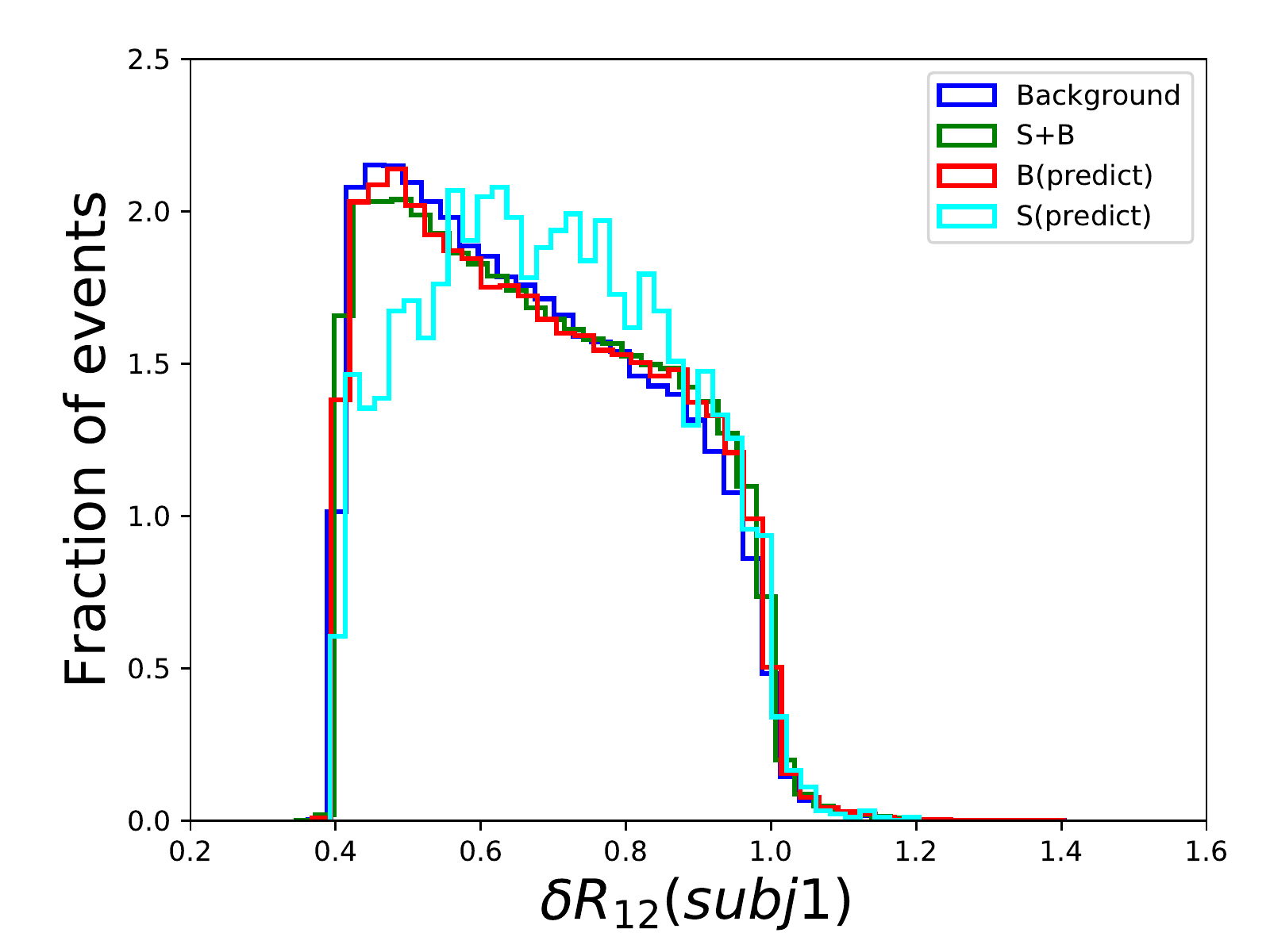}
\caption{Data Features for the Blackbox data 1.\label{fig:np2}}
\end{figure}

Using the training from the previous section and applying it to the blackbox, we found that the new physics signal is likely to be a heavy resonance of mass 3.5 GeV which further decays to two particles of mass 100 GeV and 500 GeV, as in the labeled data. We show the feature distributions in Fig.~\ref{fig:np1} and \ref{fig:np2}. Note that, due to lack of time, we have not analysed the whole blackbox, just a subsample of 200K events.

\subsubsection{Lessons Learned}
\label{sec:lessons}

Although the method outlined above is able to identify the new physics events, it is not robust enough to predict the correct range of mass of the heavy resonance. It is specific to the type of new physics it was trained on. 

This exercise and the issue of lack of generalization to new blackboxes, led to two of the authors to develop a semi-supervised algorithm, called Anomaly Awareness~\cite{Khosa:2020qrz}, with the focus on {\it robustness} respect to the types of anomalies.  

Initially we spent a significant amount of time on preliminary investigations to get the feeling of what could be the best approach. Maybe a better strategy would have been to try different methods in parallel. Different approaches have different systematic errors and sensitivities, and we would have liked to develop further the other proposals we thought of for the challenge. Alas, we had to settle for the quickest analysis to reach the deadline.  

Another point we would like to mention is the following: in the  list of observables we included correlated variables such as  $m_{j_1}$ and $m_{j_2}$. A more appropriate option would have been for example $m_{j_1}$ and $m_{j_1}-m_{j_2}$. 

 \FloatBarrier

\subsection[Factorized Topic Modeling]{Factorized Topic Modeling\footnote{Authors: Patrick Komiske, Eric Metodiev, Nilai Sarda, and Jesse Thaler.  Code will be made available at the following link:\\ \texttt{\url{https://github.com/nilais/factorized-topic-modeling}}}}
\label{sec:factorizedtopics}

In this contribution, we propose and evaluate a new technique to statistically model and discriminate between dijet configurations, called ``factorized topic modeling''.
The cross section for dijet production satisfies a leading-order factorization theorem, which implies that both jets in a dijet event are statistically independent, conditioned on their joint types.
This is an extremely powerful statement, motivated from first principles, which we convert into a statistical constraint on a generative model.
Starting from the framework of ``jet topics'' \cite{Metodiev:2018ftz}, we leverage a factorization theorem for jet substructure to build a generative model, and demonstrate a procedure for optimizing it to recover both the relative fractions of different types of jets within a mixed sample, as well as the component distribution for a given observable. 
We use factorized topic modeling in the context of anomaly detection to identify exotic jet types in the LHC Olympics R\&D dataset.
\subsubsection{Method}
\label{sec:method}

\paragraph{Review of factorization}
\label{subsec:factorization}

Factorization is the statement that the cross-section for dijet production can be decomposed into the product of independent probability functions.
Each component of the cross-section corresponds to a different physical process contributing to the observed jet pair.
Concretely, to leading order in a power expansion, the cross-section for dijet production in proton-proton collisions can be written as \cite{Ellis:1991qcd}:
\begin{equation}
\label{eq:factorization_full}
d\sigma = \sum_{ab \to cd} f_{a}(\xi_a) \otimes f_{b}(\xi_b) \otimes \mathcal{H}_{ab \to cd} \otimes \mathcal{J}_c(z_c) \otimes \mathcal{J}_d(z_d),
\end{equation}
where $f_a$ is the parton distribution function for parton $a$ inside the proton, $\xi_a$ is that parton's momentum fraction, $\mathcal{H}$ is the partonic cross section for the short-range hard scattering process $(ab \to cd)$, and $\mathcal{J}$ are the jet branching functions.
In this work, as our goal is jet tagging, we focus on the part of this equation that governs the substructure of the two jets:
\begin{equation}
\label{eq:factorization}
d\sigma \propto \sum_{cd} \mathcal{H}_{cd} \otimes \mathcal{J}_c(z_c) \otimes \mathcal{J}_d(z_d).
\end{equation}
Our goal is to translate this physical, leading-order factorization theorem into a statistical constraint on the probability distribution over jet observables.
For dijets, we consider each observation to be a pair $(\bv{x}_1, \bv{x_2})$, corresponding to the value of a given observable for the hardest and second-hardest jet in the event, respectively.
Using \Eq{eq:factorization} as a starting point, we will write down a generative model for dijet production -- more specifically, a topic model.

\paragraph{Review of topic modeling}
\label{subsec:topic}

Topic modeling was first applied to jet physics in~\cite{Metodiev:2018ftz} and has since been studied in both phenomenological and experimental contexts~\cite{Komiske:2018vkc,Aad:2019onw,Sirunyan:2019jud,Aad:2020cws}.
This body of work leverages the statistical connection between themes in text corpora and jet flavors in event samples to propose a new data-driven method for defining classes of jets.
We first consider an unfactorized topic model in a single observable $\bv{x}$.
For a mixed sample $\mathcal{M}$, this corresponds to a generative process with the following structure:
\begin{equation}
\label{prob:topic_model}
	p_{\mathcal{M}}(\bv{x}) = \sum_{k} f_{\mathcal{M}}(k) \cdot p_{k}(\bv{x}), \qquad \text{s.t.} \quad \int_\mathcal{X} d\bv{x} \, p_k(\bv{x}) = 1 \quad \forall k, \quad \sum_{k} f_\mathcal{M}(k) = 1.
\end{equation}
Each component $k$ corresponds to a jet class (i.e.~an anomalous jet or an ordinary quark/gluon jet from QCD).
The mixture components $\{p_k\}$ correspond to the distributions of any given jet observable $\bv{x}$, while the fractions $f(k)$ represent the fraction of the total sample which belongs to each component.
The goal of a topic model is to simultaneously learn the components $\{p_k\}$ and fractions $f(k)$ from a set of samples $\{\mathcal{M}_i\}$.

\paragraph{Factorized topic models}
\label{subsec:formulation}

Unlike the univariate topic model described in \Eq{prob:topic_model}, factorized topic modeling operates on pairs of observables $\bv{x}_1, \bv{x}_2$, corresponding to the leading and subleading jets in an event.
The multivariate formula for the sample distribution is then given by:
\begin{equation}
\label{eq:multivar_topic_model}
p_\mathcal{M}(\bv{x}_1, \bv{x}_2) = \sum_k f_\mathcal{M}(k) p_k(\bv{x}_1, \bv{x}_2).
\end{equation}
To specify the form for $p(\bv{x}_1, \bv{x}_2)$, we must explicitly state our constraints in a statistical language:
\begin{enumerate}
\item \emph{Sample independence}: The model assumes that, to leading order, the jet observable $\bv{x}$ depends only on the initiating parton.
In reality, there is some dependence on the process in addition to the parton flavor.
However, experimental studies have shown a high degree of empirical independence, and we suggest that these differences can be considered negligible for our model~\cite{Komiske:2018vkc}.
Defining $p^{(1)}, p^{(2)}$ as the distribution functions for the hardest and second-hardest jet, respectively, sample independence implies:
\begin{equation}
p^{(1)}_k(\bv{x}) = p^{(2)}_k(\bv{x}).
\end{equation}
\item \emph{Factorization:} The leading-order factorization theorem tells us that the two jets in an event are statistically independent, conditioned on convolution through the matrix element describing the short-range scattering.
From a statistical perspective, the factorization theorem given above is mathematically equivalent to stating that our topic model for dijets must be an \emph{mixture of products}.
In other words,
\begin{equation}
(\bv{x}_1 | k_1, k_2) \text { and } (\bv{x}_2 | k_1, k_2) \text{ are conditionally independent.}
\end{equation}
\end{enumerate}

Note that by simply replacing the structure of the sample-level probability distribution given above with these constraints, the mapping between the factorization theorem and statistical language can directly give us a topic model.
The model can be expressed as follows.
\begin{equation}
\label{eq:finalfactorizedmodel}
p_\mathcal{M}(\bv{x}_1, \bv{x}_2) = \sum_{k_1, k_2} f_\mathcal{M}(k_1, k_2) \cdot p_{k_1}(\bv{x}_1) \cdot p_{k_2}(\bv{x}_2).
\end{equation}

\paragraph{Algorithm to solve the model}
\label{subsec:algorithms}

\begin{figure}[t]
\centering
\includegraphics[width=\textwidth]{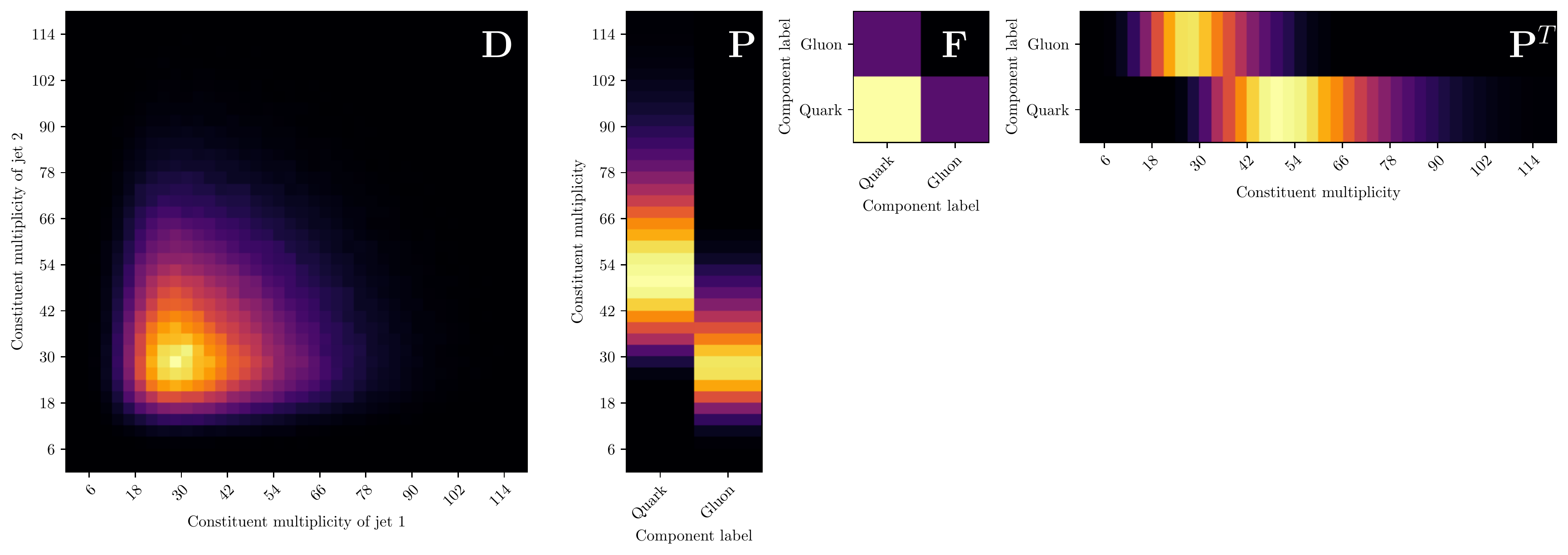}
\caption{
The paired observables from a dijet sample can be represented as a histogram, shown as the matrix $\bbm{D}$.
The generative process we describe can be visualized as the matrix product $\bbm{PFP}^\intercal$, shown as a decomposition on the right.
This example is for separating dijet events into quark and gluon categories, where the observable is jet constituent multiplicity.
}
\label{fig:matmuls}
\end{figure}

Our goal is to find the parameters of the topic model that give the best fit to the true distribution for the mixed sample $p_\mathcal{M}$.
First, we discretize the sample distribution into binned histograms, which allows us to reformulate \Eq{eq:finalfactorizedmodel} as a matrix decomposition.
Define the matrix $\bbm{D}$ to be the 2-dimensional histogram generated by jointly binning the sampled data across $\bv{x}_1$ and $\bv{x}_2$.
Similarly, let $\bbm{P}$ be the matrix whose columns are $n$-bin histograms representing each component $\bbm{p}_k$.
By rewriting the model in terms of histograms and bins, we arrive at the following non-convex program:
\begin{ftheo}
\begin{equation*}
\label{prob:disc_min}
\begin{aligned}
\min_{\substack{\bbm{F} \in \mathbbm{R}^{k \times k} \\ \bbm{P} \in \mathbbm{R}^{n \times k}}} \left\|\bbm{D} - \bbm{PFP}^\intercal\right\|^2_F, \qquad 
\text{s.t.} \quad \bbm{P}^\intercal \mathbbm{1}_n = \mathbbm{1}_k, \quad \mathbbm{1}_k^\intercal \bbm{F} \mathbbm{1}_k = 1, \quad \bbm{P,F} \geq 0,\\
\end{aligned}
\end{equation*}
\end{ftheo}
where $\mathbbm{1}_n$ is the $n$-dimensional vector of all ones, and we have taken the Frobenius norm $\|\bbm{A} - \bbm{B}\|_F = \sqrt{\sum_{ij} (\bbm{A}_{ij} - \bbm{B}_{ij})^2}$ as our loss function.
A pictorial representation of this discretization is given in \Fig{fig:matmuls}. 
While this problem is non-convex, and thus finding global optima is not guaranteed, we employed a scheme based on alternating minimization to recover locally optimal $\bbm{P}, \bbm{F}$.

\subsubsection{Results on the LHC Olympics}
\label{sec:results}

\begin{figure}[ht!]
\centering
  \centering
  \includegraphics[width=0.45\linewidth]{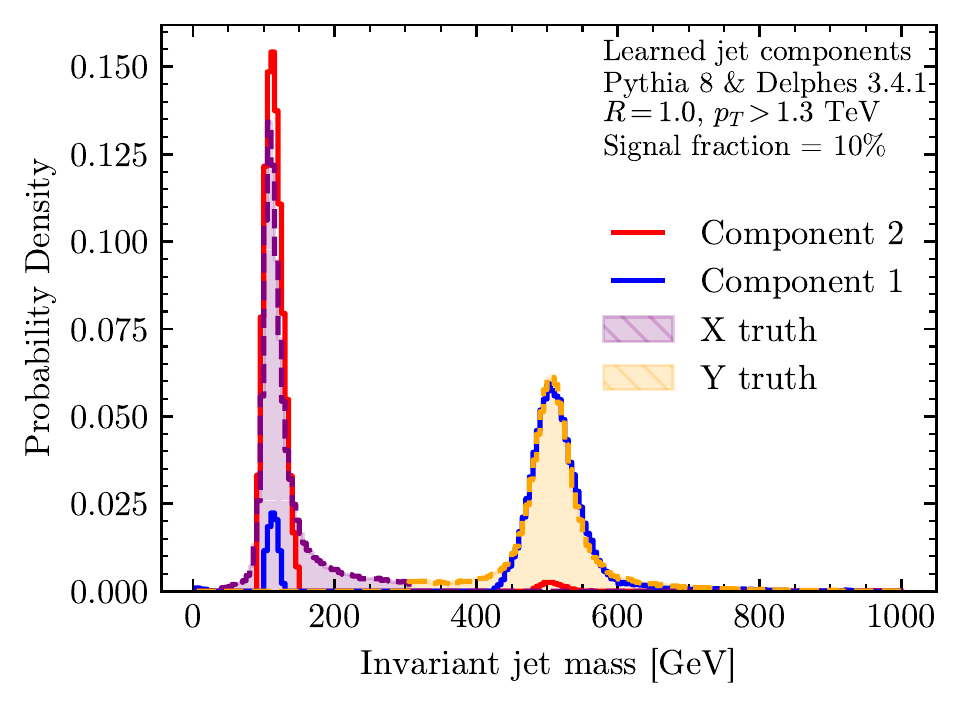}
  \includegraphics[width=0.45\linewidth]{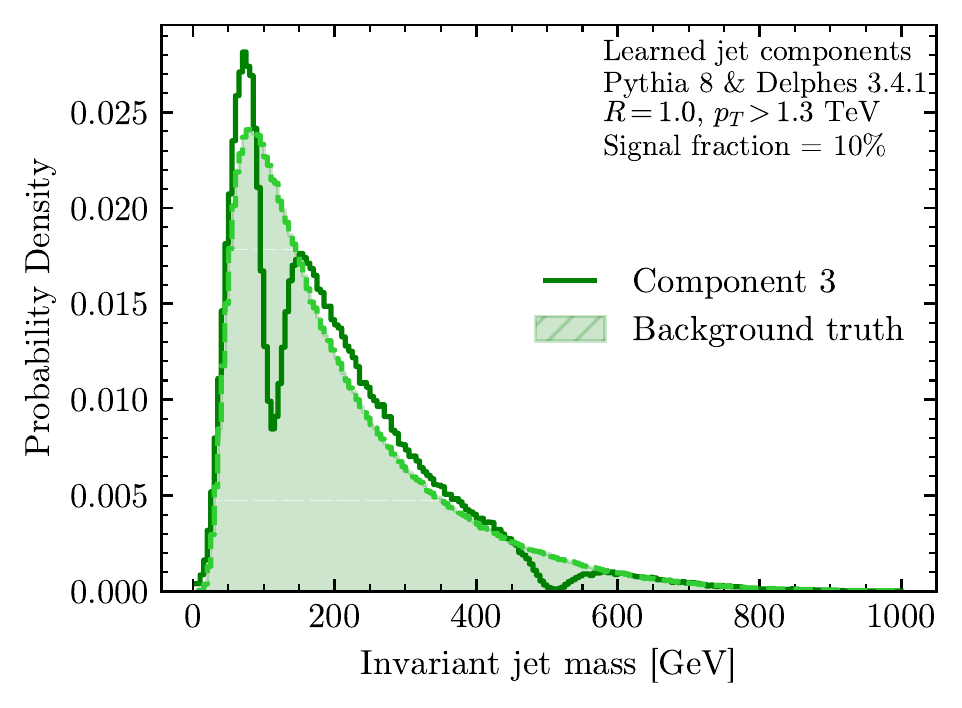}
  \includegraphics[width=0.45\linewidth]{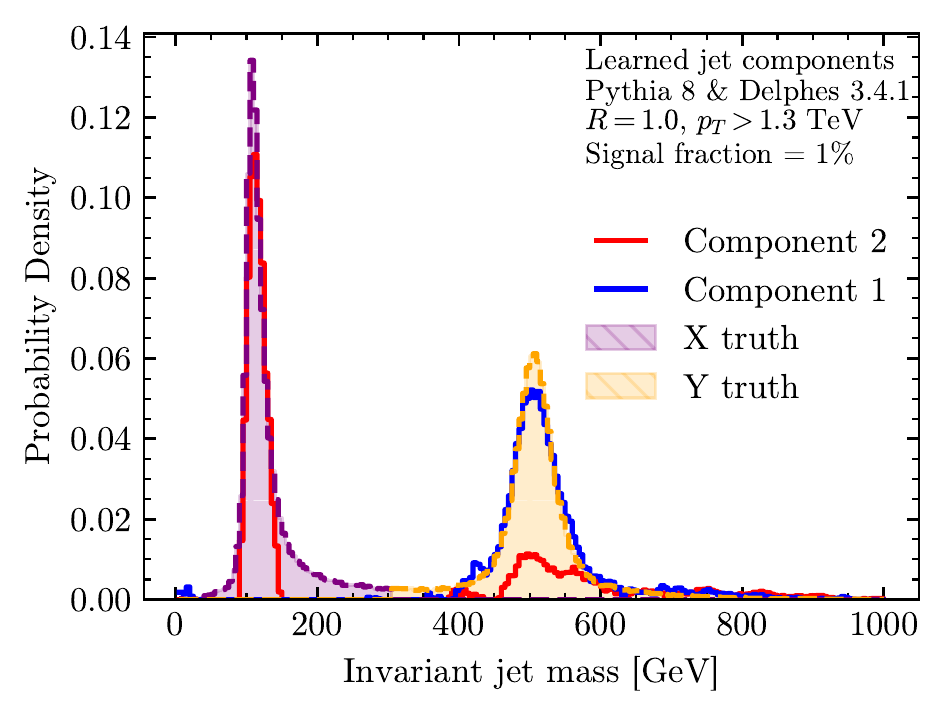}
  \includegraphics[width=0.45\linewidth]{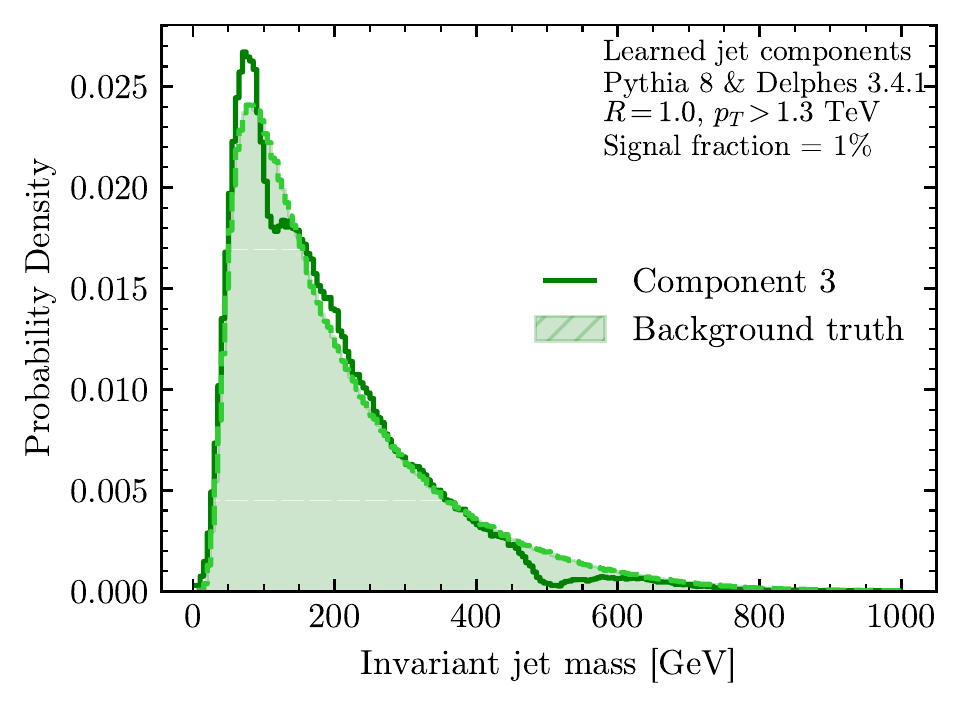}
\caption{Anomalous components at 10\% signal, Background component at 10\% signal, Anomalous components at 1\% signal, Background component at 1\% signal\label{fig:anomaly_observables}
The components retrieved from factorized topic modeling of the LHC Olympics R\&D dataset, using jet mass as our observable.
Our method shows good agreement between the learned topics and the ground truth on the jet mass observable.
We are able to recover both of the new physics resonant masses (at 100 GeV and 500 GeV) with signal fraction of 10\% (top row) and 1\% (bottom row).
The dips in the background model at the resonance masses arise because the topic finding procedure attempts to identify the most orthogonal components.
}
\end{figure}

In \Fig{fig:anomaly_observables}, we demonstrate the performance of our algorithm in recovering the jet mass distributions for dijet events from the R\&D dataset, using jet mass as our observable.
We learn a model with 3 topics, corresponding to $p_X, p_Y, p_{\text{QCD}}$, respectively.
To generate these figures, we consider a signal fraction of 10\%, and 1\% respectively, solve the topic model, and then re-weight the component distributions by subtracting the overall background distribution and renormalizing.
As the algorithms used to optimize the model returns extremal points in the polytope of all feasible solutions, the solution forces the components to be as orthogonal as possible, which is why we see characteristic dips in the background components near the $m_X$ and $m_Y$ resonance masses.

For signal fractions of both 10\% and 1\%, we are able to recover the known exotic jet masses of $m_X = 100~\text{GeV}$ and $m_Y = 500~\text{GeV}$.
As expected, the noise in the recovered distributions is noticeably larger at the lower signal fractions.
(The behavior of this model in the low-signal regime with $<0.1\%$ signal is still under investigation -- as we currently formulate it, performance degrades considerably, mostly likely due to our choice of histogram bins.)
Even in the presence of this noise, our model has significant discriminative power.
In particular, the model can infer which process any event was generated from using the likelihood ratio:
\begin{align}
\mathcal{L}(\bv{x}_1, \bv{x}_2) 
&= \frac{f(\text{signal}) \cdot p_\text{signal}(\bv{x}_1, \bv{x}_2)}{f(\text{background}) \cdot p_\text{background}(\bv{x}_1, \bv{x}_2)}\\
&= \frac{f(X, Y) \,p_X(\bv{x}_1) \, p_Y(\bv{x}_2) + f(Y, X) \, p_Y(\bv{x}_1) \, p_X(\bv{x}_2)}{f(\text{QCD, QCD})  \, p_\text{QCD}(\bv{x}_1) \, p_\text{QCD}(\bv{x}_2)}.
\end{align}
Using this likelihood ratio as a discriminant, we can test the ability of our model to classify events relative to the ground truth in the dataset.
In both cases, the model performs well, with AUCs of 0.88 and 0.81, respectively.

\subsubsection{Lessons Learned}
\label{sec:lessons}

Leveraging the leading-order factorization theorem in \Eq{eq:factorization}, we designed a statistically constrained, non-parametric, generative model to disentangle anomalous jet components from the LHC Olympics R\&D dataset.
For large enough signal fractions, our minimization algorithm finds a robust solution to Problem~\ref{prob:disc_min}, though performance degrades at lower signal fractions.
Since the input to our model is simply a 2-dimensional histogram, an interesting direction for future research could be to use this as a drop-in replacement for density estimation steps in other anomaly detection methods. 
More crucially, we see this model as a proof-of-concept for the idea of encoding physical constraints on scattering processes into a statistical language.

 \FloatBarrier

\subsection[QUAK: Quasi-Anomalous Knowledge for Anomaly Detection]{QUAK: Quasi-Anomalous Knowledge for Anomaly Detection\footnote{Authors: Sang Eon Park, Dylan Rankin, Silviu-Marian Udrescu, Mikaeel Yunus, Philip Harris.  Further details can be found in Ref.~\cite{Park:2020pak}.}}

\label{sec:quak}

Deep-learning-based anomaly detection within physics has largely focused on searching for anomalous signatures in the complete absence of a signal prior. In this scenario, two fundamental approaches have been considered: 
\begin{itemize}
\item Isolate two distinct datasets, one which may contain a signal, and one which does not; try to find a deviation between the two. 
\item Embed our knowledge of known physics processes into a simulation or a deep learning algorithm and look for events with a low likelihood of being a known physics process. 
\end{itemize}
In the first approach, colloquially referred to as classification without labels (CWoLa), conventional discrimination algorithms are used to separate the two datasets~\cite{Metodiev:2017vrx,Collins:2018epr,Collins:2019jip,Nachman:2020lpy}. Care must be taken to ensure that selection biases are mitigated so that the only discernible difference within the discrimination algorithm is the presence of an unknown physics signal. The second approach attempts to embed a complete knowledge of physics processes within a selected region into a likelihood discriminant. An excess of events with a low likelihood of being from the selected region constitutes a new physics signature. Complete knowledge of all expected physical processes within a large, high dimensional dataset can become quite complicated and can lead to reduced sensitivity~\cite{Heimel:2018mkt,Farina:2018fyg,Cerri:2018anq,Kuusela_2012}. This method broadly comprises models that utilize deep learning autoencoders. 

When comparing the two approaches, the CWoLa approach is often more sensitive, provided a signal region is present. This increase in sensitivity results from the fact that an implicit signal assumption is placed on this type of anomaly search: a signal is localized to be within a specific kinematic region. This constitutes a signal prior to the model and enhances discrimination power. For many new physics models, there are fundamental assumptions that broadly apply to all anomalies. For example, if a massive particle decays, its decay products fall within a cone determined by the particle's energy and Lorentz invariance. Neural net algorithms, on the other hand, have to learn about Lorentz invariance~\cite{Butter:2019cae}. 

By relying on one anomaly metric that measures the deviation from the background, we miss the chance to apply fundamental physical laws about how new physics may appear at the LHC, thus wasting our prior knowledge about existing physics. However, if we can incorporate this prior knowledge into the search, it should be possible to either improve the sensitivity of our search or restrict the size of the network, since additional constraints help to limit the scope of the exploration needed to construct the model.

In this section, we extend the concept of placing signal priors on anomaly searches by developing a mechanism to add signal priors without degrading the sensitivity of the pre-existing model-independent search. Through our approach, signal priors, which may or may not be accurate signal descriptions, can be embedded within an anomaly search. By inserting additional signal priors, we enhance sensitivity to signal models with characteristics similar to the embedded signal priors. Since priors are systematically added to construct information, we refer to this technique as Quasi-Anamalous Knowledge, or simply QUAK.

\subsubsection{Method}
\label{sec:method}

\begin{figure}[htbp]
\centering
\includegraphics[width=.45\linewidth]{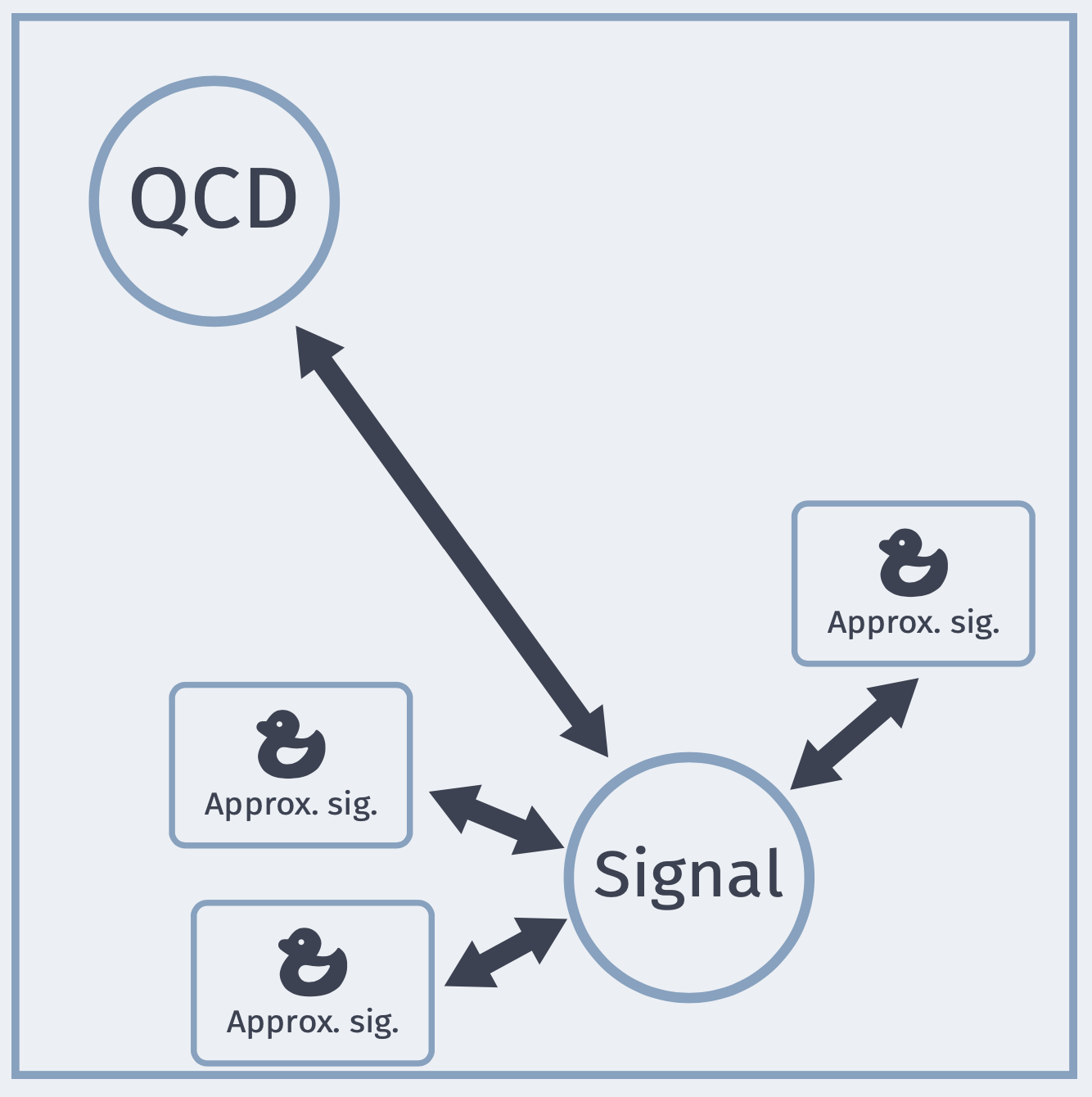}
\caption{The QUAK approach}
\label{fig:quak}
\end{figure}

In natural language processing, new models have emerged that utilize semi-supervised learning to embed these constraints on models in a natural way~\cite{chen2020big,ouali2020overview}. Semi-supervision works by training on both labeled and unlabeled data. With the unlabeled data, an unsupervised network is used. With the labeled data, a supervised classification is applied using the intermediate latent space of the unsupervised network. The unsupervised network constructs a latent space of self-organized patterns; the labeled data identifies the most useful characteristics of this space. The use of a self-supervised space has been found to be robust against variations in the data, and classifiers, in some cases, exceed that of supervised training~\cite{hendrycks2019using}. Semi-supervision has been found to be effective for anomaly detection~\cite{ruff2020deep,hendrycks2019deep}, even, very recently, within physics~\cite{Cheng:2020dal}. This paper differs from this previous approach in the construction of the network architecture and the abstraction of the latent space. 

To construct QUAK, we follow a multi-step procedure whereby we
\begin{itemize}
  \item choose a background and set of approximate signal priors that capture features of a new physics signature,
  \item train $N$ separate unsupervised probabilistic models for each signal prior or background prior,
  \item construct an $N$-dimensional ``QUAK" space consisting of the loss on each unsupervised probabilistic model, and
  \item exploit QUAK space to search for anomalies.
\end{itemize}
The construction is semi-supervised in that we use the signal priors as labels for the construction of QUAK space. 

Figure \ref{fig:quak} illustrates the concept of QUAK. By constructing an N-dimensional space, we allow for the placement of certain physics processes within specific regions of the space. A background event will have low background loss and high signal loss. An anomalous event similar to the chosen signal prior will have a low signal loss and large background loss. An anomalous event that is different from both will have large signal and background loss. By searching in the region where other proxy signals are present, we aim to isolate broadly similar anomalies, but not necessarily the same. In the following sections, we present this idea in various setups, including the LHC Olympics dataset and the MNIST dataset.

\subsubsection{Results on LHC Olympics}
\label{sec:results}

To perform the anomaly search, we first construct high-level jet features and then feed these inputs into the network for training. The high-level features consist of n-subjettiness ratios ranging from 1-prong to 4-prong and the raw jet mass of the individual jets~\cite{Thaler:2010tr,Datta:2017rhs}. Training and testing are performed with 12 variables for each event, 6 variables for each jet (4 n-subjettiness ratios, a total number of tracks, and the jet mass). 

In the construction of the unsupervised probabilistic model, an optimized scan for the studies with the LHC Olympics dataset converged on variational autoencoders (VAEs)~\cite{kingma2014autoencoding} with normalizing flows~\cite{rezende2015variational} for improved posterior approximation. Among a wide variety of normalizing flow architectures, we find Masked Autoregressive Flow~\cite{papamakarios2017masked} yields optimized results. For the training, we apply a loss metric of mean squared error reconstruction on each of the 12 variables with a KL-divergence term to regularize the sampling parameters. We choose a value of KL-divergence scale $\beta=10$~\cite{Higgins2017betaVAELB}. Additionally, we choose a latent dimension $z_{dim} = 4$, with a fully connected neural network with 3 hidden layers on either end of the VAE layer. In computing the loss within QUAK space, we remove the KL-divergence term.

With QUAK applied to the BSM search, we train multiple separate autoencoders on the QCD(background) and individual signal priors. We first utilize the single QCD autoencoder loss (1D QUAK). We progressively add additional approximate priors to the search with 2D QUAK, including one approximate signal prior, and 3D QUAK, including two approximate signal priors. To construct the ROC curve, we systematically scan the 2D space integrating from the region of minimum signal loss and maximum QCD loss. Alternative, more sophisticated approaches, such as a fit within the n-dimensional space, are not investigated here.

\begin{figure}[htbp]
\centering
\includegraphics[width=.45\linewidth]{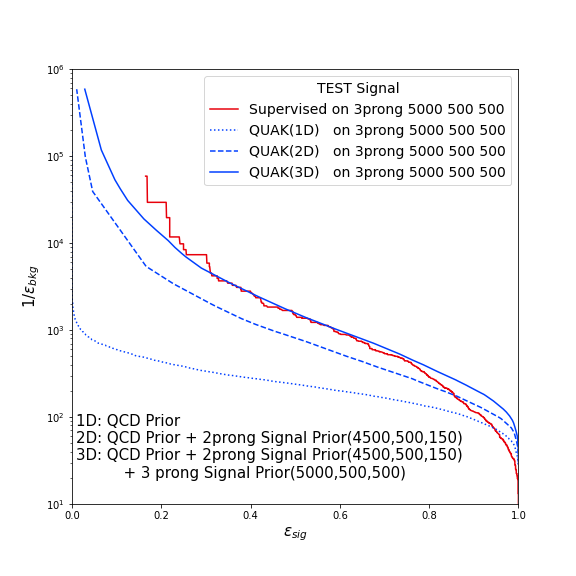}
\includegraphics[width=.45\linewidth]{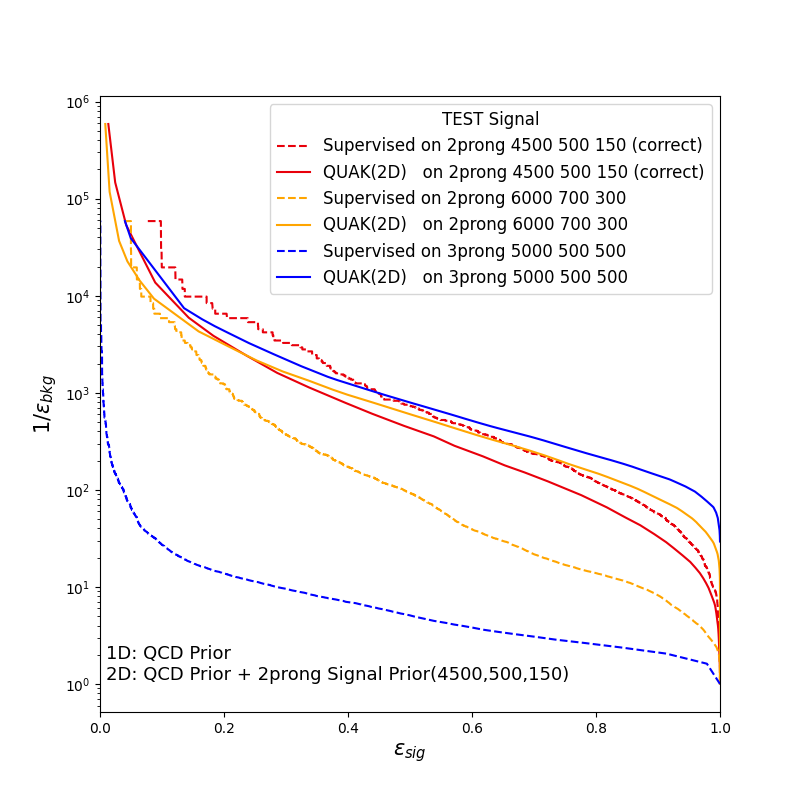}
  \caption{(Left) Receiver Operator Characteristic (ROC) for signal versus background selection for different test priors. Performance comparison of the 1D (QCD prior only), 2D (QCD prior and two prong $(m_{jj},m_{j1},m_{j2}) = (4500,500,150)$), 3D (QCD prior, two prong $(m_{jj},m_{j1},m_{j2}) = (4500,500,150)$ prior, and three prong $(m_{jj},m_{j1},m_{j2}) = (5000,500,500)$) with fully supervised training on the correct signal prior (red). Jet masses $(m_{j1}, m_{j2})$ are excluded in the training of the supervised classifier to mitigate model dependence and to allow for the potential of signal extraction through mass fits. (Right) ROC for signal versus background selection for 2D QUAK (solid) and a fixed supervised network (dashed). For both QUAK and the supervised network a signal prior of $(m_{jj},m_{j1},m_{j2}) = (4500,500,150)$ is used in the training. }
  \label{fig:supexperiment}
\end{figure}

The performance comparison of adding additional priors to the search is shown in Fig~\ref{fig:supexperiment}. By comparing dotted, dashed, and solid lines, we see that we can increase the sensitivity of the search by adding more approximate priors in training. The addition of the approximate priors approaches, and, in some places, exceeds, a fully supervised discriminator computed by training the same inputs on the known signal. Interestingly, much of the gain in discrimination of the 3-prong signal arises by adding a 2-prong signal prior.

Therefore, we observe that the addition of signal priors preserves the model-independent sensitivity of the search. Even if the signal priors are not accurate, we gain sizable performance improvement. We interpret this to mean that the added information present in the signal loss helps isolate ``signal"-like anomalies from other anomalous features present within the background. Through the construction of QUAK space, we also demonstrate that incorrect signal priors, whether they result from inaccurate simulation or different signal model choice, can still be a powerful discriminant when searching for new physics.

To contrast this with conventional new physics searches, we consider constructing a supervised classifier where we choose a signal prior and apply it to a broad range of different signal models, due to uncertainties in signal simulation and detector modeling. Nearly every signal model is inconsistent with data to a certain degree.

Figure~\ref{fig:supexperiment} compares two-dimensional QUAK, trained with QCD events and a 2-prong prior, with a supervised classifier trained with the same raw inputs and signal prior. A fully-connected network is used for both the learnable mapping from data to the latent space for the VAEs, and the supervised classifier (4 hidden layers with batch normalization~\cite{batchnorm} and dropout~\cite{hinton2012improving}). With the supervised training, we observe a general trend where the supervised classifier performs gradually worse as the test data deviates further from the 2-prong prior used to train the supervised classifier. With the 3-prong signal, we find abysmal performance with the supervised model. With QUAK, we observe relatively stable discrimination against background as the test signal further deviates from the signal prior. We conclude that QUAK incorporates signal priors in a more efficient way than supervised classifiers, and by using QUAK, we can do a more efficient scan of the whole possible BSM space. For searches where the signal prior is partially known (to within uncertainties), QUAK has the potential to mitigate loss in sensitivity.

\subsubsection{Lessons Learned}
\label{sec:lessons}
In summary, we propose the exploration of a new algorithm, QUAK, to perform model independent searches. We demonstrate this work in the context of new physics search at the LHC. We observe that the addition of approximate priors to anomaly loss allows for enhanced identification of anomalies by providing generic ``signal-like" or ``background-like" features to help in identification. With QUAK, we have presented an approach that effectively adds these priors without degrading the sensitivity of a prior-free model. QUAK is broadly applicable to a number of different problems and can help to improve both anomaly searches, and searches where large uncertainties are present on the signal modeling.

 \FloatBarrier
\subsection[Simple Supervised learning with LSTM layers]{Simple Supervised learning with LSTM layers\footnote{Authors: Zhongtian Dong.}}

\label{sec:LSTM}

\subsubsection{Method}
\label{sec:lstm}

 Recurrent neural networks have had some success in jet-tagging tasks~\cite{Guest:2018yhq}. The goal of this section is to examine how a simple model with a small number of parameters can perform on the complicated anomaly detection task.  As a first step, the raw hadron data where clustered into jets using the anti-$k_t$ algorithm with pyjet. The radius $R$ is varied during the training phase to obtain optimal test performance and is set around $R=0.7$ as a result. The input into the network is the sequence of four-momentum of jets ($p_\text{T}$, $\eta$, $\phi$, mass). The jets are ordered by their $p_\text{T}$, from largest to smallest. The length of the sequence $N$ is varied for the best performance for each data set; events that had fewer jets were be zero-padded to the same number. Typically, $N$ is chosen between $6$ and $10$. The neural network model has four hidden layers: two LSTM layers with $64$ and $128$ units, followed by two dense layers with $256$ units before a single output node. The intermediate layers have ReLU activation, and the output has a sigmoid activation. All training was done using Tensorflow through keras with the adam optimizer. $10\%$ of the R\&D data set is used as the test data set, the rest is used for training and validation. The training is done with about $30$ epochs, when the model is able to successfully identify $95\%$ of the signals in the test data.

\subsubsection{Results on LHC Olympics}
\label{sec:results}

The model performs well on the R\&D data set, which is unsurprising for a supervised learning model, but it has relatively poor performances on the black boxes. In Black Box 1,  it identifies some events as signals but with relatively low confidence, i.e. the output scores given by the model are not as high as given by the R\&D data. Compared to the actual signals presented in the data sets, the number of events identified as signals by the model is relatively large, with a higher average mass. It is possible that the model does not actually capture any real signal and incorrectly labels events as signals or backgrounds. In Black Box 2 and 3, the output results are similar to the results when running the model over the pure background data set. In retrospect, this happens perhaps for a good reason. The signal type in Black Box 3 is quite different from what is presented in the training set, which is probably the reason the neural network cannot identify them correctly.

\subsubsection{Lessons Learned}
\label{sec:lessons}

Supervised models tend to perform well on identifying specific signals comparing to a consistent background, and is probably inappropriate for anomaly detection with varying backgrounds. In general, it seems such a model is incapable of handling data that is different from the ones presented in the training set. Maybe we can still use the network structure to learn important features of the events but not doing classification tasks as it is. Rather than continuing on this type of model, I would like to study in a different direction next. Suppose we can obtain $n$ independent features for each event, and divide each feature into $2$ regions. We would have $2^n$ regions, some would be populated by "background" generic QCD events, and some other region would be not. We can focus on these regions that are rarely populated by QCD events, if many events occupy one such region in particular data sets, we may consider this an anomaly. Perhaps it is even wrong to divide into only two regions for each feature, more regions for each feature would result in a high dimensional grid, generic QCD events would rarely appear in some grid, and they can be considered as our anomaly grid. We can use machine learning to find novel features as mentioned in some previous works~\cite{Datta:2017lxt}. We may also use methods such as distance correlation to make sure features we find are independent of each other~\cite{DiscoFever}. The majority of what has been done so far are studied with dense layer networks, it would be interesting to see if we can find exotic features with more complicated network structures. Of course, there will be a lot of potential problems with this approach, one being how to make sure that data we use for training covers all of the regions that suppose to be the backgrounds, and does not falsely label signal region as background.

 \FloatBarrier

\clearpage

\section{Discussion}
\label{sec:discussion}

\begin{center}
\textit{The important thing in the Olympic Games is not to win, but to take part;\\ the important thing in Life is not triumph, but the struggle; \\the essential thing is not to have conquered but to have fought well.}\footnote{Pierre de Coubertin, founder of the International Olympic Committee, as quoted in The Olympian (1984) by Peter L. Dixon, p. 210}
\end{center}

\noindent The results of the LHCO are to be understood in a similar way. The goal is not to declare one superior method, but to foster the development of novel tools for the unsupervised detection of anomalies. With this in mind, we now turn to a discussion and comparison of the various algorithms' performance on the LHCO2020 Black Box datasets. Knowing which algorithms achieved accurate results in blinded and unblinded tests is important information, as it will provide crucial feedback for the further refinement and improvement of each approach. Also, it is important to keep in mind that an approach which did not perform well in this challenge may have its strengths elsewhere and may turn out to be better suited for a different part of phase space. 

We discuss the results in Sec.~\ref{sec:discussion_overall} and review the lessons learned --- both in terms of anomaly detection as well as in future directions for the LHCO  --- in Sec.~\ref{sec:discussion_lessons}.

\subsection{Overall Results}
\label{sec:discussion_overall}

In the following we will review the results submitted during the two LHC Olympics sessions as well as additional contributions received for this paper. As approaches were allowed to change and improve between the sessions and in preparation of this document, we chronologically walk through  results at these three stages.

\begin{figure}[h!]
\centering
\includegraphics[width=0.5\textwidth]{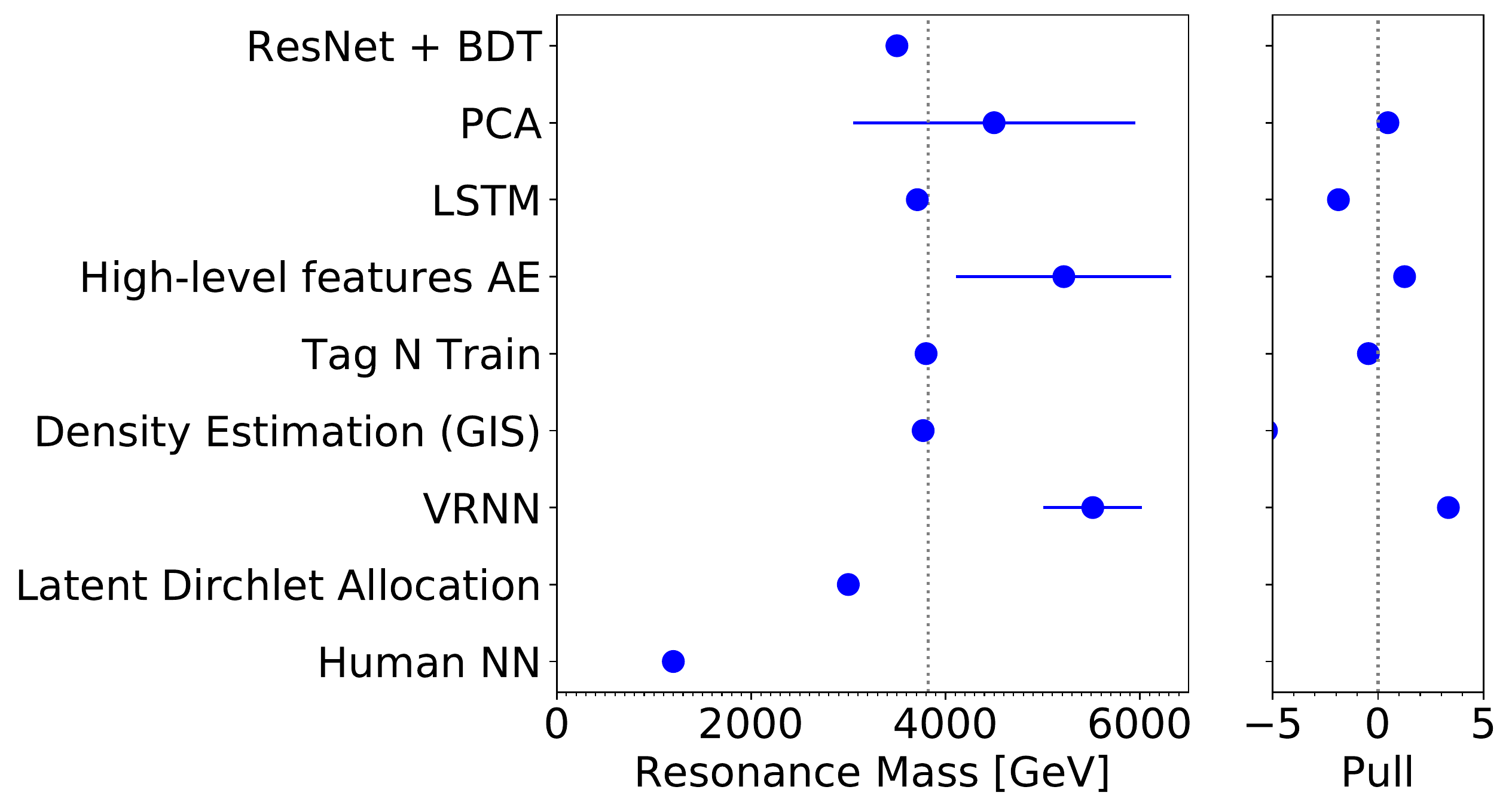}\includegraphics[width=0.5\textwidth]{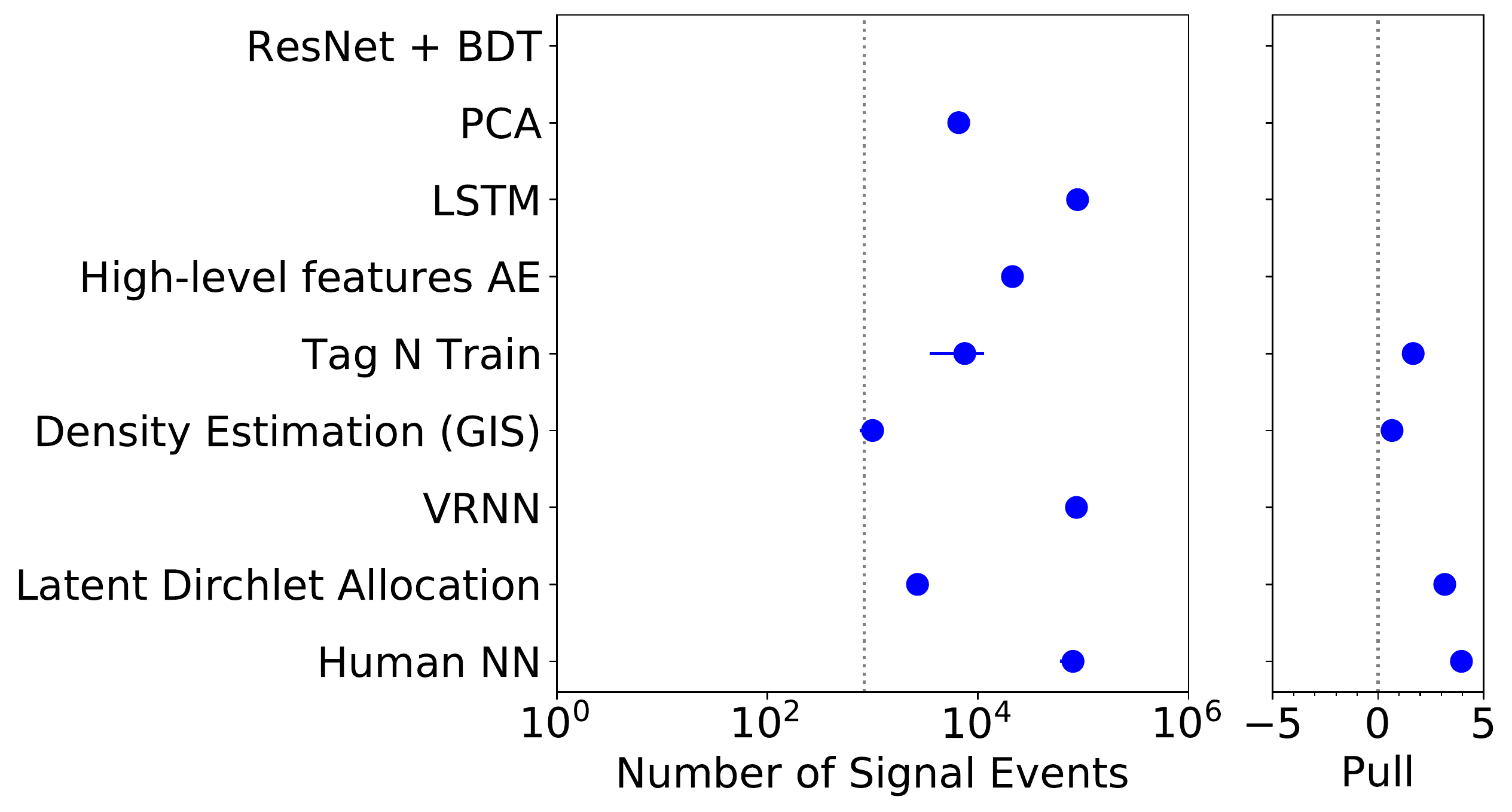}
\includegraphics[width=0.5\textwidth]{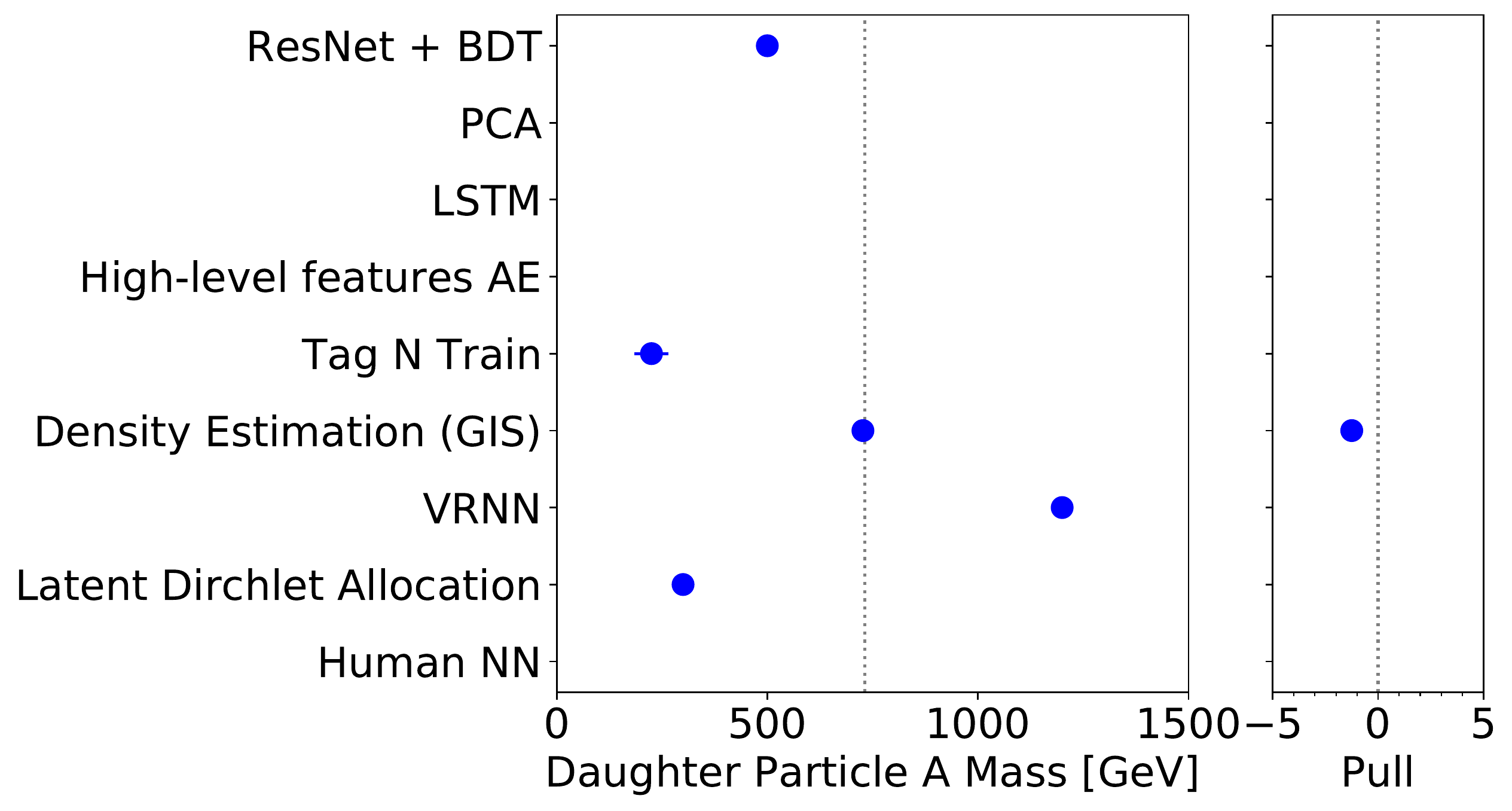}\includegraphics[width=0.5\textwidth]{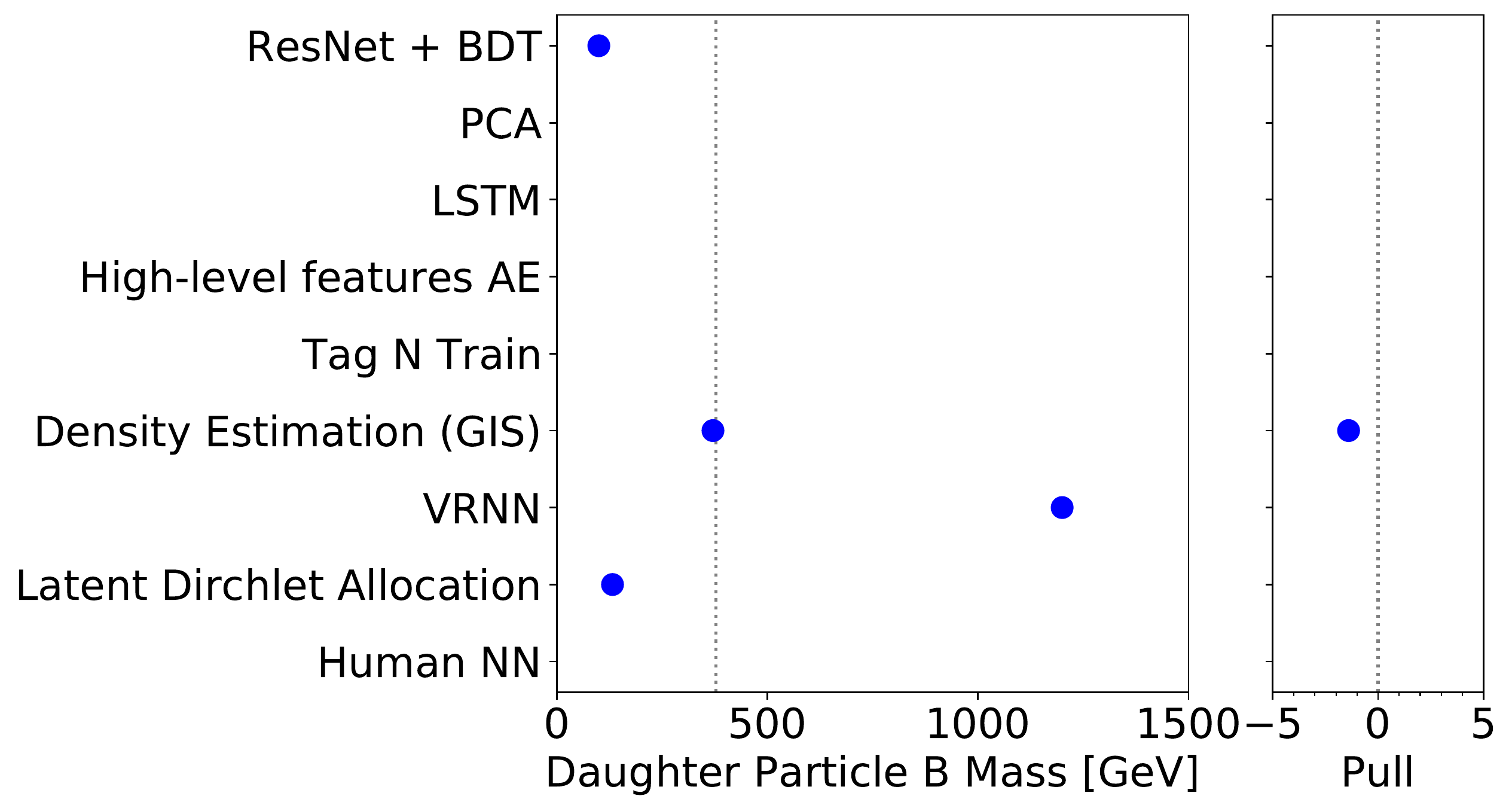}
\caption{Results of unblinding the first black box. Shown are the predicted resonance mass (top left), the number of signal events (top right), the mass of the first daughter particle (bottom left), and the mass of the second daughter particle (bottom right). Horizontal bars indicate the uncertainty (only if provided by the submitting groups). In a smaller panel the pull (answer-true)/uncertainty is given. Descriptions of the tested models are provided in the text.}
\label{fig:bb1res}
\end{figure}

As discussed in Sec.~\ref{sec:challenge_bb1}, the signal in Black Box 1 consists of 834 anomalous events with the same topology as the R\&D dataset and masses $m_{W'}=3.823$~TeV, $m_{X}=732$~GeV and $m_Y=378$~GeV and were unblinded during the LHCO session at the 2020 ML4Jets workshop~\cite{winterolympics}. Nine blind approaches were submitted and are summarised in Fig.~\ref{fig:bb1res}:
ResNet + BDT (Sec.~\ref{sec:CNNBDT}), 
PCA (Principal component analysis used as an outlier detector), 
LSTM (Sec.~\ref{sec:lstm}), 
High-level features AE (encoding kinematic and substructure variables using an autoencoder, selecting signal as events with high reconstruction MSE loss), 
Tag N Train (Sec.~\ref{sec:tnt}), 
Density Estimation (Sec.~\ref{sec:gis}), 
VRNN (Sec.~\ref{sec:vrnn}), 
Latent Dirichlet Allocation (Sec.~\ref{sec:lda}), 
and Human NN (manual search).

Of these submissions, four approaches identified the correct resonance mass either within the claimed error (PCA) or within a window of $\pm 200$~GeV (LSTM, Tag N Train, Density Estimation). Accurate predictions for the other observables were achieved only by the Density Estimation method.

Next, Black Boxes 2 and 3 were unblinded in 
Summer 2020~\cite{summerolympics}.
For Black Box 2, a resonance at 4.8~TeV (PCA),
at 4.2~TeV (VRNN, Sec.~\ref{sec:vrnn}), 
4.6~TeV (embedding clustering, Sec.~\ref{sec:ucluster}), and 5~TeV (QUAK, Sec.~\ref{sec:quak}) were predicted. 
For LDA (Sec.~\ref{sec:lda}), the absence of signal as di-jet resonance was reported. As Black Box 2 did not contain any injected signal, these results highlight a possible vulnerability of anomaly detection methods in the tail of statistical distributions.

For Black Box 3 a resonance decaying to hadrons and invisible particles (PCA), 
a resonance with a mass between 5.4 and 6.4~TeV (LDA),
at 3.1~TeV (embedding clustering), and
between 5 and 5.5~TeV (QUAK)
was reported. No signal was observed by one approach (VRNN). The true injected resonance with a mass of 4.2~TeV and two competing decay modes was not detected by any approach.

After unveiling the black boxes, further submissions and improvements to the anomaly detectors were made.
The VRNN and BuHuLaSpa (Sec.~\ref{sec:buhula}) approaches now report an enhancement at an invariant mass below 4~TeV for black box 1, while no signal
is observed for the other two black boxes. With deep ensemble anomaly detection~(Sec.~\ref{sec:CNNBDT}) a resonance at 3.5~TeV is seen
for the first black box and for Latent Dirichlet Allocation a resonance not incompatible with 3.8 TeV is observed.
Another new submission was Particle Graph Autoencoders~(Sec~\ref{sec:pga}) which detected a resonance at 3.9~TeV for the first black box.
Finally, a resonance at 3.5~TeV was seen using CWoLa hunting~(Sec.~\ref{sec:cwola}). For Black Box two and three, no additional observations of a signal
were reported after unblinding.

\subsection{Overall Lessons Learned}
\label{sec:discussion_lessons}

This large and diverse number of submissions on the blinded and unblinded datasets is very encouraging. Even better,
the resonance in the first black box was successfully detected multiple times even before unblinding. 
Of the three methods finding a mass resonance mass closest to the true value, two were based on 
building a signal-to-background likelihood ratio (Tag N Train, Density Estimation) while one 
used a signal likelihood (LSTM), and likely benefitted from the same topology between the provided
development signal and the first black box. 

However, there still is substantial room for improvement for anomaly detection in the realm of particle physics.
First, no confident statement of the absence of signal for the second black box could be made, with a number
of false positives at high values of the invariant mass. 

Second, the resonance in Black Box 3 was not detected. The structure of this 
signal was different from the shared topology of the development data and Black Box 1
which was likely to cause issues for models too finely tuned on these signals. 
Furthermore Black Box 3 featured two different decay modes which need to be combined
to achieve a significant observation. Finally, substructure offered a less useful handle here
as one decay mode involved the direct production of a pair of gluon jets.
Despite all this, the signal in Black Box 3 
still decayed as hadronic resonance with a well-defined mass-peak and visible particles in the final state.
Future developments therefore will need to both improve the sensitivity as well as the statistical interpretation
of anomaly detectors.

Beyond the reported results on the black box datasets, we also observe the use of the initial dataset for
the development of new approaches. Overall, the volume of work and results shows the 
value of targeted community studies. For anomaly detection, a new frontier would lie in the inclusion of more complex
detector effects and observables such as track and vertex information, although first a credible detection or
rejection of anomalies similar to Black Box 3 would could be desireable. While toy studies will play an important role
in developing new methods, we keenly await experimental results with these tools.

\clearpage

\section{Outlook: Anomaly Detection in Run 3, the HL-LHC and for Future Colliders}
\label{sec:outlook}

\subsection{Prospects for the (HL)-LHC}

While there are already many search results from the LHC collaborations using the full Run 2 dataset,  many more will be published in the coming years.  Notably, almost none of the methods described in this paper have been applied yet to collider data.  The ATLAS Collaboration has produced a first version of the CWoLa hunting analysis using low-dimensional features~\cite{collaboration2020dijet}, which is likely the start of a growing set of searches.  At this juncture, it is useful to consider what is possible with the full LHC dataset and what is the best way of organizing these efforts going forward.

\begin{figure}[h!]
\centering
\includegraphics[width=0.95\textwidth]{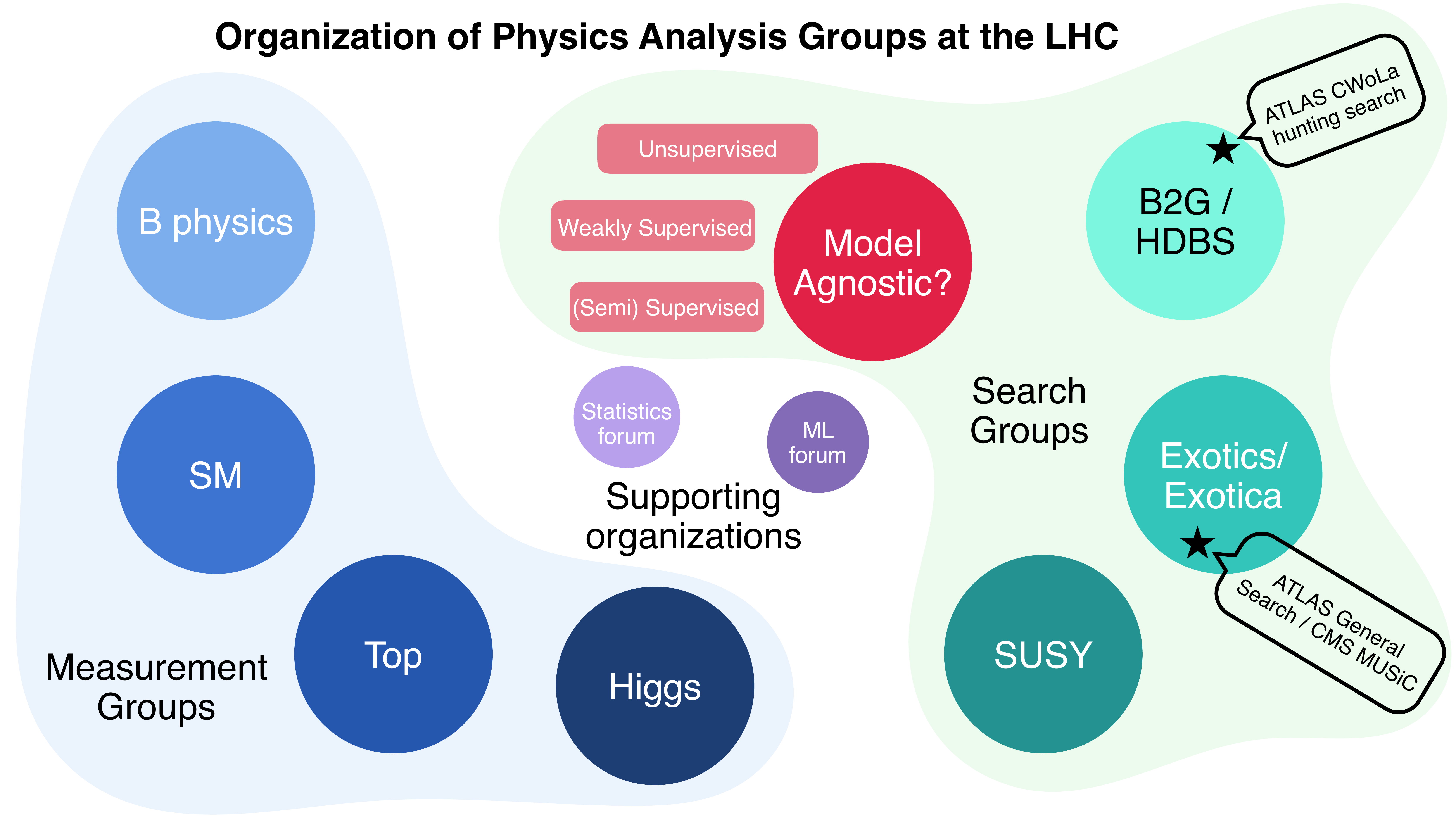}
\label{fig:groups}
\caption{The organization of physics analysis groups in ATLAS and CMS.  The large circles on the left represent analysis groups that are primarily focused on measuring properties of the Standard Model.  The group called SM is focused on the electroweak and QCD aspects of the SM that are not covered by the other groups.  The large circles on the right represent the analysis groups primarily focused on searches for new particles.  Selected supporting organizations that are connected to both measurement and search groups are depicted in smaller circles in the middle.  The ATLAS CWoLa hunting search was performed in the HDBS analysis group in ATLAS (as a `model agnostic extension of the diboson resonance search') and the ATLAS and CMS data-versus-simulation analyses are performed in the Exotics/Exotics groups.}
\end{figure}

First of all, it is clear that there is no one approach which is strictly better than every other approach.  Therefore, we envision a group of searches using complementary methodologies and targeting a variety of final states.  Currently, analyses in ATLAS and CMS are organized by physics models: there is a group focusing on supersymmetry (SUSY) models, one focused on Higgs-like particles (HDBS in ATLAS) and 3rd generation particles (B2G in CMS), and one focused on exotic particles (Exotics in ATLAS and Exotica in CMS).  It is not obvious that a model agnostic search program fits within the scope of this existing model-dependent group structure.  At the same time, the commonalities across model agnostic methods would benefit from a coherent strategy.  Therefore, a new analysis group or at least a new analysis subgroup may be required.  This is illustrated in Fig.~\ref{fig:groups}.  There are clearly strong connections with supporting groups as well, including the statistics and machine learning fora.  The analysis group home of these searches is not just a sociological question --- the technical development and physics review is primarily carried out by the analysis groups so this choice can have important implications for the success of this program.

The LHC Olympics focused on resonant new physics because there is a natural scheme for estimating backgrounds.  However, there is still a non-trivial relationship between classification and background estimation.  In particular, if the classifier is dependent on the resonant feature (e.g. the invariant mass of pairs of jets), then an artificial bump could be sculpted in the absence of any signal.  This is related to the challenge of exploring higher dimensional feature spaces, which is required to achieve the broadest sensitivity.  In some cases, automated decorrelation techniques for model-dependent searches can be adapted; in other cases, these methods would mask potential signals and so new approaches are required.  None of the methods deployed for the LHC Olympics were able to find anomalies using the full list of hadron four-vectors directly --- successful approaches all used some model-inspired dimensionality reduction.  Scaling up these approaches to high dimensional feature spaces is a key challenge for the next years and will require both methodological and computational innovation.

Exploring anomaly detection in the non-resonant case is more challenging because there is no general approach for estimating the background.  Some of the methods deployed for the LHC Olympics can achieve signal sensitivity for non-resonant signals, but significant research is required in order to combine these and possibly new approaches with background estimation strategies.  Strategies that directly compare data and background simulation are promising for final states where the background model is accurate and when the uncertainty is well-known.  A key challenge for these approaches is scaling up to high-dimensional features where the full systematic uncertainty covariance matrix may not be known.  This is a general challenge that is also faced by model-dependent approaches, where signal model uncertainties in many dimensions may not be well constrained.

Another independent dimension to consider is when in the data processing pipeline the anomaly detection happens.  The LHC Olympics is designed as an \textit{offline} analysis, where standard trigger algorithms are used to collect the data.  There is significant unexplored phase space from existing triggers, but there is also a vast phase space that is deleted in real time before it can be explored.  The development of \textit{online} anomaly detection will be a significant innovation to complement offline analysis.  Recent innovations have shown that machine learning inference can be fast enough to fit within the strict trigger latency requirements (see e.g. ~\cite{Duarte:2018ite,CERN-LHCC-2020-004}).  However, the same algorithms applied offline may not be applicable online.  For example, offline methods can make multiple passes through the dataset in order to identify anomalous regions of the phase space.  In contrast, the trigger only sees collision once before a decision to save the event or not must be made.  Even if a method could identify anomalous events within the required bandwidth, this is only a partial solution because strange collisions are only useful if we can quantify their level of strangeness.  This is one key difference between anomaly detection in high energy physics and the typical anomaly detection protocols developed in industry; we are almost never able to declare a discovery with a single collision.  Our expectation is that new physics will manifest as an `over-density' in phase space rather than being `off-manifold'. By analogy, we are not looking for flying elephants, but instead a few extra elephants than usual at the local watering hole.  The only way to know that the number of elephants is anomalous is to have a precise understanding of the usual rate of elephants.

In addition to the rich research and development program required to fully exploit the potential of these searches, there are a variety of complications involved in the interpretation of results.  The most pressing question is what to do in the case of a positive signal detection.  No fundamental particle that was not already precisely predicated by an existing theory has been discovered in decades.  Would the high energy physics community believe a significant anomaly?  It is important to start a conversation about the community plan in the case of a significant anomaly detected by one of these approaches.  If an anomaly is found before the full high-luminosity LHC (HL-LHC) dataset is recorded, then a targeted search could be conducted using an independent dataset.  What if the anomaly is only identified using the full HL-LHC dataset?  What post-hoc analysis can and should be done?  It is also important to ensure sensitivity to complex signals, where there may be multiple possible final states (as exemplified by Black Box 3).

\begin{figure}[h!]
\centering
\includegraphics[width=0.95\textwidth]{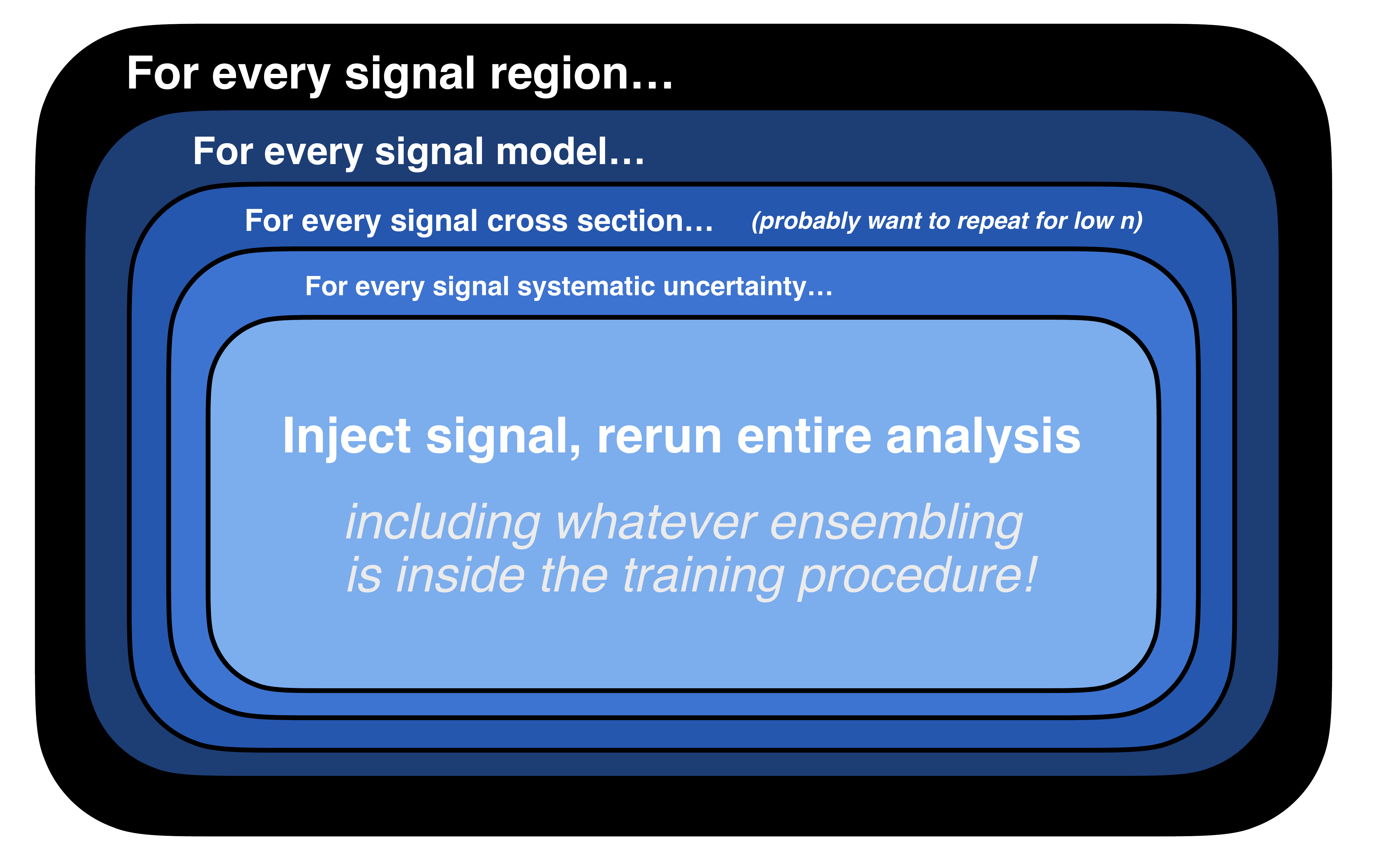}
\label{fig:trainingloop}
\caption{An illustration of the nested loops required for signal model-dependent interpretation of a model-agnostic search.  The parenthetical remark for the signal cross section refers to the fact that if the number of predicted signal events is small, one may need to repeat the injection many times due to the large statistical fluctuations in the phase space.  This is not a problem for model-dependent search where one can use all simulated signal events and scale by the predicted cross section.  Unsupervised approaches may be able to avoid certain steps if they do not change in response to modifications in the data.}
\end{figure}

In the absence of a signal detection, there is a significant challenge to quantify the sensitivity of a search.  For a model-dependent search, quantifying what was not seen is relatively straightforward for a given model, one can provide an upper limit on the cross section.  However, model agnostic methods are sensitive to many models all at once and it is challenging to define the sensitive volume.  This is particularly challenging in many dimensions.  One way to map out the sensitivity is to pick a small set of representative signal models.  Signal model dependent limits can be significantly more difficult to compute for these searches than for standard searches.  In particular, any time the anomaly classifier depends directly on the data in the signal-sensitive region, the entire analysis procedure must be repeated for every variation of the signal hypothesis.  This is represented schematically in Fig.~\ref{fig:trainingloop}.  Since the analysis selection depends on the data, the classifier must be retrained every time a different signal model cross section is injected into the data.  For example, the final exclusion plots in Ref~\cite{collaboration2020dijet} required training tens of thousands of neural networks.  Computing may become a bottleneck in the future when there is more data and higher dimensional features.  Heterogeneous High Performance Computing Centers with significant GPU resources may provide a solution to this challenge.

The dependence of the event selection on the data also complicates the usability of these data for reanalysis and reinterpretation.  One cannot simply recast published results because if a new signal was in the data, then the event selection would have been different.  If the network training and statistical analysis can be automated, then a system like RECAST~\cite{Cranmer:2010hk} may be possible whereby signal models could be submitted to collaborations for inference. Note that this is one more level of automation beyond typical analysis preservation: in addition to preserving the analysis selection, we also need to preserve the analysis optimization procedure which itself needs to be reproducible.

\subsection{Prospects for Future Colliders and Beyond}

All of the challenges described in the previous section also apply to future colliders beyond the HL-LHC.  However, a new machine opens up the possibility to imagine the co-design of accelerator/detector and analysis procedure.  What operational conditions and detector configurations are most interesting for anomaly detection?  

The methods developed for colliders may also be more broadly applicable.  Anomaly detection at other fundamental physics experiments shares many features with collider physics.  In fact, a presentation at the Summer Olympics described an anomaly detection method developed using the LHC Olympics that is now being studied for astrophysical data~\cite{streams}.

\subsection{The Role of Theory and Theorists}

While this paper is about making a search program that is model agnostic, this does not mean we should proceed without theory and without theorists.  The most powerful methods will likely employ physics-informed machine learning techniques, whereby symmetries and other physical principles are part of the learning.  These tools may allow us to find rarer signals and design procedures that are interpretable and robust.  Furthermore, there is a continuum of model independence.  Building theoretically motivated, but still relatively broad priors may achieve more powerful sensitivity to a wide class of models.

Machine learning is in general a unifying subject, where there have been many rich collaborations between experimentalists and theorists as well as between high energy physicists and machine learning researchers and practitioners.  About half of the participants in the LHC Olympics are `experimentalists' and half are `theorists'.  It is critical for the success of the anomaly detection program that model agnostic method development be a respected form of theory work and that machine learning method development and implementation be an appreciated from of experimental work.  Furthermore, barriers between theory, experiment, and computation/statistics should be as low as possible so we can make the best use of our data.  Public challenges like the LHC Olympics are an important step in this direction, but this is only one part of a bigger program of engagement.

\subsection{The Future of the LHC Olympics}

This round of the LHC Olympics was driven by a need from the community to develop and test a growing number of machine learning anomaly detection methods.  With a diverse set of submissions, we believe that this exercise has succeeded and has added value to the community.  However, there is always room for improvement.  In no particular order:

\begin{itemize}
    \item Unlike other challenges in high energy physics such as the top tagging competition~\cite{Kasieczka:2019dbj} and the challenges on the Kaggle platform like the HiggsML Challenge~\cite{pmlr-v42-cowa14}, the Flavours of Physics Challenge~\cite{flavorofphyiscs}, and the TrackML Challenge~\cite{Amrouche:2019wmx}, there was no single metric for determine a winner and therefore it was not possible to directly compare methods. (See~\cite{Rousseau:2020rnz} for a recent overview of these competitions.) This is similar to the correlation aspect of the Flavours of Physics Challenge and the efficiency-versus-fake-rate aspect of the TrackML challenge, but it even more acute for the LHC Olympics in part because the estimation of the false positive rate is non-trivial. 
    \item Without a platform like Kaggle that offers broad exposure and a monetary prize, few ML experts outside of HEP participated in the LHC Olympics.  Additionally, accessibility to non-experts could be improved.  Code to read in the data and cluster jets were provided to participants, but given that nearly every group performed additional dimensionality reduction first suggests that additional information could have been useful.
    \item One of the biggest difficulties with selecting the Black Boxes was that the anomalies should be easy enough to find that the challenge is doable, but not too easy that one could find them without new methods.  Some checks were performed before releasing the Black Boxes, but with significant work, this could have been more robust and streamlined.
\end{itemize}

There are many possibilities for variations on the LHC Olympics 2020.  Additional signal models could be considered as black boxes and more signal topologies could be studied including final state leptons, heavy flavor quarks, and long-lived particles.  We look forward to the deployment and impact of new methods developed from the LHC Olympics 2020 as well as future iterations.

\clearpage

\section{Conclusions}
\label{sec:conclusion}

Given the current lack of convincing evidence for new fundamental particles or new forces of nature from HEP experiments, it is critical that the program of dedicated searches be complemented with more model agnostic methods.  While there has been a long history of signal model agnostic methods based on binned data-simulation comparisons, there has been a recent explosion of new ideas for less model dependent approaches.  Many of the new proposals make use of machine learning to aid in the less-than-supervised exploration of collider data.  The methods presented in this paper represent a snapshot\footnote{See Ref.~\cite{livingreview} for a more updated list of papers in this area.} of the rapidly developing area of machine learning for anomaly detection in HEP. 

To address this challenge, we introduced the LHC Olympics, a community effort to develop and test anomaly detection methods in a relatively realistic setting.  A set of datasets were produced to emulate the typical setting where data are unlabeled but there is a corresponding labeled dataset for research and development.  In the LHC Olympics, three black boxes were the analog of collider data, each with a different SM background simulation and a different potential anomaly.  Many teams developed and implemented a variety of techniques on these datasets covering at least 18 different methods (some submissions compared multiple distinct methods).  

In addition to results with the R\&D dataset, many teams deployed their techniques on the black boxes.  At the Winter and Summer Olympics workshops, teams presented their results on these boxes before even knowing the nature of the signal in the datasets analyzed.  While some strategies were closer to the correct answer than others, every team followed the scientific method and gained valuable insight and experience.  In several cases, strategies were refined between the two workshops using feedback from the unveiling of the first black box.  Many of these strategies continue to be refined as they are prepared for the application to collider data in the near future.

These methods use a variety of unsupervised, semisupervised, and fully supervised machine learning methods based on neural networks and other approaches.  While there are unique advantages and disadvantages to each method, there are also common challenges across techniques, such as scaling to higher dimensions. The ultimate performance is likely to include a combination of approaches, and new method development will be required to reach the full physics potential of the data.

A data-driven revolution has started with machine learning as its catalyst.  We are well-equipped to explore the complex LHC data in new ways with immense potential for discovery.  The Run 2 data collection is over, but our exploration of these precious collisions in their natural high dimensionality is only beginning.  This LHC Olympics has been a starting point for a new chapter in collider physics that will produce exciting physics results from the current datasets as well from the datasets of the future at the LHC and beyond.

\clearpage

\section*{\label{sec::acknowledgments}Acknowledgments}

We thank the organizers and participants in the ML4Jets2020 workshop hosted at New York University and at the anomaly detection workshop hosted (virtually) by the University of Hamburg for many interesting discussions at the Winter and Summer Olympics, respectively.  B. Nachman and G. Kasieczka are grateful to the NHETC Visitor Program at Rutgers University for the generous support and hospitality during the spring of 2019 where the idea for the LHC Olympics 2020 was conceived.

A. Kahn, J. Gonski, D. Williams, and G. Brooijmans are supported by the National Science Foundation (NSF) under Grant No.~PHY-2013070.  I. Ochoa is supported by the fellowship LCF/BQ/PI20/11760025 from ``la Caixa'' Foundation (ID 100010434) and by the European Union’s Horizon 2020 research and innovation programme under the Marie Sk\l{l}odowska-Curie grant agreement No 847648.  S. E. Park, S. Udrescu, M. Yunus, P. Harris are supported by the NSF Grants \#1934700 and \#1931469.   Cloud credits for training were supported by the Internet2/NSF Grant \#190444.  V. Mikuni and F. Canelli are supported in part by the Swiss National Science Foundation (SNF) under contract No. 200020-182037.  F. F. Freitas is supported by the Center for Research and Development in Mathematics and Applications (CIDMA) through the Portuguese Foundation for Science and Technology (FCT - Funda\c{c}\~{a}o para a Ciência e a Tecnologia), references UIDB/04106/2020 and UIDP/04106/2020 and the project PTDC/FIS-PAR/31000/2017. C. K. Khosa is supported by the Royal Society, UK under the Newton International Fellowship programme (NF171488).   K. Benkendorfer was supported in part by NSF PHY REU Grant 1949923.  B. Bortolato, B. Dillon, A. Matevc, J. Kamenik, A. Smolkovic acknowledge the financial support from the Slovenian Research Agency (research core funding No. P1-0035 and J1-8137).  D. A. Faroughy is supported by SNF under contract 200021-159720.  M. Szewc would like to thank the Jozef Stefan Institute for its enormous hospitality.  P. Komiske, E. Metodiev, N. Sarda, and J. Thaler are supported by the Office of Nuclear Physics of the U.S. Department of Energy (DOE) under grant DE-SC-0011090 and by the DOE Office of High Energy Physics under grant DE-SC0012567.  N. Sarda was additionally supported by the QCRI-CSAIL Computer Research Program.  P. Komiske, E. Metodiev, N. Sarda, and J. Thaler are grateful to Benjamin Nachman and Justin Solomon for helpful conversations.  B. Nachman and J. Collins were supported by the DOE under contracts DE-AC02-05CH11231 and DE-AC02-76SF00515, respectively. P. Mart\'in-Ramiro acknowledges Berkeley LBNL, where part of this work has been developed. P. Mart\'in-Ramiro further acknowledges support from the Spanish Research Agency (Agencia Estatal de Investigaci\'on) through the contract FPA2016-78022-P and IFT Centro de Excelencia Severo Ochoa under grant SEV-2016-0597.  P. Mart\'in-Ramiro also received funding/support from the European Union’s Horizon 2020 research and innovation programme under the Marie Sk\l{l}odowska-Curie grant agreement No 690575 (RISE InvisiblesPlus).  S.~Tsan, J.~Duarte, J.-R.~Vilmant, and M. Pierini thank the University of California San Diego Triton Research and Experiential Learning Scholars (TRELS) program for supporting this research, CENIC for the 100~Gpbs networks, and Joosep Pata for helpful discussions.  They are additionally supported in part by NSF awards CNS-1730158, ACI-1540112, ACI-1541349, OAC-1826967, the University of California Office of the President, and the University of California San Diego's California Institute for Telecommunications and Information Technology/Qualcomm Institute.  J.~Duarte is supported by the DOE, Office of Science, Office of High Energy Physics Early Career Research program under Award No. DE-SC0021187. M.~Pierini is supported by the European Research Council (ERC) under the European Union's Horizon 2020 research and innovation program (Grant Agreement No. 772369). J-R.~Vilmant is partially supported by the European Research Council (ERC) under the European Union's Horizon 2020 research and innovation program (Grant Agreement No. 772369) and by the DOE, Office of Science, Office of High Energy Physics under Award No. DE-SC0011925, DE-SC0019227, and DE-AC02-07CH11359. D.~Shih is supported by DOE grant DOE-SC0010008. GK acknowledges support by the Deutsche Forschungsgemeinschaft (DFG, German Re\-search Foundation) under Germany’s Excellence Strategy – EXC 2121 ``Quantum Universe'' – 390833306.

\bibliographystyle{JHEP}
\bibliography{extra,HEPML} 

\end{document}